\documentclass[12pt]{article}
\makeatletter \@addtoreset{equation}{section} \makeatother
\renewcommand{\theequation}{\thesection.\arabic{equation}}
\newcommand{\ba}{\begin{array}}
\newcommand{\ea}{\end{array}}
\newcommand{\beq}{\begin{equation}}
\newcommand{\eeq}{\end{equation}}
\newcommand{\bea}{\begin{eqnarray}}
\newcommand{\eea}{\end{eqnarray}}
\def\bce{\begin{center}}
\def\ece{\end{center}}

\def\nonu{\nonumber}

\def\pa{\partial}

\def\be{\beta}

\def\ep{\epsilon}

\def\la{\lambda}

\def\eps6{{\displaystyle \mathop{\epsilon}^{6}}{}}
\def\g6{{\displaystyle \mathop{g}^{6}}{}}

\def\nab6{{\displaystyle \mathop{\nabla}^{6}}{}}
\def\0{{\sst{(0)}}}
\def\1{{\sst{(1)}}}
\def\2{{\sst{(2)}}}
\def\3{{\sst{(3)}}}
\def\4{{\sst{(4)}}}
\def\5{{\sst{(5)}}}
\def\6{{\sst{(6)}}}
\def\7{{\sst{(7)}}}
\def\8{{\sst{(8)}}}

\def\ba{\begin{array}}
\def\ea{\end{array}}
\def\beq{\begin{equation}}
\def\eeq{\end{equation}}
\def\be{\begin{equation}}
\def\ee{\end{equation}}

\def\la{\lambda}
\def\eps{\epsilon}

\def\ba{\begin{array}}
\def\ea{\end{array}}
\def\beq{\begin{equation}}
\def\eeq{\end{equation}}
\def\be{\begin{equation}}
\def\ee{\end{equation}}

\def\la{\lambda}
\def\eps{\epsilon}

\def\eps6{{\displaystyle \mathop{\epsilon}^{6}}{}}

\def\nab6{{\displaystyle \mathop{\nabla}^{6}}{}}

\newcommand{\bean}{\begin{eqnarray*}}
\newcommand{\eean}{\end{eqnarray*}}

\begin{document}
\thispagestyle{empty} \addtocounter{page}{-1}
   \begin{flushright}
%PUPT-2395 \\
%CALT-68-nnnn \\
%{\tt hep-th/yymmnnn}\\
\end{flushright}

\vspace*{1.3cm}
 
\centerline{ \Large \bf   
Higher Spin Currents in Wolf Space: Part I }
\vspace*{0.5cm}
%\centerline{ \Large \bf  
%in the Product of Wolf Space and $SU(2) \times U(1)$ }
%\vspace*{1.5cm}
\centerline{{\bf Changhyun Ahn 
%\footnote{On leave from the Department of Physics, Kyungpook National University, Taegu
%  702-701, Korea and 
%address until Aug. 31, 2011:
%Department of Physics, Princeton University, Jadwin Hall, 
%Princeton, NJ 08544, USA}
}} 
\vspace*{1.0cm} 
\centerline{\it 
Department of Physics, Kyungpook National University, Taegu
702-701, Korea} 
%\centerline{\it 
%Department of Physics, Princeton University, Jadwin Hall, 
%Princeton, NJ 08544, USA}
\vspace*{0.8cm} 
\centerline{\tt ahn@knu.ac.kr 
} 
\vskip2cm

\centerline{\bf Abstract}
\vspace*{0.5cm}

For the ${\cal N}=4$ superconformal coset theory described by 
$\frac{SU(N+2)}{SU(N)}$ (that contains a Wolf space) with $N=3$,  
the ${\cal N}=2$ WZW affine 
current algebra with  constraints is obtained. The $16$ generators 
of the large ${\cal N}=4$ linear superconformal algebra are described by 
those WZW affine currents explicitly. 
By factoring out four spin-$\frac{1}{2}$ currents 
and the spin-$1$ current from these $16$ generators, the remaining $11$ 
generators (spin-$2$ current, four spin-$\frac{3}{2}$ currents, 
and six spin-$1$ currents)  
corresponding to 
the large ${\cal N}=4$ nonlinear superconformal algebra are obtained.    

Based on the 
recent work by Gaberdiel and Gopakumar on the large ${\cal N}=4$ holography,
the extra $16$ currents, with spin contents 
$(1,\frac{3}{2},\frac{3}{2}, 2)$, $(\frac{3}{2}, 2, 2, 
\frac{5}{2})$, $(\frac{3}{2}, 2, 2, \frac{5}{2})$, and $(2, \frac{5}{2}, 
\frac{5}{2}, 3)$ described in terms of 
${\cal N}=2$ multiplets, 
are obtained and realized by the WZW affine currents.
As a first step towards
${\cal N}=4$ ${\cal W}$ algebra (which is NOT known so far), 
the operator product expansions (OPEs) between the above $11$ currents and 
these extra $16$ higher spin currents are found explicitly. 
It turns out that the composite fields with definite 
$U(1)$ charges, made of above $(11+16)$ currents (which commute with 
the Wolf space subgroup $SU(N=3) \times SU(2) \times U(1)$ currents),
occur in the right hand sides of these OPEs. 

\baselineskip=18pt
\newpage
\renewcommand{\theequation}
{\arabic{section}\mbox{.}\arabic{equation}}

%%%%%%%%%%%%%%%%%%%%%%%%%%%%%%%%%%%%%%%%%%%%%%%%%%%%%%%%%%%%%%%%%%%%%
%%%%%%%%%%%%%%%%%%%%%%%%%%%%%%%%%%%%%%%%%%%%%%%%%%%%%%%%%%%%%%%%%%%%%%
\section{Introduction}
%1%%%%%%%%%%%%%%%%%%%%%%%%%%%%%%%%%%%%%%%%%%%%%%%%%%%%%%%%%%%%%%%%%%%%%%
%%%%%%%%%%%%%%%%%%%%%%%%%%%%%%%%%%%%%%%%%%%%%%%%%%%%%%%%%%%%%%%%%%%%%

Gaberdiel and  Gopakumar \cite{GG1305} have proposed that 
the large ${\cal N}=4$ higher spin theory on $AdS_3$ 
based on the higher spin algebra
is dual to
the 't Hooft limit of the two dimensional 
large ${\cal N}=4$ superconformal coset theory.
The ${\cal N}=4$ higher spin algebra contains   
$8$ fields of spin $s =\frac{3}{2}, 2, \frac{5}{2}, \cdots$ and 
$7$ fields of spin $s=1$ and the exceptional superalgebra 
$D(2,1|\frac{\mu}{1-\mu})$
is the largest finite dimensional subalgebra of this ${\cal N}=4$ 
higher spin algebra.
The ${\cal N}=4$ coset theory is described by the coset 
$\frac{SU(N+2)}{SU(N)}$ with the level which is equal to the sum of 
each level of two $SU(2)$ affine algebras.
The 't Hooft coupling constant $\la$ is identified with 
the free parameter of the large ${\cal N}=4$ superconformal algebra (this 
additional parameter is due to the fact that there exist two $SU(2)$ 
affine algebras rather than one) and 
furthermore this 't Hooft parameter 
$\la$ is equal to the above $\mu$ parameter which is related 
to the mass of scalar field in the $AdS_3$ bulk theory, according to 
this proposal \cite{GG1305}. See also previous works on the 
minimal model holography \cite{GG,GG1,GG2,AGKP}.

More specifically, 
the large ${\cal N}=4$ linear superconformal algebra is generated by
spin-$2$ stress tensor, four spin-$\frac{3}{2}$ supersymmetry generators,
seven spin-$1$ currents (six of them are the generators of 
two $SU(2)$ affine algebras and one of them is $U(1)$ current) 
and four spin-$\frac{1}{2}$ currents
\cite{STVplb}.
Although the wedge algebra of the large ${\cal N}=4$ linear 
superconformal algebra is also
$D(2,1|\frac{\mu}{1-\mu})$ together with a central generator,
the large ${\cal N}=4$ linear superconformal algebra
is not a subalgebra of the extended algebra of higher spin currents
because four fermions of spin-$\frac{1}{2}$ 
do not occur in the above higher spin algebra.  
By factoring out the spin-$1$ current and four spin-$\frac{1}{2}$ currents
from the above large ${\cal N}=4$ linear superconformal algebra,
the remaining $11(=16-5)$ generators  
consist of 
the nonlinear version of the large ${\cal N}=4$ linear superconformal algebra 
\cite{GS} 
\footnote{Sometimes this algebra is called by ${\cal N}=4$ 
quasi superconformal 
algebra because this is generated quadratically (nonlinearly).
On the other hand, the small (or regular) 
${\cal N}=4$ superconformal algebra \cite{Ademolloetalplb,Ademolloetalnpb} 
can be obtained by taking one of the level
as zero and the other level as an infinity in the large ${\cal N}=4$ 
linear superconformal algebra (therefore there exists only a single $SU(2)$ affine 
algebra rather than two). Also in \cite{Schoutensnpb}, the large 
${\cal N}=4$ linear superconformal algebra was studied. See also 
\cite{Ivanov1,Ivanov2,Ivanov3,Ivanov4}.}.
The levels of two $SU(2)$ affine algebras are reduced  by one 
and the central charge appearing in the Virasoro algebra is reduced by three.
Furthermore the central term in the operator product expansions (OPE)
between the spin-$\frac{3}{2}$ currents  
is also changed and the quadratic terms in the first-order poles of the OPEs
occur.

In \cite{ST}, 
the ${\cal N}=4$ extension of Kazama-Suzuki model \cite{KS1,KS2} 
is described and 
the ${\cal N}=4$ coset model has the form 
$\mbox{Wolf} \times SU(2) \times U(1)$ where $\mbox{Wolf}$ is a Wolf space
(or quaternion-Kahler symmetric space)
\cite{Wolf,Alek,Salamon} 
\footnote{The Wolf space appeared in ${\cal N}=2$ supergravity
in four dimensions \cite{BW}.}. 
The fourteen currents of the large ${\cal N}=4$ linear 
superconformal algebra 
are expressed in terms of affine Kac-Moody currents (the remaining two currents
can be obtained from the ${\cal N}=1$ Sugawara construction) 
and the relative tensorial 
structures appearing in these expressions 
satisfy the particular identities.
They also found the nonlinear algebra by factoring out the spin-$1$ and 
spin-$\frac{1}{2}$ currents that live in the $U(1)$ 
and the coset is given by 
$\mbox{Wolf} \times SU(2)$.
The remaining three spin-$\frac{1}{2}$ currents living in
$SU(2)$ can be decoupled further.

For the simplest case where the ${\cal N}=4$ coset is given by 
$SU(2) \times U(1)$
\cite{RASS,Ahnphd} corresponding to the first entry of the table $1$ in 
\cite{ST}, 
using the above identities between the tensorial structures, 
the $16$ generators of the large ${\cal N}=4$ linear superconformal algebra
are written in terms of the affine Kac-Moody currents with constraints 
in ${\cal N}=2$ 
superspace \footnote{Although we use the ${\cal N}=2$ description in this 
paper, due to the constraints for the ${\cal N}=2$ WZW affine currents, 
effectively our description is the same as the ${\cal N}=1$ approach in 
\cite{ST}. }. 
Furthermore, following the work of \cite{ST}, by imposing on  
the above large ${\cal N}=4$ nonlinear superconformal algebra
in the ${\cal N}=4$ coset theory, the coset turns out to be a Wolf space 
itself in \cite{GK}. 

%Gukov + Martinec + Moore + Strominger

In this paper,
we would like to construct 
the $16$ generators of the large ${\cal N}=4$ linear 
superconformal  algebra
in the coset \footnote{Of course, the bosonic coset can be obtained by 
introducing the extra $SO(4N+4)$ (generated by $(4N+4)$ free fermions) 
at level $1$ in the numerator of the 
coset as in \cite{LVW,GG1305}. The number $(4N+4)$ is the dimension of the 
coset $\frac{SU(N+2)}{SU(N)}$. That is, $(N+2)^2-1 -(N^2-1)=4N+4$.}
\bea
\mbox{Wolf} \times SU(2) \times U(1) = 
\frac{SU(N+2)}{SU(N)}
\label{cosetform}
\eea 
(that is the third entry of the table $1$ in \cite{ST}) 
theory with $N=3$.
By factoring out the spin-$1$ current and four spin-$\frac{1}{2}$ currents,
the $11$ generators of the large ${\cal N}=4$ nonlinear algebra 
in the Wolf space coset 
are obtained \footnote{Let us emphasize that one can combine the 
spin-$1$ current and one spin-$\frac{1}{2}$ current and express them as
a single ${\cal N}=1$ super current which corresponds to the $U(1)$ 
factor in the left hand side of (\ref{cosetform}). However, 
the remaining three spin-$\frac{1}{2}$ currents are part of 
the ${\cal N}=1$ superconformal affine $SU(2)$ algebra. In other words, 
the superpartners of these three spin-$\frac{1}{2}$ currents, i.e. three 
spin-$1$ currents do not play an important role in 
the denominator subgroup. When one divides $SU(2) \times U(1)$ factor 
both sides of (\ref{cosetform}), then one obtains 
$\mbox{Wolf} = \frac{SU(N+2)}{SU(N) \times SU(2) \times U(1)}$ where the 
$SU(2)$ factor in the denominator refers to only the above three 
spin-$\frac{1}{2}$ currents. Their superpartners, three spin-$1$ 
currents  appear in the group $SU(N+2)$. 
\label{footnotewolf}}.
One of the findings in \cite{GG1305} is that the lowest nontrivial multiplet 
of the higher spin algebra has 
one spin-$1$ current, four spin-$\frac{3}{2}$ currents,
six spin-$2$ currents, four spin-$\frac{5}{2}$ currents and one spin-$3$
current and let us denote them by spin contents as follows:
\bea
\left(1, \frac{3}{2}, \frac{3}{2}, 2 \right), \qquad
 \left(\frac{3}{2}, 2, 2, \frac{5}{2} \right), \qquad
\left(\frac{3}{2}, 2, 2, \frac{5}{2} \right), \qquad
\left(2, \frac{5}{2}, \frac{5}{2}, 3 \right).
\label{16}
\eea
We construct these $16$ currents (\ref{16}) in terms of ${\cal N}=2$ affine 
Kac-Moody currents in the above Wolf space coset theory explicitly.
Furthermore, we calculate the various OPEs between the $11$ generators 
of the large ${\cal N}=4$ nonlinear superconformal algebra and the 
$16$ higher spin currents \footnote{The ${\cal N}=4$ $W_3$ algebra in 
different context was considered in \cite{Perret1,Perret2}. One example of 
$W$ algebra with Wolf space was found in \cite{Naka}. Recently, 
the higher spin theory with extended supersymmetries where the two dimensional 
coset theory contains Wolf space was studied in 
\cite{HLPR}.}.

By construction, the $16$ higher spin currents (anti)commute with 
the eight spin-$\frac{1}{2}$ currents (and its superpartner eight spin-$1$ 
currents) living in the effective 
${\cal N}=1$ subgroup $SU(N=3)$ of the Wolf space
coset, the three spin-$\frac{1}{2}$ currents living in the 
bosonic subgroup $SU(2)$ of the Wolf space coset and 
the remaining spin-$\frac{1}{2}$ current and its superpartner spin-$1$ 
current living in the effective ${\cal N}=1$ subgroup 
$U(1)$ of the Wolf space coset.
The $16$ currents are primary under the stress energy tensor of ${\cal N}=4$
nonlinear superconformal algebra. 

In the OPEs between the four 
spin-$\frac{3}{2}$ currents and $16$ higher spin currents,  
the right hand sides of these OPEs have special features. 
If one describes (\ref{16}) as four rows for $4 \times 4$ matrix, 
then one writes the OPEs between the above four 
spin-$\frac{3}{2}$ currents and this $4 \times 4$ matrix which 
has $16$ components.
Let us concentrate on the linear terms on the 
higher spin current appearing in the right hand side
of OPEs (There are also linear or nonlinear terms containing $11$ currents of
large ${\cal N}=4$ nonlinear superconformal algebra in the full expressions).
It turns out that there are no higher spin currents, at the linear level, 
in the second row, 
third column, second column and third row in the above each $4\times 4$ 
matrix, respectively. These vanishings of the rows and columns
are quite related to the locations of four spin-$\frac{3}{2}$ currents in the 
${\cal N}=2$ superspace multiplets.   
One can describe the $16$ currents of large ${\cal N}=4$ linear 
superconformal algebra as $4 \times 4$ matrix also as done for $16$ higher spin
currents: $(1, \frac{3}{2}, \frac{3}{2}, 2)$, $(\frac{1}{2}, 1, 1, 
\frac{3}{2})$, $(\frac{1}{2}, 1, 1, \frac{3}{2})$ and $(0, \frac{1}{2}, 
\frac{1}{2}, 1)$.  
The first, second, third and fourth  spin-$\frac{3}{2}$ currents are
 $(1,2)$, $(1,3)$, $(2,4)$ and $(3,4)$ elements of $4\times 4$ matrix
and play the role of vanishing 
second column, third column, second row and third row above respectively
\footnote{That is, when the particular spin-$\frac{3}{2}$ current acts on 
the $16$ higher spin currents described by $4\times 4$ matrix, 
one does not see any higher spin currents 
in the given row or column containing that spin-$\frac{3}{2}$ current 
in $4\times 4$ matrix. 
For example, for the first case where the spin-$\frac{3}{2}$ current 
is an $(2,4)$ element of $4\times 4$ matrix, 
the second multiplet of (\ref{16})
appears in the first- and fourth-rows in the right hand side of OPEs 
and  both 
first- and fourth-multiplets of (\ref{16}) appear in the third row
in the right hand side of OPEs:(\ref{operesult}).    
One sees the vanishing of second row in the right hand side of OPEs.
See the section 6 for more detailed descriptions.}.

In section 2, the $16$ currents of ${\cal N}=4$ linear 
superconformal algebra are obtained in the coset model (\ref{cosetform}).

In section 3, the $11$ currents of
 ${\cal N}=4$ nonlinear 
superconformal algebra are determined in the Wolf space coset (\ref{cosetform}).

In section 4, the extra $16$ currents in (\ref{16}) are obtained in the
Wolf space 
coset (\ref{cosetform}) \footnote{When we say ``the higher spin currents'' 
in 
this paper, they are given by these $16$ currents (\ref{16}). Some of the spins 
are less than $2$. In other words, 
they consist of all the extra currents besides the large ${\cal N}=4$
nonlinear algebra currents.}.

In section 5, the OPEs between the $11$ currents in section $3$
and $16$ currents in section $4$ are obtained.

In section 6, we summarize what has been done in this paper.

In Appendices $A$-$C$, some details which are necessary in sections 
$2$-$5$ are presented.
They contain the complete higher spin algebra.  

The mathematica package by Thielemans \cite{Thielemans} 
is used all the times. 
       
%%%%%%%%%%%%%%%%%%%%%%%%%%%%%%%%%%%%%%%%%%%%%%%%%%%%%%%%%%%%%%%%%%%%%
%%%%%%%%%%%%%%%%%%%%%%%%%%%%%%%%%%%%%%%%%%%%%%%%%%%%%%%%%%%%%%%%%%%%%%
\section{The large ${\cal N}=4$ (linear) 
superconformal  algebra in the coset 
minimal model    }
%2%%%%%%%%%%%%%%%%%%%%%%%%%%%%%%%%%%%%%%%%%%%%%%%%%%%%%%%%%%%%%%%%%%%%%
%%%%%%%%%%%%%%%%%%%%%%%%%%%%%%%%%%%%%%%%%%%%%%%%%%%%%%%%%%%%%%%%%%%%%

Before we construct the large ${\cal N}=4$ nonlinear superconformal algebra
using the ${\cal N}=2$ WZW affine currents,
we would like to describe its linear version first. After that,
in next section, 
the large ${\cal N}=4$ nonlinear superconformal algebra
will be constructed explicitly.

%%%%%%%%%%%%%%%%%%%%%%%%%%%%%%%%%%%%%%%%%%%
\subsection{${\cal N}=2$ WZW affine current algebra}
%%%%%%%%%%%%%%%%%%%%%%%%%%%%%%%%%%%%%%%%%%%
 
Let us consider
the particular ${\cal N}=4$ superconformal coset theory described in 
\cite{ST,GG1305} 
which 
can be described as \footnote{Usually the $U(1)$ factor in (\ref{coset})
in the denominator of Wolf space is missing for other types of Wolf
space. The details for using this $U(1)$ factor to other types are given in
\cite{SSTV}. Furthermore, 
one can see the related works on other types (orthogonal or noncompact) 
of Wolf space in 
\cite{ARV,JMS}. }
\bea
\mbox{Wolf} \times SU(2) \times U(1) = \frac{SU(N+2)}{SU(N)},
\qquad
\mbox{where} \qquad
\mbox{Wolf}= \frac{SU(N+2)}{SU(N) \times SU(2) \times U(1)}.
\label{coset}
\eea
The central charge of this coset model \cite{KS1} is given by
\bea
c_{SU(N+2)} - c_{SU(N)} & = &
\frac{3}{2} \left[ (N+2)^2-1\right] \left[1-\frac{2(N+2)}{3(k+N+2)} \right]
-\frac{3}{2} \left[ N^2-1\right] \left[1-\frac{2N}{3(k+N+2)} \right]
\nonu \\
& = & \frac{6(k+1)(N+1)}{(k+N+2)}.
\label{cosetcentral}
\eea
The level of the currents in the numerator and denominator of the
coset model (\ref{coset}) is $(k+N+2)$ respectively.
Note that the general expression of the central charge
of the large ${\cal N}=4$ linear superconformal algebra 
$c=\frac{6 k^{+} k^{-}}{(k^{+} + k^{-})}$ where
$k^{+} = (k+1)$ and $k^{-}=(N+1)$ 
are the levels of two $SU(2)$ affine algebras.
We will see that the central charge (\ref{cosetcentral})
occurs in the OPE between the spin-$2$ currents (and the OPE between 
the spin-$\frac{3}{2}$ currents) in the large ${\cal N}=4$ linear 
superconformal algebra.

Furthermore, the central charge of the Wolf space (for the time being 
we assume the contributions from the superpartners of three 
spin-$\frac{1}{2}$ fermions described in the footnote \ref{footnotewolf}) 
can be obtained as follows:
\bea
c_{\mbox{Wolf}}  & = & 
c_{SU(N+2)} -c_{SU(N)} -c_{SU(2)}-c_{U(1)} \nonu \\
& = & 
\frac{6(k+1)(N+1)}{(k+N+2)}- \frac{3}{2} \times 3 
\left[1-\frac{4}{3(k+N+2)} \right] -\frac{3}{2} 
\nonu \\
& = & \frac{6 k N}{(2+k+N)},
\label{Wolfcentral}
\eea
where the above level $(k+N+2)$ appearing in the coset 
(\ref{coset}) is taken in the $SU(2)$ factor in the 
denominator in (\ref{coset}).
In the second line of (\ref{Wolfcentral}), the result of
(\ref{cosetcentral}) is used.
Compared to the central charge in the 
large ${\cal N}=4$ linear superconformal algebra, 
this central charge will appear in the OPE between the two spin-$\frac{3}{2}$
currents in the large ${\cal N}=4$ nonlinear superconformal algebra 
in next section.  For the other central term appearing in the stress tensor,
the contribution from the central charge 
$c_{SU(2)}$ in (\ref{Wolfcentral})  will be $-\frac{3}{2}$ \footnote{
As described in \cite{GG1305},
the Wolf space can be generalized to 
the following coset model
\bea
 \frac{SU(N+M)}{SU(N) \times SU(M) \times U(1)},
\label{KS}
\eea
which appears in the standard ${\cal N}=2$ Kazama-Suzuki model \cite{KS1,KS2}.
Then the central charge 
of this coset (\ref{KS})
can be obtained from 
\bea
 c_{KS} & = &  c_{SU(N+M)} -c_{SU(N)} -c_{SU(M)}-c_{U(1)}
\nonu \\
& = & 
\frac{3}{2} \left[ (N+M)^2-1\right] \left[1-\frac{2(N+M)}{3(k+N+M)} \right]
-\frac{3}{2} \left[ N^2-1\right] \left[1-\frac{2N}{3(k+N+M)} \right]
\nonu \\ 
& - & \frac{3}{2} \left[ M^2-1\right] \left[1-\frac{2M}{3(k+N+M)} \right]
-\frac{3}{2} =  \frac{3 k M N}{(k+M+N)},
\label{KScentral}
\eea
where the level in each factor of the coset is given by $(k+N+M)$.
Of course, this will become the previous central charge (\ref{Wolfcentral}) 
for $M=2$.}.

We would like to construct the large ${\cal N}=4$ linear 
superconformal algebra for the coset theory (\ref{coset}).
For the ${\cal N}=2$ currents where the ${\cal N}=2$ superspace coordinate
$Z=(z,\theta, \bar{\theta})$ is given by the bosonic coordinate $z$ and two
Grassman coordinates $\theta$ and $\bar{\theta}$, 
the component currents are given by \footnote{One finds the explicit component
expressions in $(19)$ of \cite{RASS} or $(24)$ of \cite{HS}.}
\bea
K^m(Z) & = & 
K^m(z)+ \theta \,\, D K^m|_{\theta=\bar{\theta}=0}(z) + \bar{\theta} \,\, 
\overline{D} K^m|_{\theta=\bar{\theta}=0}(z)+ 
\theta \bar{\theta}\,\, (-1) \frac{1}{2} [ D, \overline{D} ] 
K^m|_{\theta=\bar{\theta}=0}(z),
\nonu \\ 
K^{\bar{m}}(Z) & = & 
K^{\bar{m}}(z)+ \theta \,\,  D K^{\bar{m}}|_{\theta=\bar{\theta}=0}(z) 
+\bar{\theta} \,\,
\overline{D} K^{\bar{m}}|_{\theta=\bar{\theta}=0}(z) +
\theta \bar{\theta} \,\,  (-1) \frac{1}{2} [
D, \overline{D} ] K^{\bar{m}}|_{\theta=\bar{\theta}=0}(z),
\nonu \\
J^a(Z) & = & J^a(z) +\theta \,\, D J^a|_{\theta=\bar{\theta}=0}(z) 
+\bar{\theta} \,\, \overline{D} J^a|_{\theta=\bar{\theta}=0}(z)+ 
\theta \bar{\theta} \,\, (-1) \frac{1}{2} [ D, \overline{D} ] 
J^a|_{\theta=\bar{\theta}=0}(z), 
\nonu \\ 
J^{\bar{a}}(Z) & = & 
J^{\bar{a}}(z) +\theta \,\, D J^{\bar{a}}|_{\theta=\bar{\theta}=0}(z) 
+\bar{\theta}  \,\,
\overline{D} J^{\bar{a}}|_{\theta=\bar{\theta}=0}(z) 
+\theta \bar{\theta} \,\, (-1) \frac{1}{2} [
D, \overline{D} ] J^{\bar{a}}|_{\theta=\bar{\theta}=0}(z),
\label{components}
\eea
where two complex spinor covariant derivatives $D$ and $\overline{D}$
satisfy the algebra $D \overline{D} + \overline{D} D = -\pa_z$.
The $SU(N+2)$ indices are decomposed into the $SU(N+1)$ indices and others:
the former is denoted by $m, \bar{m}$ and the latter is denoted by 
$a, \bar{a}$. We will come to this issue of the indices soon.

The nonlinear constraints, by taking $\theta, \bar{\theta}$ independent 
terms from $(2.8)$ of \cite{HS,Ahn1206}, are given by \footnote{
Then it is easy to obtain the following relations, which can be obtained 
from $(2.9)$ of \cite{Ahn1206} by putting $\theta=\bar{\theta}=0$,
\bea
\left[ D, \overline{D} \right] K^m|_{\theta=\bar{\theta}=0}(z) &= 
& -\pa K^m|_{\theta=\bar{\theta}=0}(z) +\frac{1}{(k+N+2)} 
f_{\bar{m} n}^{\;\;\;\;\bar{p}} ( \overline{D} K^n K^p-
K^n \overline{D} K^p )|_{\theta=\bar{\theta}=0}(z),
\nonu \\
\left[ D, \overline{D} \right] K^{\bar{m}}|_{\theta=\bar{\theta}=0}(z) 
&= & \pa K^{\bar{m}}|_{\theta=\bar{\theta}=0}(z)
- \frac{1}{(k+N+2)} f_{m \bar{n}}^{\;\;\;\;p} ( D K^{\bar{n}}
K^{\bar{p}}- K^{\bar{n}} D K^{\bar{p}} )|_{\theta=\bar{\theta}=0}(z),
\nonu \\
\left[ D, \overline{D} \right] J^a|_{\theta=\bar{\theta}=0}(z) &= 
& -\pa J^a|_{\theta=\bar{\theta}=0}(z) +\frac{2}{(k+N+2)} 
f_{\bar{a} b}^{\;\;\;\;\bar{m}} ( \overline{D} J^b K^m-
J^b \overline{D} K^m )|_{\theta=\bar{\theta}=0}(z),
\nonu \\
\left[ D, \overline{D} \right] J^{\bar{a}}|_{\theta=\bar{\theta}=0}(z) 
&= & \pa J^{\bar{a}}|_{\theta=\bar{\theta}=0}(z)
- \frac{2}{(k+N+2)} f_{a \bar{b}}^{\;\;\;\;m} ( D J^{\bar{b}}
K^{\bar{m}}- J^{\bar{b}} D K^{\bar{m}}  )|_{\theta=\bar{\theta}=0}(z).
\label{ConstraintsDDB}
\eea}
\bea
D K^m|_{\theta=\bar{\theta}=0}(z) & = & 
-\frac{1}{2(k+N+2)} f_{\bar{m} n}^{\;\;\;\;\bar{p}} (K^n
K^p)|_{\theta=\bar{\theta}=0}(z),
\nonu \\
D J^a|_{\theta=\bar{\theta}=0}(z) & = &  
-\frac{1}{(k+N+2)} f_{\bar{a} b}^{\;\;\;\;\bar{m}} (J^b K^m)|_{\theta=\bar{\theta}=0}(z),
\nonu \\
\overline{D} K^{\bar{m}}|_{\theta=\bar{\theta}=0}
(z)& =& -\frac{1}{2(k+N+2)} f_{m \bar{n}}^{\;\;\;\;p} (K^{\bar{n}}
K^{\bar{p}})|_{\theta=\bar{\theta}=0}(z),
\nonu \\
\overline{D} J^{\bar{a}}|_{\theta=\bar{\theta}=0}(z) & = &  
-\frac{1}{(k+N+2)}  
f_{a \bar{b}}^{\;\;\;\;m} (J^{\bar{b}} K^{\bar{m}})|_{\theta=\bar{\theta}=0}(z).
\label{Constraints}
\eea
Then the $\theta$- and $\theta \bar{\theta}$-components of 
$K^m(Z)$ and $J^a(Z)$ in (\ref{components}) 
are not independent and they can be written in terms 
of the $\theta, \bar{\theta}$-independent term and 
$\bar{\theta}$-components according to (\ref{Constraints}) and 
(\ref{ConstraintsDDB}).  
Similarly, 
 the $\bar{\theta}$- and $\theta \bar{\theta}$-components of 
$K^{\bar{m}}(Z)$ and $J^{\bar{a}}(Z)$ in (\ref{components}) 
can be written in terms 
of the $\theta, \bar{\theta}$-independent term and 
$\theta$-components \footnote{As one computes the OPEs in the component 
approach in the whole paper, for simplicity, one ignores the notation 
$|_{\theta=\bar{\theta}=0}$  acting on the  ${\cal N}=2$ superfields 
from now on \cite{Ahn1208}. 
That is, $ D K^{\bar{m}}|_{\theta=\bar{\theta}=0}(z) \equiv  
 D K^{\bar{m}}(z), \overline{D} K^m|_{\theta=\bar{\theta}=0}(z) \equiv  
 \overline{D} K^m(z),  D J^{\bar{a}}|_{\theta=\bar{\theta}=0}(z) \equiv  
 D J^{\bar{a}}(z)$, and $ \overline{D} J^a|_{\theta=\bar{\theta}=0}(z) \equiv  
 \overline{D} J^a(z)$. }. 
The on-shell current algebra in ${\cal N}=2$ superspace for the supersymmetric
WZW model on a group $SU(N+2)$ of even dimension 
can be written in terms of components of spin $s = \frac{1}{2}, 1$ 
\cite{Ahn1206}.

For $N=3$, the $24$ adjoint indices of $SU(5)$ 
are divided into $12$ unbarred 
indices and $12$ barred indices. 
Then the $8$ adjoint indices of the subgroup $SU(3)$
are given by $4,5,6$ and some combination between the index $7$ and the index
$8$ (and  their barred indices) 
\footnote{The convention of \cite{GG1305} for the 
subgroup $SU(3)$, around $(3.1)$ of \cite{GG1305}, 
is different from ours because the first $3 \times 3$ block
of $5 \times 5$ corresponds to the their subgroup $SU(3)$ (the unbarred 
indices
$1,2,4$ and some combination between the index $7$ and index $8$ and their 
barred ones will describe their $SU(3)$) while the middle 
$3 \times 3$ block of $5 \times 5$ corresponds to our subgroup 
$SU(3)$. Moreover, the adjoint representation 
${\bf 24}$ of $SU(5)$ breaks into $ {\bf 24} 
\rightarrow ({\bf 8}, {\bf 1}) \oplus ({\bf 1}, 
{\bf 3})\oplus ({\bf 1},{\bf 1}) \oplus ({\bf 3},{\bf 2}) \oplus 
({\bf \bar{3}},{\bf 2})$ under the $SU(3) \times SU(2)$. The first three 
representations  
are given by the currents with the indices $4,5,6,7, 8$ and $9$ 
(and conjugated ones). The last two 
representations are given by the currents with indices $1,2,3$ and 
$10,11,12$ (and conjugated ones). Note that the subgroup $SU(4)$ having 
the indices $1, \cdots, 8$ (and conjugated ones) cannot
be broken into $SU(3) \times SU(2)$.}.
Then the remaining indices, $1, 2, 3, 9, 10, 11, 12$ and some 
combination between the index $7$ and the index $8$
(and their barred indices)
live in the coset $\frac{SU(5)}{SU(3)}$.
We follow the convention of \cite{Ahn1206}. 
However, some currents ($K^m$ or $K^{\bar{m}}$) 
living in the subgroup $SU(4)$ of \cite{Ahn1206}
live in the coset $\frac{SU(5)}{SU(3)}$. 
In other words, one classifies the $24$ basic ${\cal N}=2$ 
WZW affine currents as follows:
\bea
K^m & = & (K^1, \, K^2, \, K^3; \, K^4, \, K^5, \, K^6, \, K^7, \, K^8):
\qquad \mbox{coset and subgroup currents},
\nonu \\
K^{\bar{m}} & = &  (K^{\bar{1}}, \, K^{\bar{2}}, \, K^{\bar{3}}; \, K^{\bar{4}}, 
\, K^{\bar{5}}, \, K^{\bar{6}}, \, K^{\bar{7}},  \, K^{\bar{8}}):
\qquad \mbox{coset and subgroup currents},
\nonu \\
J^a & = & (J^9, \, J^{10}, \, J^{11},  \, J^{12}): \qquad
\mbox{coset currents},
\nonu \\
J^{\bar{a}} & = & (J^{\bar{9}}, \, J^{\bar{10}}, \, J^{\bar{11}},  \, 
J^{\bar{12}}): \qquad \mbox{coset currents}.
\label{KKJJ}
\eea
Of course, 
one can write down  the currents having the indices $1, 2, 3$ (and their
barred indices) $J^1, J^2$ and $J^3$ (and their barred currents) 
respectively but one cannot split the currents having the indices $7$ and $8$
in terms of coset and subgroup currents at the moment. 

For the $\mbox{Wolf} \times SU(2) \times U(1) = SU(2) \times U(1)$
WZW model with $k^{+} = k+1$ and $k^{-}=1$, as described in the introduction,
the $16$ generators of the large ${\cal N}=4$ linear superconformal algebra
are expressed in terms of ${\cal N}=2$ super stress energy tensor of 
super spin-$1$,
two super spin-$\frac{1}{2}$ currents and a super spin-$0$ 
current \cite{RASS}. 
In other words,  the total $16$ currents 
are given by their spin contents
\bea
\left( 1, \frac{3}{2}, \frac{3}{2}, 2 \right), 
\qquad \left( \frac{1}{2}, 1, 1, \frac{3}{2} \right), 
\qquad \left( \frac{1}{2}, 1, 1, \frac{3}{2} \right), \qquad 
\left( 0, \frac{1}{2}, \frac{1}{2},  1 \right).
\label{16comp}
\eea
Note that the spin-$1$ current appears as a derivative of spin-$0$ 
field which is the first component field of the last ${\cal N}=2$
super current of (\ref{16comp}). See also relevant work \cite{Ahn1994}
where the $SU(2) \times U(1)$ group appears in the subgroup of the coset 
model. 

%%%%%%%%%%%%%%%%%%%%%%%%%%%%%%%%%%%%%%%%%%
\subsection{Large ${\cal N}=4$ linear superconformal algebra realization}
%%%%%%%%%%%%%%%%%%%%%%%%%%%%%%%%%%%%%%%%%%

We would like to construct the $16$ generators in terms of 
the numerator $SU(5)$ currents in the coset (\ref{coset}) explicitly.

%%%%%%%%%%%%%%%%%%%%%%%%
\subsubsection{Construction of two spin-$\frac{3}{2}$ currents}
%%%%%%%%%%%%%%%%%%%%%%%%%%

Let us consider the spin-$\frac{3}{2}$ current which is the last component
field of the second ${\cal N}=2$ super current in (\ref{16comp}). 
Among the coset indices of $1,2,3,9,10,11,12$ and some combination of 
the index $7$ and the index $8$, 
it is natural to consider the index $9$ corresponding to
the above spin-$\frac{3}{2}$ current by taking the appropriate 
derivative or spinor covariant derivatives. 
One can write down the following quantity \footnote{ We denote 
the four spin-$\frac{3}{2}$ currents
 $G_{11}(z)$, $G_{12}(z)$, $G_{21}(z)$ and  $G_{22}(z)$ here, along the line of 
\cite{BO}, for the 
$G_{+K}(z)$, $G_{+}(z)$, $G_{-}(z)$ and $G_{-K}(z)$ in \cite{npb1988}. 
That is, the vector (or fundamental) representation of $SO(4)$
is written in terms of two $SU(2)$'s fundamentals.
The three spin-$1$ currents $A^{+i}(z)$  where $i=1, 2, 3$ 
in \cite{npb1988} correspond to $A_1(z), A_2(z)$ and $A_3(z)$ 
here. The three spin-$1$ currents $A^{-i}(z)$ of spin-$1$ 
over there correspond to  $B_1(z), B_2(z)$ and 
$B_3(z)$ 
here. The four spin-$\frac{1}{2}$ currents 
$\Gamma_{+K}(z)$, $\Gamma_{+}(z)$, $\Gamma_{-}(z)$ 
and $\Gamma_{-K}(z)$ over there correspond to 
$F_{11}(z)$, $F_{12}(z)$, $F_{21}(z)$ and  $F_{22}(z)$ with two 
$SU(2)$'s. We use the 
same notations for 
the spin-$2$ and the spin-$1$ currents as $T(z)$ and $U(z)$ respectively.  }
\bea
G_{11}(z) = 
\sqrt{2} i \left(-\frac{1}{2} \left[ D, \overline{D} \right] J^9 -
\frac{(k-3)}{2 (5+k)} \pa J^9 \right)(z).
\label{g11}
\eea
With the second term of (\ref{g11}), the $G_{11}(z)$ transforms as a primary 
field under the stress energy tensor $T_{SU(5)}$ 
\footnote{Among ${\cal N}=2$ WZW affine currents, the primary currents under the
spin-$2$ stress tensor $(T_{SU(5)}-T_{SU(3))}(z)$ are given by $J^{9}(w)$ and 
$J^{\bar{9}}(w)$. 
Here $T_{SU(5)}$ can be obtained from the last component of 
${\cal N}=2$ superfield $\frac{1}{2(N+2+k)}  
[D, \overline{D} ]\left(
J^a J^{\bar{a}} + K^m K^{\bar{m}} -(f_{\bar{m} \bar{a}}^{\,\,\,\,\,\,\bar{a}}+
f_{\bar{m} \bar{n}}^{\,\,\,\,\,\,\bar{n}}) D K^{\bar{m}} -
(f_{m \bar{a}}^{\,\,\,\,\,\,\bar{a}}+
f_{m \bar{n}}^{\,\,\,\,\,\,\bar{n}}) \overline{D} K^{m} \right)(Z)$ \cite{Ahn1206} 
which is equal to $T_{SU(5)}(z) = \frac{1}{(5+k)^3} \left( K^1 K^4 K^{\bar{1}}
K^{\bar{4}} + \frac{i}{2} K^1 K^4 K^{\bar{2}} K^{\bar{7}}\right) + 
\mbox{other quartic, cubic, quadratic and linear terms}$. Similarly, 
one has $T_{SU(3)} =\frac{1}{(5+k)^3} \left( K^4 K^6 K^{\bar{4}}
K^{\bar{6}} + \frac{i}{2} K^4 K^6 K^{\bar{5}} K^{\bar{7}}\right) + 
\mbox{other quartic, cubic, quadratic and linear terms}$. The indices for the
fields in $T_{SU(3)}(z)$ are given by $4,5,6$ and $7$ and $8$ (and their 
conjugated ones). }.
The coefficient $\frac{(k-3)}{(5+k)}$ can be written as 
$\frac{(k^{+}-k^{-})}{(k^{+} + k^{-})}$ with $k^{+} = k+1$ and $k^{-}=N+1=4$ 
\footnote{ By substituting (\ref{ConstraintsDDB}) into the 
equation (\ref{g11}), one obtains
$G_{11}(z)  =   -\frac{i \sqrt{2}}{(5+k)} \,
 \sum_{(m, a)=(1,10)}^{(3,12)} \left( K^m \overline{D} J^{a}-
 \overline{D} K^m J^{a}
 \right)(z)
+ 
\frac{(3-i \sqrt{3})}{3 \sqrt{2} (5+k)} \left(
 K^7 \overline{D} J^9-
%\frac{i (3 i+\sqrt{3})}{3 \sqrt{2} (5+k)}\,
 \overline{D} K^7 J^9 \right)(z)+
\frac{(\sqrt{3}-3 i \sqrt{5})}{6 (5+k)}
 \left( K^8 \overline{D} J^9 
 - 
%\frac{(\sqrt{3}-  3 i \sqrt{5})}{6 (5+k)} \,
 \overline{D} K^8 J^9 \right)(z) 
 + 
\frac{4 i \sqrt{2} }{(5+k)} \, \pa J^9(z)$.}

Similarly,
one can construct the following primary field of spin-$\frac{3}{2}$, 
corresponding to the last component field of the third
${\cal N}=2$ super current in (\ref{16comp}),
under the $T_{SU(5)}$ as follows \footnote{Furthermore, the constraint (\ref{ConstraintsDDB}) implies that
the current (\ref{g22}) can be written in terms of
$
G_{22}(z)  =   -\frac{i \sqrt{2}}{(5+k)} \,
\sum_{(\bar{m},\bar{a})=(\bar{1},\bar{10})}^{(\bar{3},\bar{12})} 
\left( K^{\bar{m}} D J^{\bar{a}}-
D K^{\bar{m}} J^{\bar{a}} \right)(z)
 - 
\frac{(3+i \sqrt{3})}{3 \sqrt{2} (5+k)} \left(
 K^{\bar{7}} D J^{\bar{9}}-
 D K^{\bar{7}} J^{\bar{9}} \right)(z)-
\frac{(\sqrt{3}+3 i \sqrt{5})}{6 (5+k)}
 \left( K^{\bar{8}} D J^{\bar{9}} 
 - 
 D K^{\bar{8}} J^{\bar{9}} \right)(z)
 - 
\frac{4 i \sqrt{2} }{(5+k)} \, \pa J^{\bar{9}}(z)$.}:
\bea
G_{22}(z) = 
\sqrt{2} i \left(-\frac{1}{2} \left[ D, \overline{D} \right] J^{\bar{9}} +
\frac{(k-3)}{2 (5+k)} \pa J^{\bar{9}} \right)(z).
\label{g22}
\eea

%%%%%%%%%%%%%%%%%%%%%%%%
\subsubsection{Construction of three spin-$1$ currents}
%%%%%%%%%%%%%%%%%%%%%%%%%%

What about the spin-$1$ currents which are the superpartner of 
the above spin-$\frac{3}{2}$ currents $G_{11}$ and $G_{22}$?
One expects  that  two linear combinations between 
$\overline{D} J^9$ and $D J^{\bar{9}}$ provide 
the two spin-$1$ currents. Note that there are no singular terms
in the OPEs $D J^{\bar{9}}(z) \, D J^{\bar{9}}(w)$ and 
$\overline{D} J^{9}(z) \, \overline{D} J^{9}(w)$.
With the identifications
\bea
A_1(z) & = & \frac{1}{2} \left( -\overline{D} J^9 + D J^{\bar{9}} \right)(z),
\nonu \\
A_2(z) & = & \frac{i}{2} \left( \overline{D} J^9 + D J^{\bar{9}} \right)(z),
\label{a1a2}
\eea
one calculates the following OPEs
\bea
A_i(z) \, A_i(w) & = & \frac{1}{(z-w)^2} (-1) \frac{1}{2} (k+1) + \cdots,
\qquad i =1, 2,
\label{twoope}
\eea
where the second order pole can be written as $-\frac{1}{2} k^{+}$ because
of $k^{+} = (k+1)$.

Let us calculate the third spin-$1$ current by calculating the OPE
$A_1(z) \, A_2(w)$ with (\ref{a1a2}). It turns out that  
\bea
A_1(z) \, A_2(w) & = & \frac{1}{(z-w)} A_3(w) +\cdots,
\label{oneope}
\eea
where the spin-$1$ current is given by
\bea
A_3(w) & = &
  -  \frac{i}{2 (5+k)}
\sum_{(a,\bar{a})=(9,\bar{9})}^{(12,\bar{12})} J^a J^{\bar{a}}(w) 
+ 
\mbox{other quadratic and linear terms}.
\label{a3}
\eea
Furthermore one obtains 
\bea
A_i(z) \, A_j(w) & = & 
\frac{1}{(z-w)^2} (-1) \frac{1}{2} (k+1) \delta_{ij} +
\frac{1}{(z-w)} \ep_{ijk} A_k(w) + \cdots, \, \ep_{123}=1.
\label{threeope}
\eea
Therefore, the OPEs (\ref{twoope}), (\ref{oneope}), and (\ref{threeope})
provide the $SU(2)_{k^{+}}$ current algebra of ${\cal N}=4$ large superconformal
algebra.
That is,  the OPEs (\ref{threeope})
coincide with the OPE $A^{+i}(z) \, A^{+j}(w)$ appearing in $(C.3)$ of 
\cite{npb1988}.

%%%%%%%%%%%%%%%%%%%%%%%%
\subsubsection{Construction of four spin-$\frac{1}{2}$ currents and 
other two spin-$\frac{3}{2}$ currents}
%%%%%%%%%%%%%%%%%%%%%%%%%%

How does one determine the four spin-$\frac{1}{2}$ currents appearing in 
(\ref{16comp})?
Let us compute the OPE
$A_1(z) \, G_{11}(w)$ explicitly.
It turns out, from (\ref{a1a2}) and (\ref{g11}),  that 
\bea
A_1(z) \, G_{11}(w) 
 & = & \frac{1}{(z-w)^2} \, \frac{i}{2} \, 2(1-\gamma) \,
F_{21}(w) - \frac{1}{(z-w)}\, \frac{i}{2} \, G_{21}(w) +\cdots,  
\label{agope}
\eea
where the parameter is introduced
\bea
\gamma = \frac{k^{-}}{(k^{+} + k^{-})} = \frac{4}{(5+k)}
\label{gamma}
\eea
and the new 
spin-$\frac{1}{2}$ current and the spin-$\frac{3}{2}$ current are
\bea
F_{21}(w) & = &
-\frac{(3 i+\sqrt{3}) }{6 \sqrt{2}} \, K^7(w)
-\frac{1}{12} i (\sqrt{3}-3 i \sqrt{5}) \, K^8(w),
\label{f21} \\
G_{21}(w) & = &
\frac{\sqrt{2}}{5+k} \,  \sum_{(a,\bar{a})=(9,\bar{9})}^{(12,\bar{12})}
J^a D J^{\bar{a}} (w)
\nonu \\
& + &
\mbox{other cubic and quadratic terms}.
\label{g21}
\eea
The OPE (\ref{agope}) corresponds to the OPE 
$A^{+i}(z) \, G_{a}(w)$ in $(C.3)$ of \cite{npb1988} 
\footnote{More explicitly, their quantity 
$\alpha_{a}^{+ib}$ appearing in the right hand side of this OPE,
in this particular case, becomes $\alpha_{a}^{+ib} =
\alpha_{+K}^{+1-} =  \delta^{-c}  \alpha_{+K,c}^{+1}=  
2 \alpha_{+K, +}^{+1} $ which is 
equal to $ 2 \times \frac{i}{4} =\frac{i}{2}$. Here the index $a$ stands for 
$+K$, the index $b$ can be $+, -, +K$, or $-K$ and the index $i$ stands for 
$1$. Their nonzero $\delta_{ab}$ are given by $\delta_{+-} = \delta_{-+} =
\delta_{+K,-K} =\delta_{-K,+K} =\frac{1}{2}$ (not $1$). Similarly,
 $\delta^{+-} = \delta^{-+} =
\delta^{+K,-K} =\delta^{-K,+K} =2$ (not $1$). The quantity
 $\alpha_{a b}^{+i}$ are antisymmetric under $a \leftrightarrow b$. The $12$ 
nonzero independent 
quantity  $\alpha_{ab}^{\pm i}$ can be obtained from the table $3$ of 
\cite{npb1988}. }.

Let us consider the OPE $A_1(z) \, G_{22}(w)$.
It turns out, from (\ref{a1a2}) and (\ref{g22}) together with (\ref{gamma}),  
that 
\bea
A_1(z) \, G_{22}(w) 
 & = & -\frac{1}{(z-w)^2} \, \frac{i}{2} \, 2(1-\gamma) \,
F_{12}(w) + \frac{1}{(z-w)}\, \frac{i}{2} \, G_{12}(w) +\cdots,  
\label{otheragope}
\eea
where \footnote{The quantity $\alpha_{a}^{+ib}$ 
in \cite{npb1988} leads to $\alpha_{-K, -}^{+1}
=\frac{i}{4}$ and the normalization in the first-order pole in 
(\ref{otheragope})
is fixed.} one can read off the following currents
\bea
F_{12}(w) & = &
-\frac{(-3 i+\sqrt{3})}{6 \sqrt{2}} \,
K^{\bar{7}}(w)+\frac{1}{12} i (\sqrt{3}+3 i \sqrt{5}) K^{\bar{8}}(w),
\label{f12} \\
G_{12}(w) & = &
-\frac{\sqrt{2}}{5+k} \,  \sum_{(a,\bar{a})=(9,\bar{9})}^{(12,\bar{12})} 
\overline{D} J^a J^{\bar{a}} (w)
\nonu \\
& + &
\mbox{other cubic, quadratic and linear terms}.
\label{g12}
\eea
Again, this OPE  (\ref{otheragope}) 
corresponds to the OPE 
$A^{+i}(z) \, G_{a}(w)$ in $(C.3)$ of \cite{npb1988}. 

How does one determine the remaining spin-$\frac{1}{2}$ currents?
Let us consider the following OPE
$A_3(z) \, G_{11}(w)$.
It turns out, from (\ref{a3}) and (\ref{g11}),  that 
\bea
A_3(z) \, G_{11}(w) 
 & = & \frac{1}{(z-w)^2} \, \frac{i}{2} \, 2(1-\gamma) \,
F_{11}(w) - \frac{1}{(z-w)}\, \frac{i}{2} \, G_{11}(w) +\cdots,  
\label{other1agope}
\eea
where \footnote{In this case, the quantity $\alpha_{a}^{+ib}$ becomes 
$\alpha_{+K, -K}^{+3} = -\frac{i}{4}$ which agrees with the result 
of \cite{npb1988}.}
the spin-$\frac{1}{2}$ current is given by
\bea
F_{11}(w) & = & \frac{i }{\sqrt{2}} \, J^9(w).
\label{f11}
\eea
Furthermore, the following OPE,
 which can be obtained from (\ref{a3}) and (\ref{g22}),
provides other spin-$\frac{1}{2}$ current 
\bea
A_3(z) \, G_{22}(w) 
 & = & -\frac{1}{(z-w)^2} \, \frac{i}{2} \, 2(1-\gamma) \,
F_{22}(w) + \frac{1}{(z-w)}\, \frac{i}{2} \, G_{22}(w) +\cdots,  
\label{otherotheragope}
\eea
where \footnote{Furthermore, one has 
consistent expression $\alpha_{-K, +K}^{+3}= \frac{i}{4}$.}
the spin-$\frac{1}{2}$ current is
\bea
F_{22}(w) & = & -\frac{i }{\sqrt{2}} \, J^{\bar{9}}(w).
\label{f22}
\eea
The nontrivial OPEs from (\ref{f21}), (\ref{f12}), (\ref{f11}) 
and (\ref{f22}) are 
\bea
F_{11}(z) \, F_{22}(w) & = & 
-\frac{1}{(z-w)} \, \frac{1}{2} \, \frac{c}{6\gamma(1-\gamma)} +\cdots,
\nonu \\
F_{12}(z) \, F_{21}(w) & = &
-\frac{1}{(z-w)} \, \frac{1}{2} \, \frac{c}{6\gamma(1-\gamma)} +\cdots.
\label{ffope}
\eea

%%%%%%%%%%%%%%%%%%%%%%%%
\subsubsection{Construction of other three spin-$1$ currents}
%%%%%%%%%%%%%%%%%%%%%%%%%%

Let us continue to determine the remaining spin-$1$ currents.
From the OPEs $F_{11}(z) \, G_{21}(w)$ and $F_{12}(z) \, G_{22}(w)$, 
one obtains
\bea
F_{11}(z) \, G_{21}(w) & = & \frac{1}{(z-w)} \left( -i B_1 - B_2\right)(w)+
\cdots,
\nonu \\
F_{12}(z) \, G_{22}(w) & = & \frac{1}{(z-w)} \left( -i B_1 + B_2\right)(w)+
\cdots,
\label{fgope}
\eea
where the two new spin-$1$ currents are \footnote{One obtains the quantities
$\alpha_{+K, -}^{-1}=\frac{i}{4}$ and $\alpha_{+K,-}^{-2}=\frac{1}{4}$. Similarly 
one uses $\alpha_{+,-K}^{-1}=\frac{i}{4}$ and 
$\alpha_{+,-K}^{-2}=-\frac{1}{4}$. All these results are consistent with 
the results in \cite{npb1988}.  } 
obtained from the first-order poles in (\ref{fgope})
\bea
B_1(w) & = &
\frac{1}{2 (5+k)} \sum_{(m,a)=(1,10)}^{(3,12)} K^m J^a(w)
+ 
\mbox{other quadratic terms},
\nonu \\
B_2(w) & = &
\frac{i}{2 (5+k)}   \sum_{(m,a)=(1,10)}^{(3,12)} K^m J^a(w)
+ 
\mbox{other quadratic terms}.
\label{b1b2}
\eea
Now one obtains the third spin-$1$ current from (\ref{b1b2})
as follows:
\bea
B_1(z) \, B_2(w) & = & \frac{1}{(z-w)} \, B_3(w) +\cdots,
\label{b1b2b3}
\eea
where the spin-$1$ current  can be read off 
\bea
B_3(w) & = &
 -  \frac{i}{2 (5+k)} 
 \sum_{(a,\bar{a})=(9,\bar{9})}^{(12,\bar{12})} J^a J^{\bar{a}}(w) 
+ 
\mbox{other quadratic terms}.
\label{b3}
\eea
One obtains the following OPEs from (\ref{b1b2}) and (\ref{b3}) 
\bea
B_i(z) \, B_j(w) & = & -\frac{1}{(z-w)^2} 2 \delta_{ij} +
\frac{1}{(z-w)} \, \ep_{ijk} B_k(w) +\cdots.
\label{other}
\eea

Therefore, the OPEs (\ref{b1b2b3}) and (\ref{other})
provide the $SU(2)_{k^{-}}$ current algebra with $k^{-}=4$
of large ${\cal N}=4$ linear superconformal
algebra.
That is,  the OPEs (\ref{other})
coincide with the OPE $A^{-i}(z) \, A^{-j}(w)$ appearing in $(C.3)$ of 
\cite{npb1988}.

%%%%%%%%%%%%%%%%%%%%%%%%
\subsubsection{Construction of other spin-$1$ current}
%%%%%%%%%%%%%%%%%%%%%%%%%%

Finally, the undetermined spin-$1$ current
can be obtained from the OPE $F_{11}(z) \, G_{22}(w)$
\bea
F_{11}(z) \, G_{22}(w)  & = &
\frac{1}{(z-w)} \, \left( -i A_3 -i B_3 + U\right)(w) + \cdots,
\label{otherfgope}
\eea
where \footnote{In this case, one reads off $\alpha_{+K,-K}^{+3}=-\frac{i}{4}$ 
and 
$\alpha_{+K, -K}^{-3}=\frac{i}{4}$ which agree with the ones in 
\cite{npb1988}.} the first-order pole with previous expressions for 
$A_3(w)$ and $B_3(w)$ in (\ref{a3}) and (\ref{b3}) provides 
the information of $U(w)$ and the spin-$1$ current is given by
\bea
U(w)  & = & 
\frac{1}{12} (-3 i-\sqrt{3}) \overline{D} K^7(w)
-\frac{(i \sqrt{3}+3 \sqrt{5}) }{12 \sqrt{2}} \,
\overline{D} K^8(w) 
+\frac{1}{12} (-3 i+\sqrt{3})\, D K^{\bar{7}}(w)
\nonu \\
& + & \frac{(-i \sqrt{3}+3 \sqrt{5}) }{12 \sqrt{2}} \, D K^{\bar{8}}(w).
\label{uu}
\eea
From (\ref{uu}), the following OPE
can be obtained
\bea
U(z) \, U(w) =-\frac{1}{(z-w)^2} \, \frac{c}{12\gamma(1-\gamma)} 
+\cdots.
\label{uuope}
\eea
Note that one can reexpress the spin-$1$ currents using the superpartner 
of $F_{12}(z)$ and $F_{21}(z)$, i.e. $D F_{12}(z)$ and $\overline{D} F_{21}(z)$.
Then one sees that
\bea
U(z) = \frac{1}{\sqrt{2}} \left(  \overline{D} F_{21} - D F_{12} \right)(z).
\label{Usuperpartner}
\eea

%%%%%%%%%%%%%%%%%%%%%%%%
\subsubsection{Construction of spin-$2$ current}
%%%%%%%%%%%%%%%%%%%%%%%%%%

One obtains the stress tensor as a difference between the stress tensor of 
the group $SU(5)$ and the stress tensor of the subgroup $SU(3)$: 
\bea
T(z) = T_{SU(5)}(z) - T_{SU(3)}(z).
\label{tstress}
\eea
Then one can obtain the following OPEs
\bea
T(z)_{SU(5)} \, T(w)_{SU(5)} & = & \frac{1}{(z-w)^4} \, \frac{c_{SU(5)}}{2} +
\frac{1}{(z-w)^2} \, 2T(w)_{SU(5)} + \frac{1}{(z-w)} \, \pa T(w)_{SU(5)} +
\cdots,
\nonu \\
T(z)_{SU(5)} \, T(w)_{SU(3)} & = & \frac{1}{(z-w)^4} \, \frac{c_{SU(3)}}{2} +
\frac{1}{(z-w)^2} \, 2T(w)_{SU(3)} + \frac{1}{(z-w)} \, \pa T(w)_{SU(3)} +
\cdots,
\nonu \\
T(z)_{SU(3)} \, T(w)_{SU(3)} & = & \frac{1}{(z-w)^4} \, \frac{c_{SU(3)}}{2} +
\frac{1}{(z-w)^2} \, 2T(w)_{SU(3)} + \frac{1}{(z-w)} \, \pa T(w)_{SU(3)} \nonu 
\\ & + &
\cdots, 
\label{ttopes} 
\eea
where the central charges are given by 
\bea
c_{SU(5)} = \frac{12(5+3k)}{(5+k)}, \qquad 
c_{SU(3)} = \frac{12(3+k)}{(5+k)},
\qquad
c= c_{SU(5)}-c_{SU(3)} =\frac{24(1+k)}{(5+k)}.
\label{central}
\eea
Then the standard OPE between the stress tensor $T(z)$, that can be 
obtained from (\ref{ttopes}), has the following 
form 
\bea
T(z) \, T(w) & = & \frac{1}{(z-w)^4} \, \frac{c}{2} +
\frac{1}{(z-w)^2} \, 2T(w) + \frac{1}{(z-w)} \, \pa T(w) +
\cdots,
\label{opett}
\eea
where the central charge is given in (\ref{central}).

One can easily check that there are no singular terms in the OPEs between the
generators of the large ${\cal N}=4$ linear superconformal algebra and the 
subgroup stress tensor as follows \footnote{Instead, there exist 
the singular terms in 
the OPEs between $16$ currents of large ${\cal N}=4$ linear algebra
and ${\cal N}=2$ WZW affine currents (having the indices $4,5,6$ and 
some linear combination of the index $7$ and the index $8$ and their 
conjugated ones) living in 
the subgroup $SU(3)$. In next section, by factoring out spin-$1$ current and 
four spin-$\frac{1}{2}$ currents and modifying the remaining $11$ currents 
correctly, the $11$ currents of the large ${\cal N}=4$ nonlinear 
superconformal  algebra  commute with the $SU(3)$ WZW affine currents. See 
also the footnote \ref{su3currents}.}:
\bea
T_{SU(3)}(z) \, \Phi(w) & = & + \cdots, 
\label{Tphi}
\eea 
where $ \Phi(w)$ is $16$ currents of large ${\cal N}=4$ linear superconformal 
algebra.
There are no singular terms in the OPEs between the
generators of the large ${\cal N}=4$ linear superconformal algebra and the 
two subgroup spin-$\frac{3}{2}$ currents, $G_{12}^{SU(3)}(z)$ and 
$G_{21}^{SU(3)}(z)$ \footnote{That is, $
G_{12}^{SU(3)}(z) \, \Phi(w)  =  + \cdots$ and  
$ G_{21}^{SU(3)}(z) \, \Phi(w) = + \cdots$, 
where $\Phi(w)$ is defined in (\ref{Tphi}).}.

One can rearrange the $16$ currents in terms of four 
${\cal N}=2$ multiplets as 
follows:
\bea
\left( 1, \frac{3}{2}, \frac{3}{2}, 2 \right) & : &
(  -2 i \, \gamma \,  A_3 - 2i \, (1-\gamma) \, 
B_3, \quad G_{21}, \quad G_{12}, \quad T ), \nonu \\
\left(\frac{1}{2}, 1, 1, \frac{3}{2} \right) & : &
( 2F_{11}, \quad 2i(B_1-i B_2), \quad -i(A_1+i A_2) , \quad 
G_{11} + (1 - 2\gamma) \pa F_{11}  ), \nonu \\
\left(\frac{1}{2}, 1, 1, \frac{3}{2} \right) & : & ( 2F_{22}, \quad
-2i(A_1 -i A_2), \quad i(B_1+i B_2), 
\quad -G_{22} +(1-2 (1-\gamma)) \pa F_{22} ), \nonu \\
\left(0, \frac{1}{2}, \frac{1}{2}, 1 \right) & : & 
( 2\int U dz, \quad -2 F_{21}, \quad F_{12},   \quad
-i (A_3 - B_3) ).
\label{fourn2}
\eea
In the first ${\cal N}=2$ multiplet of (\ref{fourn2}),
the OPE between $G_{21}(z)$ and $G_{12}(w)$ provides 
${\cal N}=2$ $U(1)$ current of spin-$1$ and the stress tensor.
For given lowest component of second ${\cal N}=2$ multiplet in (\ref{fourn2}),
the third component can be read off from the OPE between $G_{12}(z)$ and $F_{11}(w)$. Similarly, 
the second component can be  obtained 
from the OPE between $G_{21}(z)$ and $F_{11}(w)$.
The last component can be obtained from the OPE between $G_{12}(z)$ and 
$i (B_1 -i B_2)(w)$  \footnote{In other words, one has the following OPEs:
$
G_{21}(z) \, F_{11}(w)  =  
- \frac{1}{(z-w)} i \left( B_1 - i B_2 \right)(w) 
+\cdots$, 
$G_{12}(z) \, F_{11}(w)  =  
- \frac{1}{(z-w)} i \left( A_1 + i A_2 \right)(w) 
+\cdots$, 
and 
$ G_{12}(z) \, i(B_1 - i B_2)(w)  = 
-\frac{1}{(z-w)^2} \frac{8}{(5+k)} F_{11}(w) 
+ \frac{1}{(z-w)}  \left[ - \frac{8}{(5+k)} \pa F_{11} +G_{11} \right](w) 
+\cdots$.}.
For given lowest component of third ${\cal N}=2$ multiplet in (\ref{fourn2}),
the third component can be read off from the OPE between 
$G_{12}(z)$ and $F_{22}(w)$. Similarly, 
the second component can be  obtained 
from the OPE between $G_{21}(z)$ and $F_{22}(w)$.
The last component can be obtained from the OPE between $G_{12}(z)$ and 
$i (A_1 -i A_2)(w)$ \footnote{Similarly one has 
$
G_{21}(z) \, F_{22}(w)  =  
- \frac{1}{(z-w)} i \left( A_1 - i A_2 \right)(w) 
+\cdots$, 
$G_{12}(z) \, F_{22}(w)  =  
\frac{1}{(z-w)} i \left( B_1 + i B_2 \right)(w) 
+\cdots$ 
and 
$
G_{12}(z) \, i(A_1 - i A_2)(w)  =  
\frac{1}{(z-w)^2} \frac{2(1+k)}{(5+k)} F_{22}(w) 
+ \frac{1}{(z-w)}  \left[  \frac{2(1+k)}{(5+k)} \pa F_{11} +G_{22} \right](w) 
+\cdots$. }.

Therefore, the large ${\cal N}=4$ linear 
superconformal algebra is realized by
the spin-$2$ current, (\ref{tstress}), 
the four spin-$\frac{3}{2}$ currents, (\ref{g11}), 
(\ref{g22}), (\ref{g21}) and (\ref{g12}),
the spin-$1$ currents, (\ref{a1a2}), (\ref{a3}), 
(\ref{b1b2}), (\ref{b3}) and (\ref{uu}), the spin-$\frac{1}{2}$ currents,
(\ref{f21}), (\ref{f12}), (\ref{f11}) and (\ref{f22}).
Some of the algebra consist of (\ref{twoope}), (\ref{oneope}), 
(\ref{threeope}), (\ref{agope}), (\ref{otheragope}), (\ref{other1agope}), 
(\ref{otherotheragope}), (\ref{ffope}), (\ref{fgope}), (\ref{b1b2b3}),
(\ref{other}), (\ref{otherfgope}), (\ref{uuope}) and (\ref{opett}).
The remaining OPEs can be obtained similarly \footnote{There exists 
the middle ${\cal N}=4$ linear superconformal algebra where $SU(2) \times
U(1)^4$ affine algebra is present  in \cite{AK}.}.

%%%%%%%%%%%%%%%%%%%%%%%%%%%%%%%%%%%%%%%%%%%%%%%%%%%%%%%%%%%%%%%%%%%%%
%%%%%%%%%%%%%%%%%%%%%%%%%%%%%%%%%%%%%%%%%%%%%%%%%%%%%%%%%%%%%%%%%%%%%%
\section{The large ${\cal N}=4$ nonlinear 
superconformal  algebra in the Wolf space coset 
minimal model  }
%3%%%%%%%%%%%%%%%%%%%%%%%%%%%%%%%%%%%%%%%%%%%%%%%%%%%%%%%%%%%%%%%%%%%%%
%%%%%%%%%%%%%%%%%%%%%%%%%%%%%%%%%%%%%%%%%%%%%%%%%%%%%%%%%%%%%%%%%%%%%

The large ${\cal N}=4$ nonlinear algebra can be obtained from the linear one
described in previous section by 
factoring out the four fermions and a spin-$1$ bosonic current 
\footnote{The main reason why we should decouple these five currents is that
they do not appear in the $D(2,1|\frac{\mu}{1-\mu})$ wedge subalgebra
and therefore also in the higher spin theory of \cite{GG1305}. In the CFT 
computations characterized by any OPEs obtained in this paper, because the 
number of independent currents is reduced from $16$ to $11$, there are no 
computational complications in the calculations of the OPE. In particular, we 
do not have to consider any spin-$\frac{1}{2}$ currents in 
the composite fields of any spin appearing in the singular terms 
of the OPE.  }.  

%%%%%%%%%%%%%%%%%%%%%%%%%%%%%%%%%%%%%%%%%
\subsection{Construction of spin-$2$ stress tensor}
%%%%%%%%%%%%%%%%%%%%%%%%%%%%%%%%%%%%%%%%%
 
The additional terms in the stress energy tensor 
of spin-$2$ consist of the quadratic term of spin-$1$ current 
and the $16$ quadratic terms of spin-$\frac{1}{2}$ currents with derivative.
Then the total $19$ unknown coefficient functions can be 
determined from the following regular conditions
between the $U(z)$, $F_{11}(z)$, $F_{12}(z)$, $F_{21}(z)$ and $F_{22}(z)$
and the stress energy tensor $\hat{T}= T +T_{mod}$ \cite{GS,cqg1989,npb1989}:
\bea
U(z) \, \hat{T}(w)  & = &  +\cdots, \qquad
F_{a}(z) \, \hat{T}(w)   =   +\cdots, \qquad a=11,12,21,22.
\label{hatTregular}
\eea

For $N=3$, the stress energy tensor of the large ${\cal N}=4$
nonlinear algebra can be written as  \cite{GS,cqg1989,npb1989}
\bea
\hat{T}(z) & = & T(z) + \frac{1}{(5+k)} U U(z) +\frac{1}{(5+k)}
\left( \pa F_{11} F_{22} +\pa F_{12} F_{21} + \pa F_{21} F_{12} +\pa F_{22} 
F_{11}\right)(z)
\nonu
\\
& = & 
 T_{SU(5)}(z)-T_{SU(3)}(z) 
+ \frac{1}{(5+k)} \left( U U 
+ \pa F^{a} F_{a} \right)(z).
\label{stressnonlinear} 
\eea
The general $N$ case can be obtained by putting $(5+k) \rightarrow 
(k+N+2)$ and $T \rightarrow  T_{SU(N+2)}-T_{SU(N)}$ 
in (\ref{stressnonlinear}) similarly.
One can read off the corresponding central charge appearing in the Virasoro 
algebra by calculating the OPE $\hat{T}(z) \, \hat{T}(w)$ \footnote{
The OPE between 
$U U(z)$ term  and itself contributes to $1$ for the central charge 
while the OPEs between the fermions and itself contribute to 
$2$. The OPE between $T(z)$ and the $U U(w)$ (and the OPE between 
$U U(z)$ and $T(w)$) contributes $-2$
and similarly the contributions from the $T(z)$ and fermion terms 
give to $-4$. This can be seen from the (\ref{Wolfcentral}) by realizing that
the central term coming from the three fermions in $SU(2)$
is given by $\frac{1}{2} \times 3 =\frac{3}{2}$ 
while the one from the spin-$1$ and 
spin-$\frac{1}{2}$ fields in $U(1)$ is given by $\frac{3}{2}$. Therefore, 
the sum of these is equal to $3$.  }.
Therefore, the total central charge is given by   
\bea
\hat{c} =  \frac{6(k+1)(N+1)}{(k+N+2)} -2 -4+1 +2 
=\frac{3(k+N+2k N)}{(k+N+2)} \rightarrow \frac{3(3+7k)}{(5+k)},
\label{chat}
\eea
where the $N=3$ is substituted in  the last stage.

%%%%%%%%%%%%%%%%%%%%%%%%%%%%%%%%%%%%%%%%%
\subsection{Construction of six spin-$1$ currents} 
%%%%%%%%%%%%%%%%%%%%%%%%%%%%%%%%%%%%%%%%%

Let us determine the other currents. 
For the spin-$1$ current, one can add $6$  additional terms 
coming from four fermions, $F_{11}(z)$, $F_{12}(z)$, $F_{21}(z)$ and 
$F_{22}(z)$.
The relative coefficient functions, as done in (\ref{hatTregular}), 
can be determined by 
the regular conditions   \cite{GS,cqg1989,npb1989}
\bea
U(z) \, \hat{A}_1(w)  & = &  +\cdots, \qquad
F_{a}(z) \, \hat{A}_1(w)   =   +\cdots, \qquad a=11,12,21,22.
\label{reg}
\eea
It turns out that the spin-$1$ current is given by
\bea
\hat{A}_1(z) & = & 
A_1(z)-\frac{1}{(5+k)}\left(\frac{i}{2}\, F_{11} F_{12}-
\frac{i}{2} \, F_{12} F_{11}
+\frac{i}{2}\, F_{21} F_{22}
-\frac{i}{2} \, F_{22} F_{21} \right)(z).
\label{spinonenonlinear}
\eea
One can further simplify (\ref{spinonenonlinear}) using the fact that
$F_{11} F_{12}(z) = - F_{12} F_{11}(z)$ and $F_{21} F_{22}(z) = -F_{22} F_{21}(z)$.

Similarly, the other currents of spin-$1$
can be obtained as follows:
\bea
\hat{A}_i(z) & = &
A_i(z)-\frac{1}{(5+k)} \,
\alpha_{ab}^{+i} \, F^{a} F^{b}, \qquad i =2, 3, \qquad a, b=11,12,21,22.
\label{otherspinonenonlinear} 
\eea
It is straightforward to 
calculate 
the following OPEs from (\ref{spinonenonlinear}) and 
(\ref{otherspinonenonlinear})
\bea
\hat{A}_i(z) \, \hat{A}_j(w) & = & 
-\frac{1}{(z-w)^2}  \, \frac{1}{2} \hat{k}^{+} \hat{\delta}_{ij} +
\frac{1}{(z-w)} \, \ep_{ijk} \, \hat{A}_k(w)
+ \cdots, \, \qquad \hat{\delta}_{ii}=1,
\label{opeA}
\eea
where the new level is given by $\hat{k}^{+} = k^{+}-1= k$.
The three quadratic terms appearing in (\ref{spinonenonlinear}) and 
(\ref{otherspinonenonlinear}) 
satisfy the $SU(2)_1$ current algebra of $(3.18)$ in \cite{GG1305}.
It is obvious that 
the OPEs between these three currents and $\hat{A}_1(z)$, $\hat{A}_2(z)$
and $\hat{A}_3(z)$ (that correspond to $\hat{J}(z)$ 
around the equation $(3.18)$ 
in \cite{GG1305}) 
do not have the singular terms due to the conditions 
 \cite{GS,cqg1989,npb1989}
like as (\ref{reg}):
\bea
U(z) \, \hat{A}_i(w)  & = &  +\cdots, \qquad
F_{a}(z) \, \hat{A}_i(w)   =   +\cdots, \qquad i=2,3, \qquad a=11,12,21,22.
\label{regulara2a3}
\eea
Similarly, the currents $A_1(z), A_2(z)$ and $A_3(z)$
correspond to $J(z)$ appearing around the equation $(3.18)$ 
of \cite{GG1305} \footnote{
More explicitly, 
by substituting (\ref{a1a2}), (\ref{a3}), (\ref{f11}), (\ref{f12}), 
(\ref{f21}) and (\ref{f22}) into the equations (\ref{spinonenonlinear}) and 
(\ref{otherspinonenonlinear}),
one obtains 
\bea
\hat{A}_1(z) & = &
-\frac{1}{2} \overline{D} J^9(z)+\frac{1}{2} D J^{\bar{9}}(z)
+ \left[ \frac{(3 i+\sqrt{3}) 
}{12 (5+k)} \, K^7 
+\frac{(i \sqrt{3}+3 \sqrt{5}) 
}{12 \sqrt{2} (5+k)} \, K^8 \right] J^{\bar{9}}(z)
\nonu \\
& + & \left[ \frac{(-3 i+\sqrt{3}) 
}{12 (5+k)} \, K^{\bar{7}} 
 +  \frac{(-i \sqrt{3}+3 \sqrt{5}) 
}{12 \sqrt{2} (5+k)} \, K^{\bar{8}} \right] J^9(z),
\nonu \\
\hat{A}_2(z) & = &
\frac{i}{2} \overline{D} J^9(z)+\frac{i}{2} D J^{\bar{9}}(z)
+  i \left[ \frac{(3 i+\sqrt{3}) 
}{12 (5+k)} \, K^7 
+\frac{(i \sqrt{3}+3 \sqrt{5}) 
}{12 \sqrt{2} (5+k)} \, K^8 \right] J^{\bar{9}}(z)
\nonu \\
& - & i \left[ \frac{(-3 i+\sqrt{3}) 
}{12 (5+k)} \, K^{\bar{7}} 
 +  \frac{(-i \sqrt{3}+3 \sqrt{5}) 
}{12 \sqrt{2} (5+k)} \, K^{\bar{8}} \right] J^9(z),
\nonu \\
\hat{A}_3(z) & = &
 i \left[ \frac{1}{12}  (3 i+\sqrt{3}) \, \overline{D} K^7
+\frac{ (i \sqrt{3}+3 \sqrt{5})}{12 \sqrt{2}} \,
 \overline{D} K^8 \right](z)
+ i \left[ \frac{1}{12} (-3 i + \sqrt{3}) \, D K^{\bar{7}}
\right. \nonu \\
&+ & \left. \frac{(- i \sqrt{3}  +3  \sqrt{5}) }{12 \sqrt{2}}\,
D K^{\bar{8}} \right](z) -  \frac{i}{ 2(5+k)} ( 
 2 J^9 J^{\bar{9}} +  \sum_{(a,\bar{a})=(10,\bar{10})}^{(12,\bar{12})} 
J^{a} J^{\bar{a}} )(z) 
 \nonu \\
& - &  \frac{i}{2 (5+k)}  \sum_{(m,\bar{m})=(1,\bar{1})}^{(3,\bar{3})} 
K^m K^{\bar{m}}(z).
\label{nona1a2a3}
\eea}.
Compared to (\ref{a1a2}) and (\ref{a3}), 
the currents $\hat{A}_1(z)$ and $\hat{A}_2(z)$ 
contain the quadratic parts in the 
fermions while the quadratic parts in the fermions with the indices 
$7$ and $8$ appearing in $A_3(z)$ 
disappear in the current $\hat{A}_3(z)$ \footnote{As observed in (\ref{a1a2})
and (\ref{a3}), the currents $A_1(z)$, $A_2(z)$ and $A_3(z)$ contain 
the linear spin-$1$ currents (not quadratic fermions). Then one knows its 
superpartners, $J^9(z)$, $J^{\bar{9}}(z)$ and a linear combination between 
$K^7(z), K^8(z), K^{\bar{7}}(z)$ and $K^{\bar{8}}(z)$. These are the three 
fermions in $su(2)_{\kappa}^{(1)}$ of \cite{GG1305}. Note that from 
(\ref{b1b2}) and (\ref{b3}) there are no linear spin-$1$ currents. All of 
the  expressions are written in terms of quadratic fermions. Note that 
$\hat{A}_1(z)$ and $\hat{A}_2(z)$ contain the Wolf space 
subgroup indices in their expressions 
while the $\hat{A}_3(z)$ contains both Wolf space subgroup indices and Wolf
space coset 
indices. Therefore, the $SU(2)_k$ algebra of large ${\cal N}=4$ nonlinear 
algebra comes from two currents from the Wolf space 
subgroup and one current from both
the subgroup and the coset of Wolf space. Note that  the $SU(2)_{k+1}$ 
algebra of large ${\cal N}=4$ linear 
algebra comes from three currents from the coset $\frac{SU(5)}{SU(3)}$.
This is consistent with the equation $(3.7)$ of \cite{ST}. }. 

Let us move on the other type of spin-$1$ currents.
As done before, the spin-$1$ currents, $B_1(z)$, $B_2(z)$ and 
$B_3(z)$(that correspond to $K(z)$ appearing around the equation $(3.18)$ 
of \cite{GG1305})
can  be modified under the factoring out the fermions and spin-$1$ current.
One obtains  \cite{GS,cqg1989,npb1989}
\bea
\hat{B}_i(z) & = & 
B_i(z)-\frac{1}{(5+k)} \, \alpha_{ab}^{-i}\, F^{a} F^{b}, \qquad i=1,2,3,
\qquad a, b=11,12,21,22.
\label{nonb1b2b3}
\eea
Their current algebra can be summarized by
\bea
\hat{B}_i(z) \, \hat{B}_j(w) & = & 
-\frac{1}{(z-w)^2} \frac{1}{2} \, \hat{k}^{-} \hat{\delta}_{ij} +
 \frac{1}{(z-w)} \, \ep_{ijk} \, \hat{B}_k(w)
 +\cdots,
\label{opeB}
\eea
where the new level $\hat{k}^{-} = k^{-}-1 = N=3$.
Three of four in $\tilde{K}^{\alpha \beta}(z)$ defined in the equation 
$(3.14)$ \cite{GG1305} correspond to
the present $\hat{B}_1(z)$, $\hat{B}_2(z)$ and $\hat{B}_3(z)$.
The three quadratic terms appearing in (\ref{nonb1b2b3})  
satisfy the other $SU(2)_1$ current algebra of $(3.18)$ in \cite{GG1305}
and therefore the OPEs between these three currents and the three 
currents appearing in (\ref{spinonenonlinear}) and 
(\ref{otherspinonenonlinear}) (satisfying other 
$SU(2)_1$ current algebra) do not have the singular terms
\footnote{ Furthermore, 
by substituting (\ref{b1b2}), (\ref{b3}), (\ref{f11}), (\ref{f12}), 
(\ref{f21}) and (\ref{f22}) into the equations (\ref{nonb1b2b3}), 
one has 
\bea
\hat{B}_1(z) & = &
\frac{1}{2 (5+k)} \left(\sum_{(m,a)=(1,10)}^{(3,12)} K^m J^{a}
 +  \sum_{(\bar{m},\bar{a})=(\bar{1},\bar{10})}^{(\bar{3},\bar{12})}
K^{\bar{m}} J^{\bar{a}} \right)(z),
\nonu \\
\hat{B}_2(z) & = &
\frac{i}{2 (5+k)}
 \left(\sum_{(m,a)=(1,10)}^{(3,12)} K^m J^{a}
 -  \sum_{(\bar{m},\bar{a})=(\bar{1},\bar{10})}^{(\bar{3},\bar{12})}
K^{\bar{m}} J^{\bar{a}} \right)(z),
\nonu \\
\hat{B}_3(z) & = &
 -  \frac{i}{2 (5+k)} \left( 
  \sum_{(a, \bar{a})=(10,\bar{10})}^{(12,\bar{12})} 
J^{a} J^{\bar{a}} +\sum_{(m,\bar{m})=(1,\bar{1})}^{(3,\bar{3})} K^m K^{\bar{m}}
\right)(z).
\label{b1b2b3nonlinear}
\eea}.
Compared to (\ref{b1b2}) and (\ref{b3}),
the quadratic parts having the indices $7, 8$ or $9$
disappear in (\ref{b1b2b3nonlinear}) 
\footnote{Note that all the currents $\hat{B}_1(z), \hat{B}_2(z)$
and $\hat{B}_3(z)$ contain the Wolf space coset indices only. Therefore 
the $SU(2)_3$ algebra of large ${\cal N}=4$ nonlinear superconformal algebra
comes from the currents in the Wolf space coset. This is consistent with 
the result for the structure of $J^{M-}$ in $(3.18)$ 
of \cite{GK}. Recall that 
the $SU(2)_4$ algebra of large ${\cal N}=4$ linear superconformal algebra
comes from the currents in the coset $\frac{SU(5)}{SU(3)}$.
This is consistent with the equation $(3.7)$ of \cite{ST}.
The spin-$1$ currents in (\ref{b1b2b3nonlinear}) satisfy the following 
regular conditions, as in (\ref{regulara2a3}),  as follows:
$U(z) \, \hat{B}_i(w)   =   +\cdots$ and
$F_{a}(z) \, \hat{B}_i(w)   =   +\cdots$.}.

\subsection{ Construction of four spin-$\frac{3}{2}$ currents} 
%%%%%%%%%%%%%%%%%%%%%%%%%%%%%%%%%%%%%%%%%%

One can construct four spin-$\frac{3}{2}$ currents  \cite{GS,cqg1989,npb1989}
\bea
\hat{G}_{a}(z) &= &
G_{a}(z)+\frac{2}{(5+k)}\, U F_{a}(z) 
-\frac{2}{3(5+k)^2} \, \ep_{abcd}  F^{b} F^{c} F^{d}(z)
\nonu \\
& + & \frac{4}{(5+k)} \, F^{b} \, (\alpha_{ba}^{+i} \, \hat{A}_i
- \alpha_{ba}^{-i} \, \hat{B}_i ), \qquad i=1,2,3, \quad a,b, \cdots 
=11,12,21,22.
\label{fourg}
\eea
The expressions in $(3.21)$ of \cite{GG1305} correspond to 
the extra fields in (\ref{fourg}).
The relative coefficient functions in(\ref{fourg})
are determined completely by requiring that they should commute with 
$U(1)$ current and four fermions of spin-$\frac{1}{2}$\footnote{
In other words,
$
U(z) \, \hat{G}_a(w)   =   +\cdots$ and 
$F_{a}(z) \, \hat{G}_b(w)   =   +\cdots$.}.

By substituting the expressions (\ref{f11}), (\ref{f12}), (\ref{f21}), 
(\ref{f22}), (\ref{uu}), (\ref{g11}), (\ref{g12}), (\ref{g21}), (\ref{g22}),
(\ref{nona1a2a3}) and (\ref{nonb1b2b3}) into the equations (\ref{fourg}), 
the following results can be obtained
\bea
\hat{G}_{11}(z) & = &  -\frac{i \sqrt{2}}{(5+k)} \,
 \sum_{(m,a)=(1,10)}^{(3,12)} \left( K^m \overline{D} J^{a}-
\overline{D} K^m J^{a}
 \right)(z)
\nonu \\
& + &
\mbox{other cubic and linear terms},
\label{nong11} \\
\hat{G}_{22}(z) & = &  -\frac{i \sqrt{2}}{(5+k)} \,
\sum_{(\bar{m},\bar{a})=(\bar{1},\bar{10})}^{(\bar{3},\bar{12})}
\left( K^{\bar{m}} D J^{\bar{a}}-
D K^{\bar{m}} J^{\bar{a}} \right)(z)
\nonu \\
& + &
\mbox{other cubic and linear terms},
\label{nong22}
\\
\hat{G}_{12}(z) & = &
-\frac{\sqrt{2}}{(5+k)} \,  
\sum_{(a,\bar{a})=(10,\bar{10})}^{(12,\bar{12})}
  \overline{D} J^{a} J^{\bar{a}}(z)
\nonu \\
& + &
\mbox{other cubic, quadratic and linear terms},
\label{nong12} 
\\
\hat{G}_{21}(z) & = &
\frac{\sqrt{2}}{(5+k)} \,  
\sum_{(a,\bar{a})=(10,\bar{10})}^{(12,\bar{12})}
J^{a} D J^{\bar{a}} (z)
\nonu \\
& + &
\mbox{other cubic and quadratic  terms}.
\label{nong21}
\eea
The quadratic parts having the indices $7, 8$, or $9$ in $G_{11}(z)$
disappear in $\hat{G}_{11}(z)$ and similarly,  
those having the indices  $\bar{7}, \bar{8}$, or $\bar{9}$ in $G_{22}(z)$
disappear in $\hat{G}_{22}(z)$.
Furthermore, the 
quadratic parts having the indices $7, 8$, or $9$ (and $\bar{7}, 
\bar{8}$ or $\bar{9}$) in $G_{12}(z)$
disappear in $\hat{G}_{12}(z)$. The cubic terms containing the indices 
$9, \bar{9}$ in $G_{12}(z)$ disappear. 
Similarly,  the 
quadratic parts having the indices $7, 8$, or $9$ (and $\bar{7}, 
\bar{8}$ or $\bar{9}$) in $G_{21}(z)$
disappear in $\hat{G}_{21}(z)$. The cubic terms containing the indices 
$9, \bar{9}$ in $G_{21}(z)$ disappear in $\hat{G}_{21}(z)$ 
\footnote{For the $\hat{G}_{11}(z)$ and $\hat{G}_{22}(z)$, the terms having the 
indices $\bar{7}$ and $\bar{8}$ can be rewritten using $F_{12}(z)$ 
(\ref{f12}) and the terms having the indices $7$ and $8$ can be expressed 
using the $F_{21}(z)$ (\ref{f21}). The $F_{12}(z)$ and $F_{21}(z)$ are currents of
$SU(2)$ and $U(1)$ factors in the denominator of
the Wolf space coset (\ref{coset}). See also the footnote
\ref{sect4footnote}. However, for $\hat{G}_{12}(z)$ and $\hat{G}_{21}(z)$,
it is not obvious to see whether the terms including the indices $7, 8, 
\bar{7}$ and $\bar{8}$ can be rewritten in terms of the currents with 
indices of the 
subgroup of the Wolf space coset. 
Because the currents $K^7(z), K^{\bar{7}}(z), K^8(z)$ and 
$K^{\bar{8}}$(z) can be expressed in terms of two $SU(3)$ currents, one
$SU(2)$ current and one $U(1)$ current, one realizes that any 
spin-$\frac{1}{2}$ fermion terms
including the indices $7, 8, 
\bar{7}$ and $\bar{8}$ have the subgroup $SU(3), SU(2)$ or $U(1)$ 
indices in the Wolf space coset. 
See also the footnote \ref{su3currents}. All the spin-$\frac{3}{2}$
currents have both subgroup indices and coset indices of Wolf space.}. 

The full algebra can be obtained from the OPEs between $11$ currents.
For example, from the explicit expressions in (\ref{nong12}) and 
(\ref{nong21}), the following 
OPE can be described as
\bea
\hat{G}_{12}(z) \, \hat{G}_{21}(w) & = &
\frac{1}{(z-w)^3} \, \frac{2}{3} c_{\mbox{Wolf}} +
\frac{1}{(z-w)^2} \left[ 4 i \, \gamma_A \,  \hat{A}_3 + 4i \, \gamma_B \, 
\hat{B}_3 \right](w)\nonu \\
 & + & 
\frac{1}{(z-w)} \left[ 2 \hat{T} +\frac{1}{2}  \, 4i \, \gamma_A \, \pa 
\hat{A}_3
+\frac{1}{2}   \, 4i \, \gamma_B \, \pa \hat{B}_3  +
\right. \nonu \\
& + & \left. \frac{2}{(5+k)} \left( \sum_{i=1}^3 \hat{A}_i \, \hat{A}_i 
-2 \hat{A}_3 \, \hat{B}_3 + \sum_{i=1}^{3} \hat{B}_i \, \hat{B}_i   
\right)\right](w) \nonu \\
& + & \cdots,
\label{g1221}
\eea
where the central charge in the highest pole of this OPE 
(\ref{g1221}) is $c_{\mbox{Wolf}} = \frac{18k}{(5+k)}$ which coincides 
with (\ref{Wolfcentral}) for $N=3$. 
One can easily check that the central terms from the extra three OPEs
coming from the left hand side of 
(\ref{g1221}) except $G_{12}(z) \, G_{21}(w)$ can be written as 
$-\frac{6(4+k)}{(5+k)}$ which is exactly the same as the contributions 
from $SU(2)$ and $U(1)$ in (\ref{Wolfcentral}). 
The parameters 
$\gamma_A$ and $\gamma_B$ are given by 
$\gamma_A= \frac{\hat{k}^{-}}{\hat{k}^{+} +\hat{k}^{-} +2}=
\frac{3}{(5+k)}$ and $\gamma_B= \frac{\hat{k}^{+}}
{\hat{k}^{+} +\hat{k}^{-} +2}=
\frac{k}{(5+k)}$. The stress tensor $\hat{T}(z)$ is 
given by (\ref{stressnonlinear}).
Compared to the OPE $G_{12}(z) \, G_{21}(w)$  
described in previous section, the nonlinear terms occur in the 
first-order pole in (\ref{g1221}) \footnote{Using the following 
values $\alpha_{+,-}^{+3}=
-\frac{i}{4}$, $\alpha_{+,+K}^{+2}=
-\frac{1}{4}$ and $\alpha_{-,-K}^{+2}=
-\frac{1}{4}$ (therefore, we present all the $12$ independent 
quantities with previous footnotes), this particular OPE 
coincides with $(A.7)$ of \cite{npb1989}. Note that our 
$\hat{G}_{12}(z)$ and $\hat{G}_{21}(z)$ correspond to their 
$\sqrt{2} \tilde{G}_{+}(z)$ and 
$ \sqrt{2} \tilde{G}_{-}(z)$ respectively because their
$\delta_{ab}$ is normalized by $\frac{1}{2}$ (not by $1$).}.
This is due to the fact that the OPEs between the extra terms 
in (\ref{fourg}) contribute to these nonlinear terms in the right hand side of
(\ref{g1221}).

%%%%%%%%%%%%%%%%%%%%%%%%%%%%%%%%%%%%%
\subsection{$U(1)$ charge}
%%%%%%%%%%%%%%%%%%%%%%%%%%%%%%%%%%%%%

As in $U(1)$ current of (\ref{fourn2}), 
one can describe the $U(1)$ current here.
By realizing that the $U(1)$ current appears in the second-order pole 
of the OPE 
$\hat{G}_{21}(z) \, \hat{G}_{12}(w)$,
the $U(1)$ current is identified with 
$\left(- 2 i \gamma_A \hat{A}_3 -2 i \gamma_B \hat{B}_3\right)(z)$ 
from the OPE (\ref{g1221}).
Then one obtains the following first-order poles for 
the $11$ currents as follows: 
\bea
\left(- 2 i \gamma_A \hat{A}_3 -2 i \gamma_B \hat{B}_3\right)(z) \,
\hat{A}_{\pm}(w)|_{\frac{1}{(z-w)}} & = & 
\frac{1}{(z-w)} 
\left[ \mp \frac{6}{(5+k)}\right]
\hat{A}_{\pm}(w),
\nonu \\
\left(- 2 i \gamma_A \hat{A}_3 -2 i \gamma_B \hat{B}_3\right)(z) \,
\hat{B}_{\pm}(w)|_{\frac{1}{(z-w)}} & = & 
\frac{1}{(z-w)} 
\left[ \mp \frac{2k}{(5+k)}\right]
\hat{B}_{\pm}(w),
\nonu \\
\left(- 2 i \gamma_A \hat{A}_3 -2 i \gamma_B \hat{B}_3\right)(z) \,
\hat{A}_3(w)|_{\frac{1}{(z-w)}} & = & 
0,
\nonu \\
\left(- 2 i \gamma_A \hat{A}_3 -2 i \gamma_B \hat{B}_3\right)(z) \,
\hat{B}_3(w)|_{\frac{1}{(z-w)}} & = & 
0,
\nonu \\
\left(- 2 i \gamma_A \hat{A}_3 -2 i \gamma_B \hat{B}_3\right)(z) \,
\left(
\begin{array}{c}
\hat{G}_{11} \\
\hat{G}_{22} \\
\end{array} \right) (w)|_{\frac{1}{(z-w)}} & = & 
\frac{1}{(z-w)} 
\left[ \pm \frac{(-3+k)}{(5+k)}\right]
\left(
\begin{array}{c}
\hat{G}_{11} \\
\hat{G}_{22} \\
\end{array} \right) (w),
\nonu \\
\left(- 2 i \gamma_A \hat{A}_3 -2 i \gamma_B \hat{B}_3\right)(z) \,
\left(
\begin{array}{c}
\hat{G}_{12} \\
\hat{G}_{21} \\
\end{array} \right)  (w)|_{\frac{1}{(z-w)}} & = & 
\frac{1}{(z-w)} 
\left[ \mp \frac{(3+k)}{(5+k)}\right]
\left(
\begin{array}{c}
\hat{G}_{12} \\
\hat{G}_{21} \\
\end{array} \right)(w),
\nonu \\
\left(- 2 i \gamma_A \hat{A}_3 -2 i \gamma_B \hat{B}_3\right)(z) \,
\hat{T}(w)|_{\frac{1}{(z-w)}} & = & 
0. 
\label{u1first}
\eea
For the nonzero $U(1)$ charges, we present the correspoding two 
expressions together in order to emphasize the fact that they have opposite 
$U(1)$ charges \footnote{If one takes the notations for the spin-$\frac{3}{2}$
currents in \cite{GG1305} using $(++, +-, -+,--)$, then 
one can express (\ref{u1first}) more concisely rather than column notation.}. 
We present the $U(1)$ charges for those currents in the Table $1$. 
This definite $U(1)$ charges will play an important role in 
any OPEs in this paper because the $U(1)$ charge conservation 
holds for any OPEs. For example, in the OPE between the 
spin-$\frac{3}{2}$ current and the spin-$3$ current (the most 
complicated OPE in this paper), the first-order
pole contains the spin-$\frac{7}{2}$ field. Without the $U(1)$ charge
conservation, in general, the possible spin-$\frac{7}{2}$ field
has too many terms. However,  the $U(1)$ charge assignment will 
choose the right candidate for the above spin-$\frac{7}{2}$ field
by removing the unwanted terms which do not have the correct 
$U(1)$ charge \footnote{One finds the $U(1)$ charges for the WZW affine 
currents. The fields having $U(1)$ charge $\frac{k}{(5+k)}$ are given by
$K^m(z)$ and $J^a(z)$ where $m=1, \cdots, 3$ and $a=10, \cdots, 12$.
The fields with  $U(1)$ charge $-\frac{k}{(5+k)}$ are their conjugated ones
$K^{\bar{m}}(z)$ and $J^{\bar{a}}(z)$ where $\bar{m}=\bar{1}, \cdots, \bar{3}$ 
and $\bar{a}=\bar{10}, \cdots, \bar{12}$. The fields with vanishing 
$U(1)$ charge are given by $K^m(z)$, $\overline{D} K^m(z)$ 
and $J^9(z)$ 
where $m=4, \cdots, 8$ (and their conjugated ones $K^{\bar{m}}(z)$,
$D K^{\bar{m}}(z)$ 
and $J^{\bar{9}}(z)$ where $\bar{m}=\bar{4}, \cdots, \bar{8}$. 
The $U(1)$ charges of the remaining WZW affine currents are undecided.}.     

%%%%%%%%%%%%%%%%%%%%%%%%%%%%%%%%%%%%%%%%%%%%%%%%%%%%%%%%%%%%%%%%%%%
\begin{table}[ht]
\centering % used for centering table
\begin{tabular}{|c||c| } % centered columns (4 columns)
\hline %inserts double horizontal lines
$U(1)$ charge & $11$ currents of large ${\cal N}=4$ nonlinear 
superconformal algebra  \\ [0.5ex] % inserts table
%heading
\hline \hline % inserts single horizontal line
$\frac{2k}{(5+k)}$  & $\hat{B}_{-} (\equiv \hat{B}_1 -i \hat{B}_2) $ 
\\ % inserting body of the table
\hline
$\frac{(3+k)}{(5+k)}$ &  $\hat{G}_{21}$ \\
\hline
$\frac{(-3+k)}{(5+k)}$ & $\hat{G}_{11}$  \\
\hline
$\frac{6}{(5+k)} $ & $\hat{A}_{-} (\equiv \hat{A}_1 -i \hat{A}_2) $  \\
\hline
$0$ & $\hat{A}_3, \quad \hat{B}_3, \quad \hat{T}$  \\ 
\hline
$-\frac{6}{(5+k)}$ & $ \hat{A}_{+} (\equiv \hat{A}_1 + i \hat{A}_2) $  \\ 
\hline
$-\frac{(-3+k)}{(5+k)}$ & $\hat{G}_{22}$  \\ 
\hline
$-\frac{(3+k)}{(5+k)} $ & $\hat{G}_{12}$  \\
\hline
$-\frac{2k}{(5+k)}$ & $\hat{B}_{+} (\equiv \hat{B}_1 + i \hat{B}_2 )$  \\ 
[1ex] % [1ex] adds vertical space
\hline %inserts single line
\end{tabular}
%\label{tableone} % is used to refer this table in the text
\caption{The $U(1)$ charges for the $11$ currents from (\ref{u1first}). 
The $U(1)$ charge 
conservation can be seen from the equations in Appendix (\ref{ggopenonlinear}), 
(\ref{apmfourg}), 
(\ref{bpmfourg}) and (\ref{apmbpm}). The first four currents with positive 
($ k > 3 $) 
$U(1)$ charges have their conjugated  currents with each 
opposite (negative) $U(1)$ charge.  
  } % title of Table
\end{table}
%%%%%%%%%%%%%%%%%%%%%%%%%%%%%%%%%%%%%%%%%%%%%%%%%%%%%%%%%%%%%%%%%%%%%%

Therefore, the large ${\cal N}=4$ nonlinear algebra 
is generated by 
(\ref{stressnonlinear}), (\ref{spinonenonlinear}), 
(\ref{otherspinonenonlinear}), (\ref{nonb1b2b3}) and (\ref{fourg})
and some of the algebra are given in (\ref{opeA}), (\ref{opeB}) and 
(\ref{g1221}).
The complete algebra is summarized in Appendix $A$ 
\footnote{In particular, when $\hat{k}^{+} = \hat{k}^{-}$ (or $k=3$),
the ${\cal N}=4$ nonlinear superconformal algebra becomes the 
$SO({\cal N}=4)$ Knizhnik-Bershadsky algebra \cite{Knizhnik,Bershadsky}
with central charges $c=9$ and $c_{\mbox{Wolf}}=\frac{27}{4}$ along the line 
of \cite{ST,GK}. Therefore, the higher spin currents in next section will
lead to an extension of $SO(4)$ Knizhnik-Bershadsky algebra at $c=9$.
Furthermore, according to \cite{Schoutens1987,Schoutensphd,GS}, 
the ${\cal N}=3$
linear superconformal algebra can be reduced to the $SO({\cal N}=3)$
Knizhnik-Bershadsky (nonlinear) algebra by decoupling the fermion of 
spin-$\frac{1}{2}$. The central charge is reduced with $\frac{1}{2}$.
The exact field redefinitions in order to see this in the present context 
should be done. See also \cite{BO} in the ${\cal N}=2$ 
superspace approach.}.

%%%%%%%%%%%%%%%%%%%%%%%%%%%%%%%%%%%%%%%%%%%%%%%%%%%%%%%%%%%%%%%%%%%%%
%%%%%%%%%%%%%%%%%%%%%%%%%%%%%%%%%%%%%%%%%%%%%%%%%%%%%%%%%%%%%%%%%%%%%%
\section{Higher spin currents  in the Wolf space coset 
minimal model  }
%4%%%%%%%%%%%%%%%%%%%%%%%%%%%%%%%%%%%%%%%%%%%%%%%%%%%%%%%%%%%%%%%%%%%%%
%%%%%%%%%%%%%%%%%%%%%%%%%%%%%%%%%%%%%%%%%%%%%%%%%%%%%%%%%%%%%%%%%%%%%

According to the observation of \cite{GG1305}, the $16$ lowest extra 
currents, appearing in the equation $(2.33)$ of \cite{GG1305},  
consist of single spin-$1$, four spin-$\frac{3}{2}$, six 
spin-$2$, four spin-$\frac{5}{2}$, and single spin-$3$ currents. 
Rearranging them in terms of ${\cal N}=2$ multiplets \cite{Romans},
one has four ${\cal N}=2$ supercurrents    
\bea
\left(1, \frac{3}{2}, \frac{3}{2}, 2 \right)
& : & (T^{(1)}, T_{+}^{(\frac{3}{2})}, T_{-}^{(\frac{3}{2})}, T^{(2)}), 
\nonu \\
 \left(\frac{3}{2}, 2, 2, \frac{5}{2} \right) & : & 
(U^{(\frac{3}{2})}, U_{+}^{(2)}, U_{-}^{(2)}, U^{(\frac{5}{2})} ), \nonu \\
\left(\frac{3}{2}, 2, 2, \frac{5}{2} \right) & : & 
(V^{(\frac{3}{2})}, V^{(2)}_{+}, V^{(2)}_{-}, V^{(\frac{5}{2})}),  \nonu \\
\left(2, \frac{5}{2}, \frac{5}{2}, 3 \right) & : &
 (W^{(2)}, W_{+}^{(\frac{5}{2})}, W_{-}^{(\frac{5}{2})}, W^{(3)}).
\label{new16comp}
\eea
This section will consider the particular supersymmetric Wolf space
coset minimal model
(\ref{coset}) and the higher spin currents in (\ref{new16comp}) will be 
determined \footnote{
\label{sect4footnote}
Let us emphasize that in the Wolf space coset model 
(\ref{coset}) written in terms of ${\cal N}=2$ superspace (effectively 
${\cal N}=1$ superspace due to the constraints), the higher spin currents
should commute with the ${\cal N}=2$  
WZW affine currents (of spin-$\frac{1}{2}$ fields and 
its superpartner spin-$1$ fields) living on the denominator 
$SU(3)$ (the explicit $8+8$ fields will be described in the footnote 
\ref{su3currents}), 
commute with those of three spin-$\frac{1}{2}$ currents living on
the denominator $SU(2)$ (the fields $F_{11}(z)$, $F_{22}(z)$ and 
$(F_{21}+F_{12})(z)$), and commute with 
the spin-$1$ and its superpartner spin-$\frac{1}{2}$ currents living on
the denominator $U(1)$ (i.e. the field $U(z)$ with (\ref{Usuperpartner}) 
and the field $(F_{21}-F_{12})(z)$). Once again,
the OPEs between the higher spin currents and the three $SU(2)$ currents,
[$\overline{D} F_{11}(z)$, $ D F_{22}(z)$ and 
$(\overline{D}F_{21}+ D F_{12})(z)$], which are the superpartner of above 
three spin-$\frac{1}{2}$ currents DO have singular terms. Therefore,
the $11+16=27$ 
(higher spin) currents commute with $16+3+2=21$ Wolf space 
denominator currents among 
$48$ currents in the group $SU(5)$.    }.

%%%%%%%%%%%%%%%%%%%%%%%%%
\subsection{Construction of higher spin currents of spins 
$\left(1, \frac{3}{2}, \frac{3}{2}, 2 \right)$} 
%%%%%%%%%%%%%%%%%%%%%%%%%

How does one determine the spin-$1$ current in (\ref{new16comp})?
The basic $24$ ${\cal N}=2$ WZW affine 
currents of spins $\frac{1}{2}$ are given in terms of 
$K^m(z), K^{\bar{m}}(z), J^a(z)$ and $J^{\bar{a}}(z)$ in (\ref{KKJJ}).
Then the general spin-$1$ current can be obtained from these by considering the 
quadratic expressions. 
Furthermore, the basic spin-$1$ ${\cal N}=2$ WZW affine 
currents are given by 
$\overline{D} K^m(z), D K^{\bar{m}}(z), \overline{D} J^a(z)$ and 
$ D J^{\bar{a}}(z)$.
The linear combinations of these currents should be added into the most general
spin-$1$ current $T^{(1)}(z)$ we are looking for.
Then one can write down, with $300$  coefficient functions which depend on 
the level $k$,  as
\bea
T^{(1)}(z) & = & \sum_{m,n=1}^8 c_{m,n} K^m K^n(z) +
  \sum_{m=1}^8 \sum_{\bar{n}=\bar{1}}^{\bar{8}} c_{m,\bar{n}} K^m K^{\bar{n}}(z)
+ \sum_{m=1}^8 \sum_{a=9}^{12} c_{m,a} K^m J^a(z) 
\nonu \\
& + &
  \sum_{m=1}^8 \sum_{\bar{a}=\bar{9}}^{\bar{12}} c_{m,\bar{a}} K^m J^{\bar{a}}(z)
+  \sum_{\bar{m}, \bar{n}=\bar{1}}^{\bar{8}} c_{\bar{m},\bar{n}} K^{\bar{m}} K^{\bar{n}}(z)
+\sum_{\bar{m}=\bar{1}}^{\bar{8}} \sum_{a=9}^{12} c_{\bar{m},a} K^{\bar{m}} J^a(z) 
\nonu \\
& + &  \sum_{\bar{m}=\bar{1}}^{\bar{8}} \sum_{\bar{a}=\bar{9}}^{\bar{12}} 
c_{\bar{m},\bar{a}} K^{\bar{m}} J^{\bar{a}}(z) +
 \sum_{a, b=9}^{12} c_{a,b} J^a J^b(z)
+ \sum_{a=9}^{12} \sum_{\bar{b}=\bar{9}}^{\bar{12}} c_{a,\bar{b}} J^a J^{\bar{b}}(z)
 + 
\sum_{\bar{a}=\bar{9}}^{\bar{12}} c_{\bar{a}} D J^{\bar{a}}(z)
\nonu
\\
& + &  \sum_{\bar{a}, \bar{b}=\bar{9}}^{\bar{12}} 
c_{\bar{a},\bar{b}} J^{\bar{a}} J^{\bar{b}}(z)
+ \sum_{m=1}^8 c_m \overline{D} K^m(z) +\sum_{\bar{m}=\bar{1}}^{\bar{8}} c_{\bar{m}}
D K^{\bar{m}}(z) +  \sum_{a=9}^{12} c_a \overline{D} J^a(z).
\label{generalt1} 
\eea

We would like to determine the coefficient functions appearing in 
(\ref{generalt1}) 
explicitly.
Since the regularity conditions between the spin-$1$ current $U(z)$ and the 
spin-$\frac{1}{2}$ currents, $F_{11}(z)$, $F_{12}(z)$, $F_{21}(z)$ and $F_{22}(z)$
that live in section $2$ and the spin-$1$ current 
$T^{(1)}(w)$
are preserved in this extended nonlinear 
algebra, 
the following relations, together with (\ref{uu}), (\ref{f11}), (\ref{f12}),
(\ref{f21}), (\ref{f22}) and (\ref{generalt1}), should satisfy
\bea
U(z) \, T^{(1)}(w) & = & +\cdots, 
\qquad
F_{a}(z) \, T^{(1)}(w)   =  +\cdots, \qquad a=11,12,21,22.
\label{uft1}
\eea 
Then the remaining undetermined coefficient functions 
can be fixed by the following primary field condition under the stress 
tensor (\ref{stressnonlinear})
\bea
\hat{T}(z) \, T^{(1)}(w) & = & \frac{1}{(z-w)^2} \, T^{(1)}(w) +
\frac{1}{(z-w)} \, \pa T^{(1)}(w) + \cdots.
\label{t1primary}
\eea
This OPE (\ref{t1primary})
implies the following result 
\bea
T^{(1)}(z) \, \hat{T}(w)|_{\frac{1}{(z-w)}} =0.
\label{t1tpole1}
\eea
In other words, the commutator  
$[ T_0^{(1)}, \hat{T}]$ between the zero mode of 
$T^{(1)}$ and $\hat{T}$ vanishes.
Then we are left with seven unknown coefficient functions.

All the first-order singular terms between 
the above spin-$1$ current and other six spin-$1$ currents 
can be obtained from the defining equations 
(\ref{nona1a2a3}), (\ref{b1b2b3nonlinear}) and (\ref{generalt1}) 
and by requiring that the commutators between the zero mode $T_0^{(1)}$ and six
spin-$1$ currents should vanish \cite{GG1305}
\bea
T^{(1)}(z) \, \hat{A}_i(w)|_{\frac{1}{(z-w)}} & = & 0, \qquad
T^{(1)}(z) \, \hat{B}_i(w)|_{\frac{1}{(z-w)}} =0, \qquad
\label{t1sixpole1}
\eea
all the remaining coefficient functions are completely determined except an 
overall constant.

It turns out that the lowest higher spin-$1$ 
current (\ref{generalt1}) can be obtained as follows \footnote{The first four 
terms of (\ref{t1}) 
can be written as $ -\frac{2 i \sqrt{\frac{5}{3}}}{(5+k)} U(z) + \left(-
\frac{(15 i + 5\sqrt{3} -i \sqrt{5} + \sqrt{15})}{6(5+k)} \overline{D} K^{7}+
\frac{2\sqrt{10}}{3(5+k)} \overline{D} K^8 \right)+ \left( -
\frac{(-15 i + 5\sqrt{3} +i \sqrt{5} + \sqrt{15})}{6(5+k)} D K^{\bar{7}}+
\frac{2\sqrt{10}}{3(5+k)} D K^{\bar{8}} \right)$ where the last two fields 
correspond to the Wolf space 
denominator $SU(3)$ current. This spin-$1$ current has 
both subgroup indices and coset indices of Wolf space.}:
\bea
T^{(1)}(z) & = &
-\frac{5 (3 i+\sqrt{3})}{6 (5+k)}
\overline{D} K^7(z)
 +  
\frac{(-5 i \sqrt{3}+9 \sqrt{5})}{6 \sqrt{2} (5+k)}
\overline{D} K^8(z)
 -  
\frac{5 (-3 i+\sqrt{3})}{6 (5+k)}
D K^{\bar{7}}(z)
\label{t1}
\\
& + &  
\frac{(5 i \sqrt{3}+9 \sqrt{5})}{6 \sqrt{2} (5+k)}
D K^{\bar{8}}(z)
 -  
\frac{1}{(5+k)} \left(
 \sum_{(a,\bar{a})=(10,\bar{10})}^{(12,\bar{12})} J^{a} J^{\bar{a}} 
 - \sum_{(m,\bar{m})=(1,\bar{1})}^{(3,\bar{3})} K^m K^{\bar{m}} \right)(z).
\nonu
\eea
The field contents of (\ref{t1}) look similar to those of $\hat{A}_3(z)$ in 
(\ref{nona1a2a3}) except that the former does not have the term of 
$J^9 J^{\bar{9}}(z)$.
One can easily check that this spin-$1$ current is new primary current in the 
sense that this cannot be written in terms of given six spin-$1$ currents:
$\hat{A}_i$ and $\hat{B}_i$.
Furthermore, the OPE between the spin-$1$ currents (\ref{t1}) does not
have a first-order pole \footnote{ 
\label{foot}
One obtains the OPE $T^{(1)}(z) \, T^{(1)}(w) =\frac{1}{(z-w)^2}
\left[\frac{6k}{(5+k)} \right]
+\cdots$.}: 
\bea
T^{(1)}(z) \, T^{(1)}(w)|_{\frac{1}{(z-w)}}  & = & 0. 
%\qquad
%T^{(1)}(z) \, T^{(2)}(w)|_{\frac{1}{(z-w)}}   =  0, \qquad
%T^{(1)}(z) \, U_{+}^{(2)}(w)|_{\frac{1}{(z-w)}} =0, 
%\nonu \\
%T^{(1)}(z) \, U_{-}^{(2)}(w)|_{\frac{1}{(z-w)}}  & = & 0, 
%\qquad
%T^{(1)}(z)\, V_{+}^{(2)}(w)|_{\frac{1}{(z-w)}}  =  0, \qquad
%T^{(1)}(z) \, V_{-}^{(2)}(w) |_{\frac{1}{(z-w)}} =0, 
%\nonu \\
%T^{(1)}(z) \, W^{(2)}(w)|_{\frac{1}{(z-w)}}  & = & 0, 
%\qquad
%T^{(1)}(z) \, W^{(3)}(w)|_{\frac{1}{(z-w)}}  =  0.
\label{t1t1pole1}
\eea
Therefore, this spin-$1$ current will play an important role in the 
construction of higher spin currents because this will generate, in principle, 
all the 
higher spin currents with the help of the currents in the  
${\cal N}=4$ nonlinear algebra described in previous section.

Consider the next OPE between the spin-$\frac{3}{2}$ current 
(\ref{nong21}) and
the spin-$1$ current (\ref{t1}) 
to determine the other current in 
the higher spin currents of spins $(1, \frac{3}{2}, \frac{3}{2}, 2)$.
Usually, the ${\cal N}=2$ description for this particular OPE
produces the second component of ${\cal N}=2$ superprimary current.
The result  can be expressed as
\bea
\hat{G}_{21}(z) \, T^{(1)}(w)  & = &
\frac{1}{(z-w)} \left[ \hat{G}_{21} +2 
T_{+}^{(\frac{3}{2})} \right](w) +
\cdots,
\label{g21t1}
\eea
where the new spin-$\frac{3}{2}$ current (in the sense that this 
cannot be written in terms of combination of
known currents) can be obtained
\bea
T_{+}^{(\frac{3}{2})}(w) & = &
-\frac{\sqrt{2}}{(5+k)} \sum_{(a,\bar{a})=(10,\bar{10})}^{(12,\bar{12})} 
J^{a} D J^{\bar{a}}(w) + 
\mbox{other cubic terms}. 
\label{t+3half}
\eea
The field contents of this current can be seen from the current 
$\hat{G}_{21}(z)$ (\ref{nong21}) where there are quadratic and cubic terms with
the two indices among $1$, $2$, $3$, $\bar{1}$, $\bar{2}$ and $\bar{3}$.
From the right hand side of (\ref{g21t1}), the $U(1)$ charges of 
two currents appearing in the first-order pole are the same.

There are no singular terms in the  
OPE between this spin-$\frac{3}{2}$ current (\ref{t+3half})
and the spin-$1$ current (\ref{uu}). The OPEs 
between the spin-$\frac{3}{2}$ current (\ref{t+3half}) 
and the spin-$\frac{1}{2}$ currents (\ref{f11}), (\ref{f12}), (\ref{f21}), 
and (\ref{f22}) (also the spin-$1$ current (\ref{uu})) 
do not contain any singular terms as in (\ref{uft1}):
\bea
U(z) \, T_{+}^{(\frac{3}{2})}(w) & = & +\cdots, 
\qquad
F_{a}(z) \, T_{+}^{(\frac{3}{2})}(w)   =  +\cdots, \qquad a=11,12,21,22.
\label{uft+}
\eea 
Note that the normalization of spin-$1$ current $T^{(1)}(z)$ in (\ref{t1})
(i.e. the footnote \ref{foot})
was fixed by the following OPE \footnote{We will not consider the OPEs between 
the higher spin currents themselves in this paper. Of course, they should be 
calculated to complete the full structure of the extended large ${\cal N}=4$
nonlinear algebra and will appear near future \cite{Ahn2014}. }
\bea
T^{(1)}(z) \, T_{+}^{(\frac{3}{2})}(w)  & = &
\frac{1}{(z-w)} \, 
T_{+}^{(\frac{3}{2})}(w) +
\cdots.
\label{t1t+3half}
\eea
That is, the right hand side of (\ref{t1t+3half}) can be described as
$(5+k) A(N,k)T_{+}^{(\frac{3}{2})}(w)$ with normalization $A(N,k)$. By taking 
the $U(1)$ charge under the 
$T^{(1)}$ spin-$1$ current 
to be $1$, the normalization constant is given by $A(N,k) =
\frac{1}{(5+k)}$ which appears in (\ref{t1}). 
The OPE (\ref{t1t+3half}) implies that
the $U(1)$ charge of $T^{(1)}(z)$ is equal to zero.

The next OPE between the spin-$\frac{3}{2}$ current 
(\ref{nong12}) and
the spin-$1$ current (\ref{t1}) 
can be calculated to obtain the other remaining spin-$\frac{3}{2}$ current in 
the higher spin currents of spins $(1, \frac{3}{2}, \frac{3}{2}, 2)$.
As observed before, the reason why one considers this particular OPE is 
the fact that the ${\cal N}=2$ description of this OPE 
provides the third component of 
${\cal N}=2$ superprimary current.
The result  can be expressed as
follows:
\bea
\hat{G}_{12}(z) \, T^{(1)}(w)  & = &
\frac{1}{(z-w)} \left[ -\hat{G}_{12} +2 
T_{-}^{(\frac{3}{2})} \right](w) +
\cdots,
\label{g12t1}
\eea
where the new spin-$\frac{3}{2}$ current can be written as
\bea
T_{-}^{(\frac{3}{2})}(w) & = &
-\frac{\sqrt{2}}{(5+k)} 
 \sum_{(a,\bar{a})=(10,\bar{10})}^{(12,\bar{12})} 
\overline{D} J^{a}  J^{\bar{a}}(w)+ 
\mbox{other cubic and linear terms}. 
\label{t-3half}
\eea
The field contents of this current can be seen from the current 
$\hat{G}_{12}(z)$ (\ref{nong12}) where there are quadratic and cubic terms with
the two indices among $1$, $2$, $3$, $\bar{1}$, $\bar{2}$ and $\bar{3}$.
The $U(1)$ charge of two currents in the first-order pole of (\ref{g12t1}) 
are the same. The OPE (\ref{g12t1}) is conjugated to the OPE (\ref{g21t1}).

No singular terms in the  
OPE between this spin-$\frac{3}{2}$ current (\ref{t-3half})
and the spin-$1$ current (\ref{uu}) exist. The OPEs 
between the spin-$\frac{3}{2}$ current (\ref{t-3half}) 
and the spin-$\frac{1}{2}$ currents (\ref{f11}), (\ref{f12}), (\ref{f21}), 
and (\ref{f22}) do not contain any singular terms as in (\ref{uft1}) and 
(\ref{uft+}):
\bea
U(z) \, T_{-}^{(\frac{3}{2})}(w) & = & +\cdots, 
\qquad
F_{a}(z) \, T_{-}^{(\frac{3}{2})}(w)   =  +\cdots, \qquad a=11,12,21,22.
\label{uft-}
\eea 

Consider the spin-$\frac{3}{2}$ current (\ref{nong21}) and 
the spin-$\frac{3}{2}$ current
(\ref{t-3half}) to determine the last component spin-$2$ current. 
The result can be expressed as
\bea
\hat{G}_{21}(z) \, T_{-}^{(\frac{3}{2})}(w) & = & 
\frac{1}{(z-w)^3} \, \frac{6k}{(5+k)}
+ \frac{1}{(z-w)^2} \left[ T^{(1)} +\frac{2i}{(5+k)} \left( -3\hat{A}_3 -
k \hat{B}_3 \right) \right](w)  \nonu \\
& + & \frac{1}{(z-w)}  \left[ 
 \frac{6k}{(3+7k)}\hat{T} +T^{(2)} +\frac{1}{2}\pa
 \left( T^{(1)}  + \frac{2i}{(5+k)} \left( -3  \hat{A}_3 -
k  \hat{B}_3 \right) \right) \right](w) \nonu \\
& + & \cdots,
\label{g21t-}
\eea
where the last component of the higher spin current of spins $(1, 
\frac{3}{2}, \frac{3}{2}, 2)$ can be described as
\bea
 T^{(2)}(w) & = & 
-\frac{2}{(5+k)^3} \sum_{(a,\bar{a})=(10,\bar{10})}^{(12,\bar{12})} 
J^9 J^{a} J^{\bar{9}} J^{\bar{a}} (w)
\nonu \\
&+ & \mbox{other quartic, cubic, quadratic and linear terms}.
\label{simplet2}
\eea
The complete expression  is not presented here.
In the first-order pole of (\ref{g21t-}), the new spin-$2$ current 
arises and another primary spin-$2$ current (\ref{stressnonlinear}).
The relative coefficient $\frac{1}{2}$ in the descendant field of the spin-$1$
current in the second-order pole can be obtained from the formula in the 
OPE of two (quasi) primary fields. 
Of course, this spin-$2$ current cannot be written in terms of other known 
currents. The $U(1)$ charges for the currents in the first-order pole of
(\ref{g21t-}) should be the same. 

No singular terms in the  
following OPEs exist as in (\ref{uft1}),
(\ref{uft+}) and (\ref{uft-}):
\bea
U(z) \, T^{(2)}(w) & = & +\cdots, 
\qquad
F_{a}(z) \, T^{(2)}(w)   =  +\cdots, \qquad a=11,12,21,22.
\label{uft2}
\eea 
As in (\ref{t1tpole1}), (\ref{t1sixpole1}) and (\ref{t1t1pole1}),
 the commutator between the zero mode $T_0^{(1)}$ and the 
spin-$2$ current $T^{(2)}$ vanishes \footnote{More explicitly, one has
the OPE 
$T^{(1)}(z) \, T^{(2)}(w)=\frac{1}{(z-w)^2} \left[ -\frac{6}{(5+k)} i \hat{A}_3
-\frac{2k}{(5+k)} i \hat{B}_3 + \frac{(3+k)}{(3+7k)} T^{(1)}
\right](w) +\cdots$.}:
\bea
T^{(1)}(z) \, T^{(2)}(w)|_{\frac{1}{(z-w)}}   =  0.
\label{t1t2pole1}
\eea

Therefore, the higher spin currents of spins 
$(1, \frac{3}{2}, \frac{3}{2}, 2)$
are determined completely: (\ref{t1}), (\ref{t+3half}), (\ref{t-3half}) 
and (\ref{simplet2}). 
Some of the OPEs between these higher spin currents and the currents from 
the large ${\cal N}=4$ nonlinear 
superconformal algebra are presented in (\ref{g21t1}), 
(\ref{g12t1}) and (\ref{g21t-}).
The remaining OPEs are given in Appendix $(B.1)$
and Appendix $(C.1)$.

%%%%%%%%%%%%%%%%%%%%%%%%%
\subsection{Construction of higher spin currents of spins 
$\left(\frac{3}{2}, 2, 2, \frac{5}{2} \right)$} 
%%%%%%%%%%%%%%%%%%%%%%%%%

What happens when the other spin-$\frac{3}{2}$ current $\hat{G}_{11}(z)$ 
or $\hat{G}_{22}(z)$ rather than 
$\hat{G}_{21}(z)$ or $\hat{G}_{12}(z)$ acts on the lowest 
component of previous multiplet $(1, \frac{3}{2}, \frac{3}{2}, 2)$?
Let us focus on  the OPE $\hat{G}_{11}(z) \, T^{(1)}(w)$.
From the explicit expressions (\ref{nong11}) and (\ref{t1}), 
the following OPE can be calculated easily
\bea
\hat{G}_{11}(z) \, T^{(1)}(w)  & = &
\frac{1}{(z-w)} \left[ \hat{G}_{11} +2 U^{(\frac{3}{2})} \right](w) +
\cdots,
\label{g11t1}
\eea
where the new spin-$\frac{3}{2}$ current which cannot be written in terms 
of other currents can be expressed as
\bea
U^{(\frac{3}{2})}(w) & = &
-\frac{i \sqrt{2} }{(5+k)^2}
\sum_{(a,\bar{a})=(10,\bar{10})}^{(12,\bar{12})} J^9 J^{a} J^{\bar{a}}(w)
\nonu \\
& + &  
\mbox{other cubic, quadratic and linear terms}. 
\label{u3half}
\eea
Some of the fields in (\ref{u3half}) can be seen from the $\hat{G}_{11}(w)$
but  the following three terms 
$K^2 K^{\bar{4}} J^{10}(w)$, $ K^3 K^{\bar{5}} J^{10}(w)$ and
$K^3 K^{\bar{6}} J^{11}(w)$ occur in (\ref{u3half}) newly. 
The $U(1)$ charges in the first-order pole of (\ref{g11t1})
should be the same.
As stated before, the following regularity conditions, as in 
(\ref{uft2}),  satisfy
\bea
U(z) \, U^{(\frac{3}{2})}(w) & = & +\cdots, 
\qquad
F_{a}(z) \, U^{(\frac{3}{2})}(w)   =  +\cdots, \qquad a=11,12,21,22. 
\label{ufu3half}
\eea 

Consider the next OPE between the spin-$\frac{3}{2}$ current (\ref{nong11}) 
and the 
spin-$\frac{3}{2}$ current (\ref{t+3half}). The result is as follows: 
\bea
\hat{G}_{11}(z) \, T_{+}^{(\frac{3}{2})}(w) & = & 
\frac{1}{(z-w)^2} \frac{2k}{(5+k)} \left[ i \hat{B}_{-} \right](w) \nonu 
\\
& + &
\frac{1}{(z-w)} \left[ - U_{+}^{(2)} -\frac{1}{(5+k)}  
4 \hat{A}_3  \hat{B}_{-}
+ \frac{1}{2}  \frac{2k}{(5+k)}   i  \pa \hat{B}_{-}
 \right](w) \nonu \\
& + & \cdots,
 \label{g11t+3half}
\eea
where the new spin-$2$ current is given by
\footnote{
From the OPE
\bea
\hat{G}_{21}(z) \, U^{(\frac{3}{2})}(w) & = &
\frac{1}{(z-w)^2} \frac{2k}{(5+k)} \left[ -i  \hat{B}_{-}  \right](w) +  
\frac{1}{(z-w)}  \left[ U_{+}^{(2)} 
   - \frac{1}{2}  \frac{2k}{(5+k)}  i \pa \hat{B}_{-}  \right](w) +\cdots,
\nonu 
\eea
it is natural to consider the extra spin-$2$ current in the first-order 
pole (i.e. the first-order pole subtracted by the descendant terms)
as a second component of higher spin currents of spins $(\frac{3}{2}, 2, 
2, \frac{5}{2})$.}
\bea
 U_{+}^{(2)}(w) & = & 
\frac{2 i}{(5+k)^3} \left(-K^1 J^9 J^{10} J^{\bar{9}}
+ K^1 J^{10} J^{11} J^{\bar{11}}
 +   K^1 J^{10} J^{12} J^{\bar{12}}\right)(w)
\nonu \\
& + &
\mbox{other quartic, cubic and quadratic  terms}. 
\label{u+2}
\eea
Note that the second term $\hat{A}_3 (\hat{B}_1-i \hat{B}_2)$ 
in the first order pole of (\ref{g11t+3half}) 
is primary field under the stress tensor (\ref{stressnonlinear}) 
because the current 
$\hat{A}_3(z)$ and the current 
$(\hat{B}_1-i \hat{B}_2)(z)$ are primary and commute 
with each other. Furthermore, each term of 
the spin-$2$ current (\ref{u+2}) contains 
$K^1(z)$, $K^2(z)$ or $K^3(z)$. There are no composite terms  
consisting of $J^a(z)$, $J^{\bar{a}}(z)$, 
$\overline{D} J^a(z)$ or $D J^{\bar{a}}(z)$.  

As in (\ref{ufu3half}),
the following OPEs satisfy the regular conditions
\bea
U(z) \, U_{+}^{(2)}(w) & = & +\cdots, 
\qquad
F_{a}(z) \, U_{+}^{(2)}(w)   =  +\cdots, \qquad a=11,12,21,22.
\label{ufu+2}
\eea 
Furthermore, from the equations (\ref{t1}) and (\ref{u+2}),
one checks, as in (\ref{t1t2pole1}),  
the following vanishing first-order pole
\bea
T^{(1)}(z) \, U_{+}^{(2)}(w)|_{\frac{1}{(z-w)}} =0.
\label{t1u+2poleone}
\eea

Now let us consider 
the spin-$\frac{3}{2}$ current (\ref{nong11}) acting on the spin-$\frac{3}{2}$
current (\ref{t-3half})
\bea
\hat{G}_{11}(z) \, T_{-}^{(\frac{3}{2})}(w) & = & 
\frac{1}{(z-w)^2} \frac{6}{(5+k)} \left[ i \hat{A}_{+} \right](w) \nonu \\
& + &
\frac{1}{(z-w)} \left[ - U_{-}^{(2)} + \frac{1}{2} 
 \frac{6}{(5+k)} 
i  \pa   \hat{A}_{+}
 \right](w) +\cdots,
\label{g11t-3half}
\eea
where the other spin-$2$ current which obtained by subtracting the 
descendant fields can be described as \footnote{
In this case, one has also 
\bea
\hat{G}_{12}(z) \, U^{(\frac{3}{2})}(w)  = 
\frac{1}{(z-w)^2} \frac{6}{(5+k)} \left[ i \hat{A}_{+} \right](w) +  
\frac{1}{(z-w)}  \left[ U_{-}^{(2)} 
   + \frac{1}{2}  
\frac{6}{(5+k)}  i \pa \hat{A}_{+} \right](w) +\cdots.
\nonu
\eea
After subtracting the descendant term in the first-order pole, the extra 
spin-$2$ current with plus sign is the third 
component of the higher spin currents of spins $(\frac{3}{2}, 2, 2, 
\frac{5}{2})$.
}
\bea
 U_{-}^{(2)}(w) & = & 
\frac{2 i}{(5+k)^2} \sum_{(a,\bar{a})=(10,\bar{10})}^{(12,\bar{12})} 
J^9 \overline{D} J^{a} J^{\bar{a}} (w)
\nonu \\
& + &
\mbox{other quartic, cubic, quadratic and linear terms}. 
\label{u-2}
\eea
Compared to the previous spin-$2$ current (\ref{u+2}), the first 
six terms (and the last term)
of (\ref{u-2}) contain only $J^a(z)$, $J^{\bar{a}}(z)$, $\overline{D}
J^a(z)$ or $D J^{\bar{a}}(z)$.
One realizes that the $U(1)$ charge of $U_{-}^{(2)}(z)$ is the same as the one 
of $(\hat{A}_1 + i \hat{A}_2)(z)$ from the OPE (\ref{g11t-3half}). 

As one expects in (\ref{ufu+2}), the following OPEs are satisfied  
\bea
U(z) \, U_{-}^{(2)}(w) & = & +\cdots, 
\qquad
F_{a}(z) \, U_{-}^{(2)}(w)   =  +\cdots, \qquad a=11,12,21,22. 
\label{ufu-2}
\eea 
One also checks that the first-order pole of the following OPE
vanishes as in (\ref{t1u+2poleone})
\bea
T^{(1)}(z) \, U_{-}^{(2)}(w)|_{\frac{1}{(z-w)}}  & = & 0.
\label{t1u-2poleone}
\eea

As in (\ref{g21t-}),
the last component spin-$\frac{5}{2}$ current can be calculated from the 
following OPE 
\bea
\hat{G}_{21}(z) \, U_{-}^{(2)}(w) & = &
\frac{1}{(z-w)^2} \left[ \frac{(3+2k)}{(5+k)} \hat{G}_{11}+
\frac{4(2+k)}{(5+k)} U^{(\frac{3}{2})} \right](w)
\label{g21u-2}
\\
& + & \frac{1}{(z-w)}  \left[ U^{(\frac{5}{2})}
+ \frac{1}{3} \pa \left( \frac{(3+2k)}{(5+k)}  \hat{G}_{11}+
\frac{4(2+k)}{(5+k)}  U^{(\frac{3}{2})} \right)
 \right](w)+ \cdots,
\nonu
\eea
where the new spin-$\frac{5}{2}$ current can be described as
\bea
 U^{(\frac{5}{2})}(w) & = & 
\frac{4 i \sqrt{2}}{(5+k)^4}  \left( J^9 J^{10} J^{11} J^{\bar{10}} J^{\bar{11}}
+J^9 J^{10} J^{12} J^{\bar{10}} J^{\bar{12}}
+ J^9 J^{11} J^{12} J^{\bar{11}} J^{\bar{12}} \right)(w)
\nonu \\
&+ & \mbox{other quintic, quartic, cubic, quadratic and linear terms}. 
\label{U5half}
\eea
Again, the $U(1)$ charge conservation leads to 
the fact that the three currents in (\ref{g21u-2}) in the first-order 
pole have the same 
$U(1)$ charge.
The complete expression for (\ref{U5half}) 
is not presented here.
Moreover, the following regularity conditions which are similar to 
(\ref{ufu-2}), 
satisfy
\bea
U(z) \, U^{(\frac{5}{2})}(w) & = & +\cdots, 
\qquad
F_{a}(z) \, U^{(\frac{5}{2})}(w)   =  +\cdots, \qquad a=11,12,21,22. 
\label{ufu5half}
\eea 

Therefore, the higher spin currents of spins 
$(\frac{3}{2}, 2, 2, \frac{5}{2})$
are determined completely: (\ref{u3half}), (\ref{u+2}), (\ref{u-2}) 
and (\ref{U5half}). 
Some of the OPEs between these higher spin currents and the currents from 
the large ${\cal N}=4$ nonlinear
superconformal algebra are presented in (\ref{g11t1}), 
(\ref{g11t+3half}), (\ref{g11t-3half}) and (\ref{g21u-2}).
The remaining OPEs are given in Appendix $(B.2)$
and Appendix $(C.2)$.
%%%%%%%%%%%%%%%%%%%%%%%%%
\subsection{Construction of higher spin currents of spins 
$\left(\frac{3}{2}, 2, 2, \frac{5}{2} \right)$} 
%%%%%%%%%%%%%%%%%%%%%%%%%

In this subsection, 
we consider the second case 
when the other spin-$\frac{3}{2}$ current $\hat{G}_{22}(z)$  
acts on the lowest 
component of previous multiplet $(1, \frac{3}{2}, \frac{3}{2}, 2)$:
\bea
\hat{G}_{22}(z) \, T^{(1)}(w)  & = &
\frac{1}{(z-w)} \left[ -\hat{G}_{22} +2 
V^{(\frac{3}{2})} \right](w) +
\cdots,
\label{g22t1}
\eea
where the lowest component spin-$\frac{3}{2}$ current can be expressed as
\bea
V^{(\frac{3}{2})}(w)  & = &
-\frac{i \sqrt{2} }{(5+k)^2}
\sum_{(a,\bar{a})=(10,\bar{10})}^{(12,\bar{12})} J^{a} J^{\bar{9}} J^{\bar{a}}(w)
\nonu \\
& + &
\mbox{other cubic, quadratic and linear terms}. 
\label{v3half}
\eea
The field contents of (\ref{v3half}) look similar to those of (\ref{u3half}).
The barred and unbarred indices in (\ref{u3half}) are  replaced by 
unbarred and barred ones in (\ref{v3half}) respectively. 
The $U(1)$ charges of the two currents in (\ref{g22t1}) are the same.
The OPE (\ref{g22t1}) is conjugated to the OPE (\ref{g11t1}).
As in previous results in (\ref{ufu5half}), 
the spin-$\frac{3}{2}$ current satisfies
\bea
U(z) \, V^{(\frac{3}{2})}(w) & = & +\cdots, 
\qquad
F_{a}(z) \, V^{(\frac{3}{2})}(w)   =  +\cdots, \qquad a=11,12,21,22. 
\label{ufv3half}
\eea 

Consider the next OPE 
\bea
\hat{G}_{22}(z) \, T_{+}^{(\frac{3}{2})}(w) & = & 
\frac{1}{(z-w)^2} \frac{6}{(5+k)} \left[ i \hat{A}_{-} \right](w) \nonu \\
& + &
\frac{1}{(z-w)} \left[ - V_{+}^{(2)} + \frac{1}{2}   \frac{6}{(5+k)}  
 i   \pa  \hat{A}_{-}  \right](w) +\cdots,
\label{g22t+3half}
\eea
where the new spin-$2$ current occurs in the first order pole 
of (\ref{g22t+3half}) and the result is as follows \footnote{One obtains
\bea
\hat{G}_{21}(z) \, V^{(\frac{3}{2})}(w) & = &
\frac{1}{(z-w)^2} \frac{6}{(5+k)} \left[ i  \hat{A}_{-} \right](w) + 
\frac{1}{(z-w)}  \left[ V_{+}^{(2)} 
   + \frac{1}{2}  \frac{6}{(5+k)} i \pa \hat{A}_{-}   \right](w) +\cdots.
\nonu
\eea
As before, the spin-$2$ current with plus sign appears in the first-order pole
after subtracting the descendant term.
}:
\bea
 V_{+}^{(2)}(w) & = & 
\frac{2 i}{(5+k)^2} 
\sum_{(a,\bar{a})=(10,\bar{10})}^{(12,\bar{12})}
 J^{a} J^{\bar{9}} D J^{\bar{a}}(w)
\nonu \\
& + &
\mbox{other quartic, cubic, quadratic and linear terms}. 
\label{v+2}
\eea
The field contents of (\ref{v+2}) can be obtained after 
the barred and unbarred indices in (\ref{u-2}) are  replaced by 
unbarred and barred ones respectively. 
The OPE (\ref{g22t+3half}) looks very similar to (\ref{g11t-3half}):
they are conjugated to each other under the $U(1)$ charge.

One can see immediately that the following regularity 
conditions hold, as in 
(\ref{ufu5half}) and (\ref{ufv3half}),
\bea
U(z) \, V_{+}^{(2)}(w) & = & +\cdots, 
\qquad
F_{a}(z) \, V_{+}^{(2)}(w)   =  +\cdots, \qquad a=11,12,21,22. 
\label{ufv+2}
\eea 
Furthermore, the first-order pole in the OPE $T^{(1)}(z)\, V_{+}^{(2)}(w)$
vanishes, as in (\ref{t1u-2poleone}),
\bea
T^{(1)}(z)\, V_{+}^{(2)}(w)|_{\frac{1}{(z-w)}}  =  0.
\label{t1v+2poleone}
\eea

One describes the following OPE
\bea
\hat{G}_{22}(z) \, T_{-}^{(\frac{3}{2})}(w) & = & 
\frac{1}{(z-w)^2} \frac{2k}{(5+k)} \left[ i \hat{B}_{+} \right](w) 
\nonu  \\
& + &
\frac{1}{(z-w)} \left[ - V_{-}^{(2)} + \frac{1}{2}  \frac{2k}{(5+k)}  
   i  \pa \hat{B}_{+}  + \frac{1}{(5+k)} 
4 \hat{A}_3  \hat{B}_{+}  \right](w)  \nonu \\
& + & \cdots,
\label{g22t-3half}
\eea
where the other spin-$2$ current can be described as
\footnote{
One obtains
the following OPE
\bea
\hat{G}_{12}(z) \, V^{(\frac{3}{2})}(w) & = &
\frac{1}{(z-w)^2} \frac{2k}{(5+k)} \left[ - i \hat{B}_{+} \right](w) +  
\frac{1}{(z-w)}  \left[ V_{-}^{(2)} 
   - \frac{1}{2}  \frac{2 k}{(5+k)}  i \pa \hat{B}_{+}  \right](w) +\cdots,
\nonu
\eea
where the spin-$2$ current with plus sign appears in the first-order 
pole.}
\bea
 V_{-}^{(2)}(w) & = & 
\frac{2 i}{(5+k)^3} \left( -K^{\bar{1}} J^9 J^{\bar{9}} J^{\bar{10}}
+ K^{\bar{1}} J^{11} J^{\bar{10}} J^{\bar{11}}
 +  K^{\bar{1}} J^{12} J^{\bar{10}} J^{\bar{12}} \right)(w)
\nonu \\
& + &
\mbox{other quartic, cubic and quadratic  terms}. 
\label{v-2}
\eea
The OPE (\ref{g22t-3half}) is conjugated to the OPE (\ref{g11t+3half})
under the $U(1)$ charge.
As before in (\ref{ufv+2}), one has the following OPEs
\bea
U(z) \, V_{-}^{(2)}(w) & = & +\cdots, 
\qquad
F_{a}(z) \, V_{-}^{(2)}(w)   =  +\cdots, \qquad a=11,12,21,22, 
\label{ufv-2}
\eea 
and along the line of (\ref{t1v+2poleone})
the following relation holds
\bea
T^{(1)}(z) \, V_{-}^{(2)}(w) |_{\frac{1}{(z-w)}} =0.
\label{t1v-2poleone}
\eea

Let us calculate the last component spin-$\frac{5}{2}$ current,
as in (\ref{g21u-2}), 
\bea
\hat{G}_{21}(z) \, V_{-}^{(2)}(w) & = &
\frac{1}{(z-w)^2} \left[ -\frac{(6+k)}{(5+k)} \hat{G}_{22}+
\frac{2(7+k)}{(5+k)} V^{(\frac{3}{2})} \right](w)
\label{g21v-2}
\\
& + & \frac{1}{(z-w)}  \left[ V^{(\frac{5}{2})}
+\frac{1}{3} \pa \left(- \frac{(6+k)}{(5+k)}  \hat{G}_{22}+
\frac{2(7+k)}{(5+k)}  V^{(\frac{3}{2})}
 \right) \right](w)+ \cdots,
\nonu
\eea
where the new spin-$\frac{5}{2}$ current after subtracting the descendant 
fields can be described as
\bea
 V^{(\frac{5}{2})}(w) & = & 
-\frac{4 i \sqrt{2}}{(5+k)^4} \left( J^{10} J^{11} J^{\bar{9}}  J^{\bar{10}} 
J^{\bar{11}} + J^{10} J^{12} J^{\bar{9}}  J^{\bar{10}} 
J^{\bar{12}} + J^{11} J^{12} J^{\bar{9}}  J^{\bar{11}} 
J^{\bar{12}}\right)(w)
\nonu \\
& + &
\mbox{other quintic, 
quartic, cubic, quadratic and linear terms}. 
\label{v5half}
\eea
Furthermore one has the regularity conditions, similar to (\ref{ufv3half}), 
(\ref{ufv+2}) and (\ref{ufv-2}),  as follows: 
\bea
U(z) \, V^{(\frac{5}{2})}(w) & = & +\cdots, 
\qquad
F_{a}(z) \, V^{(\frac{5}{2})}(w)   =  +\cdots, \qquad a=11,12,21,22. 
\label{ufv5half}
\eea 

Therefore, the higher spin currents of spins 
$(\frac{3}{2}, 2, 2, \frac{5}{2})$
are determined completely: (\ref{v3half}), (\ref{v+2}), (\ref{v-2}) 
and (\ref{v5half}). 
Some of the OPEs between these higher spin currents and the currents from 
the large ${\cal N}=4$ nonlinear
superconformal algebra are presented in (\ref{g22t1}), 
(\ref{g22t+3half}), (\ref{g22t-3half}) and (\ref{g21v-2}).
The remaining OPEs are given in Appendix $(B.2)$
and Appendix $(C.2)$.

%%%%%%%%%%%%%%%%%%%%%%%%%
\subsection{Construction of higher spin currents of spins 
$\left(2, \frac{5}{2}, \frac{5}{2}, 3 \right)$} 
%%%%%%%%%%%%%%%%%%%%%%%%%

In this subsection, the final ${\cal N}=2$ multiplet can be determined.
One way to see the presence of new spin-$2$ current is to calculate the 
following  OPE, together with (\ref{nong22}) and (\ref{u3half}),
\bea
\hat{G}_{22}(z) \, U^{(\frac{3}{2})}(w) & = &
-\frac{1}{(z-w)^3} \, \frac{6k}{(5+k)} +
\frac{1}{(z-w)^2} \left[ \frac{2i}{(5+k)} ( 3 \hat{A}_3 -
k \hat{B}_3) + T^{(1)} \right](w) \nonu \\
& + &  
\frac{1}{(z-w)}  \left[ -W^{(2)} 
  + \frac{1}{2} \pa \left(  
\frac{2i}{(5+k)} ( 3 \hat{A}_3 -
k \hat{B}_3) + T^{(1)}
%\frac{3i}{(5+k)} \pa \hat{A}_3 -
%\frac{i k}{(5+k)} \pa \hat{B}_3 +\frac{1}{2} \pa T^{(1)}
\right)
\right](w) \nonu \\
& + & \cdots,
\label{g22u3half} 
\eea
where the new spin-$2$ current can be written as
\bea
 W^{(2)}(w) & = & 
\frac{4}{(5+k)^3} \left( J^9 J^{10} J^{\bar{9}} J^{\bar{10}}
+ J^9 J^{11} J^{\bar{9}} J^{\bar{11}}
+ J^9 J^{12} J^{\bar{9}} J^{\bar{12}} \right.
\nonu \\
& + & 
\left. J^{10} J^{11} J^{\bar{10}} J^{\bar{11}}
+J^{10} J^{12} J^{\bar{10}} J^{\bar{12}}
+J^{11} J^{12} J^{\bar{11}} J^{\bar{12}} \right)(w)
\nonu \\
&+ & \mbox{other quartic, cubic, quadratic and linear terms}.
\label{w2simple}
\eea
The full expression of (\ref{w2simple}) is not presented here.
As in (\ref{t1v-2poleone}) and (\ref{ufv5half}),
the above spin-$2$ current satisfies
\bea
U(z) \, W^{(2)}(w) & = & +\cdots, 
\qquad
F_{a}(z) \, W^{(2)}(w)   =  +\cdots, \qquad a=11,12,21,22, 
\label{ufw2}
\eea 
and 
\bea
T^{(1)}(z) \, W^{(2)}(w)|_{\frac{1}{(z-w)}}  & = & 0. 
\label{t1w2polepone}
\eea

Let us consider the following OPE
\bea
\hat{G}_{21}(z) \, W^{(2)}(w)  & = & 
\frac{1}{(z-w)^2} \frac{1}{(5+k)} \left[ 2(2+k) \hat{G}_{21} +
(-3+k)  T_{+}^{(\frac{3}{2})}\right](w) 
\label{g21w2}
\\
& + &
\frac{1}{(z-w)} \frac{1}{(5+k)} \left[ W_{+}^{(\frac{5}{2})} +
 \frac{1}{3} \pa \left( 2(2+k)  \hat{G}_{21} +
 (-3+k)   T_{+}^{(\frac{3}{2})} \right)
\right](w) +\cdots,
\nonu
\eea
where the new spin-$\frac{5}{2}$ current is given by
\bea
 W_{+}^{(\frac{5}{2})}(w) & = & 
\frac{2 i \sqrt{2}}{(5+k)^4} \left( K^1 K^4 K^7 K^{\bar{1}} K^{\bar{4}}
+ K^1 K^5 K^7 K^{\bar{1}} K^{\bar{5}} \right)
\nonu \\
& + & 
 \mbox{other quintic, 
quartic, cubic, quadratic and linear terms}. 
\label{w+5half}
\eea
The $U(1)$ charges of three currents in (\ref{g21w2})
are the same.
One can check that the spin-$\frac{5}{2}$ current commute with 
the subgroup currents, as in (\ref{ufw2}),  as follows:
\bea
U(z) \, W_{+}^{(\frac{5}{2})}(w) & = & +\cdots, 
\qquad
F_{a}(z) \, W_{+}^{(\frac{5}{2})}(w)   =  +\cdots, \qquad a=11,12,21,22. 
\label{uw+5half}
\eea 

Let us consider the following OPE
\bea
\hat{G}_{12}(z) \, W^{(2)}(w)  & = & 
\frac{1}{(z-w)^2} \frac{1}{(5+k)} \left[ 2(2+k) \hat{G}_{12} -
(-3+k)  T_{-}^{(\frac{3}{2})}\right](w) 
\label{g12w2} 
\\
& + &
\frac{1}{(z-w)} \frac{1}{(5+k)} \left[ W_{-}^{(\frac{5}{2})} +
 \frac{1}{3} \pa \left( 2(2+k)  \hat{G}_{12} -
 (-3+k)   T_{-}^{(\frac{3}{2})} \right)
\right](w) +\cdots,
\nonu
\eea
where the new spin-$\frac{5}{2}$ current can be obtained
\bea
 W_{-}^{(\frac{5}{2})}(w) & = & 
-\frac{2 i \sqrt{2}}{(5+k)^4} \left( K^1 K^4 K^{\bar{1}} K^{\bar{4}} K^{\bar{7}}
+ K^1 K^5 K^{\bar{1}} K^{\bar{5}} K^{\bar{7}} \right)
\nonu \\
& + & 
 \mbox{other quintic, 
quartic, cubic, quadratic and linear terms}. 
\label{w-5half}
\eea
The two OPEs (\ref{g21w2}) and (\ref{g12w2}) are conjugated to each other.
Also one obtains the following regularity conditions (i.e. 
the higher spin current should commute with the subgroup currents)
as in (\ref{uw+5half})
\bea
U(z) \, W_{-}^{(\frac{5}{2})}(w) & = & +\cdots, 
\qquad
F_{a}(z) \, W_{-}^{(\frac{5}{2})}(w)   =  +\cdots, \qquad a=11,12,21,22. 
\label{uw-5half}
\eea 

Finally, the highest higher spin current of spin-$3$ \footnote{
\label{Nahmformula}
The OPE of two quasi-primary fields, of spins $h_i$ and 
$h_j$ respectively, takes the form 
\cite{Bowcock,BFKNRV,BS,Nahm1,Nahm2,Ahn1211,AP1310} 
\bea
&& \Phi_i(z) \; \Phi_j(w)  =  \frac{1}{(z-w)^{h_i+h_j}} \, \gamma_{ij} \nonu \\
&& +  \sum_k C_{ijk} \sum_{n=0}^{\infty} \frac{1}{(z-w)^{h_i+h_j-h_k-n}}
\left[\frac{1}{n!} \prod_{x=0}^{n-1} \frac{(h_i-h_j+h_k+x)}{(2h_k+x)} 
\right]
\pa^n \Phi_k(w).
\label{PhiPhi}
\eea
The $\gamma_{ij}$ corresponds to a metric on the space of quasi-primary fields.
The structure constant $C_{ijk}$
appears in the three-point function.
The index $k$ specifies all the quasi-primary fields occurring
in the right hand side of (\ref{PhiPhi}).
The relative coefficient functions 
$\left[\frac{1}{n!} \prod_{x=0}^{n-1} \frac{(h_i-h_j+h_k+x)}{(2h_k+x)} 
\right]$
in the descendant fields 
depend on the spins
and number of derivatives.
For the fixed $C_{ijk}$, 
the relative coefficient functions in the descendant fields
using (\ref{PhiPhi})
can be obtained. 
It is quite nontrivial to rearrange those expressions in terms of 
determined (and known) higher spin currents.
}, 
can be obtained from the following OPE
\bea
\hat{G}_{21}(z) \, W_{-}^{(\frac{5}{2})}(w)  & = & 
\frac{1}{(z-w)^3}  \left[ \frac{8 i (1+3k)}{(5+k)^2} \hat{A}_3 +
\frac{80 i k}{3(5+k)^2} \hat{B}_3 + \frac{8(-3+k)}{3(5+k)} T^{(1)}
\right](w) 
\nonu \\
& + & \frac{1}{(z-w)^2} \left[ 
-\frac{4(15+65k+22k^2)}{3(3+7k)(5+k)} \hat{T}
+\frac{4(-3+k)}{3(5+k)} T^{(2)}
+\frac{4(4+k)}{(5+k)} W^{(2)}
\right.
\nonu \\
& - & \frac{16(-1+k)}{3(5+k)^2}  \hat{A}_1 \, \hat{A}_1 
-  \frac{16(-1+k)}{3(5+k)^2}  \hat{A}_2 \, \hat{A}_2 
+  \frac{4(13-4k)}{3(5+k)^2}  \hat{A}_3 \, \hat{A}_3 
\nonu \\
& - &    \frac{4}{3(5+k)}  \hat{B}_1 \, \hat{B}_1 -
 \frac{4}{3(5+k)}  \hat{B}_2 \, \hat{B}_2 +
 \frac{4(-5+2k)}{3(5+k)^2}  \hat{B}_3 \, \hat{B}_3 \nonu \\
& + &  \left. 
 \frac{4 i }{(5+k)} T^{(1)} \, \hat{B}_3   
+ \frac{8(-4+k)}{3(5+k)^2}  \hat{A}_3 \, \hat{B}_3 
 -
 \frac{4 i}{(5+k)} T^{(1)} \, \hat{A}_3
\right](w)
\nonu \\
& + &
\frac{1}{(z-w)}  \left[ \frac{1}{4} \, \pa 
\, \{ \hat{G}_{21} \, W_{-}^{(\frac{5}{2})} \}_{-2}
+\frac{24 i(1+3k)}{(19+23k)(5+k)} \left( \hat{T} \, \hat{A}_3-
\frac{1}{2} \pa^2 \hat{A}_3 \right)
\right.
\nonu \\
& + &  \frac{80i k }{(19+23k)(5+k)}
 \left( \hat{T} \, \hat{B}_3-
\frac{1}{2} \pa^2 \hat{B}_3 \right)
+\frac{8(-3+k)}{(19+23k)}
 \left( \hat{T} \, T^{(1)}-
\frac{1}{2} \pa^2 T^{(1)} \right)
\nonu \\
& + & \left. W^{(3)} 
\right](w) +\cdots,
\label{aboveg21w-5half}
\eea
where the new spin-$3$ current can be described as
\bea
W^{(3)} & = & -\frac{4(-27+23k+2k^2)}{5+k)^5(19+23k)} \left( K^1 K^2 K^4
K^{\bar{1}} K^{\bar{2}} K^{\bar{4}} +
 K^1 K^3 K^4
K^{\bar{1}} K^{\bar{3}} K^{\bar{4}}
\right)
\nonu \\
& + &  \mbox{other sextic, quintic, 
quartic, cubic, quadratic and linear terms}. 
\label{w3}
\eea
Note that because
the relative coefficient functions for the descendant fields of 
the spin-$1$ currents appearing in the third-order pole in 
(\ref{aboveg21w-5half})
vanish, there are no descendant fields in the second-order pole. 
In the first-order pole in (\ref{aboveg21w-5half}), 
the total derivative terms with the coefficient $\frac{1}{4}$ are descendant 
fields of the spin-$2$ currents in the second-order pole. 
The remaining three terms containing $\hat{T}(w)$ are quasi-primary fields of
spin-$3$ under the stress tensor (\ref{stressnonlinear})
\footnote{One has the following regularity conditions between the spin-$3$ current 
and four spin-$\frac{1}{2}$
currents and spin-$1$ current, as in (\ref{uw-5half}) 
\bea
U(z) \, W^{(3)}(w) & = & +\cdots, 
\qquad
F_{a}(z) \, W^{(3)}(w)   =  +\cdots, \qquad a=11,12,21,22. 
\nonu
\eea 
Furthermore, compared to (\ref{t1w2polepone}), the first-order pole between 
the spin-$1$ current and the spin-$3$ current
doesn't vanish
\bea
T^{(1)}(z) \, W^{(3)}(w)|_{\frac{1}{(z-w)}}  \neq  0.
\nonu
\eea}.

Therefore, the higher spin currents of spins 
$(2, \frac{5}{2}, \frac{5}{2}, 3)$
are determined completely: (\ref{w2simple}), (\ref{w+5half}), 
(\ref{w-5half}) and (\ref{w3}). 
Some of the OPEs between these higher spin currents and the currents from 
the large ${\cal N}=4$ nonlinear
superconformal algebra are presented in (\ref{g22u3half}), 
(\ref{g21w2}), (\ref{g12w2}) and (\ref{aboveg21w-5half}).
The remaining OPEs are given in Appendix $(B.3)$
and Appendix $(C.3)$.

%%%%%%%%%%%%%%%%%%%%%%%%%%%%%%%%%%%%%%%%%%%%%%%%%%%%%%%%%%%%%%%%%%%%%
%%%%%%%%%%%%%%%%%%%%%%%%%%%%%%%%%%%%%%%%%%%%%%%%%%%%%%%%%%%%%%%%%%%%%%
\section{Extension of large ${\cal N}=4$ nonlinear 
superconformal  algebra in the coset 
minimal model    }
%2%%%%%%%%%%%%%%%%%%%%%%%%%%%%%%%%%%%%%%%%%%%%%%%%%%%%%%%%%%%%%%%%%%%%%
%%%%%%%%%%%%%%%%%%%%%%%%%%%%%%%%%%%%%%%%%%%%%%%%%%%%%%%%%%%%%%%%%%%%%

In section $3$, the $11$ currents of
large ${\cal N}=4$ nonlinear superconformal algebra 
were constructed and in section $4$, the $16$ 
higher spin currents were found.
For the extension of  large ${\cal N}=4$ nonlinear 
superconformal  algebra, one should calculate 
the OPEs between the $11$ currents in section $3$ and 
the $16$ currents in section $4$
as follows:
\bea
\left(
\begin{array}{c}
 \hat{T} \\
\hat{A}_1, \hat{A}_2, \hat{A}_3 \\
\hat{B}_1, \hat{B}_2, \hat{B}_3 \\
\hat{G}_{11}, \hat{G}_{12}, \hat{G}_{21}, \hat{G}_{22}
 \\
\end{array}
\right)(z)  \,
\left(
\begin{array}{cccc} 
T^{(1)}, T_{+}^{(\frac{3}{2})}, T_{-}^{(\frac{3}{2})}, 
T^{(2)} \\
U^{(\frac{3}{2})}, U_{+}^{(2)}, U_{-}^{(2)}, U^{(\frac{5}{2})} 
 \\
V^{(\frac{3}{2})}, V_{+}^{(2)}, 
V_{-}^{(2)}, V^{(\frac{5}{2})} 
\\
W^{(2)}, W_{+}^{(\frac{5}{2})}, 
W_{-}^{(\frac{5}{2})}, W^{(3)} \\
\end{array} \right)(w).  
\label{opediagram}
\eea
In other words, the OPEs between $11$ currents in the left hand side of
(\ref{opediagram}) and the $16$ currents in the right hand side of 
(\ref{opediagram}) \footnote{
\label{su3currents} One can easily check that 
the OPEs between $(11+16)$ currents and four quantities 
$\frac{1}{12\sqrt{2}} \left(
4(3 i +\sqrt{3}) K^7 +  (-i \sqrt{6} - 3 \sqrt{10}) K^8 \right)(w)$,
$\frac{1}{12\sqrt{2}} \left(
4(-3 i +\sqrt{3}) K^{\bar{7}} +  i (\sqrt{6} + 3 i 
\sqrt{10}) K^{\bar{8}} \right)(w)$,  $\frac{1}{12\sqrt{2}} \left(
4(3 i +\sqrt{3}) \overline{D} K^7 +  
(-i \sqrt{6} - 3 \sqrt{10}) \overline{D} K^8 \right)(w)$,
and $\frac{1}{12\sqrt{2}} \left(
4(-3 i +\sqrt{3}) D K^{\bar{7}} +  i (\sqrt{6} + 3 i 
\sqrt{10}) D K^{\bar{8}} \right)(w)$ do not have any singular terms. 
The first and second fields look similar to $F_{21}(z)$ and $F_{12}(z)$. 
The only difference appears in the coefficient of $K^{7}(z)$ and 
$K^{\bar{7}}(z)$. 
These are 
${\cal N}=2$ 
WZW affine currents for the denominator $SU(3)$ in the Wolf space 
coset model.
Furthermore, the OPEs of above $(11+16)$ currents and [$K^m(z)$, $K^{\bar{m}}$,
$\overline{D} K^m(z)$, and $D K^{\bar{m}}(z)$] where $m =4, \cdots, 6$ and 
$\bar{m}=\bar{4}, \cdots, \bar{6}$ do not have any singular terms. Therefore,
these eight currents (and its superpartners)
consist of the Wolf space denominator $SU(3)$ ${\cal N}=2$ WZW affine 
currents. } are needed.  

%%%%%%%%%%%%%%%%%%%%%%%%%
\subsection{$U(1)$ charges of higher spin currents} 
%%%%%%%%%%%%%%%%%%%%%%%%%

It is straightforward to calculate the various $U(1)$ charges under 
the $U(1)$ current introduced in previous section. The result is
as follows: 
\bea
\left(- 2 i \gamma_A \hat{A}_3 -2 i \gamma_B \hat{B}_3\right)(z) \,
T^{(1)}(w)|_{\frac{1}{(z-w)}} & = & 
0, \nonu \\
\left(- 2 i \gamma_A \hat{A}_3 -2 i \gamma_B \hat{B}_3\right)(z) \,
T_{\pm}^{(\frac{3}{2})} (w)|_{\frac{1}{(z-w)}} & = & 
\frac{1}{(z-w)} 
\left[  \pm \frac{(3+k)}{(5+k)}\right]
T_{\pm}^{(\frac{3}{2})} (w),
\nonu \\
\left(- 2 i \gamma_A \hat{A}_3 -2 i \gamma_B \hat{B}_3\right)(z) \,
T^{(2)}(w)|_{\frac{1}{(z-w)}} & = & 
0, 
\nonu \\
\left(- 2 i \gamma_A \hat{A}_3 -2 i \gamma_B \hat{B}_3\right)(z) \,
\left(
\begin{array}{c}
U^{(\frac{3}{2})} \\
V^{(\frac{3}{2})} \\
\end{array} \right) (w)|_{\frac{1}{(z-w)}} & = & 
\frac{1}{(z-w)} 
\left[  \pm \frac{(-3+k)}{(5+k)}\right]
\left(
\begin{array}{c}
U^{(\frac{3}{2})} \\
V^{(\frac{3}{2})} \\
\end{array} \right)(w),
\nonu \\
\left(- 2 i \gamma_A \hat{A}_3 -2 i \gamma_B \hat{B}_3\right)(z) \,
\left(
\begin{array}{c}
U_{+}^{(2)} \\
V_{-}^{(2)} \\
\end{array} \right)
(w)|_{\frac{1}{(z-w)}} & = & 
\frac{1}{(z-w)} 
\left[  \pm \frac{2k}{(5+k)}\right]
\left(
\begin{array}{c}
U_{+}^{(2)} \\
V_{-}^{(2)} \\
\end{array} \right)
(w),
\nonu \\
\left(- 2 i \gamma_A \hat{A}_3 -2 i \gamma_B \hat{B}_3\right)(z) \,
\left(
\begin{array}{c}
U_{-}^{(2)} \\
V_{+}^{(2)} \\
\end{array} \right)
(w)|_{\frac{1}{(z-w)}} & = & 
\frac{1}{(z-w)} 
\left[ \mp \frac{6}{(5+k)}\right]
\left(
\begin{array}{c}
U_{-}^{(2)} \\
V_{+}^{(2)} \\
\end{array} \right)(w),
\nonu \\
\left(- 2 i \gamma_A \hat{A}_3 -2 i \gamma_B \hat{B}_3\right)(z) \,
\left(
\begin{array}{c}
U^{(\frac{5}{2})} \\
V^{(\frac{5}{2})} \\
\end{array} \right)
(w)|_{\frac{1}{(z-w)}} & = & 
\frac{1}{(z-w)} 
\left[ \pm  \frac{(-3+k)}{(5+k)}\right]
\left(
\begin{array}{c}
U^{(\frac{5}{2})} \\
V^{(\frac{5}{2})} \\
\end{array} \right) (w),
\nonu \\
\left(- 2 i \gamma_A \hat{A}_3 -2 i \gamma_B \hat{B}_3\right)(z) \,
W^{(2)}(w)|_{\frac{1}{(z-w)}} & = & 
0,
\nonu \\
\left(- 2 i \gamma_A \hat{A}_3 -2 i \gamma_B \hat{B}_3\right)(z) \,
W_{\pm}^{(\frac{5}{2})} (w)|_{\frac{1}{(z-w)}} & = & 
\frac{1}{(z-w)} 
\left[  \pm \frac{(3+k)}{(5+k)}\right]
W_{\pm}^{(\frac{5}{2})} (w),
\nonu \\
\left(- 2 i \gamma_A \hat{A}_3 -2 i \gamma_B \hat{B}_3\right)(z) \,
W^{(3)}(w)|_{\frac{1}{(z-w)}} & = & 
0.
\label{u1second}
\eea
We present these $U(1)$ charges in the Table $2$ explicitly 
\footnote{If one  uses the notations $(++,+-,-+,--)$ for the fermion 
currents and $(+,-,3)$ for the bosonic currents which tansform as triplet 
under the $SU(2)$, then the expression (\ref{u1second}) will be simpler.}. 
Then the $U(1)$ charges of  all the composite fields 
coming from the $(11+16)$ currents  can be determined by 
the assignments in Table $1$ and Table $2$.
The necessary $U(1)$ assignments in this paper are presented in 
Table $3$-Table $7$.

%\bea
%\frac{2k}{(5+k)} & : & (\hat{B}_1 - i \hat{B}_2), \qquad U_{+}^{(2)}, 
%\nonu \\
%\frac{(3+k)}{(5+k)} & : & \hat{G}_{21}, \qquad T_{+}^{(\frac{3}{2})}, \qquad
%W_{+}^{(\frac{5}{2})},
%\nonu \\
%\frac{(-3+k)}{(5+k)} & : & \hat{G}_{11}, \qquad U^{(\frac{3}{2})}, \qquad
%U^{(\frac{5}{2})},
%\nonu \\
%\frac{6}{(5+k)} & : & (\hat{A}_1 - i \hat{A}_2), \qquad V_{+}^{(2)},
%\nonu \\
%0 & : & \hat{A}_3, \qquad \hat{B}_3, \qquad T^{(1)}, \qquad T^{(2)}, \qquad
%W^{(2)}, \qquad W^{(3)},
%\nonu \\
%-\frac{6}{(5+k)} & : & 
% (\hat{A}_1 + i \hat{A}_2), \qquad U_{-}^{(2)},
%\nonu \\
%-\frac{(-3+k)}{(5+k)} & : & \hat{G}_{22}, \qquad V^{(\frac{3}{2})}, \qquad
%V^{(\frac{5}{2})},
%\nonu \\
%-\frac{(3+k)}{(5+k)} & : & \hat{G}_{12}, \qquad T_{-}^{(\frac{3}{2})}, \qquad
%W_{-}^{(\frac{5}{2})},
%\nonu \\
%-\frac{2k}{(5+k)} & : &  (\hat{B}_1 + i \hat{B}_2), \qquad V_{-}^{(2)}.
%\nonu
%\eea

%%%%%%%%%%%%%%%%%%%%%%%%%
\subsection{Structure of OPEs} 
%%%%%%%%%%%%%%%%%%%%%%%%%

The $16$ higher spin currents are primary fields under the stress energy 
tensor $\hat{T}(z)$ (\ref{stressnonlinear}).
Then the remaining nontrivial 
OPEs between the $16$ currents and the $11$ currents
are given by 1) the OPEs between the six spin-$1$ currents and the 
$16$ higher spin currents described in Appendix $B$ and 2)
the OPEs between the four spin-$\frac{3}{2}$ currents and 
the $16$ higher spin currents described in Appendix $C$.
It is nontrivial to extract all the structures in the right hand sides of
above OPEs. Because the left hand sides of these OPEs are known in terms 
${\cal N}=2$ WZW affine currents, we would like to express them in terms of
known currents. If one cannot write them with the known currents, then 
one should obtain the new primary fields appearing in the right hand sides
along the line of footnote \ref{Nahmformula}.  
One should write down the correct terms, which preserve the right 
$U(1)$ charge described before via Tables $1$-$7$,  
with arbitrary coefficients at the specific 
pole (with fixed spin) in the given OPE. 

For example, in the OPEs between the spin-$1$ currents of 
large ${\cal N}=4$ nonlinear algebra and the largest spin-$3$ higher spin
current, by dimensional analysis of spin, the first-order pole contains 
the spin-$3$ fields in Appendices (\ref{A+w3})-(\ref{B-w3}).
One can rearrange these OPEs in terms of descendant fields with known 
coefficients and (quasi) primary fields at first-order pole terms. 
The $U(1)$ charges of composite spin-$3$ fields are presented in Table 
$5$. The $U(1)$ charge of the left hand side of Appendix 
(\ref{A+w3}) is the sum of
the $U(1)$ charge of $(\hat{A}_1 +i \hat{A}_2)(z)$ and the one of 
$W^{(3)}(w)$. The former is given by $-\frac{6}{(5+k)}$ 
and the latter is given by $0$ from Table $2$. 
Therefore, the total $U(1)$ charge, $-\frac{6}{(5+k)}$, should appear in the 
right hand side of Appendix (\ref{A+w3}). Then from the first row of Table $5$, 
there exist $26$ possible composite spin-$3$ fields with $U(1)$ charge 
$-\frac{6}{(5+k)}$.    
Because the left hand side of Appendix 
(\ref{A+w3}) is written in terms of ${\cal N}=2$
WZW affine currents explicitly, let us subtract the above $26$ terms with 
arbitrary $k$-dependent coefficients from the left hand side of Appendix
(\ref{A+w3}).
In order to vanish these quantities, one should solve this equation.
This is equivalent to solve many linear equations with respect to the above 
undetermined $k$-dependent coefficients. 
It turns out that the unknown coefficient functions can be obtained completely
and they are given in Appendix 
(\ref{A+w3}). Of course, when the order of any product
is changed, the extra derivative terms arise either in Table $6$ or in the OPE
of Appendix (\ref{A+w3}). 

What about the OPEs between the spin-$\frac{3}{2}$ currents of
large ${\cal N}=4$ nonlinear superconformal algebra and the above spin-$3$ 
higher spin current? They are presented in Appendix (\ref{g11w3})-(\ref{g22w3}).
One can have the composite spin-$\frac{7}{2}$ fields in the first-order pole of
Appendix (\ref{g11w3}) by dimensional  analysis as before.
The total $U(1)$ charge of this OPE should be preserved. 
Because the $U(1)$ charge of the left hand side is given by 
$\frac{(-3+k)}{5+k}$ from the Table $2$, the possible $68$ 
composite fields with 
this $U(1)$ charge are presented in the first row of Table $7$. 
As performed in previous paragraph, 
it turns out that  the unknown 
$41$ coefficient functions can be obtained completely
and they are given in Appendix (\ref{g11w3}).

In these examples, there are no new (quasi) primary fields in the 
first-order pole. As in (\ref{aboveg21w-5half}), when one cannot solve the 
linear equations completely due to the appearance of new (quasi) primary fields,
one should resort to the procedure described in the footnote 
\ref{Nahmformula}. That is, one should check that any singular terms 
consist of a couple of descendant fields (with determined coefficient functions
using the formula in the footnote \ref{Nahmformula}) and a couple of (quasi) 
primary fields. 

As in an abstract, the result from Appendices $B$ and $C$ shows that 
the right hand sides of all the OPEs contains the composite fields which 
can be obtained from the known $(11+16)$ currents. There exist no new 
primary fields. One expects that the new primary fields will appear in the 
OPEs between the $16$ currents and themselves \cite{Ahn2014}.

%%%%%%%%%%%%%%%%%%%%%%%%%%%%%%%%%%%%%%%%%%%%%%%%%%%%%%%%%%%%%%%%%%%
\begin{table}[ht]
%\caption{The $U(1)$ charges for the $16$ currents} % title of Table
\centering % used for centering table
\begin{tabular}{|c||c||c| } % centered columns (4 columns)
\hline %inserts double horizontal lines
$U(1)$ charge & $16$ higher spin currents & $11$ currents in Table 1
\\ [0.5ex] % inserts table
%heading
\hline \hline % inserts single horizontal line
$\frac{2k}{(5+k)}$  & $ \quad U_{+}^{(2)}$ &  $\hat{B}_{-}$ 
\\ % inserting body of the table
\hline
$\frac{(3+k)}{(5+k)}$ &  $ T_{+}^{(\frac{3}{2})}, \quad
W_{+}^{(\frac{5}{2})}$ &  $\hat{G}_{21}$ \\
\hline
$\frac{(-3+k)}{(5+k)}$ & $\quad U^{(\frac{3}{2})}, \quad
U^{(\frac{5}{2})}$ &  $\hat{G}_{11}$  \\
\hline
$\frac{6}{(5+k)} $ & $V_{+}^{(2)}$ &  $\hat{A}_{-}$  \\
\hline
$0$ & $ T^{(1)}, \quad T^{(2)}, \quad
W^{(2)}, \quad W^{(3)}$ & $\hat{A}_3$, \qquad $\hat{B}_3$, \qquad 
$\hat{T}$  \\ 
\hline
$-\frac{6}{(5+k)}$ & $ U_{-}^{(2)}$ & $\hat{A}_{+}$  \\ 
\hline
$-\frac{(-3+k)}{(5+k)}$ & $ V^{(\frac{3}{2})}, \quad
V^{(\frac{5}{2})}$ & $\hat{G}_{22}$  \\ 
\hline
$-\frac{(3+k)}{(5+k)} $ & $ T_{-}^{(\frac{3}{2})}, \quad
W_{-}^{(\frac{5}{2})}$ & $\hat{G}_{12}$  \\
\hline
$-\frac{2k}{(5+k)}$ & $V_{-}^{(2)}$ & 
$\hat{B}_{+}$  \\ 
[1ex] % [1ex] adds vertical space
\hline %inserts single line
\end{tabular}
%\label{table:nonlin} % is used to refer this table in the text
\caption{The $U(1)$ charges for the $16$ currents from (\ref{u1second}).
For convenience, the $U(1)$ charges for the $11$ currents of
large ${\cal N}=4$ nonlinear superconformal algebra  
in Table $1$ are presented also.
The fields appearing in the first-order poles  of 
(\ref{b+t2}), 
(\ref{a-u3half}), 
[(\ref{a+t+3half}) and (\ref{G11t1})], 
(\ref{a+t2}), (\ref{b+u+2}), (\ref{a+t2}), 
(\ref{b+t+3half}),
[(\ref{b+u3half}) and (\ref{G12t1})] 
and (\ref{b+t2}) correspond to the above ones  located at each row, respectively. 
} % title of Table
\centering % used for centering table
\end{table}
%%%%%%%%%%%%%%%%%%%%%%%%%%%%%%%%%%%%%%%%%%%%%%%%%%%%%%%%%%%%%%%%%%%%%%

%%%%%%%%%%%%%%%%%%%%%%%%%%%%%%%%%%%%%%%%%%%%%%%%%%%%%%%%%%%%%%%%%%%%
\begin{table}[ht]
\centering % used for centering table
\begin{tabular}{|c||c| } % centered columns (4 columns)
\hline %inserts double horizontal lines
$U(1)$ charge & Composite fields of spin-$2$  
\\ [0.5ex] % inserts table
%heading
\hline \hline % inserts single horizontal line
%$\frac{4k}{(5+k)}$  & $(\hat{B}_1 - i \hat{B}_2)(\hat{B}_1 - i \hat{B}_2)$ 
%\\ % inserting body of the table
%\hline
$\frac{2(3+k)}{(5+k)}$ &  $\hat{A}_{-}
\hat{B}_{-}$ \\
\hline
$\frac{2k}{(5+k)}$ & $T^{(1)}\hat{B}_{-}, \quad
U_{+}^{(2)}, \quad \pa \hat{B}_{-}, \quad 
\hat{A}_3 \hat{B}_{-}, \quad \hat{B}_{-} 
\hat{B}_3$  \\
\hline
$\frac{2(-3+k)}{(5+k)} $ & $\hat{A}_{+}\hat{B}_{-}$  \\
\hline
%$\frac{12}{(5+k)}$ & $\hat{A}_{-}\hat{A}_{-}$  \\ 
%\hline
$\frac{6}{(5+k)}$ & $ T^{(1)}\hat{A}_{-}, \quad
V_{+}^{(2)}, \quad \pa \hat{A}_{-}, \quad 
\hat{A}_{-} \hat{A}_3, \quad \hat{A}_{-} 
\hat{B}_3$  \\ 
\hline
$0$ & $T^{(1)} \hat{A}_3, \quad T^{(1)} \hat{B}_3, \quad
\hat{T}, \quad T^{(2)}, \quad W^{(2)}, \quad T^{(1)} T^{(1)}, 
\quad \pa \hat{A}_3, \quad \pa \hat{B}_3, \quad \pa T^{(1)} $  \\ 
 & $\hat{A}_{+}\hat{A}_{-}, \quad
\hat{A}_3 \hat{A}_3, \quad \hat{A}_3 \hat{B}_3, \quad 
\hat{B}_{+}\hat{B}_{-}, \quad \hat{B}_3 
\hat{B}_3$  \\ 
\hline
$-\frac{6}{(5+k)} $ & $ T^{(1)}\hat{A}_{+}, \quad
U_{-}^{(2)}, \quad \pa \hat{A}_{+}, \quad 
\hat{A}_{+}\hat{A}_3, \quad \hat{A}_{+}
\hat{B}_3$  \\
\hline
%$-\frac{12}{(5+k)}$ & $(\hat{A}_1+i\hat{A}_2)(\hat{A}_1+i\hat{A}_2)$  \\ 
%\hline
$-\frac{2(-3+k)}{(5+k)}$ & $\hat{A}_{-}\hat{B}_{+}$  \\ 
\hline
$-\frac{2k}{(5+k)}$ & $T^{(1)}\hat{B}_{+}, \quad
V_{-}^{(2)}, \quad \pa \hat{B}_{+}, \quad 
\hat{A}_3 \hat{B}_{+}, \quad \hat{B}_{+} 
\hat{B}_3$  \\ 
\hline
$-\frac{2(3+k)}{(5+k)}$ & $\hat{A}_{+}
\hat{B}_{+}$  \\ 
%\hline
%$-\frac{4k}{(5+k)}$ & $(\hat{B}_1 + i \hat{B}_2)
%(\hat{B}_1 + i \hat{B}_2)$  \\ 
[1ex] % [1ex] adds vertical space
\hline %inserts single line
\end{tabular}
%\label{table:nonlin} % is used to refer this table in the text
\caption{The $U(1)$ charges for the spin-$2$ fields which can be obtained 
from the $U(1)$ charges in Table $2$.
 The fields appearing in the first-order pole in 
[(\ref{a-u+2}) and (\ref{G21t+3half})], 
[(\ref{G11t+3half}) and (\ref{G21u3half})], 
[(\ref{a+u+2}), (\ref{b-u-2}) and (\ref{G11u3half})],  
(\ref{G22t+3half}), 
[(\ref{a-u-2}),  
(\ref{G12t+3half}) and (\ref{G22u3half})], 
[(\ref{a+w2}) and (\ref{G12u3half})], 
(\ref{a-u+2}), 
(\ref{b+w2}) and
(\ref{b+u-2})  
correspond to the above ones respectively. One sees that 
the fields located at the first four rows have their conjugated ones 
in the last four rows. } % title of Table
\end{table}
%%%%%%%%%%%%%%%%%%%%%%%%%%%%%%%%%%%%%%%%%%%%%%%%%%%%%%%%%%%%%%%%%%%%%%%

%%%%%%%%%%%%%%%%%%%%%%%%%%%%%%%%%%%%%%%%%%%%%%%%%%%%%%%%%%%%%%%%%%%%%%%
\begin{table}[ht]
\centering % used for centering table
\begin{tabular}{|c||c| } % centered columns (4 columns)
\hline %inserts double horizontal lines
$U(1)$ charge & Composite fields of spin-$\frac{5}{2}$  
\\ [0.5ex] % inserts table
%heading
\hline \hline % inserts single horizontal line
$\frac{3(1+k)}{(5+k)}$  & $\hat{B}_{-} \hat{G}_{21}, \quad
\hat{B}_{-} T_{+}^{(\frac{3}{2})}
$ 
\\ % inserting body of the table
\hline
$\frac{3(-1+k)}{(5+k)}$ &  $
\hat{B}_{-} \hat{G}_{11}, \quad
\hat{B}_{-} U^{(\frac{3}{2})}
$ \\
\hline
$\frac{(9+k)}{(5+k)}$ & $\hat{A}_{-} \hat{G}_{21}, \quad
\hat{A}_{-} T_{+}^{(\frac{3}{2})}$  \\
\hline
$\frac{(3+k)}{(5+k)} $ & $W_{+}^{(\frac{5}{2})}, \quad \hat{A}_{-} 
\hat{G}_{11}, \quad \hat{A}_{-} U^{\frac{3}{2}}, \quad
\hat{A}_3 \hat{G}_{21}, \quad \hat{A}_3 T_{+}^{(\frac{3}{2})}, \quad
\hat{B}_{-} \hat{G}_{22}$  \\
& $\hat{B}_{-} V^{\frac{3}{2}}, \quad \hat{B}_3 \hat{G}_{21},
\quad \hat{B}_3 T_{+}^{(\frac{3}{2})}, \quad T^{(1)} \hat{G}_{21}, \quad
T^{(1)} T_{+}^{(\frac{3}{2})}, \quad \pa \hat{G}_{21}, \quad \pa
T_{+}^{(\frac{3}{2})} $ \\
\hline
$\frac{(-3+k)}{(5+k)}$ & $\hat{A}_{+}\hat{G}_{21}, \quad
\hat{A}_{+}T_{+}^{(\frac{3}{2})}, \quad U^{(\frac{5}{2})},
\quad \hat{A}_3 \hat{G}_{11}, \quad \hat{A}_3 U^{(\frac{3}{2})}, \quad
\hat{B}_{-} \hat{G}_{12}$  \\ 
& $\hat{B}_{-} T_{-}^{(\frac{3}{2})}, \quad \hat{B}_3 \hat{G}_{11},
\quad \hat{B}_3 U^{(\frac{3}{2})}, \quad T^{(1)} \hat{G}_{11}, \quad
T^{(1)} U^{(\frac{3}{2})}, \quad \pa \hat{G}_{11}, 
\quad \pa U^{(\frac{3}{2})}$ \\
\hline
$\frac{(-9+k)}{(5+k)}$ & $\hat{A}_{+}\hat{G}_{11}, \quad
\hat{A}_{+}U^{(\frac{3}{2})} $  \\ 
\hline
$-\frac{(-9+k)}{(5+k)}$ & $\hat{A}_{-} \hat{G}_{22}, \quad
\hat{A}_{-} V^{(\frac{3}{2})}  $  \\ 
\hline
$-\frac{(-3+k)}{(5+k)}$ & $
\hat{A}_{-} \hat{G}_{12}, \quad
\hat{A}_{-} T_{-}^{(\frac{3}{2})}, \quad V^{(\frac{5}{2})},
\quad \hat{A}_3 \hat{G}_{22}, \quad \hat{A}_3 V^{(\frac{3}{2})}, \quad
\hat{B}_{+} \hat{G}_{21}$  \\ 
& $\hat{B}_{+} T_{+}^{(\frac{3}{2})}, \quad \hat{B}_3 \hat{G}_{22},
\quad \hat{B}_3 V^{(\frac{3}{2})}, \quad T^{(1)} \hat{G}_{22}, \quad
T^{(1)} V^{(\frac{3}{2})}, \quad \pa \hat{G}_{22}, 
\quad \pa V^{(\frac{3}{2})}
$  
\\ 
\hline
$-\frac{(3+k)}{(5+k)}$ & $
W_{-}^{(\frac{5}{2})}, \quad \hat{A}_{+}
\hat{G}_{22}, \quad \hat{A}_{+}V^{\frac{3}{2}}, \quad
\hat{A}_3 \hat{G}_{12}, \quad \hat{A}_3 T_{-}^{(\frac{3}{2})}, \quad
\hat{B}_{+} \hat{G}_{11}$  \\
& $\hat{B}_{+} U^{\frac{3}{2}}, \quad \hat{B}_3 \hat{G}_{12},
\quad \hat{B}_3 T_{-}^{(\frac{3}{2})}, \quad T^{(1)} \hat{G}_{12}, \quad
T^{(1)} T_{-}^{(\frac{3}{2})}, \quad \pa \hat{G}_{12}, \quad \pa
T_{-}^{(\frac{3}{2})}
$  \\
\hline
$-\frac{(9+k)}{(5+k)}$ & 
$\hat{A}_{+}\hat{G}_{12}, \quad
\hat{A}_{+}T_{-}^{(\frac{3}{2})}$  \\ 
\hline
$-\frac{3(-1+k)}{(5+k)}$ & 
$\hat{B}_{+} \hat{G}_{22}, \quad
\hat{B}_{+} V^{(\frac{3}{2})}$  \\ 
\hline
$-\frac{3(1+k)}{(5+k)} $ & $ 
\hat{B}_{+} \hat{G}_{12}, \quad
\hat{B}_{+} T_{-}^{(\frac{3}{2})}$  \\ 
[1ex] % [1ex] adds vertical space
\hline %inserts single line
\end{tabular}
%\label{table:nonlin} % is used to refer this table in the text
\caption{The $U(1)$ charges for the spin-$\frac{5}{2}$ 
fields which can be obtained from the $U(1)$ charges in Table $2$.
 The fields appearing in the first-order pole in 
(\ref{G21u+2}), 
(\ref{G11u+2}), 
(\ref{G12u-2}), 
(\ref{G22u+2}), 
[(\ref{G11t2}), (\ref{G12u+2}), (\ref{G21u-2}) and (\ref{G11w2})], 
(\ref{G11u-2}), 
(\ref{G11u-2}), 
[(\ref{G22t2}), (\ref{G12v+2}) and (\ref{G21v-2})], 
[(\ref{G12t2}), (\ref{G22u-2}) and (\ref{G12w2})], 
(\ref{G12u-2}), 
(\ref{G11u+2})  
and 
(\ref{G21u+2}) correspond to the above 
ones respectively. Similarly, 
 the fields appearing in the first-order pole in 
(\ref{B-w+5half}), 
(\ref{B-u5half}), 
(\ref{A-w+5half}), 
[(\ref{A-u5half}) and (\ref{B-v5half})], 
[(\ref{A+w+5half}) and (\ref{B-w-5half})], 
(\ref{A+u5half}), 
(\ref{A-v5half}), 
[(\ref{B+w+5half}) and (\ref{A-w-5half})], 
[(\ref{B+u5half}) and (\ref{A+v5half})], 
(\ref{A-w+5half}), 
(\ref{B+v5half})  and 
(\ref{B-w+5half}) 
correspond to the above 
ones respectively.
The fields in the first six rows
are conjugated to each other in those in the last six rows.
} % title of Table
\end{table}
%%%%%%%%%%%%%%%%%%%%%%%%%%%%%%%%%%%%%%%%%%%%%%%%%%%%%%%%%%%%%%%%%%%%%%%%%%

%%%%%%%%%%%%%%%%%%%%%%%%%%%%%%%%%%%%%%%%%%%%%%%%%%%%%%%%%%%%%%%%%%%%%%%%%%%
\begin{table}[ht]
\centering % used for centering table
\begin{tabular}{|c||c| } % centered columns (4 columns)
\hline %inserts double horizontal lines
$U(1)$ charge & Composite fields of spin-$3$  
\\ [0.5ex] % inserts table
%heading
\hline \hline % inserts single horizontal line
$\frac{(6+2k)}{(5+k)}$  & $
\hat{A}_3  \hat{A}_{-} \hat{B}_{-}, \hat{B}_3  
\hat{A}_{-} \hat{B}_{-}, T^{(1)}  
\hat{A}_{-} \hat{B}_{-}$, 
 $\hat{A}_{-} U_{+}^{(2)}, \hat{A}_{-} \pa 
\hat{B}_{-}, \hat{B}_{-} V_{+}^{(2)}, 
\hat{B}_{-} \pa \hat{A}_{-} $, \\
& $\hat{G}_{21} \hat{G}_{21}, \hat{G}_{21} T_{+}^{(\frac{3}{2})}, 
T_{+}^{(\frac{3}{2})}  T_{+}^{(\frac{3}{2})} $, \\
\hline
$\frac{2k}{(5+k)}$ &  $ T^{(1)} U_{+}^{(2)}, T^{(1)} 
\pa \hat{B}_{-}, \hat{A}_3  U_{+}^{(2)},  \hat{A}_3 
\pa \hat{B}_{-},  \hat{B}_{-} \hat{T},   
\hat{B}_{-} T^{(2)}$,   $ \hat{B}_{-} W^{(2)},  \hat{B}_{-} 
\pa \hat{A}_3,  \hat{B}_{-} \pa \hat{B}_3$, \\
& $\hat{B}_{-} 
\pa T^{(1)}, \hat{B}_3  U_{+}^{(2)}$,  $  \hat{B}_3 \pa \hat{B}_{-}, 
\pa  U_{+}^{(2)},   \pa^2 \hat{B}_{-}, 
\hat{G}_{21} \hat{G}_{11}, \hat{G}_{21} U^{(\frac{3}{2})}, 
\hat{G}_{11} T_{+}^{(\frac{3}{2})}$,  \\
& $ T_{+}^{(\frac{3}{2})}  U^{(\frac{3}{2})}$,  
 $ \hat{A}_{-} \hat{A}_{+} \hat{B}_{-}, 
\hat{A}_3 \hat{A}_3  \hat{B}_{-},  \hat{A}_3 \hat{B}_3  
\hat{B}_{-}$, $   \hat{B}_{-}    \hat{B}_{-}   
\hat{B}_{+},  \hat{B}_3 \hat{B}_3  \hat{B}_{-},  
T^{(1)} \hat{A}_3   \hat{B}_{-}$, \\
&   $T^{(1)} \hat{B}_3   \hat{B}_{-}, T^{(1)} T^{(1)} 
\hat{B}_{-} $, \\
\hline
$\frac{(-6+2k)}{(5+k)}$ &   $\hat{A}_3  \hat{A}_{+} 
\hat{B}_{-}, \hat{B}_3  
\hat{A}_{+} \hat{B}_{-}, T^{(1)}  
\hat{A}_{+} 
\hat{B}_{-}$, 
$\hat{B}_{-} U_{-}^{(2)}, \hat{B}_{-} \pa 
\hat{A}_{+}, \hat{A}_{+} U_{+}^{(2)}, 
\hat{A}_{+} \pa \hat{B}_{-} $, \\
& $\hat{G}_{11} \hat{G}_{11}, \hat{G}_{11} U^{(\frac{3}{2})}, 
U^{(\frac{3}{2})}  U^{(\frac{3}{2})} $
\\
\hline
$\frac{6}{(5+k)}$ &   $ T^{(1)} V_{+}^{(2)}, T^{(1)} \pa 
\hat{A}_{-}, \hat{A}_{-} \hat{T}, 
\hat{A}_{-} T^{(2)}, \hat{A}_{-} W^{(2)}$,
 $ \hat{A}_{-} \pa \hat{A}_3,  
\hat{A}_{-} \pa \hat{B}_3,  \hat{A}_{-}\pa T^{(1)}$, \\ 
& $\hat{A}_3 V_{+}^{(2)}, \pa
V_{+}^{(2)},  \pa^2 \hat{A}_{-}$,  
$\hat{G}_{21} \hat{G}_{22}, \hat{G}_{21} V^{(\frac{3}{2})}, 
\hat{G}_{22} T_{+}^{(\frac{3}{2})}, T_{+}^{(\frac{3}{2})} V^{(\frac{3}{2})},
 \hat{A}_{-} \hat{A}_{+}  
\hat{A}_{-}$, \\
& $  \hat{A}_{-} \hat{A}_3 \hat{A}_3,  
\hat{A}_{-} \hat{A}_3 \hat{B}_3,  
\hat{A}_{-} \hat{B}_{+} 
\hat{B}_{-}$,  $\hat{A}_{-} \hat{B}_3 \hat{B}_3,  \hat{B}_3 V_{+}^{(2)},  
\hat{A}_{-} T^{(1)} \hat{A}_3,  
\hat{A}_{-} T^{(1)} \hat{B}_3$,  \\
& $\hat{A}_{-} T^{(1)} T^{(1)}, \hat{A}_3 \pa  
\hat{A}_{-}, \hat{B}_3 \pa  
\hat{A}_{-}$ \\
\hline
$0$ & $W^{(3)}, T^{(1)} \hat{T},  T^{(1)} T^{(2)},
 T^{(1)} W^{(2)},  T^{(1)} \pa \hat{A}_3,  T^{(1)} \pa \hat{B}_3,  T^{(1)}
\pa T^{(1)}, T^{(1)} T^{(1)} T^{(1)}$, \\
& $\hat{A}_{+} V_{+}^{(2)},  \hat{A}_{+}\pa 
\hat{A}_{-}, \hat{A}_{-}  U_{-}^{(2)},  
\hat{A}_{-} \pa 
\hat{A}_{+}$,  $\hat{A}_3 \hat{T},\hat{A}_3 T^{(2)}, \hat{A}_3 W^{(2)}$, 
 $T_{+}^{(\frac{3}{2})} T_{-}^{(\frac{3}{2})}$
\\ 
& $\hat{A}_3 \pa 
\hat{A}_3, \hat{A}_3 \pa \hat{B}_3, \hat{A}_3 \pa T^{(1)},
\hat{B}_{+}  U_{+}^{(2)}$,  $  \hat{B}_{+} \pa 
\hat{B}_{-}, \hat{B}_{-}  V_{-}^{(2)},  
\hat{B}_{-} \pa 
\hat{B}_{+}, \hat{B}_3 \hat{T}, 
\hat{B}_3 T^{(2)}$, \\
& $ \hat{B}_3 W^{(2)}, \hat{B}_3 \pa 
\hat{A}_3, \hat{B}_3 \pa \hat{B}_3, \hat{B}_3 \pa T^{(1)},
\pa \hat{T}, \pa T^{(2)}, \pa W^{(2)}, T^{(1)} \pa T^{(1)},
\pa^2 \hat{A}_3, \pa^2 \hat{B}_3, \pa^2 T^{(1)}$, \\
& $\pa T^{(1)}  \hat{A}_3, \pa T^{(1)}  \hat{B}_3, \hat{G}_{11} 
\hat{G}_{22}, \hat{G}_{11} V^{(\frac{3}{2})}, \hat{G}_{12} \hat{G}_{21},
\hat{G}_{12} T_{+}^{(\frac{3}{2})}, \hat{G}_{21} T_{-}^{(\frac{3}{2})},
\hat{G}_{22}  U^{(\frac{3}{2})},  U^{(\frac{3}{2})}  V^{(\frac{3}{2})}$, \\
& $\hat{A}_{+}\hat{A}_{-} 
\hat{A}_3, \hat{A}_{+}\hat{A}_{-} 
\hat{B}_3, \hat{A}_3 \hat{A}_3 \hat{A}_3, \hat{A}_3 \hat{A}_3 \hat{B}_3,
\hat{A}_3 \hat{B}_3 \hat{B}_3$,  $ \hat{B}_{+} \hat{B}_{-} 
\hat{B}_3, \hat{B}_{+} \hat{B}_{-} 
\hat{A}_3, \hat{B}_3 \hat{B}_3 \hat{B}_3$, \\
& $T^{(1)}  \hat{A}_{+}\hat{A}_{-},
T^{(1)} \hat{A}_3 \hat{A}_3, T^{(1)} \hat{A}_3 \hat{B}_3, T^{(1)}
 \hat{B}_{+} \hat{B}_{-}$, 
$T^{(1)} \hat{B}_3 \hat{B}_3, T^{(1)} T^{(1)} \hat{A}_3, T^{(1)} T^{(1)} 
\hat{B}_3$, \\
 \hline % inserts single horizontal line
$-\frac{6}{(5+k)}$ &   $ T^{(1)} U_{-}^{(2)}, T^{(1)} \pa 
\hat{A}_{+}, \hat{A}_{+} \hat{T}, 
\hat{A}_{+} T^{(2)}, 
\hat{A}_{+} W^{(2)}$,
 $ \hat{A}_{+} \pa \hat{A}_3,  
\hat{A}_{+} \pa \hat{B}_3,  
\hat{A}_{+}\pa T^{(1)}$, \\
& $\hat{A}_3 U_{-}^{(2)}, \pa
U_{-}^{(2)},  \pa^2 \hat{A}_{+}$,  
$\hat{G}_{12} \hat{G}_{11}, \hat{G}_{12} U^{(\frac{3}{2})}, 
\hat{G}_{11} T_{-}^{(\frac{3}{2})}, T_{-}^{(\frac{3}{2})} U^{(\frac{3}{2})},
 \hat{A}_{+} \hat{A}_{-}  
\hat{A}_{+}$, \\
& $  \hat{A}_{+} \hat{A}_3 \hat{A}_3,  
\hat{A}_{+} \hat{A}_3 \hat{B}_3,  
\hat{A}_{+} \hat{B}_{-} 
\hat{B}_{+}$,  $\hat{A}_{+} 
\hat{B}_3 \hat{B}_3,  \hat{B}_3 U_{-}^{(2)}$, \\
&  
$\hat{A}_{+} T^{(1)} \hat{A}_3,  
\hat{A}_{+} T^{(1)} \hat{B}_3$,  $\hat{A}_{+} T^{(1)} T^{(1)}, \hat{A}_3 \pa  
\hat{A}_{+}, \hat{B}_3 \pa  \hat{A}_{+}$ \\
\hline
$-\frac{(-6+2k)}{(5+k)}$ & $
\hat{A}_3  \hat{A}_{-} 
\hat{B}_{+}, \hat{B}_3  
\hat{A}_{-} 
\hat{B}_{+}, T^{(1)}  
\hat{A}_{-} \hat{B}_{+}$, 
 $\hat{B}_{+} V_{+}^{(2)}, 
\hat{B}_{+} \pa \hat{A}_{-}$, \\ 
& $\hat{A}_{-} V_{-}^{(2)}, 
\hat{A}_{-} \pa 
\hat{B}_{+} $, 
 $\hat{G}_{22} \hat{G}_{22}, \hat{G}_{22} V^{(\frac{3}{2})}, 
V^{(\frac{3}{2})}  V^{(\frac{3}{2})}
 $  \\
\hline
$-\frac{2k}{(5+k)} $ & $T^{(1)} V_{-}^{(2)}, T^{(1)} 
\pa \hat{B}_{+}, \hat{A}_3  V_{-}^{(2)},  \hat{A}_3 
\pa \hat{B}_{+},    \hat{B}_{+} T^{(2)}$,   $ \hat{B}_{+} W^{(2)},  \hat{B}_{+} 
\pa \hat{A}_3,  \hat{B}_{+} \pa \hat{B}_3$, 
$\hat{B}_{+} \pa T^{(1)}$, \\
& $ \hat{B}_3  V_{-}^{(2)},  \hat{B}_{+} 
\hat{T}$, 
$  \hat{B}_3 \pa \hat{B}_{+}, \pa  V_{-}^{(2)}, 
\hat{G}_{12} \hat{G}_{22}, \hat{G}_{12} V^{(\frac{3}{2})}, 
\hat{G}_{22} T_{-}^{(\frac{3}{2})},  T_{-}^{(\frac{3}{2})}  V^{(\frac{3}{2})}$,   
$ \hat{A}_{+} \hat{A}_{-} \hat{B}_{+}$, \\
& $ \hat{A}_3 
\hat{A}_3  \hat{B}_{+},   \pa^2 
\hat{B}_{+},\hat{A}_3 \hat{B}_3  
\hat{B}_{+}$,  $   \hat{B}_{+}    \hat{B}_{+}   
\hat{B}_{-},  \hat{B}_3 \hat{B}_3  
\hat{B}_{+}$,  $T^{(1)} \hat{A}_3   
\hat{B}_{+}$,    $T^{(1)} \hat{B}_3   \hat{B}_{+}, T^{(1)} T^{(1)} 
\hat{B}_{+} $, \\
\hline
$-\frac{(6+2k)}{(5+k)}$
& $\hat{A}_3  \hat{A}_{+} 
\hat{B}_{+}, \hat{B}_3  
\hat{A}_{+} \hat{B}_{+}, T^{(1)}  
\hat{A}_{+} \hat{B}_{+}$, 
 $\hat{A}_{+} V_{-}^{(2)}, \hat{A}_{+} 
\pa \hat{B}_{+}$, \\
& $\hat{B}_{+} U_{-}^{(2)}, 
\hat{B}_{+} \pa \hat{A}_{+} $,  
$\hat{G}_{12} \hat{G}_{12}, \hat{G}_{12} T_{-}^{(\frac{3}{2})}, 
T_{-}^{(\frac{3}{2})}  T_{-}^{(\frac{3}{2})}$ \\
[1ex] % [1ex] adds vertical space
\hline %inserts single line
\end{tabular}
%\label{table:nonlin} % is used to refer this table in the text
\caption{The $U(1)$ charges for the spin-$3$ 
fields which can be obtained from the 
$U(1)$ charges in Table $2$. 
The fields appearing in the first-order pole in Appendix (\ref{g21w+5half}), 
[(\ref{g21u5half}) and (\ref{g11w+5half})], 
(\ref{g11u5half}), [(\ref{A-w3}), 
(\ref{g21v5half}) and (\ref{g22w+5half})], 
[(\ref{g22u5half}), (\ref{g11v5half}), (\ref{g12w+5half}) and
(\ref{g21w-5half})], 
[(\ref{A+w3}), 
(\ref{g12u5half}) and 
(\ref{g11w-5half})], 
(\ref{g22v5half}), 
[(\ref{B+w3}), (\ref{g12v5half}) and (\ref{g22w-5half})] 
and (\ref{g12w-5half})
correspond to the above ones respectively.
 } % title of Table
\end{table}
%%%%%%%%%%%%%%%%%%%%%%%%%%%%%%%%%%%%%%%%%%%%%%%%%%%%%%%%%%%%%%%%%%%%%%%%%%

%%%%%%%%%%%%%%%%%%%%%%%%%%%%%%%%%%%%%%%%%%%%%%%%%%%%%%%%%%%%%%%%%%%
\begin{table}[ht]
\centering % used for centering table
\begin{tabular}{|c||c| } % centered columns (4 columns)
\hline %inserts double horizontal lines
$U(1)$ charge & Composite fields of spin-$\frac{7}{2}$  
\\ [0.5ex] % inserts table
%heading
\hline \hline % inserts single horizontal line
$\frac{(3+k)}{(5+k)}$  & 
 $ \hat{A}_{-} \hat{A}_{+}  \hat{G}_{21},
\hat{A}_{-}\hat{A}_{+} T_{+}^{(\frac{3}{2})},
\hat{A}_{-} U^{(\frac{5}{2})},
$ 
 $ \hat{A}_{-} \hat{A}_3 \hat{G}_{11}, \hat{A}_{-}
\hat{A}_3 U^{(\frac{3}{2})}, \hat{A}_{-}\hat{B}_{-}
\hat{G}_{12} $, \\
& $ \hat{A}_{-} \hat{B}_{-} 
T_{-}^{(\frac{3}{2})}, \hat{A}_{-} \hat{B}_3 \hat{G}_{11},
\hat{A}_{-} \hat{B}_3 U^{(\frac{3}{2})},
\hat{A}_{-} \pa \hat{G}_{11}$, 
 $\hat{A}_{-} \pa U^{(\frac{3}{2})},
\hat{B}_{+}\hat{B}_{-}\hat{G}_{21},
\hat{B}_{+}\hat{B}_{-} T_{+}^{(\frac{3}{2})} $, \\
& $\hat{B}_{-} V^{(\frac{5}{2})}, \hat{B}_{-}
\hat{A}_3 \hat{G}_{22}, \hat{B}_{-} \hat{A}_3 V^{(\frac{3}{2})},
\hat{B}_{-} \hat{B}_3 \hat{G}_{22}$, \\
& $\hat{B}_{-} \hat{B}_3 V^{(\frac{3}{2})},
\hat{B}_{-} T^{(1)} \hat{G}_{22},
\hat{B}_{-} T^{(1)} V^{(\frac{3}{2})},
 \hat{B}_{-} \pa \hat{G}_{22}
$, \\
& $\hat{B}_{-} \pa  V^{(\frac{3}{2})},  
\hat{A}_3 W_{+}^{(\frac{5}{2})}, \hat{A}_3 \hat{A}_3 \hat{G}_{21},
\hat{A}_3 \hat{A}_3 T_{+}^{(\frac{3}{2})},
\hat{A}_3 \hat{B}_3 \hat{G}_{21}, \hat{A}_3 
\hat{B}_3 T_{+}^{(\frac{3}{2})}$, \\
& $\hat{A}_3 \pa \hat{G}_{21}, \hat{A}_3 \pa T_{+}^{(\frac{3}{2})},
\hat{B}_3 W_{+}^{(\frac{5}{2})},  \hat{B}_3 \hat{B}_3 \hat{G}_{21},
\hat{B}_3 \hat{B}_3 T_{+}^{(\frac{3}{2})}, \hat{B}_3 T^{(1)} 
\hat{G}_{21},  \hat{B}_3 T^{(1)} 
T_{+}^{(\frac{3}{2})}   $, \\
& $\hat{B}_3 \pa \hat{G}_{21}, \hat{B}_3 \pa T_{+}^{(\frac{3}{2})},
T^{(1)}  W_{+}^{(\frac{5}{2})}, T^{(1)} \hat{A}_{-} \hat{G}_{11},
T^{(1)} \hat{A}_{-} U^{(\frac{3}{2})}, T^{(1)} \hat{A}_3  
\hat{G}_{21}$, \\
& $ T^{(1)} \hat{A}_3 T_{+}^{(\frac{3}{2})}, T^{(1)} T^{(1)} \hat{G}_{21},
T^{(1)} T^{(1)} T_{+}^{(\frac{3}{2})}, T^{(1)} \pa \hat{G}_{21}, T^{(1)} 
\pa T_{+}^{(\frac{3}{2})}, \pa  W_{+}^{(\frac{5}{2})}, \pa \hat{A}_{-}  
\hat{G}_{11}$, \\
& $\pa \hat{A}_{-}  U^{(\frac{3}{2})},
\pa \hat{A}_3  \hat{G}_{21}, \pa \hat{A}_3  T_{+}^{(\frac{3}{2})},
\pa \hat{B}_{-}  
\hat{G}_{22}, \pa \hat{B}_{-}  V^{(\frac{3}{2})},
\pa \hat{B}_3  \hat{G}_{21}$, \\ 
& $
\pa \hat{B}_3  T_{+}^{(\frac{3}{2})}, \pa T^{(1)}  \hat{G}_{21}, \pa T^{(1)} 
T_{+}^{(\frac{3}{2})}, \pa^2 \hat{G}_{21}, \pa^2 T_{+}^{(\frac{3}{2})}, \hat{G}_{21}
\hat{T}, \hat{G}_{21} T^{(2)}, \hat{G}_{21} W^{(2)}, T_{+}^{(\frac{3}{2})} 
\hat{T}$, 
\\
& $ T_{+}^{(\frac{3}{2})} T^{(2)},  T_{+}^{(\frac{3}{2})} W^{(2)}, \hat{G}_{11}
V_{+}^{(2)}, U^{(\frac{3}{2})} V_{+}^{(2)},
 \hat{G}_{22}
U_{+}^{(2)}, V^{(\frac{3}{2})} U_{+}^{(2)}$ \\
\hline
$-\frac{(3+k)}{(5+k)} $ & $ 
\hat{A}_{+}\hat{A}_{-} \hat{G}_{12},
\hat{A}_{+}\hat{A}_{-} T_{-}^{(\frac{3}{2})},
\hat{A}_{+} V^{(\frac{5}{2})},
$, $ \hat{A}_{+} \hat{A}_3 \hat{G}_{22}, \hat{A}_{+}
\hat{A}_3 V^{(\frac{3}{2})}, \hat{A}_{+}\hat{B}_{+}
\hat{G}_{21} $, \\
& $\hat{A}_{+}\hat{B}_{+} 
T_{+}^{(\frac{3}{2})}, \hat{A}_{+} \hat{B}_3 \hat{G}_{22},
\hat{A}_{+} \hat{B}_3 V^{(\frac{3}{2})},
\hat{A}_{+} \pa \hat{G}_{22}$,  $\hat{A}_{+} \pa V^{(\frac{3}{2})},
\hat{B}_{-}\hat{B}_{+}\hat{G}_{12},
\hat{B}_{-}\hat{B}_{+} T_{-}^{(\frac{3}{2})} $, \\
& $\hat{B}_{+} U^{(\frac{5}{2})}, \hat{B}_{+}
\hat{A}_3 \hat{G}_{11}, \hat{B}_{+} \hat{A}_3 U^{(\frac{3}{2})},
\hat{B}_{+} \hat{B}_3 \hat{G}_{11}$, \\
& $\hat{B}_{+} \hat{B}_3 U^{(\frac{3}{2})},
\hat{B}_{+} T^{(1)} \hat{G}_{11},
\hat{B}_{+} T^{(1)} U^{(\frac{3}{2})},
 \hat{B}_{+} \pa \hat{G}_{11}
$, \\
& $\hat{B}_{+} \pa  U^{(\frac{3}{2})},  
\hat{A}_3 W_{-}^{(\frac{5}{2})}, \hat{A}_3 \hat{A}_3 \hat{G}_{12},
\hat{A}_3 \hat{A}_3 T_{-}^{(\frac{3}{2})},
\hat{A}_3 \hat{B}_3 \hat{G}_{12}, \hat{A}_3 
\hat{B}_3 T_{-}^{(\frac{3}{2})}$, \\
& $\hat{A}_3 \pa \hat{G}_{12}, \hat{A}_3 \pa T_{-}^{(\frac{3}{2})},
\hat{B}_3 W_{-}^{(\frac{5}{2})},  \hat{B}_3 \hat{B}_3 \hat{G}_{12},
\hat{B}_3 \hat{B}_3 T_{-}^{(\frac{3}{2})}, \hat{B}_3 T^{(1)} 
\hat{G}_{12},  \hat{B}_3 T^{(1)} 
T_{-}^{(\frac{3}{2})}   $, \\
& $\hat{B}_3 \pa \hat{G}_{12}, \hat{B}_3 \pa T_{-}^{(\frac{3}{2})},
T^{(1)}  W_{-}^{(\frac{5}{2})}, T^{(1)} \hat{A}_{+} \hat{G}_{22},
T^{(1)} \hat{A}_{+} V^{(\frac{3}{2})}, T^{(1)} \hat{A}_3  
\hat{G}_{12}$, \\
& $ T^{(1)} \hat{A}_3 T_{-}^{(\frac{3}{2})}, T^{(1)} T^{(1)} \hat{G}_{12},
T^{(1)} T^{(1)} T_{-}^{(\frac{3}{2})}, T^{(1)} \pa \hat{G}_{12}, T^{(1)} 
\pa T_{-}^{(\frac{3}{2})}, \pa  W_{-}^{(\frac{5}{2})}, \pa \hat{A}_{+}  
\hat{G}_{22}$, \\
& $\pa \hat{A}_{+}  V^{(\frac{3}{2})},
\pa \hat{A}_3  \hat{G}_{12}, \pa \hat{A}_3  T_{-}^{(\frac{3}{2})},
\pa \hat{B}_{+}  
\hat{G}_{11}, \pa \hat{B}_{+}  U^{(\frac{3}{2})},
\pa \hat{B}_3  \hat{G}_{12}$, \\ 
& $
\pa \hat{B}_3  T_{-}^{(\frac{3}{2})}, \pa T^{(1)}  \hat{G}_{12}, \pa T^{(1)} 
T_{-}^{(\frac{3}{2})}, \pa^2 \hat{G}_{12}, \pa^2 T_{-}^{(\frac{3}{2})}, \hat{G}_{12}
\hat{T}, \hat{G}_{12} T^{(2)}, \hat{G}_{12} W^{(2)}, T_{-}^{(\frac{3}{2})} 
\hat{T}$, 
\\
& $ T_{-}^{(\frac{3}{2})} T^{(2)},  T_{-}^{(\frac{3}{2})} W^{(2)}, \hat{G}_{22}
U_{-}^{(2)}, V^{(\frac{3}{2})} U_{-}^{(2)},
 \hat{G}_{11}
V_{-}^{(2)}, U^{(\frac{3}{2})} V_{-}^{(2)}
$  \\ 
[1ex] % [1ex] adds vertical space
\hline %inserts single line
\end{tabular}
%\label{table:nonlin} % is used to refer this table in the text
\caption{The $U(1)$ charges for the spin-$\frac{7}{2}$ 
fields.  The fields appearing in the first-order pole in 
Appendix (\ref{g21w3}) and (\ref{g12w3}) correspond to the above ones 
respectively. The fields in the first row are conjugated to those in the 
second row.  } % title of Table
\end{table}
%%%%%%%%%%%%%%%%%%%%%%%%%%%%%%%%%%%%%%%%%%%%%%%%%%%%%%%%%%%%%%%%%

%%%%%%%%%%%%%%%%%%%%%%%%%%%%%%%%%%%%%%%%%%%%%%%%%%%%%%%%%%%%%%%%%%
\begin{table}[ht]
\centering % used for centering table
\begin{tabular}{|c||c| } % centered columns (4 columns)
\hline %inserts double horizontal lines
$U(1)$ charge & Composite fields of spin-$\frac{7}{2}$  

\\ [0.5ex] % inserts table
%heading
\hline \hline % inserts single horizontal line
%$\frac{(3+k)}{(5+k)}$  & $
%$
%\\ % inserting body of the table
%\hline
$\frac{(-3+k)}{(5+k)}$ &  $ \hat{A}_{-}\hat{A}_{+}
\hat{G}_{11},  \hat{A}_{-}\hat{A}_{+} 
U^{(\frac{3}{2})}, 
 \hat{B}_{+}\hat{B}_{-} 
\hat{G}_{11}$,   $\hat{B}_{+}\hat{B}_{-} 
U^{(\frac{3}{2})}, 
 \hat{B}_{-}\hat{A}_{+}
\hat{G}_{22},  \hat{B}_{-}\hat{A}_{+} 
V^{(\frac{3}{2})}$,
\\
& $\hat{B}_{-} W_{-}^{(\frac{5}{2})}, \hat{B}_{-} \hat{A}_3
\hat{G}_{12},  \hat{B}_{-} \hat{A}_3 
T_{-}^{(\frac{3}{2})}, 
 \hat{B}_{-} \hat{B}_3 \hat{G}_{12}
$, 
 $\hat{B}_{-} \hat{B}_3 T_{-}^{(\frac{3}{2})}, 
\hat{B}_{-} T^{(1)} \hat{G}_{12}$, \\
& $\hat{B}_{-} T^{(1)} T_{-}^{(\frac{3}{2})},
\hat{B}_{-} \pa \hat{G}_{12}$, 
 $ \hat{B}_{-} \pa T_{-}^{(\frac{3}{2})}, \hat{A}_{+}
W_{+}^{(\frac{5}{2})}, \hat{A}_{+} \hat{A}_3 \hat{G}_{21},
\hat{A}_{+} \hat{A}_3 T_{+}^{(\frac{3}{2})}$ \\
 & $ \hat{A}_{+} \hat{B}_3 \hat{G}_{21},
\hat{A}_{+} \hat{B}_3 T_{+}^{(\frac{3}{2})},
 \hat{A}_{+} T^{(1)} \hat{G}_{21},
\hat{A}_{+} T^{(1)} T_{+}^{(\frac{3}{2})}$,
 $\hat{A}_{+} \pa \hat{G}_{21}, \hat{A}_{+} \pa
T_{+}^{(\frac{3}{2})}$, \\
& $\hat{A}_3 U^{(\frac{5}{2})}, \hat{A}_3 \hat{A}_3 \hat{G}_{11},
 \hat{A}_3 \hat{A}_3 U^{(\frac{3}{2})},  \hat{A}_3 \hat{B}_3 \hat{G}_{11}$, 
 $ \hat{A}_3 \hat{B}_3 U^{(\frac{3}{2})},  \hat{A}_3 T^{(1)} \hat{G}_{11}$,
\\
 & 
$\hat{A}_3 T^{(1)} U^{(\frac{3}{2})}, \hat{A}_3 \pa \hat{G}_{11}, \hat{A}_3
\pa U^{(\frac{3}{2})},  \hat{B}_3 U^{(\frac{5}{2})}, \hat{B}_3 \hat{B}_3
\hat{G}_{11} $,  $ \hat{B}_3 \hat{B}_3 U^{(\frac{3}{2})},
 \hat{B}_3 T^{(1)} \hat{G}_{11}$, \\
& $\hat{B}_3 T^{(1)} U^{(\frac{3}{2})}, \hat{B}_3 \pa \hat{G}_{11}, \hat{B}_3
\pa U^{(\frac{3}{2})}, T^{(1)} U^{(\frac{5}{2})},
T^{(1)} T^{(1)} \hat{G}_{11}
$, \\
& $T^{(1)} T^{(1)} U^{(\frac{3}{2})}, T^{(1)} \pa \hat{G}_{11}, T^{(1)} \pa 
 U^{(\frac{3}{2})}, \pa \hat{A}_{+} \hat{G}_{21}, \pa 
\hat{A}_{+} T_{+}^{(\frac{3}{2})}$, \\
& $ \pa \hat{B}_{-} \hat{G}_{12}, \pa 
\hat{B}_{-} T_{-}^{(\frac{3}{2})}, \pa \hat{B}_3 
\hat{G}_{11}, \pa \hat{B}_3 U^{(\frac{3}{2})}, 
\pa T^{(1)} 
\hat{G}_{11}, \pa T^{(1)} U^{(\frac{3}{2})}
$, \\
& $ 
\pa^2 \hat{G}_{11}, \pa^2  U^{(\frac{3}{2})},  \pa  U^{(\frac{5}{2})}, 
\hat{G}_{21} U_{-}^{(2)}, T_{+}^{(\frac{3}{2})} U_{-}^{(2)},
\hat{G}_{12} U_{+}^{(2)}, T_{-}^{(\frac{3}{2})} U_{+}^{(2)},
\hat{G}_{11} \hat{T}, \hat{G}_{11} T^{(2)}$, \\
& $\hat{G}_{11} W^{(2)}, \hat{G}_{11} \pa \hat{A}_3,
   U^{(\frac{3}{2})} \hat{T}, 
 U^{(\frac{3}{2})} T^{(2)},  U^{(\frac{3}{2})} W^{(2)},  U^{(\frac{3}{2})}
\pa \hat{A}_3 $ \\
\hline
$-\frac{(-3+k)}{(5+k)}$ & $
 \hat{A}_{+}\hat{A}_{-}
\hat{G}_{22},  \hat{A}_{+}\hat{A}_{-} 
V^{(\frac{3}{2})}, 
 \hat{B}_{-}\hat{B}_{+}
\hat{G}_{22}$, 
   $\hat{B}_{-}\hat{B}_{+} 
V^{(\frac{3}{2})}, 
 \hat{B}_{+}\hat{A}_{-}
\hat{G}_{11},  \hat{B}_{+}\hat{A}_{-} 
U^{(\frac{3}{2})}$,
\\
& $\hat{B}_{+} W_{+}^{(\frac{5}{2})}, \hat{B}_{+} \hat{A}_3
\hat{G}_{21},  \hat{B}_{+} \hat{A}_3 
T_{+}^{(\frac{3}{2})}, 
 \hat{B}_{+} \hat{B}_3 \hat{G}_{21}
$, 
 $\hat{B}_{+} \hat{B}_3 T_{+}^{(\frac{3}{2})}, 
\hat{B}_{+} T^{(1)} \hat{G}_{21}$, \\
&
 $\hat{B}_{+} T^{(1)} T_{+}^{(\frac{3}{2})},
\hat{B}_{+} \pa \hat{G}_{21}$, 
 $ \hat{B}_{+} \pa T_{+}^{(\frac{3}{2})}, \hat{A}_{-}
W_{-}^{(\frac{5}{2})}, \hat{A}_{-} \hat{A}_3 \hat{G}_{12},
\hat{A}_{-} \hat{A}_3 T_{-}^{(\frac{3}{2})}$ \\
 & $ \hat{A}_{-} \hat{B}_3 \hat{G}_{12},
\hat{A}_{-} \hat{B}_3 T_{-}^{(\frac{3}{2})},
 \hat{A}_{-} T^{(1)} \hat{G}_{12},
\hat{A}_{-} T^{(1)} T_{-}^{(\frac{3}{2})}$ \\
& $\hat{A}_{-} \pa \hat{G}_{12}, \hat{A}_{-} \pa
T_{-}^{(\frac{3}{2})}, \hat{A}_3 V^{(\frac{5}{2})}, \hat{A}_3 \hat{A}_3 \hat{G}_{22},
 \hat{A}_3 \hat{A}_3 V^{(\frac{3}{2})},  \hat{A}_3 \hat{B}_3 \hat{G}_{22}$, \\
& $ \hat{A}_3 \hat{B}_3 V^{(\frac{3}{2})},  \hat{A}_3 T^{(1)} \hat{G}_{22},
\hat{A}_3 T^{(1)} V^{(\frac{3}{2})}, \hat{A}_3 \pa \hat{G}_{22}, \hat{A}_3
\pa V^{(\frac{3}{2})},  \hat{B}_3 V^{(\frac{5}{2})}, \hat{B}_3 \hat{B}_3
\hat{G}_{22} $, \\
& $ \hat{B}_3 \hat{B}_3 V^{(\frac{3}{2})},
 \hat{B}_3 T^{(1)} \hat{G}_{22},
\hat{B}_3 T^{(1)} V^{(\frac{3}{2})}, \hat{B}_3 \pa \hat{G}_{22}, \hat{B}_3
\pa V^{(\frac{3}{2})}, T^{(1)} V^{(\frac{5}{2})},
T^{(1)} T^{(1)} \hat{G}_{22}
$, \\
& $T^{(1)} T^{(1)} V^{(\frac{3}{2})}, T^{(1)} \pa \hat{G}_{22}, T^{(1)} \pa 
 V^{(\frac{3}{2})}, \pa \hat{A}_{-} \hat{G}_{12}, \pa 
\hat{A}_{-} T_{-}^{(\frac{3}{2})}$, \\
& $ \pa \hat{B}_{+} \hat{G}_{21}, \pa 
\hat{B}_{+} T_{+}^{(\frac{3}{2})}, \pa \hat{B}_3 
\hat{G}_{22}, \pa \hat{B}_3 V^{(\frac{3}{2})}, 
\pa T^{(1)} 
\hat{G}_{22}, \pa T^{(1)} V^{(\frac{3}{2})}
$, \\
& $ 
\pa^2 \hat{G}_{22}, \pa^2  V^{(\frac{3}{2})},  \pa  V^{(\frac{5}{2})}, 
\hat{G}_{12} V_{+}^{(2)}, T_{-}^{(\frac{3}{2})} V_{+}^{(2)},
\hat{G}_{21} V_{-}^{(2)}, T_{+}^{(\frac{3}{2})} V_{-}^{(2)},
\hat{G}_{22} \hat{T}, \hat{G}_{22} T^{(2)}$, \\
& $\hat{G}_{22} W^{(2)}, \hat{G}_{22} \pa \hat{A}_3, 
   V^{(\frac{3}{2})} \hat{T}, 
 V^{(\frac{3}{2})} T^{(2)},  V^{(\frac{3}{2})} W^{(2)},  V^{(\frac{3}{2})}
\pa \hat{A}_3
 $  \\
%\hline
%$-\frac{(3+k)}{(5+k)} $ & $ $  \\ 
[1ex] % [1ex] adds vertical space
\hline %inserts single line
\end{tabular}
%\label{table:nonlin} % is used to refer this table in the text
\caption{The $U(1)$ charges for the spin-$\frac{7}{2}$ 
fields. The fields appearing in the first-order pole in 
Appendix (\ref{g11w3}) and (\ref{g22w3}) correspond to the above ones
respectively. The fields in two rows are conjugated to 
each other.} % title of Table
\end{table}
%%%%%%%%%%%%%%%%%%%%%%%%%%%%%%%%%%%%%%%%%%%%%%%%%%%%%%%%%%%%%%%%%%

%%%%%%%%%%%%%%%%%%%%%%%%%%%%%%%%%%%%%%%%%%%%%%%%%%%%%%%%%%%%%%%%%%%%%%%%%%%%%%%
%%%%%%%%%%%%%%%%%%%%%%%%%%%%%%%%%%%%%%%%%%%%%%%%%%%%%%%%%%%%%%%%%%%%%%%%%%%%%%%%
\section{Conclusions and outlook }
%5%%%%%%%%%%%%%%%%%%%%%%%%%%%%%%%%%%%%%%%%%%%%%%%%%%%%%%%%%%%%%%%%%%%%%%%%%%%%%%%
%%%%%%%%%%%%%%%%%%%%%%%%%%%%%%%%%%%%%%%%%%%%%%%%%%%%%%%%%%%%%%%%%%%%%%%%%%%%%%%%

As in an abstract,
the $16$ higher spin currents given in (\ref{16}) and $11$ currents of 
large ${\cal N}=4$ nonlinear algebra are found explicitly in the Wolf space
coset model $\frac{SU(5)}{SU(3) \times SU(2) \times U(1)}$. 
Part of this extended large ${\cal N}=4$ nonlinear algebra
are described in Appendices $A$, $B$ and $C$. 
If one says the last statement of the abstract precisely, 
the $(11+16)$ currents commute with the ${\cal N}=2$ (effectively ${\cal N}=1$)
WZW affine currents
living in the ${\cal N}=1$ subgroup $SU(3) \times U(1)$ and the bosonic 
subgroup $SU(2)$.  
The regular conditions of these currents with 
the four spin-$\frac{1}{2}$ currents and 
spin-$1$ current were very crucial. Three of four spin-$\frac{1}{2}$ currents
live in the above bosonic subgroup 
$SU(2)$ while one of them and spin-$1$ current 
live in the ${\cal N}=1$ subgroup $U(1)$.
The two $SU(2)$ affine algebras are embedded in the ${\cal N}=4$ 
coset theory in nontrivial way.
The $SU(2)_3$ algebra is realized by the Wolf space coset currents while
the $SU(2)_k$ algebra is realized by two currents from the Wolf space 
subgroup   and one remaining current from both Wolf space subgroup and 
Wolf space coset. 

Compared to the work of \cite{Ahn1206} where the ${\cal N}=2$ ${\cal W}_5$ 
algebra with spin contents of 
$(1, \frac{3}{2}, \frac{3}{2}, 2)$, $(2, \frac{5}{2}, \frac{5}{2}, 3)$,
$(3, \frac{7}{2}, \frac{7}{2}, 4)$ and $(4, \frac{9}{2}, \frac{9}{2}, 5)$ 
is realized on the coset ${\bf CP}^4 =\frac{SU(5)}{SU(4) \times U(1)}$, 
the present Wolf space coset model has more higher spin currents in the sense
that there exist the additional second and third multiplets of (\ref{16}).
This is because the dimension of the subgroup in Wolf space coset is less than
the one of the above ${\bf CP}^4$ model and there exists more room for 
construction of higher spin currents.

From the OPEs between the $11$ currents and $16$ higher spin currents,
one realizes that some of the OPEs have 
special features. Let us consider the OPEs between 
four spin-$\frac{3}{2}$ currents
and $16$ currents. In the right hand side of these OPEs, we only 
focus on the linear higher spin current terms. See also Appendix $C$ where
one obtains all the detailed results.  
The first result is as follows:
\bea
\hat{G}_{11} & \times & 
\left(
\begin{array}{cccc}
T^{(1)}, & T_{+}^{(\frac{3}{2})}, & T_{-}^{(\frac{3}{2})}, & T^{(2)}  \\
U^{(\frac{3}{2})}, & U_{+}^{(2)}, & U_{-}^{(2)}, & U^{(\frac{5}{2})}  \\
V^{(\frac{3}{2})}, & V^{(2)}_{+}, & V^{(2)}_{-}, & V^{(\frac{5}{2})}  \\
W^{(2)}, & W_{+}^{(\frac{5}{2})}, & W_{-}^{(\frac{5}{2})}, & W^{(3)}  
\end{array} \right)
\nonu \\
& \rightarrow &
\left(
\begin{array}{cccc}
U^{(\frac{3}{2})}, & U_{+}^{(2)}, & U_{-}^{(2)}, & U^{(\frac{3}{2})},U^{(\frac{5}{2})}  \\
0,  & 0, & 0,  & 0  \\
T^{(1)}, W^{(2)}, & T_{+}^{(\frac{3}{2})},  W_{+}^{(\frac{5}{2})}, 
& T_{-}^{(\frac{3}{2})}, W_{-}^{(\frac{5}{2})}, & T^{(1)}, 
T^{(2)},  W^{(2)},  W^{(3)}    \\
U^{(\frac{3}{2})}, & U_{+}^{(2)}, & U_{-}^{(2)}, & U^{(\frac{3}{2})},U^{(\frac{5}{2})} 
\end{array} \right).
\label{operesult}
\eea
There are no higher spin currents, denoted by zeros in (\ref{operesult})
corresponding to the row containing $G_{11}(w)$ when one writes  
(\ref{fourn2}) in terms of $4\times 4$ matrix, 
in the OPEs between the current $\hat{G}_{11}(z)$
and the four currents living in the second multiplet of (\ref{new16comp}). 
These four currents appear in first and last rows 
in the final expression of (\ref{operesult}).  
Also the current $U^{(\frac{3}{2})}(w)$ appears in the last columns of these rows
by conformal dimensional analysis.
Finally, the first and last multiplets of (\ref{new16comp}) appear in the 
third row in (\ref{operesult}).
Of course, the currents $T^{(1)}(w)$ and $W^{(2)}(w)$ can appear in the last 
column of this row.

Similarly, it turns out that the following result for the 
spin-$\frac{3}{2}$ current and $16$ currents is given by
\bea
\hat{G}_{12} & \times & 
\left(
\begin{array}{cccc}
T^{(1)}, & T_{+}^{(\frac{3}{2})}, & T_{-}^{(\frac{3}{2})}, & T^{(2)}  \\
U^{(\frac{3}{2})}, & U_{+}^{(2)}, & U_{-}^{(2)}, & U^{(\frac{5}{2})}  \\
V^{(\frac{3}{2})}, & V^{(2)}_{+}, & V^{(2)}_{-}, & V^{(\frac{5}{2})}  \\
W^{(2)}, & W_{+}^{(\frac{5}{2})}, & W_{-}^{(\frac{5}{2})}, & W^{(3)}  
\end{array} \right)
\nonu \\
& \rightarrow&
\left(
\begin{array}{cccc}
T_{-}^{(\frac{3}{2})}, & T^{(1)}, T^{(2)}, & 0, & T_{-}^{(\frac{3}{2})}   \\
U_{-}^{(2)}, & U^{(\frac{3}{2})},U^{(\frac{5}{2})}, & 0, & U_{-}^{(2)}  \\
V_{-}^{(2)}, & V^{(\frac{3}{2})}, V^{(\frac{5}{2})}, & 0, & V_{-}^{(2)}  \\
T_{-}^{(\frac{3}{2})}, W_{-}^{(\frac{5}{2})},  & T^{(1)}, T^{(2)}, W^{(2)}, W^{(3)}, 
& 0, & T_{-}^{(\frac{3}{2})}, W_{-}^{(\frac{5}{2})}  
\end{array} \right).
\label{operesult1}
\eea
There are no higher spin currents, denoted by zeros in (\ref{operesult1})
corresponding to the column containing $G_{12}(w)$ in 
(\ref{fourn2}), 
in the OPEs between the current $\hat{G}_{12}(z)$
and the four currents living in the third component of each  
multiplet of (\ref{new16comp}).
These four currents appear in the first and the last columns 
in the final expression of (\ref{operesult1}).  
Also the current $T_{-}^{(\frac{3}{2})}(w)$ 
appears in the last rows of these columns.
Finally, the first and last components in each multiplet 
of (\ref{new16comp}) appear in the 
second column in (\ref{operesult1}).
Of course, the currents $T^{(1)}(w)$ and $W^{(2)}(w)$ can appear in the last 
row of this column.

One has the following result
\bea
\hat{G}_{21} & \times & 
\left(
\begin{array}{cccc}
T^{(1)}, & T_{+}^{(\frac{3}{2})}, & T_{-}^{(\frac{3}{2})}, & T^{(2)}  \\
U^{(\frac{3}{2})}, & U_{+}^{(2)}, & U_{-}^{(2)}, & U^{(\frac{5}{2})}  \\
V^{(\frac{3}{2})}, & V^{(2)}_{+}, & V^{(2)}_{-}, & V^{(\frac{5}{2})}  \\
W^{(2)}, & W_{+}^{(\frac{5}{2})}, & W_{-}^{(\frac{5}{2})}, & W^{(3)}  
\end{array} \right)
\nonu \\
& \rightarrow&
\left(
\begin{array}{cccc}
T_{+}^{(\frac{3}{2})}, & 0, & T^{(1)}, T^{(2)},  & T_{+}^{(\frac{3}{2})}   \\
U_{+}^{(2)}, & 0, & U^{(\frac{3}{2})},U^{(\frac{5}{2})}, & U_{+}^{(2)}  \\
V_{+}^{(2)}, & 0,  & V^{(\frac{3}{2})}, V^{(\frac{5}{2})},  & V_{+}^{(2)}  \\
T_{+}^{(\frac{3}{2})}, W_{+}^{(\frac{5}{2})},  & 0, & T^{(1)}, T^{(2)}, W^{(2)}, W^{(3)}, 
 & T_{+}^{(\frac{3}{2})}, W_{+}^{(\frac{5}{2})}  
\end{array} \right).
\label{operesult2} 
\eea
There are no higher spin currents, denoted by zeros in (\ref{operesult2})
corresponding to the column containing $G_{21}(w)$ in
(\ref{fourn2}), 
in the OPEs between the current $\hat{G}_{21}(z)$
and the four currents living in the second component of each  
multiplet of (\ref{new16comp}).
These four currents appear in the first and the last columns 
in the final expression of (\ref{operesult2}).  
Also the current $T_{+}^{(\frac{3}{2})}(w)$ 
appears in the last rows of these columns.
Finally, the first and last components in each multiplet 
of (\ref{new16comp}) appear in the 
third column in (\ref{operesult2}).
Of course, the currents $T^{(1)}(w)$ and $W^{(2)}(w)$ can appear in the last 
row of this column.

Finally, one obtains the following result 
\bea
\hat{G}_{22} & \times & 
\left(
\begin{array}{cccc}
T^{(1)}, & T_{+}^{(\frac{3}{2})}, & T_{-}^{(\frac{3}{2})}, & T^{(2)}  \\
U^{(\frac{3}{2})}, & U_{+}^{(2)}, & U_{-}^{(2)}, & U^{(\frac{5}{2})}  \\
V^{(\frac{3}{2})}, & V^{(2)}_{+}, & V^{(2)}_{-}, & V^{(\frac{5}{2})}  \\
W^{(2)}, & W_{+}^{(\frac{5}{2})}, & W_{-}^{(\frac{5}{2})}, & W^{(3)}  
\end{array} \right)
\nonu \\
& \rightarrow &
\left(
\begin{array}{cccc}
V^{(\frac{3}{2})}, & V_{+}^{(2)}, & V_{-}^{(2)}, & V^{(\frac{3}{2})},V^{(\frac{5}{2})}  \\
T^{(1)}, W^{(2)}, & T_{+}^{(\frac{3}{2})},  W_{+}^{(\frac{5}{2})}, 
& T_{-}^{(\frac{3}{2})}, W_{-}^{(\frac{5}{2})}, & T^{(1)},T^{(2)},  W^{(2)},  W^{(3)}    \\
0,  & 0, & 0,  & 0  \\
V^{(\frac{3}{2})}, & V_{+}^{(2)}, & V_{-}^{(2)}, & V^{(\frac{3}{2})},V^{(\frac{5}{2})} 
\end{array} \right).
\label{operesult3}
\eea
There are no higher spin currents, denoted by zeros in (\ref{operesult3})
corresponding to the row containing $G_{22}(w)$ in
(\ref{fourn2}), 
in the OPEs between the current $\hat{G}_{22}(z)$
and the four currents living in the third multiplet of (\ref{new16comp}). 
These four currents appear in first and last rows in  
the final expression of (\ref{operesult3}).  
Also the current $V^{(\frac{3}{2})}(w)$ appears in the last columns of these rows.
Finally, the first and last multiplets of (\ref{new16comp}) appear in the 
second row in (\ref{operesult}).
Of course, the currents $T^{(1)}(w)$ and $W^{(2)}(w)$ can appear in the last 
column of this row.

We present some future interesting research directions.

%%%%%%%%%%%%%%%%%%%%%%%%%%%%%%%%%%%%%%%%%%%%%%%%%%%%%%%%%%%%%%%%
$\bullet$ The OPEs between the higher spin currents themselves.
%%%%%%%%%%%%%%%%%%%%%%%%%%%%%%%%%%%%%%%%%%%%%%%%%%%%%%%%%%%%%%%%

Because the $16$ higher spin currents are found explicitly, it is natural 
to calculate the OPEs between them \cite{Ahn2014}. The next higher spin 
currents are specified by the following spin contents by adding one more 
spin to the spin contents in (\ref{16}):
\bea
\left(2, \frac{5}{2}, \frac{5}{2}, 3 \right), \qquad
 \left(\frac{5}{2}, 3, 3, \frac{7}{2} \right), \qquad
\left(\frac{5}{2}, 3, 3, \frac{7}{2} \right), \qquad
\left(3, \frac{7}{2}, \frac{7}{2}, 4 \right).
\label{other16}
\eea 
One expects that
some of the new higher spin currents in (\ref{other16}) should arise   
in the OPEs between the higher spin currents in (\ref{16}).
For $N=3$, there exist finite higher spin currents consisting of 
(\ref{16}) and (\ref{other16}) besides the higher spin 
currents in the short representation \cite{GG1305}.
One expects that the most complicated OPE is given by the OPE 
between the spin-$3$ current and itself.
It would be interesting to study the complete structure of these higher spin 
currents. 
Eventually, it is an open problem, after the work of \cite{Ahn2014} is done,
to obtain the two dimensional boundary
conformal field theory (in the context of higher spin theory) 
dual to the string theory in $AdS_3 \times {\bf S}^3
\times {\bf S}^3 \times {\bf S}^1$ compactification \cite{BPS,EFGT,dPS,GMMS}, 
as suggested in \cite{GG1305}.  
See also \cite{Gaberdieltalk} in the context of alternating spin chain 
\cite{BSZ,OSS}.

%%%%%%%%%%%%%%%%%%%%%%%%%%%%%%%%%%%%%%%%%%%%%%%%%%%%%%%%%%%%
$\bullet$ For general $N$, how the extended currents arise?
%%%%%%%%%%%%%%%%%%%%%%%%%%%%%%%%%%%%%%%%%%%%%%%%%%%%%%%%%%%%

Eventually one should have the higher spin currents for general $N$.
In doing this, the first step is to write down the $16$ currents in the 
large ${\cal N}=4$ linear superconformal algebra  
or $11$ currents in the large ${\cal N}=4$ nonlinear superconformal algebra
for general $N$. 
Some of the information are given in \cite{ST,GK}.
One expects that the calculations of OPEs for general 
$N$, by hand, are based on the component OPEs
along the line of \cite{BBSS1,BBSS2,Ahn1111,Ahn1305,AH1308}. 
Then one should know the various identities between the products of $f$ symbols
in the complex basis.
One can extract some tensor structures from  the OPE results in 
Appendices $A, B$ and $C$ because the fermionic currents transform as 
$({\bf 2},{\bf 2})$ under the two $SU(2)$'s from the observation of 
\cite{GG1305}. Of course, it is quite nontrivial and complicated to
obtain the full algebra like as Appendices $A, B$ and $C$ but at least 
one should find the currents themselves (not the whole algebra).

%%%%%%%%%%%%%%%%%%%%%%%%%%%%%%%%%%%%%%%%%%%%%%%%%%%%%%%%%%%%%%%%%%%%%
$\bullet$ Inserting the four spin-$\frac{1}{2}$ and spin-$1$ currents
%%%%%%%%%%%%%%%%%%%%%%%%%%%%%%%%%%%%%%%%%%%%%%%%%%%%%%%%%%%%%%%%%%%%%

What happens when we substitute $\hat{T}(z)$, $\hat{A}_1(z)$, [$\hat{A}_2(z)$ 
and 
$\hat{A}_3(z)$], [$\hat{B}_1(z)$, $\hat{B}_2(z)$ and $\hat{B}_3(z)$],
and [$\hat{G}_{11}(z)$, $\hat{G}_{12}(z)$, $\hat{G}_{21}(z)$ and 
$\hat{G}_{22}(z)$] using 
the equation (\ref{stressnonlinear}), (\ref{spinonenonlinear}), 
(\ref{otherspinonenonlinear}),
(\ref{nonb1b2b3}) and (\ref{fourg}) 
in the large ${\cal N}=4$ nonlinear 
superconformal algebra? 
Then one expects that one obtains the large ${\cal N}=4$ linear superconformal
algebra. See also the work of \cite{GK}.
It would be interesting to study the other OPEs by reintroducing the four
spin-$\frac{1}{2}$ currents and a spin-$1$ current.

%%%%%%%%%%%%%%%%%%%%%%%%%%%%%%%%%%%%%%%
$\bullet$ Other group realization.
%%%%%%%%%%%%%%%%%%%%%%%%%%%%%%%%%%%%%%%

In the classification of ${\cal N}=4$ coset theory in \cite{ST}, 
there exists a coset model 
\bea
\mbox{Wolf} \times SU(2) \times U(1) = \frac{SO(N+4)}{SO(N) \times SU(2)} 
\times U(1).
\label{cosetcoset}
\eea
As described in \cite{AP1310}, the central charge for this coset model is
given by 
$c =\frac{6(k+1)(N+1)}{(k+N+2)}$ which is equal to the one in 
(\ref{cosetcentral}).
The first nontrivial case is for $N=4$. At least, the large 
${\cal N}=4$ (non)linear superconformal algebra should exist. 
But the spin contents
for the higher spin currents will be different from those in the coset model
(\ref{coset})
in this paper.
Therefore, it would be interesting to study the higher spin currents in the 
above Wolf space for (fixed) general $N$. The corresponding ${\cal N}=2$
current algebra in ${\cal N}=2$ superspace for the supersymmetric WZW model,
with level $k$, on a group $G =SO(N+4)$ of even dimension 
(for the odd dimension one can introduce the extra $U(1)$ in the both
numerator and the denominator of the coset),  
can be obtained from \cite{Ahn1206}. 
In general, the component expression of ${\cal N}=2$ current algebra in the 
above coset model (\ref{cosetcoset}) 
looks different. 
The previous works \cite{Ahn1106,Ahn1202,AP1301} 
in this direction will be helpful.
Note that the similar coset $\frac{SO(N+4)}{SO(N+2) \times SO(2)}$
appears in the Kazama-Suzuki model. In this case, one can also construct 
the higher spin currents of even spins as well.

%%%%%%%%%%%%%%%%%%%%%%%%%%%%%%%%%%%%%%%%%%%%%%%%%%%%%%%%%%%%%%%%%%%%%%%
$\bullet$ The general Kazama-Suzuki model and nonabelian generalization 
%%%%%%%%%%%%%%%%%%%%%%%%%%%%%%%%%%%%%%%%%%%%%%%%%%%%%%%%%%%%%%%%%%%%%%%

As observed in \cite{GG1305}, the Wolf space can be generalized to the 
Kazama-Suzuki coset model in (\ref{KS})
\bea
\frac{SU(N+M)}{SU(N) \times SU(M) \times U(1)}.
\label{gencoset}
\eea 
In this coset model, 
the central charge $c =\frac{3k M N}{(M+N+k)}$ in (\ref{KScentral})
depends on $k, M$ and $N$. 
Then in the stringy limit where the three quantities $k, M$ and $N$
go to infinity simultaneously, the central charge  behaves $N^2$ which 
is appropriate for a stringy model.
It would be interesting to study the higher spin currents in this generalized 
coset model. 
For general $M$ and $N$, this coset model has at least ${\cal N}=2$ 
supersymmetry.
Note that the coset (\ref{gencoset}) for 
$M=2$ case is the coset model considered in this 
paper. The immediate question is how to construct the $M^2$ fermions which 
can be decoupled from the algebra.

%$\bullet$ ${\cal N}=3$ linear and nonlinear superconformal algebra.

$\bullet$ Extension of ${\cal N}=2$ $U(3)$ or $U(2|1)$ 
Knizhnik-Bershadsky algebras and others. 

In \cite{AIS},  the two ${\cal N}=2$ superconformal 
algebras were found. Decoupling of four spin-$\frac{1}{2}$ currents 
appropriately, the  $U(3)$ \cite{Knizhnik,Bershadsky} 
or $U(2|1)$ \cite{DTH} 
Knizhnik-Bershadsky algebras were reproduced.
Furthermore, the ${\cal N}=2$ WZW affine current superalgebra has been 
found. Then it is an open problem to obtain the extension of 
unitary Knizhnik-Bershadsky algebras in the higher spin current context.

For ${\cal N} > 4$, there exists a coset construction. For example, 
in \cite{GK1}, the ${\cal N}=7$ and ${\cal N}=8$ nonlinear superconformal 
algebras were studied. 
See also \cite{BG,FL,Bowcock1992}.
It would be interesting to see whether these algebras 
can be extended or not in the context of minimal model holography even though
the cosets of these models 
do not have any group parameters like as the rank of
the group.  

It would be interesting to see whether one can 
express the $(11+16)$ currents in terms of 
${\cal N}=2$ superspace as done in \cite{Ahn1206} or even further ${\cal N}=4$
superspace in \cite{Schoutensnpb}.
For the $16$ currents of the large ${\cal N}=4$ linear superconformal algebra
in \cite{STVplb}, the explicit field redefinitions  in \cite{STVplb} provide
that  
these $16$ currents are transformed into those currents in 
$O(4)$ extended superconformal
algebra given in \cite{Schoutensnpb}. Then the $16$ currents presented in 
section $2$ can be combined into one single ${\cal N}=4$ superfield in 
\cite{Schoutensnpb}. It is not obvious how one can express the $11$ currents 
of large ${\cal N}=4$ nonlinear superconformal algebra
in terms of a single ${\cal N}=4$ superfield after
decoupling one spin-$1$ current and four spin-$\frac{1}{2}$ currents.     

We clarify the difference between the $16$ higher spin currents in this paper
and the ones in \cite{GG1305}. One can find the particular basis where 
the $16$ higher spin currents transform under the two $SU(2)$ currents characterized by $\hat{A}_i(z)$ and $\hat{B}_i(z)$ in very simple form 
\footnote{We would like to thank the 
referee for raising this issue.}.
For example, one redefines the four spin-$\frac{3}{2}$ currents 
appearing in (\ref{g11t1}), (\ref{g12t1}), (\ref{g21t1}) and (\ref{g22t1})
by absorbing the spin-$\frac{3}{2}$ currents $\hat{G}_a(w)$ of
large ${\cal N}=4$ nonlinear algebra.
Then in this new basis, there are no spin-$\frac{3}{2}$ current 
$\hat{G}_a(w)$ dependences in the right hand side of Appendix 
(\ref{b+t+3half}) and (\ref{b+u3half}).
Furthermore, the OPEs between $\hat{G}_a(z)$ and these redefined 
(primary) spin-$\frac{3}{2}$
currents transforming as $({\bf 2},{\bf 2})$ under the 
$SU(2) \times SU(2)$ 
do not contain the third-order singular terms.

For the spin-$2$ currents in this new basis, 
one can write each (redefined) hatted spin-$2$ current
in terms of each unhatted spin-$2$ current found in (\ref{new16comp}) 
with coefficient $1$ 
and other composite spin-2 fields
with undetermined coefficients 
in Table $3$. Then the relative coefficients can be determined by 
the vanishing of second-order pole in the OPEs between the six 
spin-$1$ currents, $\hat{A}_i(z)$ and $\hat{B}_i(z)$, and 
each hatted spin-$2$ current, $\hat{T}^{(2)}(w)$, $\hat{U}_{\pm}^{(2)}(w)$,
$\hat{V}_{\pm}^{(2)}(w)$ and $\hat{W}^{(2)}(w)$.  
These hatted spin-$2$ currents should be again 
primary under the stress energy 
tensor $\hat{T}(z)$ and be 
$({\bf 3},{\bf 1})$ and $({\bf 1},{\bf 3})$ under 
the $SU(2) \times SU(2)$.

For the spin-$\frac{5}{2}$ currents, 
as before, each hatted spin-$\frac{5}{2}$ current (should be 
$({\bf 2}, {\bf 2})$ under the $SU(2) \times SU(2)$) 
can be written as
each unhatted one in (\ref{new16comp}) 
and other spin-$\frac{5}{2}$ composite fields with 
undetermined coefficients appearing in 
Table $4$. 
By requiring that  the OPEs between $\hat{G}_a(z)$ and these 
redefined spin-$\frac{5}{2}$
currents do not contain the third-order singular terms 
and the OPEs between the above six 
spin-$1$ currents and 
those hatted spin-$\frac{5}{2}$ currents
do not contain the second-order singular 
terms (and also they should be primary
under the stress energy tensor $\hat{T}(z)$),
most of the relative coefficients are fixed.
The remaining ones are determined by 
the condition that 
the OPEs between the spin-$1$ current $T^{(1)}(z)$ and these 
hatted spin-$\frac{5}{2}$ currents do not contain the second-order 
singular terms 
\cite{Ahn2014}.

For the spin-$3$ current, 
the hatted spin-$3$ current can be written as
the unhatted one in (\ref{new16comp}) 
and other spin-$3$ composite fields with 
undetermined coefficients appearing in 
Table $5$. 
By requiring that  
the OPEs between $\hat{A}_i(z)$ (and $\hat{B}_i(z)$) and the 
hatted spin-$3$
current do not contain any singular terms, the relative coefficients 
can be fixed. 

The higher spin-$1$ current remains unchanged because it 
commutes with  the above six spin-$1$ currents in 
$SU(2) \times SU(2)$(i.e., The OPEs do not contain any singular terms)
and both the spin-$1$ and the spin-$3$ currents, $\hat{T}^{(1)}(z)$
and $\hat{W}^{(3)}(z)$, play 
the role of $({\bf 1}, {\bf 1})$ in $SU(2) \times SU(2)$ respectively
as in \cite{GG1305}. 

Now we present the final $16$ higher spin currents in this new basis
explicitly as follows:
\bea
\hat{T}^{(1)}(z) & = & T^{(1)}(z), \nonu \\
\hat{T}_{\pm}^{(\frac{3}{2})}(z)  & = &  T_{\pm}^{(\frac{3}{2})}(z) \pm \frac{1}{2} 
\left( \begin{array}{c}
\hat{G}_{21} \\
\hat{G}_{12} \end{array} \right)(z),
\qquad
%\hat{T}_{-}^{(\frac{3}{2})}  & = &  T_{-}^{(\frac{3}{2})} -\frac{1}{2} 
%\hat{G}_{12},
%\nonu \\
\left( 
\begin{array}{c} 
\hat{U}^{(\frac{3}{2})} \\
\hat{V}^{(\frac{3}{2})} \end{array} 
\right)(z)   =   \left( 
\begin{array}{c} 
U^{(\frac{3}{2})} \\
V^{(\frac{3}{2})} \end{array} \right)(z) \pm \frac{1}{2} 
\left(
\begin{array}{c} 
\hat{G}_{11} \\
\hat{G}_{22} \end{array} \right)(z),
\nonu \\
%\hat{V}^{(\frac{3}{2})}  & = &  V^{(\frac{3}{2})} -\frac{1}{2} 
%\hat{G}_{22},
%\nonu \\
\hat{T}^{(2)}(z)  & = &  T^{(2)}(z) + \frac{i}{k} T^{(1)} \hat{A}_3(z)
+  \frac{i}{3} T^{(1)} \hat{B}_3(z) - \frac{3(3+k)(13+9k)}{20(2+k)(3+7k)} 
T^{(1)} T^{(1)}(z) \nonu \\
& + &  \frac{1}{(5+k)}  \left[ -
\frac{3i(-3+3k+2k^2)}{(2+k)(3+7k)}  \pa \hat{A}_3
+  \frac{i(-27+k)k}{5(3+7k)} 
\pa \hat{B}_3  \right. \nonu \\
& - &   \frac{3(-3+3k+2k^2)}{(2+k)(3+7k)} 
\hat{A}_{+} \hat{A}_{-} 
+  \frac{k(-27+k)}{5(3+7k)}
\hat{B}_3 \hat{B}_3
\nonu \\
& - &  \left. \frac{3(-3+3k+2k^2)}{(2+k)(3+7k)}  
\hat{A}_3 \hat{A}_3
+2 
\hat{A}_3 \hat{B}_3 +      
\frac{k(-27+k)}{5(3+7k)} \hat{B}_{+} \hat{B}_{-} \right](z),
\nonu \\
\left(
\begin{array}{c}
\hat{U}_{+}^{(2)} \\
\hat{V}_{-}^{(2)} 
\end{array} \right)(z)
  & = &  \left( 
\begin{array}{c} 
U_{+}^{(2)} \\
V_{-}^{(2)} \end{array} \right)(z) 
\mp \frac{i}{3} T^{(1)} \hat{B}_{\mp}(z)
\pm \frac{2}{(5+k)} \hat{A}_3 \hat{B}_{\mp}(z),
\nonu \\
\left(
\begin{array}{c}
\hat{U}_{-}^{(2)} \\
\hat{V}_{+}^{(2)} \end{array}
\right)(z) & = & \left(
\begin{array}{c} 
U_{-}^{(2)} \\
V_{+}^{(2)} \end{array} \right)(z) 
\pm \frac{i}{k} T^{(1)} \hat{A}_{\pm}(z)
\pm \frac{2}{(5+k)} \hat{A}_{\pm} \hat{B}_3(z),
\nonu \\
%\hat{V}_{+}^{(2)}  & = &  V_{+}^{(2)} -\frac{i}{k} T^{(1)} \hat{A}_{-}
%-\frac{2}{(5+k)} \hat{A}_{-} \hat{B}_3,
%\nonu \\
%\hat{V}_{-}^{(2)}  & = &  V_{-}^{(2)} +\frac{i}{3} T^{(1)} \hat{B}_{+}
%-\frac{2}{(5+k)} \hat{A}_3 \hat{B}_{+},
%\nonu \\
\hat{W}^{(2)}(z)  & = &  W^{(2)}(z) - \frac{i}{k} T^{(1)} \hat{A}_3(z)
+  \frac{i}{3} T^{(1)} \hat{B}_3(z) - \frac{3(13+9k)}{20(2+k)} 
T^{(1)} T^{(1)}(z) \nonu \\
& + & \frac{1}{(5+k)} \left[ \frac{3i}{(2+k)}  \pa \hat{A}_3
+  \frac{ik}{5} 
\pa \hat{B}_3 +  \frac{3}{(2+k)} 
\hat{A}_{+} \hat{A}_{-} \right. \nonu \\
&+ & \left.   \frac{3}{(2+k)}  
\hat{A}_3 \hat{A}_3
-2 
\hat{A}_3 \hat{B}_3 +  \frac{k}{5} \hat{B}_{+} \hat{B}_{-}
+  \frac{k}{5}
\hat{B}_3 \hat{B}_3 \right](z),
\nonu \\
\hat{U}^{(\frac{5}{2})}(z) & = & U^{(\frac{5}{2})}(z) +
\frac{1}{(5+k)(-25+41k+38k^2)}
\left[ 16ik(8+k)
\hat{A}_{+} \hat{G}_{21}  \right. \nonu \\
& + &  2i(-25+169k+54k^2)
\hat{A}_{+} T_{+}^{(\frac{3}{2})}
\nonu \\
& + &  2i(5+2k)(-5+23k) \hat{A}_3 \hat{G}_{11} +
 4i(5+2k)(-5+23k) \hat{A}_3 U^{(\frac{3}{2})}
\nonu \\
& - &  6i(-1+k)(1+k)(5+2k) 
\hat{B}_{-} \hat{G}_{12}
+   12i(-1+k)(1+k)(5+2k) 
\hat{B}_{-} T_{-}^{(\frac{3}{2})}
\nonu \\
& + & 
 12i(-1+k)(1+k)(5+2k)
\hat{B}_3 U^{(\frac{3}{2})}
+  (-3+k)(17+10k)(5+k) T^{(1)} \hat{G}_{11}
\nonu \\
& + &   2i(8+k)(-5+5k+6k^2) \hat{B}_3 \hat{G}_{11}
+  \frac{2}{3}(-3+k)(50+71k+14k^2)  \pa \hat{G}_{11}
\nonu \\
& + &  \left.
 \frac{4}{3} (-125-204k-9k^2+14k^3) \pa U^{(\frac{3}{2})} \right](z),
\nonu \\
\hat{V}^{(\frac{5}{2})}(z) & = & V^{(\frac{5}{2})}(z) + 
\frac{1}{(5+k)(-25+41k+38k^2)} \left[-
16ik(8+k) 
\hat{A}_{-} \hat{G}_{12} \right.  \nonu \\
& + &  32ik(8+k) 
\hat{A}_{-} T_{-}^{(\frac{3}{2})}
-   2i(5+2k)(-5+23k) \hat{A}_3 \hat{G}_{22} +
 32ik(8+k) \hat{A}_3 V^{(\frac{3}{2})}
\nonu \\
& + &  2i(8+k)(-5+5k+6k^2) 
\hat{B}_{+} \hat{G}_{21}
+   2i(-11+8k+12k^2)(5+k) 
\hat{B}_{+} T_{+}^{(\frac{3}{2})}
\nonu \\
& + & 
4i(8+k)(-5+5k+6k^2) 
\hat{B}_3 V^{(\frac{3}{2})}
-  (-3+k)(17+10k)(5+k) T^{(1)} \hat{G}_{22}
\nonu \\
& - &   2i(8+k)(-5+5k+6k^2) \hat{B}_3 \hat{G}_{22}
+  \frac{2}{3}(-3+k)(50+71k+14k^2)  
\pa \hat{G}_{22}
\nonu \\
& - &  \left.  \frac{4}{3}(1+k)(-175+53k+14k^2) \pa V^{(\frac{3}{2})}
\right](z),
\nonu \\
\hat{W}_{\pm}^{(\frac{5}{2})}(z) & = & W_{\pm}^{(\frac{5}{2})}(z) +
\frac{1}{(5+k)(-25+41k+38k^2)} \left[ \pm 16ik(8+k)
\hat{A}_{\mp}
\left( \begin{array}{c} 
 \hat{G}_{11} \\
\hat{G}_{22} 
\end{array} \right) \right.
 \nonu \\
& + &  2i(-25+169k+54k^2)
\hat{A}_{\mp} \left(
\begin{array}{c}
 U^{(\frac{3}{2})} \\
V^{(\frac{3}{2})} 
\end{array} \right) 
\nonu \\
& \pm &  2i(-25-23k+30k^2) \hat{A}_3 \left( 
\begin{array}{c} 
\hat{G}_{21} \\
\hat{G}_{12} \end{array} \right) - 
 32ik(8+k) \hat{A}_3 T_{\pm}^{(\frac{3}{2})}
\nonu \\
& \pm &  6i(-1+k)(1+k)(5+2k) 
\hat{B}_{\mp} \left(
\begin{array}{c} 
\hat{G}_{22} \\
\hat{G}_{11} \end{array}
\right)
\nonu \\
&- &   2i(-11+8k+12k^2)(5+k)
\hat{B}_{\mp} \left( 
\begin{array}{c} 
V^{(\frac{3}{2})} \\
U^{(\frac{3}{2})} \end{array}
\right)
\nonu \\
& \pm & 
 2i(8+k)(-5+5k+6k^2) 
\hat{B}_3 \left(
\begin{array}{c} 
\hat{G}_{21} \\
\hat{G}_{12} 
\end{array} \right)
+   12i(-1+k)(1+k)(5+2k) 
\hat{B}_3 T_{\pm}^{(\frac{3}{2})} 
\nonu \\
& \pm &  (-3+k)(17+10k)(5+k) T^{(1)} \left( 
\begin{array}{c} \hat{G}_{21} \\
\hat{G}_{12} \end{array} \right)
\nonu \\
& + &   \frac{2}{3}(8+k)(-25-7k+14k^2)  \pa 
\left( \begin{array}{c} 
\hat{G}_{21} \\
\hat{G}_{12} \end{array} \right)
\nonu \\
& \pm &   \left. 
\frac{4}{3}(-3+k)(50+71k+14k^2) \pa  T_{\pm}^{(\frac{3}{2})} \right](z),
\nonu \\
\hat{W}^{(3)}(z) & = & 
W^{(3)}(z) 
 + \frac{1}{(k+5)} \left[ \frac{4}{(k+5)}   \hat{A}_{+}\pa 
\hat{A}_{-}
 -   \frac{6}{(k+5)} 
\hat{A}_{-} \pa 
\hat{A}_{+}
 +   \frac{6}{(5 + k)} \hat{A}_3 \pa 
\hat{A}_3 \right. 
\nonu \\
& - &
i \hat{A}_3 \pa T^{(1)}
 -   
\pa \hat{T} -  \pa W^{(2)}
+ i \hat{B}_3 \pa T^{(1)}
-    
\frac{8 i}{(k+5)}
\hat{A}_{+}\hat{A}_{-} 
\hat{A}_3
 -   
T^{(1)} \hat{A}_3 \hat{A}_3 
\nonu \\
& - &   \left.  T^{(1)}  \hat{A}_{+}\hat{A}_{-} 
-   
\frac{8 i}{(k+5)}
\hat{A}_3 \hat{A}_3 \hat{A}_3
-   \frac{8 i}{(k+5)} 
\hat{A}_3 \hat{B}_3 \hat{B}_3 \right](z)
\nonu \\
&+ & \frac{1}{(786 k^3+3727 k^2+2920 k-1925)(k+5)} \nonu \\
&\times & \left[ - 
\frac{3 (7860 k^5+100436 k^4+512137 k^3+950183 k^2+359535 k-296975)}{
2  (23 k+19)}
T^{(1)} \hat{T} \right. \nonu \\
& - &  \frac{i}{2} (534 k^4+6689 k^3+23191 k^2+16723 k-11065) 
T^{(1)} \pa \hat{B}_3
\nonu \\
&+ & \frac{i}{2} (4890 k^2+7823 k-2975)(5+k)
\hat{A}_{+} V_{+}^{(2)}  \nonu \\
&- & \frac{i}{2} (1746 k^3+17365 k^2+24460 k-7175)\hat{A}_{-}  U_{-}^{(2)} 
 \nonu \\
& - &  \frac{i (519762 k^5+3980167 k^4+8204637 k^3
+4284093 k^2-1220395 k-739200)}{
 (7 k+3) (23 k+19)}
\nonu \\
& \times & \hat{A}_3 \hat{T}
+  \frac{3}{2} i (466 k^3-819 k^2-4260 k-175)
\hat{A}_3 T^{(2)} \nonu \\
&+ & 
\frac{i}{2} (4890 k^2+7823 k-2975)(5+k)
\hat{A}_3 W^{(2)} \nonu \\
&+ &  
\frac{1}{(k+5)}(2022 k^4+28451 k^3+82512 k^2+19815 k-59900) 
 \hat{A}_3 \pa \hat{B}_3 
 \nonu \\
& - & \frac{i}{2} (1098 k^3+4607 k^2+2629 k-2700)(k+5)
\hat{B}_{+}  U_{+}^{(2)} 
\nonu \\
& + &   \frac{1}{(k+5)} (312 k^4+5584 k^3+19017 k^2+9450 k-11575)
  \hat{B}_{+} \pa 
\hat{B}_{-} \nonu \\
& + & \frac{i}{2} (1098 k^4+6953 k^3+10756 k^2-1235 k-5800) 
\hat{B}_{-}  V_{-}^{(2)} \nonu \\
& - &  \frac{1}{(k+5)} (312 k^4+7156 k^3+26471 k^2+15290 k-15425)
\hat{B}_{-} \pa 
\hat{B}_{+} \nonu \\
& - &  
\frac{i}{
 (7 k+3) (23 k+19)} 
(101016 k^6 +  1081846 k^5+4027699 k^4
\nonu \\
& + & 6039421 k^3
+2381493 k^2 -  1777935 k-769500)
\hat{B}_3 \hat{T} \nonu \\
& + &  \frac{i}{2}  (1098 k^4+3809 k^3-4152 k^2-12915 k+1900)
\hat{B}_3 T^{(2)} \nonu 
\\
& + &  \frac{i}{2} (1098 k^3+4607 k^2+2629 k-2700)(5+k)
 \hat{B}_3 W^{(2)} \nonu \\
& + &  \frac{2}{(k+5)} (2034 k^4+13661 k^3+29267 k^2+7775 k-18825)
\hat{B}_3 \pa 
\hat{A}_3 \nonu \\
&- & \frac{1}{(k+5)}(1098 k^4+11669 k^3+33118 k^2+16285 k-17350)
\hat{B}_3 \pa \hat{B}_3
\nonu \\
& - &  \frac{i}{3(k+5)} (2994 k^4-475 k^3-53632 k^2-46735 k+40500)
 \pa^2 \hat{B}_3
\nonu \\
& - & 
\frac{1}{6} (1398 k^3+2433 k^2-4957 k-3500)(5+k)
\pa^2 T^{(1)} 
\nonu \\
& - & \frac{1}{4}(1398 k^3+2433 k^2-4957 k-3500)(5+k) \hat{G}_{11} 
(\hat{G}_{22} 
%\nonu \\
%& + & 
%\frac{1}{4} (1398 k^3+2433 k^2-4957 k-3500)(5+k)
-  V^{(\frac{3}{2})}) 
\nonu \\
& + &  
\frac{1}{4} (1398 k^4+15711 k^3+37024 k^2-4925 k-32900)
\hat{G}_{12} \hat{G}_{21} \nonu \\
& + &
\frac{1}{4} (1398 k^3+2433 k^2-4957 k-3500)(5+k) 
(\hat{G}_{12} T_{+}^{(\frac{3}{2})} 
%\nonu \\
%&+ & 
%\frac{1}{4} (1398 k^3+2433 k^2-4957 k-3500)(5+k)
+ \hat{G}_{21} T_{-}^{(\frac{3}{2})} 
%\nonu \\
%& + & 
%\frac{1}{4} (1398 k^3+2433 k^2-4957 k-3500)(5+k)
+\hat{G}_{22}  U^{(\frac{3}{2})} )
\nonu \\
& - & 
\frac{i}{(k+5)} (1098 k^4+13589 k^3+60394 k^2+59365 k-27850)
 \hat{A}_{+}\hat{A}_{-} 
\hat{B}_3
\nonu \\
&- & 
\frac{i}{(k+5)} (1098 k^4+7301 k^3+30578 k^2+36005 k-12450) 
\hat{A}_3 \hat{A}_3 \hat{B}_3 
\nonu \\
& - & 
i (1098 k^3+4607 k^2+2629 k-2700)
(2 \hat{A}_3 \hat{B}_{+} \hat{B}_{-}
 %\nonu \\
%& - & 
%i (1098 k^3+4607 k^2+2629 k-2700)
+ \hat{A}_3 \hat{B}_{+} \hat{B}_3
%\nonu \\
%& - & i (1098 k^3+4607 k^2+2629 k-2700)
+ \hat{B}_{-} \hat{B}_{+} \hat{B}_3
%\nonu \\
%& - & i (1098 k^3+4607 k^2+2629 k-2700)
+\hat{B}_3 \hat{B}_3 \hat{B}_3)
\nonu \\
& - &  
\frac{1}{2}(534 k^4+5117 k^3+15737 k^2+10883 k-7215)
T^{(1)}
( \hat{B}_{+} \hat{B}_{-} +\hat{B}_3 \hat{B}_3) 
\nonu \\ 
%& - & \frac{1}{2} (534 k^4+5117 k^3+15737 k^2+10883 k-7215) 
%T^{(1)} \hat{B}_3 \hat{B}_3 \nonu \\
& + & \left.  
\frac{1}{24} (4086 k^4+50329 k^3+275267 k^2+523139 k+300835)
T^{(1)} T^{(1)} T^{(1)} \right](z).
\label{16new} 
%\hat{W}_{-}^{(\frac{5}{2})} & = & W_{-}^{(\frac{5}{2})} -
%\frac{16ik(8+k)}{(5+k)(-25+41k+38k^2)} 
%\hat{A}_{+} \hat{G}_{22} + \frac{2i(-25+169k+54k^2)}{(5+k)(-25+41k+38k^2)} 
%\hat{A}_{+} V^{(\frac{3}{2})}
%\nonu \\
%& - &  \frac{2i(-25-23k+30k^2)}{
%(5+k)(-25+41k+38k^2)} \hat{A}_3 \hat{G}_{12} -
% \frac{32ik(8+k))}{
%(5+k)(-25+41k+38k^2)} \hat{A}_3 T_{-}^{(\frac{3}{2})}
%\nonu \\
%& - &  \frac{6i(-1+k)(1+k)(5+2k)}{(5+k)(-25+41k+38k^2)} 
%\hat{B}_{+} \hat{G}_{11}
%-   \frac{2i(-11+8k+12k^2)}{(-25+41k+38k^2)} 
%\hat{B}_{+} U^{(\frac{3}{2})}
%\nonu \\
%& - & 
% \frac{2i(8+k)(-5+5k+6k^2)}{(5+k)(-25+41k+38k^2)} 
%\hat{B}_3 \hat{G}_{12}
%-  \frac{12i(-1+k)(1+k)(5+2k)}{(5+k)(-25+41k+38k^2)} 
%\hat{B}_3 T_{-}^{(\frac{3}{2})} 
%\nonu \\
%& - &   \frac{(-3+k)(17+10k)}{
%(-25+41k+38k^2)} T^{(1)} \hat{G}_{12}
%+  \frac{2(8+k)(-25-7k+14k^2)}{3(5+k)(-25+41k+38k^2)}  \pa \hat{G}_{12}
%\nonu \\
%& - &   \frac{4(-3+k)(50+71k+14k^2)}{
%3(5+k)(-25+41k+38k^2)} \pa T_{-}^{(\frac{3}{2})}
%\nonu 
\eea
The corresponding algebras between the $10$ currents, $\hat{A}_{\pm}(z)$, 
$\hat{A}_3(z)$, $\hat{B}_{\pm}(z)$, $\hat{B}_3(z)$ and $\hat{G}_a(z)$, from 
large ${\cal N}=4$ nonlinear algebra and $16$ higher spin currents, 
$\hat{T}^{(1)}(w)$, $\hat{T}_{\pm}^{(\frac{3}{2})}(w)$, $\hat{T}^{(2)}(w)$, 
$\hat{U}^{(\frac{3}{2})}(w)$, $\hat{U}_{\pm}^{(2)}(w)$, $\hat{U}^{(\frac{5}{2})}(w)$,
$\hat{V}^{(\frac{3}{2})}(w)$, $\hat{V}_{\pm}^{(2)}(w)$, $\hat{V}^{(\frac{5}{2})}(w)$,
$\hat{W}^{(2)}(w)$, $\hat{W}_{\pm}^{(\frac{5}{2})}(w)$ and $ \hat{W}^{(3)}(w)$, 
in
(\ref{16new})  are given at the end of 
Appendices $B$ and $C$.

\vspace{.7cm}

%%%%%%%%%%%%%%%%%%%%%%%%%%%%%%%%%%%%%%%%%%%%%%%%%%%%%%%%%%%%%%
%%%%%%%%%%%%%%%%%%%%%%%%%%%%%%%%%%%%%%%%%%%%%%%%%%%%%%%%%%%%%%%
\centerline{\bf Acknowledgments}
%%%%%%%%%%%%%%%%%%%%%%%%%%%%%%%%%%%%%%%%%%%%%%%%%%%%%%%%%%%%%%%
%%%%%%%%%%%%%%%%%%%%%%%%%%%%%%%%%%%%%%%%%%%%%%%%%%%%%%%%%%%%%%%

CA would like to thank M.R. Gaberdiel for intensive discussions 
and R. Gopakumar and H. Kim for discussions. 
CA would like to thank his previous collaborators K. Schoutens and 
A. Sevrin for earlier works.
This work was supported by the Mid-career Researcher Program through
the National Research Foundation of Korea (NRF) grant 
funded by the Korean government (MEST) (No. 2012-045385/2013-056327).
%We thank the Galileo Galilei Institute for Theoretical Physics 
%for the hospitality and the INFN for partial support during 
%the completion of this work. 
CA acknowledges warm hospitality from 
the School of  Liberal Arts (and Institute of Convergence Fundamental
Studies), Seoul National University of Science and Technology.

\newpage

\appendix

\renewcommand{\thesection}{\large \bf \mbox{Appendix~}\Alph{section}}
\renewcommand{\theequation}{\Alph{section}\mbox{.}\arabic{equation}}

%%%%%%%%%%%%%%%%%%%%%%%%%%%%%%%%%%%%%%%%%%%%%%%%%%%%%%%%%%%%%%%%%%%%%
%%%%%%%%%%%%%%%%%%%%%%%%%%%%%%%%%%%%%%%%%%%%%%%%%%%%%%%%%%%%%%%%%%%%%
\section{The OPEs of the large ${\cal N}=4$ nonlinear 
superconformal algebra from WZW affine currents}
%C%%%%%%%%%%%%%%%%%%%%%%%%%%%%%%%%%%%%%%%%%%%%%%%%%%%%%%%%%%%%%%%%%%%
%%%%%%%%%%%%%%%%%%%%%%%%%%%%%%%%%%%%%%%%%%%%%%%%%%%%%%%%%%%%%%%%%%%%

In section $3$, the stress tensor is given by (\ref{stressnonlinear})
and the OPE between this stress tensor and itself
can be described as
\bea
\hat{T}(z) \, \hat{T}(w) & = & \frac{1}{(z-w)^4} \, \frac{\hat{c}}{2} +
\frac{1}{(z-w)^2} \, 2 \hat{T}(w) + \frac{1}{(z-w)} \, 
\pa \hat{T}(w) +
\cdots,
\label{thatthat}
\eea
where 
the central charge $\hat{c}$ is given by (\ref{chat}).

The $10$ currents which are given in (\ref{nong11})-(\ref{nong21}), 
(\ref{nona1a2a3}) and 
(\ref{b1b2b3nonlinear}) are primary fields under this stress tensor 
(\ref{stressnonlinear}) 
as follows:
\bea
\hat{T}(z) \, \Phi(w) & = & 
\frac{1}{(z-w)^2} \, h \, \Phi(w) + \frac{1}{(z-w)} \, 
\pa \Phi(w) +
\cdots,
\label{primarynonlinear}
\eea
where $h =\frac{3}{2}$ for $\Phi=\hat{G}_a$ and $h =1$ for 
$\Phi=\hat{A}_i, \hat{B}_i$.

Let us present the nontrivial remaining OPEs between these $10$ currents 
as follows: 

%%%%%%%%%%%%%%%%%%%%%%%%%%%%%%%%%%%%%%%%%%%%%%%%%%%%%%%
$\bullet$ spin-$\frac{3}{2}$ spin-$\frac{3}{2}$ OPEs
%%%%%%%%%%%%%%%%%%%%%%%%%%%%%%%%%%%%%%%%%%%%%%%%%%%%%%%

From the explicit expressions in (\ref{nong11})-(\ref{nong21}), 
the following OPEs can be 
obtained
\bea
\hat{G}_{a}(z) \, \hat{G}_{b}(w) & = &
\frac{1}{(z-w)^3} \, \frac{2}{3} c_{\mbox{Wolf}} \, \hat{\delta}_{ab} -
\frac{1}{(z-w)^2} \frac{16}{(5+k)} 
\left[ \hat{k}^{-} \, \alpha_{ab}^{+i} \,  \hat{A}_i + \hat{k}^{+} \, 
\alpha_{ab}^{-i} \, 
\hat{B}_i \right](w)\nonu \\
 & + & 
\frac{1}{(z-w)} \left[ 2 \hat{T} \,  \hat{\delta}_{ab}
-\frac{1}{2}  \,    \frac{16}{(5+k)} 
 ( \hat{k}^{-} \, 
\alpha_{ab}^{+i} \,  \pa \hat{A}_i + \hat{k}^{+} \, 
\alpha_{ab}^{-i} \, \pa 
\hat{B}_i ) +
\right. \nonu \\
& - & \left. \frac{16}{(5+k)} (  
\alpha^{+i} \,  \hat{A}_i -
\alpha^{-i} \, 
\hat{B}_i)_{c(a }  (  
\alpha^{+j} \,  \hat{A}_j -
\alpha^{-j} \, 
\hat{B}_j )_{b)}^{\,\,c}   \right](w) +  \cdots,
\label{ggopenonlinear}
\eea
where $\hat{\delta}_{ab}$ is normalized to be $1$.
The central charge $c_{\mbox{Wolf}}$ appearing above  is given by 
$c_{\mbox{Wolf}} = \frac{18k}{(5+k)}$ (\ref{Wolfcentral}) for $N=3$.
In general, this central charge is different from the one in the OPE 
(\ref{thatthat}) of
stress energy tensor and itself.
Note that compared to the linear case,
the nonlinear terms of (\ref{ggopenonlinear})
are nontrivial.   

%%%%%%%%%%%%%%%%%%%%%%%%%%%%%%%%%%%%%%%%%%%%
$\bullet$ spin-$1$ spin-$\frac{3}{2}$ OPEs
%%%%%%%%%%%%%%%%%%%%%%%%%%%%%%%%%%%%%%%%%%%%

From the relations (\ref{nona1a2a3}) and the above 
spin-$\frac{3}{2}$ currents, one obtains the following OPEs between them: 
\bea
\hat{A}_i(z) \, \hat{G}_{a}(w) 
 & = &  \frac{1}{(z-w)} \, \alpha_{ab}^{+ i} \, \hat{G}^b(w) +\cdots.  
\label{agnonlinear}
\eea
Compared to the linear case, the right hand sides of (\ref{agnonlinear}) 
look similar to the one in linear case 
by ignoring the spin-$\frac{1}{2}$ currents
over there. 
Note that the OPEs $\hat{A}_{+}(z) \, \hat{G}_{11}(w)$,
 $\hat{A}_{+}(z) \, \hat{G}_{12}(w)$,
$\hat{A}_{-}(z) \, \hat{G}_{21}(w)$ and
$\hat{A}_{-}(z) \, \hat{G}_{22}(w)$
do not contain the singular terms.
This can be understood from the $U(1)$ charge assignment described in Table 
$1$.
Furthermore, one has, from (\ref{agnonlinear}), 
\bea
\hat{A}_{\mp}(z) \, 
\left(
\begin{array}{c}
\hat{G}_{11} \\
\hat{G}_{22} \\
\end{array} \right)(w) 
 & = &  \mp \frac{1}{(z-w)}\, i \, 
\left(
\begin{array}{c}
\hat{G}_{21} \\
\hat{G}_{12} \\
\end{array} \right)  (w) +\cdots,  
\nonu \\
\hat{A}_{\mp}(z) \, 
\left(
\begin{array}{c}
\hat{G}_{12} \\
\hat{G}_{21} \\
\end{array} \right)(w) 
 & = &   \pm \frac{1}{(z-w)}\, i \, 
\left(
\begin{array}{c}
\hat{G}_{22} \\
\hat{G}_{11} \\
\end{array} \right) (w) +\cdots.  
\label{apmfourg}
\eea
One can easily check that the $U(1)$ charge from Table 
$1$ is preserved in (\ref{apmfourg}).

For the other spin-$1$ currents (\ref{nonb1b2b3}), the similar OPEs 
can be described as
\bea
\hat{B}_i(z) \, \hat{G}_{a}(w) & = & 
\frac{1}{(z-w)} \, \alpha_{ab}^{-i} \, \hat{G}^{b}(w) +\cdots.
\label{bgnonlinear}
\eea
Similarly, the OPEs $\hat{B}_{-}(z) \, \hat{G}_{11}(w)$,
 $\hat{B}_{+}(z) \, \hat{G}_{12}(w)$,
$\hat{B}_{-}(z) \, \hat{G}_{21}(w)$ and
$\hat{B}_{+}(z) \, \hat{G}_{22}(w)$
do not contain the singular terms due to the $U(1)$ charge conservation. 
Instead, one has the following singular OPEs, from (\ref{bgnonlinear}),
\bea
\hat{B}_{\pm}(z) \, 
\left(
\begin{array}{c}
\hat{G}_{11} \\
\hat{G}_{22} \\
\end{array} \right) (w) & = & 
\pm \frac{1}{(z-w)} \, i \, 
\left(
\begin{array}{c}
\hat{G}_{12} \\
\hat{G}_{21} \\
\end{array} \right) (w) +\cdots,
\nonu \\
\hat{B}_{\mp} (z) \, 
\left(
\begin{array}{c}
\hat{G}_{12} \\
\hat{G}_{21} \\
\end{array} \right) (w) & = & 
\pm \frac{1}{(z-w)} \, i \, 
\left(
\begin{array}{c}
\hat{G}_{11} \\
\hat{G}_{22} \\
\end{array} \right) (w) +\cdots.
\label{bpmfourg}
\eea
In this case also, 
the $U(1)$ charge from Table 
$1$ is preserved in (\ref{bpmfourg}).

%%%%%%%%%%%%%%%%%%%%%%%%%%%%%%%%%%%%%%%%%
$\bullet$ spin-$1$ spin-$1$ OPEs
%%%%%%%%%%%%%%%%%%%%%%%%%%%%%%%%%%%%%%%%

Finally, the OPEs between the spin-$1$ currents in (\ref{nona1a2a3}) 
and (\ref{nonb1b2b3}) are as follows:
\bea
\hat{A}_i(z) \, \hat{A}_j(w) & = & 
-\frac{1}{(z-w)^2} \, \frac{1}{2} \hat{k}^{+} \, \hat{\delta}_{ij} +
 \frac{1}{(z-w)} \, \ep_{ijk} \, \hat{A}_k(w)
 \cdots,
\nonu \\
\hat{B}_i(z) \, 
\hat{B}_j(w) & = & -\frac{1}{(z-w)^2} \, \frac{1}{2} \, \hat{k}^{-} 
\, \hat{\delta}_{ij} +
\frac{1}{(z-w)} \, \ep_{ijk} \, \hat{B}_k(w) 
\cdots,
\label{aboverelation}
\eea
where $\hat{k}^{+}=k$ and $\hat{k}^{-}=3$.
One also has, from (\ref{aboverelation}),  the following OPEs
\bea
\hat{A}_{+}(z) \, \hat{A}_{-}(w) & = & 
\frac{1}{(z-w)^2} \,  \hat{k}^{+}  -\frac{1}{(z-w)} 2 i
\hat{A}_3(w) +\cdots,
\nonu \\
\hat{A}_{\pm}(z) \, \hat{A}_3(w) & = & 
\pm \frac{1}{(z-w)}  i \hat{A}_{\pm} (w) +\cdots,
\nonu \\
 \hat{B}_{+}(z) \, \hat{B}_{-}(w) & = & 
\frac{1}{(z-w)^2} \,  \hat{k}^{-}  -\frac{1}{(z-w)} 2 i
\hat{B}_3(w) +\cdots,
\nonu \\
\hat{B}_{\pm} (z) \, \hat{B}_3(w) & = & 
\pm \frac{1}{(z-w)}  i
\hat{B}_{\pm} (w) +\cdots.
\label{apmbpm}
\eea
The $U(1)$ charge from Table 
$1$ is preserved in (\ref{apmbpm}) also.

Therefore, the equations (\ref{ggopenonlinear}), (\ref{agnonlinear}) 
and (\ref{bgnonlinear}) correspond to the ones in $(A.7)$ 
in \cite{npb1989} (or $(4.3)$ in \cite{cqg1989}).
There are also (\ref{thatthat}) and (\ref{primarynonlinear}).

%%%%%%%%%%%%%%%%%%%%%%%%%%%%%%%%%%%%%%%%%%%%%%%%%%%%%%%%%%%%%%%%%%%%%
%%%%%%%%%%%%%%%%%%%%%%%%%%%%%%%%%%%%%%%%%%%%%%%%%%%%%%%%%%%%%%%%%%%%%
\section{The OPEs between the large ${\cal N}=4$ nonlinear algebra
currents and the higher spin currents-I}
%C%%%%%%%%%%%%%%%%%%%%%%%%%%%%%%%%%%%%%%%%%%%%%%%%%%%%%%%%%%%%%%%%%%%
%%%%%%%%%%%%%%%%%%%%%%%%%%%%%%%%%%%%%%%%%%%%%%%%%%%%%%%%%%%%%%%%%%%%

In this Appendix, we present the OPEs between the six spin-$1$ currents 
in section $3$ and the $16$ higher spin currents in section $4$.

%%%%%%%%%%%%%%%%%%
\subsection{ The OPEs between six spin-$1$ currents and the higher spin 
current of spins $(1, \frac{3}{2}, \frac{3}{2}, 2)$
}
%%%%%%%%%%%%%%%%%%%%
%\label{spinonefirstmultiplet}

It is easy to see that there are no singular terms in the OPEs 
$\hat{A}_{\mp}(z) \, T_{\pm}^{(\frac{3}{2})}(w)$ and 
$\hat{B}_{\mp}(z) \, T_{\pm}^{(\frac{3}{2})}(w)$ due to the $U(1)$
charge conservation. 
Furthermore, one has  the following OPEs,
\bea
\hat{A}_{\pm} (z) \, T_{\pm}^{(\frac{3}{2})}(w) 
& = & \frac{1}{(z-w)} \, i \, 
\left(
\begin{array}{c}
-U^{(\frac{3}{2})}     \\ 
V^{(\frac{3}{2})}    \\
\end{array} \right)(w) +\cdots,
\label{a+t+3half} \\
\hat{A}_3(z) \, T_{\pm}^{(\frac{3}{2})}(w) 
& = & \pm \frac{1}{(z-w)} \, \frac{i}{2} \, 
T_{\pm}^{(\frac{3}{2})}(w) +\cdots=\hat{B}_3(z) \, T_{\pm}^{(\frac{3}{2})}(w), 
\nonu \\
\hat{B}_{\pm} (z) \, T_{\pm}^{(\frac{3}{2})}(w) 
& = & \frac{1}{(z-w)}  i 
\left(
\begin{array}{c}
\hat{G}_{22}- V^{(\frac{3}{2})}  \\
 \hat{G}_{11}+ U^{(\frac{3}{2})} \\
\end{array} \right)(w) 
 +\cdots.
\label{b+t+3half}
\eea
One can check the $U(1)$ charge conservation in these OPEs from Table $2$. 

As before, one obtains the following OPEs  
as follows \footnote{The spin-$2$ current $T^{(2)}(z)$ is 
not to be confused with the spin-$2$ stress tensor $\hat{T}(z)$ in 
(\ref{stressnonlinear}).}: 
\bea
\hat{A}_{\pm}(z)   \, T^{(2)}(w) & = &
-\frac{1}{(z-w)^2} \, \left[\frac{3(-3+3k+2k^2)}{(3+7k)(5+k)}\right]
 \hat{A}_{\pm} (w) +\frac{1}{(z-w)}\, i  
\left(
\begin{array}{c}
U^{(2)}_{-} \\
 V^{(2)}_{+} \\
\end{array} \right)(w) 
\nonu \\
& + & \cdots,
\label{a+t2} \\
\hat{A}_3(z) \, T^{(2)}(w) & = &
\frac{1}{(z-w)^2} \, \left[ -\frac{3(-3+3k+2k^2)}{(3+7k)(5+k)}
\, \hat{A}_3 + \frac{k}{(5+k)} \hat{B}_3 +
\frac{i}{2} T^{(1)} \right](w) +\cdots,
\nonu \\
\hat{B}_{\pm} (z) \, T^{(2)}(w) & = &
\frac{1}{(z-w)^2} \, \left[\frac{(-27+k)k}{(3+7k)(5+k)} \right]
\hat{B}_{\pm} (w) \nonu \\
& + & \frac{1}{(z-w)}\,  
\left[ i 
\left(
\begin{array}{c}
V_{-}^{(2)} \\
U_{+}^{(2)} \\
\end{array} \right)
  \mp \frac{4 i}{(5+k)} \hat{A}_3   \hat{B}_{\pm} 
\right](w) +\cdots,
\label{b+t2} \\
\hat{B}_3(z) \, T^{(2)}(w) & = &
\frac{1}{(z-w)^2} \left[ \frac{(-27+k)k}{(3+7k)(5+k)}
\, \hat{B}_3 + \frac{3}{(5+k)} \hat{A}_3 +\frac{i}{2} 
T^{(1)}\right](w) +\cdots.
\nonu
\eea
The $U(1)$ charge conservation in these OPEs from Tables $2$ and $3$ 
can be checked 
again.

%%%%%%%%%%%%%%%%%%
\subsection{The OPEs between six spin $1$ currents and two higher spin 
currents of spins $(\frac{3}{2}, 2, 2, \frac{5}{2})$}
%%%%%%%%%%%%%%%%%%%%

One has  the following OPEs between the spin-$1$ currents and the 
spin-$\frac{3}{2}$ currents:
\bea
\hat{A}_{\mp} (z)   \, 
\left(
\begin{array}{c} 
U^{(\frac{3}{2})} \\
V^{(\frac{3}{2})} \\
\end{array} \right)(w) & = & 
\mp \frac{1}{(z-w)} \, i \, T_{\pm}^{(\frac{3}{2})}(w)+\cdots, 
\label{a-u3half} \\
\hat{A}_3(z) \, 
\left(
\begin{array}{c} 
U^{(\frac{3}{2})} \\
V^{(\frac{3}{2})} \\
\end{array} \right)(w)
 & = & 
\frac{1}{(z-w)} \, \frac{i}{2} \, 
\left(
\begin{array}{c} 
-U^{(\frac{3}{2})} \\
V^{(\frac{3}{2})} \\
\end{array} \right)(w) +\cdots =
-\hat{B}_3(z) \, 
\left(
\begin{array}{c} 
U^{(\frac{3}{2})} \\
V^{(\frac{3}{2})} \\
\end{array} \right)(w), 
\nonu \\
\hat{B}_{\pm} (z) \, 
\left(
\begin{array}{c} 
U^{(\frac{3}{2})} \\
V^{(\frac{3}{2})} \\
\end{array} \right)(w)
 & = & 
\frac{1}{(z-w)} i \left[ 
\pm T_{\mp}^{(\frac{3}{2})} -
\left(
\begin{array}{c} 
\hat{G}_{12} \\
\hat{G}_{21} \\
\end{array} \right)  \right](w)+\cdots. 
\label{b+u3half}
\eea
One can check the $U(1)$ charge conservation from Table $2$.

As performed before, one has the OPEs between the spin-$1$ currents and the 
spin-$2$ currents,
\bea
\hat{A}_{+} (z) \, 
\left(
\begin{array}{c} 
U_{+}^{(2)} \\
V_{-}^{(2)} 
\end{array} \right)
(w) & = & 
\frac{1}{(z-w)} \frac{2}{(5+k)} \left[ \mp i \hat{A}_{+} \hat{B}_{\mp} 
\right](w) +\cdots,
\label{a+u+2} \\
\hat{A}_{-} (z) \, 
\left(
\begin{array}{c} 
U_{+}^{(2)} \\
V_{-}^{(2)} 
\end{array} \right)(w) & = & 
\frac{1}{(z-w)} \frac{2}{(5+k)} \left[ \pm i \hat{A}_{-}  \hat{B}_{\mp} 
\right](w) +\cdots,
\label{a-u+2} \\
\hat{A}_3(z) \, 
\left(
\begin{array}{c} 
U_{+}^{(2)} \\
V_{-}^{(2)} 
\end{array} \right)(w)
 & = & \pm 
\frac{1}{(z-w)^2} \frac{k}{(5+k)} \left[\hat{B}_{\mp} \right](w) +\cdots,
\nonu \\
\hat{B}_{\pm}  (z) \, 
\left(
\begin{array}{c} 
U_{+}^{(2)} \\
V_{-}^{(2)} 
\end{array} \right)
(w) & = & 
\frac{1}{(z-w)^2} \left[ \pm \frac{6}{(5+k)} \hat{A}_3 \mp
i T^{(1)} \right](w)
\label{b+u+2} \\
& + & \frac{1}{(z-w)}  \left[-\frac{2 i(15+23k+4k^2)}
{(3+7k)(5+k)} \hat{T} +i T^{(2)} +i
W^{(2)} 
\right. \nonu \\
& + &  \left. \frac{2i}{(5+k)}\left( -\hat{A}_i \,\hat{A}_i
+
  2 \hat{A}_3 \, \hat{B}_3-
\hat{B}_i \, \hat{B}_i  \right) \right](w) +\cdots,
\nonu \\
\hat{B}_3(z) \, 
\left(
\begin{array}{c} 
U_{+}^{(2)} \\
V_{-}^{(2)} 
\end{array} \right)
(w)
 & = & 
   \frac{1}{(z-w)} \,  i\, 
\left(
\begin{array}{c} 
U_{+}^{(2)} \\
-V_{-}^{(2)} 
\end{array} \right)
(w) +\cdots.
\nonu
\eea
One can check the $U(1)$ charge conservation in these OPEs from Tables 
$2$ and $3$.

Furthermore, the remaining OPEs can be rewritten as 
\bea
\hat{A}_{\mp} (z) \, 
\left(
\begin{array}{c} 
U_{-}^{(2)} \\
V_{+}^{(2)} 
\end{array} \right)
(w) & = &
\frac{1}{(z-w)^2} \, \left[ \pm \frac{2k}{(5+k)} \hat{B}_3  \pm i
T^{(1)} \right](w) \nonu \\
& + & \frac{1}{(z-w)} \left[ \frac{6 i k}
{(3+7k)} \hat{T} + i T^{(2)} - i W^{(2)}
\right](w)+\cdots,
\label{a-u-2} \\
\hat{A}_3(z) \, 
\left(
\begin{array}{c} 
U_{-}^{(2)} \\
V_{+}^{(2)} 
\end{array} \right)
(w) & = &
\frac{1}{(z-w)} \, i 
\left(
\begin{array}{c} 
-U_{-}^{(2)} \\
V_{+}^{(2)} 
\end{array} \right)
(w) +\cdots,
\nonu \\
\hat{B}_{\pm} (z) \, 
\left(
\begin{array}{c} 
U_{-}^{(2)} \\
V_{+}^{(2)} 
\end{array} \right)
(w) & = &
\frac{1}{(z-w)} \, \frac{2}{(5+k)} \left[-i \hat{A}_{\pm} 
\hat{B}_{\pm} \right](w) +\cdots,
\label{b+u-2} \\
\hat{B}_{\mp} (z) \, 
\left(
\begin{array}{c} 
U_{-}^{(2)} \\
V_{+}^{(2)} 
\end{array} \right)(w) & = &
\frac{1}{(z-w)} \, \frac{2}{(5+k)} \left[i \hat{A}_{\pm} 
 \hat{B}_{\mp} \right](w) +\cdots,
\label{b-u-2}
\\
\hat{B}_3(z) \, 
\left(
\begin{array}{c} 
U_{-}^{(2)} \\
V_{+}^{(2)} 
\end{array} \right)(w)
 & = & \pm
\frac{1}{(z-w)^2} \, \frac{3}{(5+k)} \left[ \hat{A}_{\pm} \right](w) +\cdots.
\nonu
\eea
One can check the $U(1)$ charge conservation in these OPEs from Tables 
$2$ and $3$.

The following OPEs between the spin-$1$ currents and the spin-$\frac{5}{2}$ 
current can be obtained as
\bea
\hat{A}_{+} (z) \, U^{(\frac{5}{2})}(w) & = & 
\frac{1}{(z-w)} 
 \frac{1}{(5+k)} \left[ 2 \hat{A}_{+} ( \hat{G}_{11} 
+  U^{(\frac{3}{2})}) \right](w) +\cdots,
\label{A+u5half} \\
\hat{A}_{-} (z) \, U^{(\frac{5}{2})}(w) & = & 
\frac{1}{(z-w)^2} \left[ \frac{2i(9+k)}{3(5+k)} \hat{G}_{21} 
+ \frac{2i(19+5k)}{3(5+k)} T_{+}^{(\frac{3}{2})} \right](w) 
\nonu \\
& + & \frac{1}{(z-w)} \left[ -i W_{+}^{(\frac{5}{2})} -
\frac{4i}{3(5+k)} \pa \hat{G}_{21} -\frac{4i}{3(5+k)} \pa 
T_{+}^{(\frac{3}{2})} \right. \nonu \\
&+ & \left.  \frac{1}{(5+k)} \left( -2 \hat{A}_{-}  (\hat{G}_{11} 
+  U^{(\frac{3}{2})}) - 2\hat{B}_{-}  
V^{(\frac{3}{2})}\right)\right](w) +  \cdots,
\label{A-u5half} \\
\hat{A}_3(z) \, U^{(\frac{5}{2})}(w) & = & 
\frac{1}{(z-w)^2} \left[ \frac{2i(3+2k)}{3(5+k)} \hat{G}_{11} 
+ \frac{8i(2+k)}{3(5+k)} U^{(\frac{3}{2})} \right](w) 
 -  \frac{1}{(z-w)} \, \frac{i}{2} 
\, U^{(\frac{5}{2})} +\cdots,
\nonu \\
\hat{B}_{+} (z) \, U^{(\frac{5}{2})}(w) & = & 
\frac{1}{(z-w)^2} \left[ -\frac{4i(1+k)}{3(5+k)} \hat{G}_{12} 
+ \frac{8i(2+k)}{3(5+k)} T_{-}^{(\frac{3}{2})} \right](w) 
\nonu \\
& + & \frac{1}{(z-w)} \left[ -i W_{-}^{(\frac{5}{2})} -
\frac{8i}{3(5+k)} \pa \hat{G}_{12} +\frac{4i}{3(5+k)} \pa 
T_{-}^{(\frac{3}{2})} \right. 
\label{B+u5half}
\\
&+ &  \left.
\frac{1}{(5+k)} \left( 2 \hat{A}_{+} \hat{G}_{22} -4 \hat{A}_3 \, 
(\hat{G}_{12} -T_{-}^{(\frac{3}{2})}) 
+2  \hat{B}_{+} (\hat{G}_{11} -  U^{(\frac{3}{2})})
 \right) \right](w) +\cdots,
\nonu \\
\hat{B}_{-} (z) \, U^{(\frac{5}{2})}(w) & = & 
-\frac{1}{(z-w)} 
\frac{1}{(5+k)} \left[ 2 \hat{B}_{-}  \hat{G}_{11} 
 \right](w) +\cdots,
\label{B-u5half} \\
\hat{B}_3(z) \, U^{(\frac{5}{2})}(w) & = & 
\frac{1}{(z-w)^2} \left[ \frac{2i(6+k)}{3(5+k)} \hat{G}_{11} 
+ \frac{4i(2+k)}{3(5+k)} U^{(\frac{3}{2})} \right](w) 
 +  \frac{1}{(z-w)} \, \frac{i}{2}  U^{(\frac{5}{2})}(w)  
 +\cdots.
\nonu
\eea
One can check the $U(1)$ charge conservation in these OPEs from Tables 
$2$ and $4$.

%%%%%%%%%%%%%%%%%%
%\subsection{The OPEs between six spin $1$ currents and the higher spin 
%current of spins $(\frac{3}{2}, 2, 2, \frac{5}{2})$}
%%%%%%%%%%%%%%%%%%%%

Then the following nontrivial remaining OPEs hold as follows:
\bea
\hat{A}_{+} (z) \, V^{(\frac{5}{2})}(w) & = &
\frac{1}{(z-w)^2} \left[-\frac{2i(9+k)}{3(5+k)} \hat{G}_{12}
+ \frac{4i(7+k)}{3(5+k)} T_{-}^{(\frac{3}{2})}\right](w) 
\nonu \\
&+& \frac{1}{(z-w)} \left[ i W_{-}^{(\frac{5}{2})}+
\frac{4i}{3(5+k)} \pa \hat{G}_{12} + \frac{4i}{3(5+k)}
\pa T_{-}^{(\frac{3}{2})} \right. 
\label{A+v5half}
\\
& + & \left.  \frac{1}{(5+k)} \left(- 2 \hat{A}_{+}
 (V^{(\frac{3}{2})}+ \hat{G}_{22})
 - 2 \hat{B}_{+} \hat{G}_{11} 
+   4 \hat{B}_3 
\, T_{-}^{(\frac{3}{2})}  \right)\right](w) +\cdots,
\nonu \\
\hat{A}_{-} (z) \, V^{(\frac{5}{2})}(w) & = &
\frac{1}{(z-w)} 
   \frac{1}{(5+k)} \left[ 2 \hat{A}_{-} \hat{G}_{22}
 \right](w) +\cdots,
\label{A-v5half} \\
\hat{A}_3(z) \, V^{(\frac{5}{2})}(w) & = &
\frac{1}{(z-w)^2} \left[-\frac{2i(3+2k)}{3(5+k)} \hat{G}_{22}
+ \frac{2i(7+k)}{3(5+k)} V^{(\frac{3}{2})}\right](w) 
+ \frac{1}{(z-w)}\, \frac{i}{2}\, 
V^{(\frac{5}{2})}(w), 
\nonu \\
&+& \cdots, \nonu \\
\hat{B}_{+} (z) \, V^{(\frac{5}{2})}(w) & = &
 \frac{1}{(z-w)}  \frac{1}{(5+k)} \left[ - 2 \hat{B}_{+} 
( \hat{G}_{22}
- V^{(\frac{3}{2})}) \right](w) +\cdots,
\label{B+v5half} \\
\hat{B}_{-} (z) \, V^{(\frac{5}{2})}(w) & = &
\frac{1}{(z-w)^2} \left[\frac{4i(8+k)}{3(5+k)} \hat{G}_{21}
+ \frac{4i(11+2k)}{3(5+k)} T_{+}^{(\frac{3}{2})}\right](w) 
\nonu \\
&+& \frac{1}{(z-w)} \left[ i W_{+}^{(\frac{5}{2})}
%+\frac{2i}{3(5+k)} \pa \hat{G}_{12} 
- \frac{4i}{3(5+k)}
\pa T_{+}^{(\frac{3}{2})} \right. \nonu \\
& + &  \left. \frac{1}{(5+k)} \left(- 2 \hat{A}_{-} 
 U^{(\frac{3}{2})} + 2 \hat{B}_{-} ( \hat{G}_{22}
- V^{(\frac{3}{2})})  \right)\right](w) +  \cdots,
\label{B-v5half} \\
\hat{B}_3(z) \, V^{(\frac{5}{2})}(w) & = &
\frac{1}{(z-w)^2} \left[-\frac{2i(6+k)}{3(5+k)} \hat{G}_{22}
+ \frac{4i(7+k)}{3(5+k)} V^{(\frac{3}{2})}\right](w) 
- \frac{1}{(z-w)} \, \frac{i}{2} V^{(\frac{5}{2})}(w) \nonu \\
& + & \cdots.
\nonu
\eea
One can check the $U(1)$ charge conservation in these OPEs from Tables  
$2$ and $4$.

%%%%%%%%%%%%%%%%%%
\subsection{ The OPEs between six spin $1$ currents and the higher spin 
current of spins $(2, \frac{5}{2}, \frac{5}{2}, 3)$}
%%%%%%%%%%%%%%%%%%%%

The nontrivial OPEs between the spin-$1$ currents and the spin-$2$ current 
can be described as
\bea
\hat{A}_{\pm}(z) \, W^{(2)}(w) & = &
\frac{1}{(z-w)^2} \, \frac{3}{(5+k)} \hat{A}_{\pm}(w)
-\frac{1}{(z-w)} i 
\left(
\begin{array}{c} 
U_{-}^{(2)} \\
V_{+}^{(2)} 
\end{array} \right)
(w) +  \cdots,
\label{a+w2} \\
\hat{A}_3(z) \, W^{(2)}(w) & = &
\frac{1}{(z-w)^2} \left[ \frac{3}{(5+k)} \hat{A}_3-
\frac{k}{(5+k)} \hat{B}_3 - \frac{i}{2} T^{(1)} \right](w)
+\cdots,
\nonu \\
\hat{B}_{\pm}(z) \, W^{(2)}(w) & = &
\frac{1}{(z-w)^2} \, \frac{k}{(5+k)} \hat{B}_{\pm} (w)
+\frac{1}{(z-w)} i 
\left(
\begin{array}{c} 
V_{-}^{(2)} \\
U_{+}^{(2)} 
\end{array} \right)(w)
+  \cdots,
\label{b+w2} \\
\hat{B}_3(z) \, W^{(2)}(w) & = &
\frac{1}{(z-w)^2} \left[ -
\frac{3}{(5+k)} \hat{A}_3 + \frac{k}{(5+k)}\hat{B}_3 +
\frac{i}{2} T^{(1)} \right](w)
+\cdots.
\nonu 
\eea

One can write the OPEs between the spin-$1$ currents and the 
spin-$\frac{5}{2}$ currents  as follows:
\bea
\hat{A}_{+}(z) \, W_{+}^{(\frac{5}{2})}(w) & = &
\frac{1}{(z-w)^2} \left[ \frac{2i}{3} \hat{G}_{11} +
\frac{10i(3+k)}{3(5+k)} U^{(\frac{3}{2})} \right](w) 
\nonu \\
& + & \frac{1}{(z-w)} \left[ \frac{4i}{3(5+k)} \pa \hat{G}_{11}
+ \frac{4i}{3(5+k)} \pa U^{(\frac{3}{2})} -i 
U^{(\frac{5}{2})} \right. 
 \label{A+w+5half}
\\
& + & \left. \frac{1}{(5+k)} \left( 2 \hat{A}_{+} ( \hat{G}_{21}
+  T_{+}^{(\frac{3}{2})}) -2  \hat{B}_{-} 
T_{-}^{(\frac{3}{2})}  \right) \right](w)+\cdots, 
\nonu \\
\hat{A}_{\mp}(z) \, W_{\pm}^{(\frac{5}{2})}(w) & = &
-\frac{1}{(z-w)} \frac{2}{(5+k)} \hat{A}_{\mp} \left(
\begin{array}{c} 
  \hat{G}_{21}
+  T_{+}^{(\frac{3}{2})} \\
  \hat{G}_{12}
-  T_{-}^{(\frac{3}{2})}
\end{array}
 \right)(w)+\cdots, 
\label{A-w+5half} \\
\hat{A}_3(z) \, W_{+}^{(\frac{5}{2})}(w) & = &
\frac{1}{(z-w)^2} \left[ \frac{2i(-4+k)}{3(5+k)} \hat{G}_{21} -
\frac{2i(9+k)}{3(5+k)} T_{+}^{(\frac{3}{2})} \right](w) 
\nonu \\
& + & \frac{1}{(z-w)} \, \frac{i}{2}  W_{+}^{(\frac{5}{2})}(w) +\cdots, 
\nonu \\
\hat{B}_{+}(z) \, W_{+}^{(\frac{5}{2})}(w) & = &
\frac{1}{(z-w)^2} \left[ \frac{4i(1+k)}{3(5+k)} \hat{G}_{22} -
\frac{4i(9+2k)}{3(5+k)} V^{(\frac{3}{2})} \right](w) 
\nonu \\
& + & \frac{1}{(z-w)} \left[ \frac{4i}{3(5+k)} \pa \hat{G}_{22}
- \frac{4i}{3(5+k)} \pa V^{(\frac{3}{2})} +i 
V^{(\frac{5}{2})} \right. \nonu \\
& + &  \left. \frac{1}{(5+k)} \left( - 
2 \hat{A}_{-} (\hat{G}_{12}
+   T_{-}^{(\frac{3}{2})}) 
- 
2 \hat{B}_{+} T_{+}^{(\frac{3}{2})}  \right) \right](w)+\cdots, 
\label{B+w+5half} \\
\hat{B}_{\mp}(z) \, W_{\pm}^{(\frac{5}{2})}(w) & = &
\frac{1}{(z-w)} 
\frac{1}{(5+k)} \left[ \pm 2 \hat{B}_{\mp} T_{\pm}^{(\frac{3}{2})}  
\right](w)+\cdots,
\label{B-w+5half} \\
\hat{B}_3(z) \, W_{+}^{(\frac{5}{2})}(w) & = &
\frac{1}{(z-w)^2} \left[ \frac{2i(8+k)}{3(5+k)} \hat{G}_{21} +
\frac{4i(3+k)}{3(5+k)} T_{+}^{(\frac{3}{2})} \right](w) 
+  \frac{1}{(z-w)} \, \frac{i}{2}  W_{+}^{(\frac{5}{2})}(w) +\cdots. 
\nonu
\eea

In this case, the four remaining OPEs  can be rewritten as
\bea
\hat{A}_{-}(z) \, W_{-}^{(\frac{5}{2})}(w) & = &
\frac{1}{(z-w)^2} \left[ -\frac{2i}{3} \hat{G}_{22} +
\frac{10i(3+k)}{3(5+k)} V^{(\frac{3}{2})} \right](w) 
\nonu \\
& + & \frac{1}{(z-w)} \left[ -\frac{4i}{3(5+k)} \pa \hat{G}_{22}
- \frac{4i}{3(5+k)} \pa V^{(\frac{3}{2})} +i 
V^{(\frac{5}{2})} \right. 
\label{A-w-5half}
 \\
& + &  \left. \frac{1}{(5+k)} \left(  
%2( \hat{A}_1  + i \hat{A}_2) T_{-}^{(\frac{3}{2})}
2  \hat{A}_{-} \hat{G}_{12}
+ 4 \hat{A}_3 \, V^{(\frac{3}{2})}  
 -  2 \hat{B}_{+}  \hat{G}_{21}
 -     4 \hat{B}_3 \,
V^{(\frac{3}{2})}  
\right) \right](w)+\cdots, 
\nonu \\
\hat{A}_3(z) \, W_{-}^{(\frac{5}{2})}(w) & = &
\frac{1}{(z-w)^2} \left[ -\frac{2i(-4+k)}{3(5+k)} \hat{G}_{12} -
\frac{2i(9+k)}{3(5+k)} T_{-}^{(\frac{3}{2})} \right](w) 
 -  \frac{1}{(z-w)} \, \frac{i}{2}  
W_{-}^{(\frac{5}{2})}(w) \nonu \\
& + & \cdots, 
\nonu \\
\hat{B}_{-}(z) \, W_{-}^{(\frac{5}{2})}(w) & = &
\frac{1}{(z-w)^2} \left[ -\frac{4i(1+k)}{3(5+k)} \hat{G}_{11} -
\frac{4i(9+2k)}{3(5+k)} U^{(\frac{3}{2})} \right](w) 
\nonu \\
& + & \frac{1}{(z-w)} \left[ -\frac{4i}{3(5+k)} \pa \hat{G}_{11}
+ \frac{4i}{3(5+k)} \pa U^{(\frac{3}{2})} -i 
U^{(\frac{5}{2})} \right. 
\label{B-w-5half}
\\
& + &  \left. \frac{1}{(5+k)} \left( - 2 \hat{A}_{+} \hat{G}_{21} 
+4 \hat{A}_3 \, U^{(\frac{3}{2})}
+ 2 \hat{B}_{-} \hat{G}_{12} 
- 
 4 \hat{B}_3 
\, U^{(\frac{3}{2})} \right) \right](w)+\cdots,
\nonu \\
\hat{B}_3(z) \, W_{-}^{(\frac{5}{2})}(w) & = &
\frac{1}{(z-w)^2} \left[ -\frac{2i(8+k)}{3(5+k)} \hat{G}_{12} +
\frac{4i(3+k)}{3(5+k)} T_{-}^{(\frac{3}{2})} \right](w) 
-  \frac{1}{(z-w)} \, \frac{i}{2}  W_{-}^{(\frac{5}{2})}(w) \nonu \\
& + & \cdots. 
\nonu
\eea

Finally the six nontrivial OPEs between the spin-$1$ currents and the 
spin-$3$ current  can be described as 
\bea
\hat{A}_{+}(z) \, W^{(3)}(w) & = &
\frac{1}{(z-w)^3} \left[- \frac{2(35-20k+33k^2)}{(5+k)^2(19+23k)}
\right]  \hat{A}_{+} (w) 
\nonu \\
& + &
\frac{1}{(z-w)^2} \left[ 
 -\frac{(13+2k)}{(5+k)} i U_{-}^{(2)} 
-\frac{6(1+41k+12k^2)}{(5+k)^2(19+23k)} i \hat{A}_{+} \hat{A}_3
\right. \nonu \\
& -&  \frac{2(76+311k+63k^2)}{(5+k)^2(19+23k)} i \hat{A}_{+} 
\hat{B}_3 +\frac{8}{(5+k)^2} i \hat{A}_{+} \hat{B}_3 
\nonu \\
& - & \left.
\frac{(-32 +143k + 3k^2)}{(5+k)^2(19+23k)}  \pa  \hat{A}_{+}
-  \frac{(-139-7k+8k^2)}{(5+k)(19+23k)} 
T^{(1)} \hat{A}_{+}
%\frac{(1+k)}{(5+k)} U_{-}^{(2)} -
%\frac{(1+k)}{(5+k)} V_{+}^{(2)} + \frac{16}{(5+k)^2} 
%\hat{A}_1 \, \hat{B}_3 
\right](w) \nonu \\
%\frac{1}{(z-w)^2} \left[
% \frac{2i(1+k)}{(5+k)} U_{-}^{(2)}  + 
%\frac{16 i}{(5+k)^2} 
%(\hat{A}_1 + i \hat{A}_2 ) \hat{B}_3 
%\right](w) \nonu \\
& + & 
\frac{1}{(z-w)} 
\left[ \frac{1}{(5+k)} 
 T^{(1)}  \pa \hat{A}_{+} -  
\frac{8(48+97k+29k^2)}{(3+7k)(19+23k)(5+k)}  
\hat{A}_{+} \hat{T} 
\right. \nonu \\
& + & \frac{4}{(5+k)} \hat{A}_{+}  T^{(2)}
+  \frac{2}{(5+k)^2} i \hat{A}_{+} \pa \hat{A}_3
+ \frac{2k}{(5+k)^2} i  \hat{A}_{+} \pa \hat{B}_3 
\nonu \\
& - & 
\frac{1}{(5+k)}  \hat{A}_{+}  \pa  T^{(1)}
-\frac{2}{(5+k)^2} i  \hat{A}_3  \pa  \hat{A}_{+}  
-\frac{4}{(5+k)}  \hat{B}_3 U_{-}^{(2)}
\nonu \\
&- & \frac{2(-2+k)}{(5+k)^2} i \hat{B}_3   \pa \hat{A}_{+}  
-\frac{1}{(5+k)} i \pa  U_{-}^{(2)} -
\frac{2}{(5+k)} i \hat{G}_{11} \hat{G}_{12}
\nonu \\
&- & \frac{4}{(5+k)^2}   \hat{A}_{+} 
 \hat{A}_{+} 
 \hat{A}_{-}  +\frac{16}{(5+k)^2}  \hat{A}_{+}  
\hat{A}_3 \hat{B}_3
- \frac{8}{(5+k)^2}  \hat{A}_{+}  \hat{B}_3 \hat{B}_3 
\label{A+w3}
\\
& - & \left. \frac{4}{(5+k)^2}  \hat{A}_{-}
 \hat{A}_{+} 
 \hat{A}_{+}
-   \frac{8}{(5+k)^2}  \hat{A}_3
 \hat{A}_{+} \hat{A}_3
- \frac{4}{(5+k)^2}   \hat{B}_{-} 
 \hat{A}_{+}   \hat{B}_{+} 
%\frac{1}{(5+k)}\left[ 
%\frac{12 k}{(3+7k)} ( \hat{A}_1 + i \hat{A}_2) \hat{T}
%+ 2 (\hat{A}_1 + i \hat{A}_2 ) T^{(2)} \right. \nonu \\
%& - &  2 (\hat{A}_1 + i \hat{A}_2) W^{(2)}
%+\frac{4 i(-4+k)}{(5+k)} ( \hat{A}_1 + i \hat{A}_2 ) \pa \hat{B}_3 
%\nonu \\
%&- &  4 \hat{A}_3 \, U_{-}^{(2)}  -
%\frac{4i(-4+k)}{(5+k)} \hat{B}_3  \pa ( \hat{A}_1 + i \hat{A}_2) 
%- 2 i \pa U_{-}^{(2)}
%\nonu \\
%& - & \left. 2  \pa T^{(1)} (\hat{A}_1 + i \hat{A}_2) + 
%2  T^{(1)}  \pa ( \hat{A}_1 + i \hat{A}_2)
%
\right](w) +  \cdots, 
\nonu \\
\hat{A}_{-}(z) \, W^{(3)}(w) & = &
\frac{1}{(z-w)^3} \left[ \frac{2(35-20k+33k^2)}{(5+k)^2(19+23k)}
\right]  \hat{A}_{-} (w) 
\nonu \\
& + &
\frac{1}{(z-w)^2} \left[ 
 \frac{(15+4k)}{(5+k)} i V_{+}^{(2)} 
-\frac{6(1+41k+12k^2)}{(5+k)^2(19+23k)} i \hat{A}_{-}  \hat{A}_3
\right. \nonu \\
& -&  \frac{2(76+311k+63k^2)}{(5+k)^2(19+23k)} i \hat{A}_{-} 
\hat{B}_3 -\frac{8}{(5+k)^2} i \hat{A}_{-} \hat{B}_3 
\nonu \\
& + & \left.
\frac{(-32 +143k + 3k^2)}{(5+k)^2(19+23k)}  \pa \hat{A}_{-} 
-   \frac{(-139-7k+8k^2)}{(5+k)(19+23k)} 
T^{(1)} \hat{A}_{-} 
%\frac{(1+k)}{(5+k)} U_{-}^{(2)} -
%\frac{(1+k)}{(5+k)} V_{+}^{(2)} + \frac{16}{(5+k)^2} 
%\hat{A}_1 \, \hat{B}_3 
\right](w) \nonu \\
%
%\frac{1}{(z-w)^2} \left[
%\right](w) \nonu \\
& + & 
\frac{1}{(z-w)} 
\left[-\frac{1}{(5+k)} 
 T^{(1)}  \pa  \hat{A}_{-}  -\frac{2}{(5+k)}  
\hat{A}_{-} W^{(2)}  \right. \nonu \\
& + & 
\frac{4(96+251k+127k^2)}{(3+7k)(19+23k)(5+k)}  
\hat{A}_{-}  \hat{T}
- \frac{2}{(5+k)} \hat{A}_{-}  T^{(2)}
+  \frac{18}{(5+k)^2} i \hat{A}_{-}  \pa \hat{A}_3
\nonu \\
& - & \frac{2k}{(5+k)^2} i \hat{A}_{-}  \pa \hat{B}_3 
+  
\frac{1}{(5+k)}  \hat{A}_{-} \pa  T^{(1)}
+\frac{6}{(5+k)^2} i  \hat{A}_3  \pa \hat{A}_{-}  
\nonu \\
&
+ & \frac{4}{(5+k)}  \hat{A}_3 V_{+}^{(2)}
-  
\frac{1}{(5+k)} i \pa  V_{+}^{(2)} +
\frac{2}{(5+k)} i \hat{G}_{21} \hat{G}_{22}
\nonu \\
&+ & \frac{8}{(5+k)^2}   \hat{A}_{+}  
 \hat{A}_{-} 
 \hat{A}_{-}  +\frac{8}{(5+k)^2}  \hat{A}_{-}  
\hat{A}_3 \hat{A}_3
- \frac{2(-2+k)}{(5+k)^2}  \hat{A}_{-}  \hat{A}_3 \hat{B}_3 
\nonu \\
& + & \frac{8}{(5+k)^2}  \hat{A}_{-} 
 \hat{B}_3 \hat{B}_3
+   \frac{2(-10+k)}{(5+k)^2}  \hat{A}_3
 \hat{A}_{-} \hat{B}_3
\nonu \\
& + & \left. \frac{4}{(5+k)^2}   \hat{B}_{-} 
 \hat{A}_{-}   \hat{B}_{+} 
-  \frac{4}{(5+k)}   \hat{B}_3  V_{+}^{(2)} 
 \right](w) +  \cdots, 
%\nonu
%\frac{1}{(z-w)^2} \left[ \frac{2i(1+k)}{(5+k)} V_{+}^{(2)}  -
%\frac{16 i}{(5+k)^2} 
%(\hat{A}_1 - i \hat{A}_2 ) \hat{B}_3 \right](w) \nonu \\
%& + & 
%\frac{1}{(z-w)} \frac{1}{(5+k)}\left[ 
%\frac{12 k}{(3+7k)} ( \hat{A}_1 - i \hat{A}_2) \hat{T}
%+ 2 (\hat{A}_1 - i \hat{A}_2 ) T^{(2)} \right. \nonu \\
%& - &  2 (\hat{A}_1 - i \hat{A}_2) W^{(2)}
%-\frac{4 i(-4+k)}{(5+k)} ( \hat{A}_1 - i \hat{A}_2 ) \pa \hat{B}_3 
%\nonu \\
%&+ &  4 \hat{A}_3 \, V_{+}^{(2)}  +
%\frac{4i(-4+k)}{(5+k)} \hat{B}_3  \pa ( \hat{A}_1 - i \hat{A}_2) 
%- 2 i \pa V_{+}^{(2)}
%\nonu \\
%& + & \left. 2  \pa T^{(1)} (\hat{A}_1 - i \hat{A}_2) - 
%2  T^{(1)}  \pa ( \hat{A}_1 - i \hat{A}_2)
%\right](w) + \cdots. 
%\nonu
\label{A-w3} \\
\hat{A}_3(z) \, W^{(3)}(w) & = &
\frac{1}{(z-w)^3} 
\left[ \frac{2(-3+k) k}{(5+k)^2} \hat{B}_3 +
\frac{(1+k)}{(5+k)} i T^{(1)}
%\left[ \frac{4(-3+k) k}{(5+k)^2} \hat{B}_3 +
%\frac{2i(1+k)}{(5+k)} T^{(1)}\right](w) \nonu \\
\right](w) \nonu \\
& + &
\frac{1}{(z-w)^2} \left[  
-\frac{2(-57+350k+699k^2)}{(3+7k)(19+23k)(5+k)} i \hat{T}
-\frac{8}{(5+k)} i T^{(2)}
\right. \nonu \\
&- & \frac{4(-2+k)}{(5+k)^2} i \hat{A}_1 \hat{A}_1 
-\frac{4(-2+k)}{(5+k)^2} i \hat{A}_2 \hat{A}_2
-\frac{2(-73+69k+82k^2)}{(5+k)^2(19+23k)} i \hat{A}_3 \hat{A}_3
\nonu \\
& + & \frac{4(-76-173k+3k^2)}{(5+k)^2(19+23k)}
\hat{A}_3 \hat{B}_3 -\frac{2(-1+k)}{(5+k)^2} i \hat{B}_1 \hat{B}_1 
-\frac{2(-1+k)}{(5+k)^2} i \hat{B}_2 \hat{B}_2
\nonu \\
&+ & \frac{2}{(5+k)^2} i \hat{B}_3 \hat{B}_3 
-\frac{(-3+k) k}{(5+k)^2} \pa \hat{B}_3 -\frac{(1+k)}{2(5+k)} i \pa T^{(1)}
+\frac{2(4+k)}{(5+k)} i W^{(2)} 
\nonu \\
%-\frac{1}{2} \frac{4(-3+k) k}{(5+k)^2} \pa 
%\hat{B}_3 -
%\frac{1}{2}
%\frac{2i(1+k)}{(5+k)} \pa T^{(1)}
&-& \left. \frac{2(-79-15k+4k^2)}{(5+k)(19+23k)} T^{(1)} \hat{A}_3
-\frac{2}{(5+k)} T^{(1)} \hat{B}_3
\right](w)  + \cdots, 
\nonu \\
\hat{B}_{+}(z) \, W^{(3)}(w) & = &
\frac{1}{(z-w)^3} \left[-\frac{4k(-167+101k)}{3(5+k)^2(19+23k)} \right] 
\hat{B}_{+} (w) \nonu \\
& + &  \frac{1}{(z-w)^2} 
\left[ 
\frac{(7+4k)}{(5+k)} i V_{-}^{(2)}
-\frac{2(79+253k+82k^2)}{(5+k)^2(19+23k)} i \hat{A}_3 \hat{B}_{+}
\right. \nonu \\
& -&  \frac{2(1+k)}{(5+k)^2}  i \hat{A}_3 
\hat{B}_{+} 
-\frac{2k(181+17k)}{(5+k)^2(19+23k)} i \hat{B}_{+}  \hat{B}_3 
\nonu \\
& - &  \left. \frac{k(-877 +151k)}{3(5+k)^2(19+23k)} \pa \hat{B}_{+}
-   \frac{(-101+39k+8k^2)}{(5+k)(19+23k)} 
T^{(1)} \hat{B}_{+} \right](w) \nonu \\
& + &  \frac{1}{(z-w)} \left[- \frac{1}{(5+k)} 
 T^{(1)}  \pa \hat{B}_{+} -  
\frac{8(-27+k)k}{(3+7k)(19+23k)(5+k)}  
\hat{B}_{+} \hat{T}
\right. \nonu \\
& - &  \frac{4}{(5+k)} \hat{B}_{+} T^{(2)}
-  \frac{2(2+k)}{(5+k)^2} i \hat{B}_{+}  \pa \hat{A}_3
+ \frac{14k}{3(5+k)^2} i \hat{B}_{+} \pa \hat{B}_3 
\nonu \\
& + & 
\frac{1}{(5+k)}  \hat{B}_{+} \pa  T^{(1)}
+\frac{4}{(5+k)}  \hat{A}_3 V_{-}^{(2)}
+ 
\frac{1}{(5+k)} i \pa  V_{-}^{(2)} 
\nonu \\
& + & \frac{5k}{3(5+k)^2}  \hat{B}_{+} 
 \hat{B}_{+}  \hat{B}_{-} 
+  \frac{4}{(5+k)^2}   \hat{A}_{-} 
 \hat{A}_{+}  \hat{B}_{+} 
+  \frac{2(-8+k)}{(5+k)^2}  \hat{A}_3 
  \hat{B}_{+} \hat{B}_3
\nonu \\
& +&  \frac{2k}{(5+k)^2}  \hat{B}_{+} \hat{B}_3 \hat{B}_3 
-\frac{5k}{3(5+k)^2}  \hat{B}_{-} 
\hat{B}_{+} \hat{B}_{+}
-  \frac{2(-8+k)}{(5+k)^2}  \hat{B}_3
 \hat{A}_3 \hat{B}_{+} 
\nonu \\
& - & \left. \frac{2k}{(5+k)^2}    \hat{B}_3
 \hat{B}_{+}  \hat{B}_3
+  
\frac{2}{(5+k)} i \hat{G}_{12} \hat{G}_{22}
\right](w) +   \cdots, 
%\nonu \\
%\frac{1}{(z-w)^2} \left[ -\frac{8i}{(5+k)} V_{-}^{(2)} +
%\frac{4(1+k)}{(5+k)^2} i \hat{A}_3 (\hat{B}_1 +i \hat{B}_2)\right](w) \nonu \\
%& + & 
%\frac{1}{(z-w)} \left[ \frac{4(-4+k)}{(5+k)^2} i   \hat{A}_3 \pa 
%(\hat{B}_1+i \hat{B}_2)
%+\frac{4(3+4k)}{(3+7k)(5+k)}  (\hat{B}_1+i \hat{B}_2) \hat{T}
%\right. \nonu \\
%& - & \frac{2}{(5+k)}  (\hat{B}_1+i \hat{B}_2) T^{(2)} 
%-\frac{2}{(5+k)}  (\hat{B}_1+i \hat{B}_2) W^{(2)} 
%\nonu \\
%& + & \frac{4}{(5+k)^2} i  (\hat{B}_1+i \hat{B}_2) \pa \hat{B}_3
%+  \frac{4}{(5+k)} \hat{B}_3  V_{-}^{(2)} +\frac{2i}{(5+k)} \pa 
% V_{-}^{(2)} \nonu \\
%& + &  \frac{2}{(5+k)} \pa T^{(1)}    (\hat{B}_1+i \hat{B}_2)
%- \frac{2}{(5+k)}  T^{(1)}  \pa  (\hat{B}_1+i \hat{B}_2)
%\nonu \\
%&+& \frac{2(-1+k)}{(5+k)^2}  (\hat{A}_1+i \hat{A}_2)  (\hat{A}_1-i \hat{A}_2)
% (\hat{B}_1+i \hat{B}_2) 
%\nonu \\
%& - & \frac{2(-3+k)}{(5+k)^2}  (\hat{A}_1-i \hat{A}_2)  (\hat{A}_1+i \hat{A}_2)
% (\hat{B}_1+i \hat{B}_2) +\frac{4}{(5+k)^2} \hat{A}_3 \hat{A}_3 
% (\hat{B}_1+i \hat{B}_2) 
%\nonu \\
%&+ & \frac{4}{(5+k)^2}   (\hat{B}_1+i \hat{B}_2)  (\hat{B}_1+i \hat{B}_2)
% (\hat{B}_1-i \hat{B}_2)
%+ \frac{4}{(5+k)^2}   (\hat{B}_1+i \hat{B}_2) \hat{B}_3 \hat{B}_3
%\nonu \\
%& - & \left. \frac{8}{(5+k)^2} \hat{B}_3 \hat{A}_3    (\hat{B}_1+i \hat{B}_2)
%\right](w) + \cdots, 
\label{B+w3} \\
\hat{B}_{-}(z) \, W^{(3)}(w) & = &
\frac{1}{(z-w)^3} \left[\frac{4k(-167+101k)}{3(5+k)^2(19+23k)} \right] 
\hat{B}_{-} (w) \nonu \\
& + &  \frac{1}{(z-w)^2} 
\left[- 
\frac{(15+4k)}{(5+k)} i U_{+}^{(2)}
-\frac{2(79+253k+82k^2)}{(5+k)^2(19+23k)} i \hat{A}_3 \hat{B}_{-}
\right. \nonu \\
& +&  \frac{2(1+k)}{(5+k)^2}  i \hat{A}_3 
\hat{B}_{-} 
-\frac{2k(181+17k)}{(5+k)^2(19+23k)} i \hat{B}_{-} \hat{B}_3 
\nonu \\
& + &  \left. \frac{k(-877 +151k)}{3(5+k)^2(19+23k)} \pa \hat{B}_{-}
-    \frac{(-101+39k+8k^2)}{(5+k)(19+23k)} 
T^{(1)} \hat{B}_{-} \right](w) \nonu \\
& + & \frac{1}{(z-w)} \left[- \frac{2k}{(5+k)^2} 
 i \hat{A}_3  \pa \hat{B}_{-} +\frac{4}{(5+k)} \hat{A}_3  
U_{+}^{(2)} 
+\frac{4(-3+2k)}{3(5+k)^2} \pa^2 \hat{B}_{-}
\right. \nonu \\
& + & 
\frac{4(57+199k+90k^2)}{(3+7k)(19+23k)(5+k)}  
\hat{B}_{-} \hat{T}
+ \frac{2}{(5+k)} \hat{B}_{-} T^{(2)}
\nonu \\
& + & \frac{2k}{(5+k)^2} i \hat{B}_{-} \pa \hat{A}_3
- \frac{2(-6+k)}{3(5+k)^2} i  \hat{B}_{-} \pa \hat{B}_3 
-  
\frac{1}{(5+k)}  \hat{B}_{-} \pa  T^{(1)}
\nonu \\
& - & \frac{4}{(5+k)}  \hat{B}_3 U_{+}^{(2)}
+ 
\frac{1}{(5+k)} i \pa  U_{+}^{(2)} 
+  \frac{2k}{(5+k)^2} i \hat{B}_3 \pa \hat{B}_{-}
\nonu \\
& - & 
\frac{2}{(5+k)} i \hat{G}_{11} \hat{G}_{21}
+\frac{4}{(5+k)^2}  \hat{A}_3 \hat{A}_3
 \hat{B}_{-}
 -\frac{2}{(5+k)}  \hat{B}_{-} W^{(2)}
\nonu \\
&- & \frac{8}{(5+k)^2}   \hat{A}_3  
 \hat{B}_{-} \hat{B}_3 
+\frac{4}{(5+k)^2}  \hat{B}_{-}
 \hat{B}_3 \hat{B}_3
+ \frac{4}{(5+k)^2} \hat{B}_{+} \hat{B}_{-} \hat{B}_{-}  
\nonu \\
&+ & \left. \frac{1}{(5+k)}  i T^{(1)}
\hat{B}_{-} \hat{B}_3 
-\frac{1}{(5+k)}   i \hat{B}_3 T^{(1)}
 \hat{B}_{-}  
\right](w) +   \cdots,
\label{B-w3} \\
\hat{B}_3(z) \, W^{(3)}(w) & = &
\frac{1}{(z-w)^3} 
\left[ -\frac{6(-3+k) }{(5+k)^2} \hat{A}_3 -
\frac{4}{(5+k)} i T^{(1)}
\right](w) \nonu \\
& + &
\frac{1}{(z-w)^2} \left[  
-\frac{2(399+1234k+859k^2+184k^3)}{(3+7k)(19+23k)(5+k)} i \hat{T}
+\frac{2(1+k)}{(5+k)} i T^{(2)}
\right. \nonu \\
&- & \frac{4(2+k)}{(5+k)^2} i \hat{A}_1 \hat{A}_1 
-\frac{4(2+k)}{(5+k)^2} i \hat{A}_2 \hat{A}_2
-\frac{2(1+2k)}{(5+k)^2} i \hat{A}_3 \hat{A}_3
+\frac{2(4+k)}{(5+k)} i W^{(2)} 
\nonu \\
& + & \frac{4(46+15k+5k^2)}{(5+k)^2(19+23k)} i
\hat{A}_3 \hat{B}_3 -\frac{2(7+k)}{(5+k)^2} i \hat{B}_1 \hat{B}_1 
-\frac{2(7+k)}{(5+k)^2} i \hat{B}_2 \hat{B}_2
\nonu \\
&- & \frac{2(133+361k+40k^2)}{(5+k)^2(19+23k)} i \hat{B}_3 \hat{B}_3 
-\frac{2(-41+31k+4k^2)}{(5+k)(19+23k)} T^{(1)} \hat{B}_3
\nonu \\
&+& \left.  \frac{2}{(5+k)} T^{(1)} \hat{A}_3
+\frac{3(-3+k)}{(5+k)^2} \pa \hat{A}_3 
 +  \frac{2}{(5+k)} i \pa T^{(1)}
\right](w)  + \cdots.
\nonu
%\frac{1}{(z-w)^2} \left[ -\frac{8i}{(5+k)} U_{+}^{(2)} -
%\frac{4(1+k)}{(5+k)^2} i \hat{A}_3 (\hat{B}_1 -i \hat{B}_2)\right](w) \nonu \\
%& + & 
%\frac{1}{(z-w)} \left[ 
%\frac{4(3+4k)}{(3+7k)(5+k)}  (\hat{B}_1-i \hat{B}_2) \hat{T}
%-\frac{2}{(5+k)}  (\hat{B}_1-i \hat{B}_2) T^{(2)} 
%\right. \nonu \\
%& - & \frac{2}{(5+k)}  (\hat{B}_1-i \hat{B}_2) W^{(2)}
%+\frac{4(-1+k)}{(5+k)^2} i  (\hat{B}_1-i \hat{B}_2) \pa \hat{A}_3
%\nonu \\
%& - & \frac{4}{(5+k)} \hat{B}_3   U_{+}^{(2)}
%+ \frac{2i}{(5+k)} \hat{B}_3 \pa  U_{+}^{(2)} + \frac{2}{(5+k)^2}
%\pa^2 (\hat{B}_1-i \hat{B}_2)
%\nonu \\
%&- & \frac{2}{(5+k)} \pa T^{(1)}   (\hat{B}_1-i \hat{B}_2)
%+\frac{4}{(5+k)^2}  (\hat{A}_1+i \hat{A}_2)
% (\hat{A}_1-i \hat{A}_2) (\hat{B}_1-i \hat{B}_2)
%\nonu \\
%& + & \frac{4}{(5+k)^2} \hat{A}_3 \hat{A}_3  (\hat{B}_1-i \hat{B}_2)
%+\frac{4(-4+k)}{(5+k)^2} \hat{A}_3  (\hat{B}_1-i \hat{B}_2) \hat{B}_3  
%\nonu \\
%&- & \frac{2}{(5+k)^2}  (\hat{B}_1+i \hat{B}_2)
% (\hat{B}_1-i \hat{B}_2)  (\hat{B}_1-i \hat{B}_2) 
%\nonu \\
%& + & \frac{6}{(5+k)^2}
% (\hat{B}_1-i \hat{B}_2)  (\hat{B}_1+i \hat{B}_2)  (\hat{B}_1-i \hat{B}_2)
%+ \frac{4}{(5+k)^2}
% (\hat{B}_1-i \hat{B}_2)  \hat{B}_3  \hat{B}_3
%\nonu \\
%& - & \frac{4(-2+k)}{(5+k)^2} \hat{B}_3 \hat{A}_3  (\hat{B}_1-i \hat{B}_2)
%+\frac{2}{(5+k)} T^{(1)} i (\hat{B}_1-i \hat{B}_2) \hat{B}_3 
%\nonu \\
%& - & \left. \frac{2}{(5+k)} i \hat{B}_3 T^{(1)}  (\hat{B}_1-i \hat{B}_2) 
%\right](w) + \cdots, 
%\nonu
\eea 
One can check the $U(1)$ charge conservation in these OPEs from Tables 
$2, 3$ and $5$.

As done in the OPEs between the four spin-$\frac{3}{2}$ currents and 
$16$ higher spin currents, (\ref{operesult})-(\ref{operesult3}), one obtains the following results between the spin-$1$
currents and the $16$ higher spin currents by focusing on the linear terms of
higher spin currents in the right hand side of the OPEs 
\bea
\hat{A}_{+} & \times & 
\left(
\begin{array}{cccc}
T^{(1)}, & T_{+}^{(\frac{3}{2})}, & T_{-}^{(\frac{3}{2})}, & T^{(2)}  \\
U^{(\frac{3}{2})}, & U_{+}^{(2)}, & U_{-}^{(2)}, & U^{(\frac{5}{2})}  \\
V^{(\frac{3}{2})}, & V^{(2)}_{+}, & V^{(2)}_{-}, & V^{(\frac{5}{2})}  \\
W^{(2)}, & W_{+}^{(\frac{5}{2})}, & W_{-}^{(\frac{5}{2})}, & W^{(3)}  
\end{array} \right)
%\nonu \\
 \rightarrow 
\left(
\begin{array}{cccc}
0 & U^{(\frac{3}{2})}, & 0, & U_{-}^{(2)}  \\
0,  & 0, & 0,  & 0  \\
T_{-}^{(\frac{3}{2})}, & T^{(1)}, 
T^{(2)},  W^{(2)}, & 0, &  T_{-}^{(\frac{3}{2})}, W_{-}^{(\frac{5}{2})}   \\
U_{-}^{(2)}, & U^{(\frac{3}{2})},U^{(\frac{5}{2})}, & 0, &  U_{-}^{(2)}
\end{array} \right),
\nonu \\
\hat{A}_{-} & \times & 
\left(
\begin{array}{cccc}
T^{(1)}, & T_{+}^{(\frac{3}{2})}, & T_{-}^{(\frac{3}{2})}, & T^{(2)}  \\
U^{(\frac{3}{2})}, & U_{+}^{(2)}, & U_{-}^{(2)}, & U^{(\frac{5}{2})}  \\
V^{(\frac{3}{2})}, & V^{(2)}_{+}, & V^{(2)}_{-}, & V^{(\frac{5}{2})}  \\
W^{(2)}, & W_{+}^{(\frac{5}{2})}, & W_{-}^{(\frac{5}{2})}, & W^{(3)}  
\end{array} \right)
%\nonu \\
 \rightarrow 
\left(
\begin{array}{cccc}
0, & 0, & V^{(\frac{3}{2})},  & V_{+}^{(2)}  \\
T_{+}^{(\frac{3}{2})}, & 0,  & T^{(1)}, 
T^{(2)},  W^{(2)},  &  T_{+}^{(\frac{3}{2})}, W_{+}^{(\frac{5}{2})}   \\
0, & 0, & 0, & 0 \\ 
V_{+}^{(2)}, &0, & V^{(\frac{3}{2})},V^{(\frac{5}{2})}, &  V_{+}^{(2)}
\end{array} \right),
\nonu  \\
\hat{A}_{3} & \times & 
\left(
\begin{array}{cccc}
T^{(1)}, & T_{+}^{(\frac{3}{2})}, & T_{-}^{(\frac{3}{2})}, & T^{(2)}  \\
U^{(\frac{3}{2})}, & U_{+}^{(2)}, & U_{-}^{(2)}, & U^{(\frac{5}{2})}  \\
V^{(\frac{3}{2})}, & V^{(2)}_{+}, & V^{(2)}_{-}, & V^{(\frac{5}{2})}  \\
W^{(2)}, & W_{+}^{(\frac{5}{2})}, & W_{-}^{(\frac{5}{2})}, & W^{(3)}  
\end{array} \right)
%\nonu \\
 \rightarrow 
\left(
\begin{array}{cccc}
0, & T_{+}^{(\frac{3}{2})}, & T_{-}^{(\frac{3}{2})},  & T^{(1)}  \\
U^{(\frac{3}{2})}, & 0,  & U_{-}^{(2)}, 
  &  U^{(\frac{3}{2})}, U^{(\frac{5}{2})}   \\
 V^{(\frac{3}{2})}, & V_{+}^{(2)}, & 0, &  V^{(\frac{3}{2})},V^{(\frac{5}{2})} \\ 
T^{(1)}, & T_{+}^{(\frac{3}{2})},W_{+}^{(\frac{5}{2})}, & 
T_{-}^{(\frac{3}{2})},W_{-}^{(\frac{5}{2})}, &  T^{(1)}, T^{(2)}, W^{(2)}
\end{array} \right),
\nonu  \\
\hat{B}_{+} & \times & 
\left(
\begin{array}{cccc}
T^{(1)}, & T_{+}^{(\frac{3}{2})}, & T_{-}^{(\frac{3}{2})}, & T^{(2)}  \\
U^{(\frac{3}{2})}, & U_{+}^{(2)}, & U_{-}^{(2)}, & U^{(\frac{5}{2})}  \\
V^{(\frac{3}{2})}, & V^{(2)}_{+}, & V^{(2)}_{-}, & V^{(\frac{5}{2})}  \\
W^{(2)}, & W_{+}^{(\frac{5}{2})}, & W_{-}^{(\frac{5}{2})}, & W^{(3)}  
\end{array} \right)
%\nonu \\
 \rightarrow 
\left(
\begin{array}{cccc}
0 & V^{(\frac{3}{2})}, & 0, & V_{-}^{(2)}  \\
T_{-}^{(\frac{3}{2})}, & T^{(1)}, 
T^{(2)},  W^{(2)}, & 0, &  T_{-}^{(\frac{3}{2})}, W_{-}^{(\frac{5}{2})}   \\
 0, & 0, & 0, & 0 \\
V_{-}^{(2)}, & V^{(\frac{3}{2})},V^{(\frac{5}{2})}, & 0, &  V_{-}^{(2)}
\end{array} \right),
\nonu \\
\hat{B}_{-} & \times & 
\left(
\begin{array}{cccc}
T^{(1)}, & T_{+}^{(\frac{3}{2})}, & T_{-}^{(\frac{3}{2})}, & T^{(2)}  \\
U^{(\frac{3}{2})}, & U_{+}^{(2)}, & U_{-}^{(2)}, & U^{(\frac{5}{2})}  \\
V^{(\frac{3}{2})}, & V^{(2)}_{+}, & V^{(2)}_{-}, & V^{(\frac{5}{2})}  \\
W^{(2)}, & W_{+}^{(\frac{5}{2})}, & W_{-}^{(\frac{5}{2})}, & W^{(3)}  
\end{array} \right)
%\nonu \\
 \rightarrow 
\left(
\begin{array}{cccc}
0, &0, & U^{(\frac{3}{2})},   & U_{+}^{(2)}  \\
0, & 0, & 0, & 0 \\
T_{+}^{(\frac{3}{2})}, & 0, & T^{(1)}, 
T^{(2)},  W^{(2)}, &  T_{+}^{(\frac{3}{2})}, W_{+}^{(\frac{5}{2})}   \\ 
U_{+}^{(2)}, & 0, & U^{(\frac{3}{2})},U^{(\frac{5}{2})}, &  U_{+}^{(2)}
\end{array} \right),
\nonu \\
\hat{B}_{3} & \times & 
\left(
\begin{array}{cccc}
T^{(1)}, & T_{+}^{(\frac{3}{2})}, & T_{-}^{(\frac{3}{2})}, & T^{(2)}  \\
U^{(\frac{3}{2})}, & U_{+}^{(2)}, & U_{-}^{(2)}, & U^{(\frac{5}{2})}  \\
V^{(\frac{3}{2})}, & V^{(2)}_{+}, & V^{(2)}_{-}, & V^{(\frac{5}{2})}  \\
W^{(2)}, & W_{+}^{(\frac{5}{2})}, & W_{-}^{(\frac{5}{2})}, & W^{(3)}  
\end{array} \right)
%\nonu \\
 \rightarrow 
\left(
\begin{array}{cccc}
0, & T_{+}^{(\frac{3}{2})}, & T_{-}^{(\frac{3}{2})},  & T^{(1)}  \\
U^{(\frac{3}{2})}, & U_{+}^{(2)}, & 0, 
  &  U^{(\frac{3}{2})}, U^{(\frac{5}{2})}   \\
 V^{(\frac{3}{2})}, & 0, & V_{-}^{(2)}, &  V^{(\frac{3}{2})},V^{(\frac{5}{2})} \\ 
T^{(1)}, & T_{+}^{(\frac{3}{2})}, W_{+}^{(\frac{5}{2})}, & 
T_{-}^{(\frac{3}{2})},W_{-}^{(\frac{5}{2})}, &  T^{(1)}, T^{(2)}, W^{(2)}
\end{array} \right).
\nonu
\eea
Note that the zeros of the right hand sides for the OPEs between 
$\hat{A}_{\pm}$ and $\hat{B}_{\pm}$ acting on the $16$ higher spin currents
are located at 
the row and column containing the corresponding spin-$1$ current in
(\ref{fourn2}). 

%%%%%%%%%%%%%%%%%%
\subsection{The OPEs between six spin $1$ currents and the higher spin 
currents in different basis}
%%%%%%%%%%%%%%%%%%%%

In other basis described in section $6$, 
the OPEs between the six spin-$1$ currents and the 
four (redefined an hatted)
spin-$\frac{3}{2}$ currents in (\ref{16new}) can be summarized by
\bea
\hat{A}_{\pm} (z) \, \hat{T}_{\pm}^{(\frac{3}{2})}(w) 
& = &  \mp \frac{1}{(z-w)} \, i \, 
\left(
\begin{array}{c}
\hat{U}^{(\frac{3}{2})}     \\ 
\hat{V}^{(\frac{3}{2})}    \\
\end{array} \right)(w) +\cdots,
%\label{newa+t+3half} 
\nonu
\\
\hat{A}_3(z) \, \hat{T}_{\pm}^{(\frac{3}{2})}(w) 
& = & \pm \frac{1}{(z-w)} \, \frac{i}{2} \, 
\hat{T}_{\pm}^{(\frac{3}{2})}(w) +\cdots=\hat{B}_3(z) \, \hat{T}_{\pm}^{(\frac{3}{2})}(w), 
\nonu \\
\hat{B}_{\pm} (z) \, \hat{T}_{\pm}^{(\frac{3}{2})}(w) 
& = &  \mp \frac{1}{(z-w)}  i 
\left(
\begin{array}{c}
 \hat{V}^{(\frac{3}{2})}  \\
 \hat{U}^{(\frac{3}{2})} \\
\end{array} \right)(w) 
 +\cdots,
%\label{newb+t+3half}
\nonu
\\
\hat{A}_{\mp} (z)   \, 
\left(
\begin{array}{c} 
\hat{U}^{(\frac{3}{2})} \\
\hat{V}^{(\frac{3}{2})} \\
\end{array} \right)(w) & = & 
\mp \frac{1}{(z-w)} \, i \, \hat{T}_{\pm}^{(\frac{3}{2})}(w)+\cdots, 
%\label{newa-u3half} 
\nonu
\\
\hat{A}_3(z) \, 
\left(
\begin{array}{c} 
\hat{U}^{(\frac{3}{2})} \\
\hat{V}^{(\frac{3}{2})} \\
\end{array} \right)(w)
 & = & \mp
\frac{1}{(z-w)} \, \frac{i}{2} \, 
\left(
\begin{array}{c} 
\hat{U}^{(\frac{3}{2})} \\
\hat{V}^{(\frac{3}{2})} \\
\end{array} \right)(w) +\cdots =
-\hat{B}_3(z) \, 
\left(
\begin{array}{c} 
\hat{U}^{(\frac{3}{2})} \\
\hat{V}^{(\frac{3}{2})} \\
\end{array} \right)(w), 
\nonu \\
\hat{B}_{\pm} (z) \, 
\left(
\begin{array}{c} 
\hat{U}^{(\frac{3}{2})} \\
\hat{V}^{(\frac{3}{2})} \\
\end{array} \right)(w)
 & = & 
\pm \frac{1}{(z-w)} i 
 \hat{T}_{\mp}^{(\frac{3}{2})}(w)+\cdots. 
%\label{newb+u3half}
\label{newope1ope3half}
\eea
The four spin-$\frac{3}{2}$ currents, $\hat{T}_{\pm}^{(\frac{3}{2})}(z)$,
$\hat{U}^{(\frac{3}{2})}(z)$ and $\hat{V}^{(\frac{3}{2})}(z)$,
 in (\ref{newope1ope3half}) 
transform nontrivially under the 
two $SU(2)$ currents.
It is easy to see that there are no $\hat{G}_a(w)$-dependences in the 
right hand side of (\ref{newope1ope3half}).
Furthermore, 
if one writes the above four spin-$\frac{3}{2}$ currents,
$\hat{U}^{(\frac{3}{2})}(w)$, $\hat{T}_{-}^{(\frac{3}{2})}(w)$, 
$\hat{T}_{+}^{(\frac{3}{2})}(w)$
and $\hat{V}^{(\frac{3}{2})}(w)$ 
as $11$-,$12$-,$21$-,$22$-components 
of $({\bf 2}, {\bf 2} )$ in $SU(2) \times SU(2)$ 
respectively, then
the half of (\ref{newope1ope3half}) will lead to  
the equation Appendix (\ref{agnonlinear}) where $\hat{G}_a(w)$ is replaced 
with the above four quantities and the remaining equations 
of Appendix (\ref{newope1ope3half})  become the equation 
Appendix (\ref{bgnonlinear})
with same replacement of $\hat{G}_a(w)$.

The OPEs between the six spin-$1$ currents and the 
six spin-$2$ currents in (\ref{16new}) can be described by
\bea
\hat{A}_{+}(z) \, 
\left(
\begin{array}{c} 
\hat{U}_{-}^{(2)} \\
\hat{T}^{(2)} -\hat{W}^{(2)} \\
\hat{V}_{+}^{(2)} 
\end{array} \right)
(w) & = &
\frac{1}{(z-w)} \,  
\left(
\begin{array}{c} 
0 \\
2 i \hat{U}_{-}^{(2)} \\
\{  \hat{A}_{+} \, \hat{V}_{+}^{(2)}\}_{-1} 
\end{array} \right)
(w) +\cdots,
\nonu \\
\hat{A}_{-}(z) \, 
\left(
\begin{array}{c} 
\hat{U}_{-}^{(2)} \\
\hat{T}^{(2)} -\hat{W}^{(2)} \\
\hat{V}_{+}^{(2)} 
\end{array} \right)
(w) & = &
\frac{1}{(z-w)} \,  
\left(
\begin{array}{c} 
\{  \hat{A}_{-} \, \hat{U}_{-}^{(2)}\}_{-1} 
\\
2 i \hat{V}_{+}^{(2)} \\
0 
\end{array} \right)
(w) +\cdots,
\nonu \\
%\hat{A}_{\pm}(z)   \, (\hat{T}^{(2)}-\hat{W}^{(2)})(w) & = &
%\frac{1}{(z-w)}\, 2 i  
%\left(
%\begin{array}{c}
%\hat{U}^{(2)}_{-} \\
% \hat{V}^{(2)}_{+} \\
%\end{array} \right)(w) 
%+  \cdots,
%\label{newa+t2} \\
%\nonu \\
%\hat{A}_{\mp} (z) \, 
%\left(
%\begin{array}{c} 
%\hat{U}_{-}^{(2)} \\
%\hat{V}_{+}^{(2)} 
%\end{array} \right)
%(w) & = &
% \frac{1}{(z-w)} \left[ ... \right](w)+\cdots,
%\label{newa-u-2} 
\nonu \\
\hat{A}_3(z) \, 
\left(
\begin{array}{c} 
\hat{U}_{-}^{(2)} \\
\hat{T}^{(2)} -\hat{W}^{(2)} \\
\hat{V}_{+}^{(2)} 
\end{array} \right)
(w) & = &
\frac{1}{(z-w)} \, i 
\left(
\begin{array}{c} 
-\hat{U}_{-}^{(2)} \\
0 \\
\hat{V}_{+}^{(2)} 
\end{array} \right)
(w) +\cdots,
\nonu \\
%\hat{B}_{\pm} (z) \, (\hat{T}^{(2)} +\hat{W}^{(2)})(w) & = &
% \frac{1}{(z-w)}\,  
% 2 i 
%\left(
%\begin{array}{c}
%V_{-}^{(2)} \\
%U_{+}^{(2)} \\
%\end{array} \right)(w) +\cdots.
%\label{newb+t2} 
%\nonu \\
%\hat{B}_{\pm}  (z) \, 
%\left(
%\begin{array}{c} 
%\hat{U}_{+}^{(2)} \\
%\hat{V}_{-}^{(2)} 
%\end{array} \right)
%(w) & = & 
%\label{newb+u+2} 
% \frac{1}{(z-w)}  \left[ ... ... \right](w) +\cdots,
%\nonu \\
\hat{B}_{+}(z) \, 
\left(
\begin{array}{c} 
\hat{V}_{-}^{(2)} \\
\hat{T}^{(2)} +\hat{W}^{(2)} \\
\hat{U}_{+}^{(2)} \\
\end{array} \right)
(w)
 & = & 
   \frac{1}{(z-w)} \,  
\left(
\begin{array}{c} 
0 \\
2 i \hat{V}_{-}^{(2)} \\
\{  \hat{B}_{+} \, \hat{U}_{+}^{(2)}\}_{-1}
 \\
\end{array} \right)
(w) +\cdots,
\nonu \\
\hat{B}_{-}(z) \, 
\left(
\begin{array}{c} 
\hat{V}_{-}^{(2)} \\
\hat{T}^{(2)} +\hat{W}^{(2)} \\
\hat{U}_{+}^{(2)} \\
\end{array} \right)
(w)
 & = & 
   \frac{1}{(z-w)} \,  
\left(
\begin{array}{c} 
\{  \hat{B}_{-} \, \hat{V}_{-}^{(2)}\}_{-1}
 \\
2 i \hat{U}_{+}^{(2)} \\
0 \\
\end{array} \right)
(w) +\cdots,
\nonu \\
\hat{B}_3(z) \, 
\left(
\begin{array}{c} 
\hat{V}_{-}^{(2)} \\
\hat{T}^{(2)} +\hat{W}^{(2)} \\
\hat{U}_{+}^{(2)} \\
\end{array} \right)
(w)
 & = & 
   \frac{1}{(z-w)} \,  i\, 
\left(
\begin{array}{c} 
-\hat{V}_{-}^{(2)} \\
0 \\
\hat{U}_{+}^{(2)} \\
\end{array} \right)
(w) +\cdots.
\label{newope1ope2}
\eea
The first half of (\ref{newope1ope2}) implies that 
the spin-$2$ currents,
$\hat{U}_{-}^{(2)}(w)$,  $(\hat{T}^{(2)}-\hat{W}^{(2)})(w)$ 
and $\hat{V}_{+}^{(2)}(w)$ transform under the 
$SU(2)$ current $\hat{A}_i(z)$ nontrivially and the OPEs between the other 
$SU(2)$ current $\hat{B}_i(z)$ and these three spin-$2$ currents 
do not contain any singular 
terms (i.e., singlet under the second $SU(2)$ realized by $\hat{B}_i(z)$).
Note that the OPE between the $\hat{A}_{+}(z)$ and the top component 
$\hat{U}_{-}^{(2)}(w)$ does not produce any singular term,
 the OPE between the $\hat{A}_{+}(z)$ and the middle component 
$(\hat{T}^{(2)}-\hat{W}^{(2)})(w)$ produces the above top component
$\hat{U}_{-}^{(2)}(w)$
and  the OPE between the $\hat{A}_{+}(z)$ and the bottom component 
$\hat{V}_{+}^{(2)}(w)$ produces the above middle component 
$(\hat{T}^{(2)}-\hat{W}^{(2)})(w)$
and other nonlinear 
terms which will be determined soon.

Similarly, 
the remaining equations of (\ref{newope1ope2}) imply that 
the spin-$2$ currents, $\hat{V}_{-}^{(2)}(w)$,
$(\hat{T}^{(2)}+\hat{W}^{(2)})(w)$
and $\hat{U}_{+}^{(2)}(w)$   transform under the 
$SU(2)$ current $\hat{B}_i(z)$ nontrivially and the OPEs between 
these three spin-$2$ currents
and the other 
$SU(2)$ current $\hat{A}_i(z)$ do not contain the singular terms
(that is singlet under the first $SU(2)$ realized by $\hat{A}_i(z)$).
One can also analyze the raising and lowering operators
of the second $SU(2)$ 
acting on the three representations of $SU(2)$ as above. 
In summary, the above six spin-$2$ currents transform as
$({\bf 3},{\bf 1}) \oplus ({\bf 1},{\bf 3})$ under the $SU(2) \times SU(2)$.
All the second-order poles appeared in previous basis are disappeared in this
new basis (\ref{newope1ope2}).

The nonlinear terms appearing in (\ref{newope1ope2})
are given by
\bea
\{  \hat{A}_{+} \, \hat{V}_{+}^{(2)}\}_{-1}(w) & = &  
\{  \hat{A}_{-} \, \hat{U}_{-}^{(2)}\}_{-1}(w)
=   i   
(\hat{T}^{(2)} - \hat{W}^{(2)})(w)
+  \frac{6 i k}{(3+7k)} \tilde{\hat{T}}(w), 
\nonu \\
\{  \hat{B}_{+} \, \hat{U}_{+}^{(2)}\}_{-1}(w) & = &  
\{  \hat{B}_{-} \, \hat{V}_{-}^{(2)}\}_{-1}(w)
=   i   
(\hat{T}^{(2)} + \hat{W}^{(2)})(w)
-  \frac{2 i (3+4k)}{(3+7k)} \tilde{\hat{T}}(w),
\nonu 
\eea
where the spin-$2$ primary field (under the $\hat{T}(z)$) is given by 
\bea
\tilde{\hat{T}}(w) & = &
 \left[ \hat{T} -
\frac{3  (9 k+13)}{20 (k+2)}
\hat{T}^{(1)} \hat{T}^{(1)} 
+
\frac{ 1  }{(k+2)}
\sum_{i=1}^3 \hat{A}_i \hat{A}_i +   
\frac{1}{5} \sum_{i=1}^3 \hat{B}_i \hat{B}_i \right](w).
\label{tildehatt}
\eea
The presence of $\hat{T}^{(1)} \hat{T}^{(1)}(w)$ makes 
the fourth-order pole in the OPE between $\hat{T}(z)$ and 
$\tilde{\hat{T}}(w)$ to vanish.
The third-order pole in the OPE between $\hat{T}(z)$
and each term of (\ref{tildehatt}) vanishes. 

%\bea
%\hat{A}_{\pm}(z) \, \hat{W}^{(2)}(w) & = &
%-\frac{1}{(z-w)} i 
%\left(
%\begin{array}{c} 
%\hat{U}_{-}^{(2)} \\
%\hat{V}_{+}^{(2)} 
%\end{array} \right)
%(w) +  \cdots,
%\label{newa+w2} 
%\nonu \\
%\hat{B}_{\pm}(z) \, \hat{W}^{(2)}(w) & = &
%\frac{1}{(z-w)} i 
%\left(
%\begin{array}{c} 
%\hat{V}_{-}^{(2)} \\
%\hat{U}_{+}^{(2)} 
%\end{array} \right)(w)
%+  \cdots,
%\label{newb+w2} 
%\eea

Finally, the nontrivial OPEs between the six spin-$1$ currents and the 
four spin-$\frac{5}{2}$ currents in (\ref{16new}) can be given by
\bea
\hat{A}_{\mp} (z) \, 
\left(
\begin{array}{c}
\hat{U}^{(\frac{5}{2})} \\
\hat{V}^{(\frac{5}{2})} \end{array} \right) (w) & = & 
\mp \frac{1}{(z-w)} i \hat{W}_{\pm}^{(\frac{5}{2})}(w) +  \cdots,
%\label{newA-u5half} 
\nonu \\
\hat{A}_3(z) \, \left(
\begin{array}{c} 
\hat{U}^{(\frac{5}{2})} \\
\hat{V}^{(\frac{5}{2})} 
\end{array} \right) (w) & = & 
\mp
   \frac{1}{(z-w)} \, \frac{i}{2} 
\, \left( \begin{array}{c} 
\hat{U}^{(\frac{5}{2})} \\
\hat{V}^{(\frac{5}{2})}
\end{array} \right) +\cdots=-\hat{B}_3(z) \, \left(
\begin{array}{c} 
\hat{U}^{(\frac{5}{2})} \\
\hat{V}^{(\frac{5}{2})} 
\end{array} \right) (w),
\nonu \\
\hat{B}_{\pm} (z) \, 
\left(
\begin{array}{c} 
\hat{U}^{(\frac{5}{2})} \\
\hat{V}^{(\frac{5}{2})} 
\end{array} \right)  (w) & = & 
\mp \frac{1}{(z-w)} i \hat{W}_{\mp}^{(\frac{5}{2})}(w) +\cdots,
\nonu 
%\hat{B}_3(z) \, \left(
%\begin{array}{c} 
%\hat{U}^{(\frac{5}{2})} \\
%\hat{V}^{(\frac{5}{2})} 
%\end{array} \right) (w) & = & 
%  \frac{1}{(z-w)} \, \frac{i}{2}  
%\left( \begin{array}{c}
%\hat{U}^{(\frac{5}{2})} \\
%-\hat{V}^{(\frac{5}{2})} 
%\end{array} \right) (w)  
% +\cdots.
%\nonu
%\bea
%\hat{A}_{+} (z) \, V^{(\frac{5}{2})}(w) & = &
% \frac{1}{(z-w)} \left[ ... \right](w) +\cdots,
%\nonu \\
%\hat{A}_3(z) \, \hat{V}^{(\frac{5}{2})}(w) & = &
% \frac{1}{(z-w)}\, \frac{i}{2}\, 
%\hat{V}^{(\frac{5}{2})}(w) + \cdots, 
%\nonu \\
%\hat{B}_{-} (z) \, \hat{V}^{(\frac{5}{2})}(w) & = &
% \frac{1}{(z-w)} \left[ ... \right](w) +  \cdots,
%\label{B-v5half} 
%\\
%\hat{B}_3(z) \, \hat{V}^{(\frac{5}{2})}(w) & = &
%- \frac{1}{(z-w)} \, \frac{i}{2} \hat{V}^{(\frac{5}{2})}(w) +  \cdots.
%\nonu
%\eea
\\
\hat{A}_{\pm}(z) \, \hat{W}_{\pm}^{(\frac{5}{2})}(w) & = &
\mp
 \frac{1}{(z-w)} i \left( \begin{array}{c} 
\hat{U}^{(\frac{5}{2})} \\
\hat{V}^{(\frac{5}{2})} \\
\end{array} \right)(w)+\cdots, 
\nonu \\
\hat{A}_3(z) \, \hat{W}_{\pm}^{(\frac{5}{2})}(w) & = &
\pm  \frac{1}{(z-w)} \, \frac{i}{2}  \hat{W}_{\pm}^{(\frac{5}{2})}(w) +\cdots
=\hat{B}_3(z) \, \hat{W}_{\pm}^{(\frac{5}{2})}(w), 
\nonu \\
\hat{B}_{\pm}(z) \, \hat{W}_{\pm}^{(\frac{5}{2})}(w) & = &
\pm
 \frac{1}{(z-w)} i \left( \begin{array}{c} 
\hat{V}^{(\frac{5}{2})} \\
\hat{U}^{(\frac{5}{2})} \\
\end{array}  \right)(w)+\cdots.
%\label{B+w+5half} 
%\hat{B}_3(z) \, \hat{W}_{\pm}^{(\frac{5}{2})}(w) & = &
%\pm \frac{1}{(z-w)} \, \frac{i}{2}  \hat{W}_{\pm}^{(\frac{5}{2})}(w) +\cdots. 
\label{newope1ope5half}
\eea
The four spin-$\frac{5}{2}$ currents, $\hat{W}_{\pm}^{(\frac{5}{2})}(z)$,
$\hat{U}^{(\frac{5}{2})}(z)$ and $\hat{V}^{(\frac{5}{2})}(z)$,
 in (\ref{newope1ope5half}) 
transform nontrivially under the 
two $SU(2)$ currents as in (\ref{newope1ope3half}).
There are sign changes in the OPEs between $\hat{B}_{\pm}(z)$
and these spin-$\frac{5}{2}$ currents.
All the second-order poles appeared in previous basis are disappeared in this
new basis (\ref{newope1ope5half}).
%\bea
%\hat{A}_{-}(z) \, \hat{W}_{-}^{(\frac{5}{2})}(w) & = &
% \frac{1}{(z-w)} \left[ ... \right](w)+\cdots, 
%\nonu \\
%\hat{A}_3(z) \, \hat{W}_{-}^{(\frac{5}{2})}(w) & = &
% -  \frac{1}{(z-w)} \, \frac{i}{2}  
%\hat{W}_{-}^{(\frac{5}{2})}(w) \nonu \\
%& + & \cdots, 
%\nonu \\
%\hat{B}_{-}(z) \, \hat{W}_{-}^{(\frac{5}{2})}(w) & = &
% \frac{1}{(z-w)} \left[ ... \right](w)+\cdots,
%\nonu \\
%\hat{B}_3(z) \, \hat{W}_{-}^{(\frac{5}{2})}(w) & = &
%-  \frac{1}{(z-w)} \, \frac{i}{2}  \hat{W}_{-}^{(\frac{5}{2})}(w) \nonu \\
%& + & \cdots. 
%\nonu
%\eea

Note that the OPEs between the six spin-$1$ currents and the 
spin-$1$ current $\hat{T}^{(1)}(w) =T^{(1)}(w)$ 
do not have any singular terms 
and similarly the OPEs between those six spin-$1$ currents 
and the spin-$3$ current $\hat{W}^{(3)}(w)$ do not contain 
any singular terms. 
This reflects the fact that these two higher spin currents transform
as $({\bf 1},{\bf 1})$ under the $SU(2) \times SU(2)$ respectively. 
In other words, $\hat{A}_i(z) \, \hat{T}^{(1)}(w) = 
\hat{B}_i(z) \, \hat{T}^{(1)}(w) =\hat{A}_i(z) \, \hat{W}^{(3)}(w) = 
\hat{B}_i(z) \, \hat{W}^{(3)}(w) = +\cdots$.
The OPEs between the four spin-$\frac{3}{2}$ currents 
$\hat{G}_a(z)$ and the $16$ higher spin currents
in this new basis will be given at the end of Appendix $C$. 

%%%%%%%%%%%%%%%%%%%%%%%%%%%%%%%%%%%%%%%%%%%%%%%%%%%%%%%%%%%%%%%%%%%%%
%%%%%%%%%%%%%%%%%%%%%%%%%%%%%%%%%%%%%%%%%%%%%%%%%%%%%%%%%%%%%%%%%%%%%
\section{The OPEs between the large ${\cal N}=4$ nonlinear 
algebra currents and the higher spin currents-II}
%C%%%%%%%%%%%%%%%%%%%%%%%%%%%%%%%%%%%%%%%%%%%%%%%%%%%%%%%%%%%%%%%%%%%
%%%%%%%%%%%%%%%%%%%%%%%%%%%%%%%%%%%%%%%%%%%%%%%%%%%%%%%%%%%%%%%%%%%%

In this Appendix, we present the OPEs between the four 
spin-$\frac{3}{2}$ currents 
in section $3$ and the $16$ higher spin currents in section $4$.

%%%%%%%%%%%%%%%%%%
\subsection{The OPEs between four spin $\frac{3}{2}$ currents 
and the higher spin 
current of spins $(1, \frac{3}{2}, \frac{3}{2}, 2)$
}
%\label{spin3halffirstmultiplet}
%%%%%%%%%%%%%%%%%%%%

From the explicit expressions (\ref{nong11})-(\ref{nong21}) 
and (\ref{t1}), the following OPEs between the 
spin-$\frac{3}{2}$ currents and the 
spin-$1$ current  can be derived 
\bea
\left(
\begin{array}{c}
\hat{G}_{11} \\
\hat{G}_{22}
\end{array} \right)(z) \, T^{(1)}(w) & = &
\frac{1}{(z-w)} \left(
\begin{array}{c}
 \hat{G}_{11} + 2 U^{(\frac{3}{2})}
\\
 -\hat{G}_{22} + 2 V^{(\frac{3}{2})}
\end{array}
\right)(w) +
\cdots, 
\label{G11t1} \\
\left(
\begin{array}{c}
\hat{G}_{12} \\
\hat{G}_{21} 
\end{array} \right) (z) \, T^{(1)}(w) & = &
\frac{1}{(z-w)} \left(
\begin{array}{c} 
-\hat{G}_{12} + 2 T_{-}^{(\frac{3}{2})} \\
\hat{G}_{21} + 2 T_{+}^{(\frac{3}{2})}
\end{array}
\right)(w) +
\cdots. 
\label{G12t1} 
\eea
One can check the $U(1)$ charge conservation in these OPEs from Table 
$2$.

For the spin-$\frac{3}{2}$ currents (\ref{t+3half}) and (\ref{t-3half}), 
one obtains the 
following OPEs between the spin-$\frac{3}{2}$ currents and the 
spin-$\frac{3}{2}$ currents as follows:
\bea
\left(
\begin{array}{c}
\hat{G}_{11} \\
\hat{G}_{22}
\end{array} \right)(z) \, 
T_{\pm}^{(\frac{3}{2})} (w) & = & 
\frac{1}{(z-w)^2} \frac{2k}{(5+k)} \left[ i \hat{B}_{\mp} \right](w)  
\label{G11t+3half}
\\
& + &
\frac{1}{(z-w)} \left[ -\left(
\begin{array}{c} 
 U_{+}^{(2)} \\
V_{-}^{(2)} 
\end{array}
\right) +\frac{1}{(5+k)} \left( 
\mp 4 \hat{A}_3  \hat{B}_{\mp}
+ k i \pa  \hat{B}_{\mp}
\right) \right](w) +\cdots,
 \nonu \\
\left(
\begin{array}{c}
\hat{G}_{12} \\
\hat{G}_{21} 
\end{array} \right)(z) \, T_{\pm}^{(\frac{3}{2})}(w) & = & 
\mp \frac{1}{(z-w)^3}\, \frac{6k}{(5+k)} +
 \frac{1}{(z-w)^2} \left[ \frac{2i}{(5+k)} \left( -3 \hat{A}_3 - 
k \hat{B}_3 \right) +  T^{(1)}\right](w) \nonu \\
& + &
\frac{1}{(z-w)} \left[ \mp \frac{6k}{(3+7k)}\hat{T} 
\mp T^{(2)} +\frac{i}{(5+k)} \left( 
-3  \pa \hat{A}_3 - k \pa \hat{B}_3
\right) \right.
\nonu \\
& + & \left. \frac{1}{2} \pa T^{(1)} \right](w)  + \cdots,
\label{G12t+3half} \\
\left(
\begin{array}{c}
\hat{G}_{21} \\
\hat{G}_{12} \\
\end{array} \right)(z) \, T_{\pm}^{(\frac{3}{2})}(w) & = & 
\frac{1}{(z-w)} \frac{2}{(5+k)} \left[ 
\mp \hat{A}_{\mp} 
 \hat{B}_{\mp}  \right](w) +\cdots,
\label{G21t+3half} \\
\left(
\begin{array}{c}
\hat{G}_{22} \\
\hat{G}_{11} \\
\end{array} \right)(z) \, T_{\pm}^{(\frac{3}{2})}(w) & = & 
\frac{1}{(z-w)^2} \frac{6}{(5+k)} \left[ i \hat{A}_{\mp} \right](w) \nonu \\
& + &
\frac{1}{(z-w)} \left[ - 
\left(
\begin{array}{c} 
V_{+}^{(2)} \\
U_{-}^{(2)} \\
\end{array} \right)
+\frac{3}{(5+k)}  
 i  \pa  \hat{A}_{\mp}  \right](w) +\cdots.
\label{G22t+3half}
\eea
The $U(1)$ charge conservation in these OPEs can be checked from Tables 
$2$ and $3$.

For the last component spin-$2$ current,
one calculates the following OPEs between the spin-$\frac{3}{2}$ currents 
and the spin-$2$ current as follows:
\bea
\hat{G}_{11}(z) \, T^{(2)}(w) & = &
\frac{1}{(z-w)^2} \left[ \frac{(12-11k+5k^2)}{(3+7k)(5+k)}
\hat{G}_{11} +\frac{(-3+k)}{(5+k)} U^{(\frac{3}{2})} \right](w)
\nonu \\
& + & \frac{1}{(z-w)} \left[ U^{(\frac{5}{2})} +
\frac{k(-39+5k)}{(9+21k)(5+k)} \pa \hat{G}_{11} +
\frac{(-7+k)}{3(5+k)} \pa   U^{(\frac{3}{2})} \right. \nonu \\
& + & \left. \frac{2}{(5+k)} 
\left( 2 i \hat{A}_3 \, (\hat{G}_{11} +  
U^{(\frac{3}{2})}) -i \hat{B}_{-} T_{-}^{(\frac{3}{2})}  \right) \right](w) +\cdots,
\label{G11t2} \\
\left(
\begin{array}{c}
\hat{G}_{12} \\
\hat{G}_{21} \\
\end{array} \right) (z) \, T^{(2)}(w) & = &
\frac{1}{(z-w)^2} \left[ -\frac{(-3+38k+9k^2)}{(3+7k)(5+k)}
\left(
\begin{array}{c}
\hat{G}_{12} \\
\hat{G}_{21} \\
\end{array} \right)  
\pm \frac{(11+3k)}{(5+k)} T_{\mp}^{(\frac{3}{2})} \right](w)
\label{G12t2}
\\
& + & \frac{1}{(z-w)} \left[ -
\frac{(3+22k+3k^2)}{(3+7k)(5+k)} \pa 
\left(
\begin{array}{c}
\hat{G}_{12} \\
\hat{G}_{21} \\
\end{array} \right) \pm
\pa   T_{\mp}^{(\frac{3}{2})} \right. \nonu \\
& + & \left. \frac{2}{(5+k)} 
\left( \mp i \hat{A}_{\pm}  \left(
\begin{array}{c} 
\hat{G}_{22} \\
\hat{G}_{11} \\
\end{array} \right) +  i \hat{A}_{\pm} \left(
\begin{array}{c}
 V^{(\frac{3}{2})} \\
 U^{(\frac{3}{2})} \\
\end{array} \right)
+
i \hat{B}_{\pm} \left( 
\begin{array}{c} 
U^{(\frac{3}{2})} \\
 V^{(\frac{3}{2})} \\
\end{array} \right)
\right) \right](w) \nonu \\
& + & \cdots,
\nonu \\
\hat{G}_{22}(z) \, T^{(2)}(w) & = &
\frac{1}{(z-w)^2} \left[ \frac{(12-11k+5k^2)}{(3+7k)(5+k)}
\hat{G}_{22} -\frac{(-3+k)}{(5+k)} V^{(\frac{3}{2})} \right](w)
\nonu \\
& + & \frac{1}{(z-w)} \left[ V^{(\frac{5}{2})} +
\frac{k(-39+5k)}{(9+21k)(5+k)} \pa \hat{G}_{22} 
-\frac{(1+k)}{3(5+k)} \pa   V^{(\frac{3}{2})} \right. 
\label{G22t2}
\\
& + & \left. \frac{2}{(5+k)} 
\left( - i  \hat{A}_{-} 
T_{-}^{(\frac{3}{2})} -2 i \hat{A}_3 \, \hat{G}_{22} +
i \hat{B}_{+} \hat{G}_{21} + 2i \hat{B}_3  \, V^{(\frac{3}{2})} 
\right) \right](w) +  \cdots.
\nonu
\eea
The $U(1)$ charge conservation in these OPEs can be checked from Tables 
$2, 3$ and $4$.

%%%%%%%%%%%%%%%%%%
\subsection{The OPEs between four spin $\frac{3}{2}$ currents 
and two higher spin 
currents of spins $(\frac{3}{2}, 2, 2, \frac{5}{2})$}
%%%%%%%%%%%%%%%%%%%%

From (\ref{nong11})-(\ref{nong21}),
(\ref{u3half}) and  (\ref{v3half}), 
the following OPEs between the spin-$\frac{3}{2}$ currents and the 
spin-$\frac{3}{2}$ currents can be obtained 
\bea
\left(
\begin{array}{c}
\hat{G}_{11} \\
\hat{G}_{22} \\
\end{array} \right)(z) \, 
\left(
\begin{array}{c}
U^{(\frac{3}{2})} \\
V^{(\frac{3}{2})} \\
\end{array} \right)(w) & = &
\frac{1}{(z-w)} \frac{2}{(5+k)} \left[ \pm
\hat{A}_{\pm}  \hat{B}_{\mp} \right](w) +\cdots,
\label{G11u3half} \\
\left(
\begin{array}{c}
\hat{G}_{12} \\
\hat{G}_{21} \\
\end{array} \right)(z) \, 
\left(
\begin{array}{c}
U^{(\frac{3}{2})} \\
V^{(\frac{3}{2})} \\
\end{array} \right) (w) & = &
\frac{1}{(z-w)^2} \frac{6}{(5+k)} \left[ i  \hat{A}_{\pm} \right](w) \nonu  \\
&  + & 
\frac{1}{(z-w)}  \left[ 
\left(
\begin{array}{c} 
U_{-}^{(2)} \\
V_{+}^{(2)} \\
\end{array} \right)
   + \frac{3}{(5+k)}  i \pa \hat{A}_{\pm} \right](w) +\cdots,
\label{G12u3half} \\
\left(
\begin{array}{c}
\hat{G}_{21} \\
\hat{G}_{12} \\
\end{array} \right) (z) \, 
\left(
\begin{array}{c} 
U^{(\frac{3}{2})} \\
V^{(\frac{3}{2})} \\
\end{array} \right)(w) & = &
\frac{1}{(z-w)^2} \frac{2k}{(5+k)} \left[ -i  \hat{B}_{\mp} \right](w) \nonu \\
& + & 
\frac{1}{(z-w)}  \left[ \left( 
\begin{array}{c}
 U_{+}^{(2)} \\
  V_{-}^{(2)} \\
\end{array} \right) 
   - \frac{k}{(5+k)}  i \pa \hat{B}_{\mp} \right](w) +\cdots,
\label{G21u3half} \\
\left(
\begin{array}{c}
\hat{G}_{22} \\
\hat{G}_{11} \\
\end{array} \right)(z) \, 
\left(
\begin{array}{c}
U^{(\frac{3}{2})} \\
V^{(\frac{3}{2})} \\
\end{array} \right) (w) & = &
\mp \frac{1}{(z-w)^3} \, \frac{6k}{(5+k)} \nonu \\
& + & 
\frac{1}{(z-w)^2} \left[ \frac{2i}{(5+k)} ( 3 \hat{A}_3 -
k \hat{B}_3) + T^{(1)} \right](w) \nonu \\
& + &  
\frac{1}{(z-w)}  \left[ \mp W^{(2)} 
  + \frac{3i}{(5+k)} \pa \hat{A}_3 -
\frac{i k}{(5+k)} \pa \hat{B}_3 +\frac{1}{2} \pa T^{(1)}
\right](w) \nonu \\
& + & \cdots.
\label{G22u3half} 
\eea
The $U(1)$ charge conservation in these OPEs can be checked from Tables 
$2$ and $3$.

For the spin-$2$ currents (\ref{u+2}) and (\ref{v-2}),
one calculates the nontrivial  OPEs between the spin-$\frac{3}{2}$ currents
and the spin-$2$ currents 
\bea
\left(
\begin{array}{c}
\hat{G}_{11} \\
\hat{G}_{22} \\
\end{array} \right) (z) \, \left(
\begin{array}{c} 
 U_{+}^{(2)} \\
 V_{-}^{(2)} \\
\end{array} \right) (w) & = &
-\frac{1}{(z-w)} \frac{2}{(5+k)} \left[
i  \hat{B}_{\mp} \left(
\begin{array}{c} \hat{G}_{11} +   U^{(\frac{3}{2})} \\
   \hat{G}_{22} -   V^{(\frac{3}{2})} \\
\end{array} \right)
 \right](w)+ \cdots,
\label{G11u+2} \\
\hat{G}_{12}(z) \, U_{+}^{(2)}(w) & = &
\frac{1}{(z-w)^2} \left[ \frac{(6+k)}{(5+k)} \hat{G}_{11}+
\frac{2(7+k)}{(5+k)} U^{(\frac{3}{2})} \right](w)
\nonu \\
& + & \frac{1}{(z-w)} \left[ - U^{(\frac{5}{2})}
+\frac{(6+k)}{3(5+k)} \pa \hat{G}_{11}+
\frac{2(11+k)}{3(5+k)} \pa U^{(\frac{3}{2})} \right.
\nonu \\
&+&  \frac{2}{(5+k)} \left( -i 
\hat{A}_{+} T_{+}^{(\frac{3}{2})}
-2 i \hat{A}_3 \, U^{(\frac{3}{2})} - i \hat{B}_{-} (\hat{G}_{12}
 -T_{-}^{(\frac{3}{2})}) 
\right. 
\nonu \\
& + & \left. \left.  
     2i \hat{B}_3 
\, U^{(\frac{3}{2})} \right) \right](w) +\cdots,
\label{G12u+2} \\
\left(
\begin{array}{c}
\hat{G}_{21} \\
\hat{G}_{12} \\ 
\end{array} \right) (z) \, 
\left(
\begin{array}{c}
U_{+}^{(2)} \\
V_{-}^{(2)} \\
\end{array} \right) (w) & = &
\mp\frac{1}{(z-w)} \frac{2}{(5+k)} \left[
i \hat{B}_{\mp}  T_{\pm}^{(\frac{3}{2})}   \right](w)+ \cdots,
\label{G21u+2} \\
\left(
\begin{array}{c}
\hat{G}_{22} \\
\hat{G}_{11} \\
\end{array} \right)  (z) \, 
\left(
\begin{array}{c}
U_{+}^{(2)} \\
V_{-}^{(2)} \\
\end{array} \right)
(w) & = &
\frac{1}{(z-w)^2} \left[ \mp \frac{(8+k)}{(5+k)} 
 \left(
\begin{array}{c}
\hat{G}_{21} \\
\hat{G}_{12} \\
\end{array} \right)-
\frac{2(7+k)}{(5+k)} T_{\pm}^{(\frac{3}{2})} \right](w)
\nonu \\
& + & \frac{1}{(z-w)} \left[ \pm  W_{\pm}^{(\frac{5}{2})}
\mp \frac{(8+k)}{3(5+k)} \pa \left( 
\begin{array}{c} 
\hat{G}_{21} \\
\hat{G}_{12} \\
\end{array} \right) -
\frac{2(9+k)}{3(5+k)} \pa T_{\pm}^{(\frac{3}{2})} \right.
\nonu \\
&+& \left. \frac{2}{(5+k)} \left( \pm i 
\hat{A}_{\mp} \left( 
\begin{array}{c}
 U^{(\frac{3}{2})} \\
 V^{(\frac{3}{2})} \\
\end{array} \right)
\mp 2 i \hat{B}_3 \, T_{\pm}^{(\frac{3}{2})}  \right) \right](w) +\cdots.
\label{G22u+2}
\eea
It is easy to check that 
the spin-$\frac{5}{2}$ fields in (\ref{G12u+2}) and (\ref{G22u+2}) 
after subtracting the descendant 
fields (the derivative terms for the fields in the second-order pole with 
coefficient $\frac{1}{3}$) 
in the first-order pole are primary under the 
stress energy tensor (\ref{stressnonlinear}). 
The $U(1)$ charge conservation in these OPEs can be checked from Tables 
$2$ and $4$.

Similarly, from the expression (\ref{u-2}) and (\ref{v+2}), 
the following remaining OPEs can be described as
\bea
\left(
\begin{array}{c}
\hat{G}_{11} \\
\hat{G}_{22} \\
\end{array} \right) (z) \, \left(
\begin{array}{c} 
U_{-}^{(2)} \\
V_{+}^{(2)} \\
\end{array} \right)
(w) & = &
\frac{1}{(z-w)} \frac{2}{(5+k)} \left[
i \hat{A}_{\pm} \left(
\begin{array}{c} 
\hat{G}_{11} +  U^{(\frac{3}{2})} \\
\hat{G}_{22} -  V^{(\frac{3}{2})} \\
\end{array} \right)   
\right](w)+ \cdots,
\label{G11u-2} \\
\left(
\begin{array}{c}
\hat{G}_{12} \\
\hat{G}_{21} \\
\end{array} \right)(z) \, 
\left(
\begin{array}{c} 
U_{-}^{(2)} \\
V_{+}^{(2)} \\
\end{array} \right) (w) & = & -
 \frac{1}{(z-w)}  \frac{2}{(5+k)} \left[i 
\hat{A}_{\pm} \left(
\begin{array}{c} 
\hat{G}_{12} - T_{-}^{(\frac{3}{2})}  \\
\hat{G}_{21} + T_{+}^{(\frac{3}{2})}
\\
\end{array} \right)  
\right](w) +\cdots,
\label{G12u-2} \\
\hat{G}_{21}(z) \, U_{-}^{(2)}(w) & = &
\frac{1}{(z-w)^2} \left[ \frac{(3+2k)}{(5+k)} \hat{G}_{11}+
\frac{4(2+k)}{(5+k)} U^{(\frac{3}{2})} \right](w)
\label{G21u-2}
\\
& + & \frac{1}{(z-w)}  \left[ U^{(\frac{5}{2})}
+ \frac{(3+2k)}{3(5+k)} \pa \hat{G}_{11}+
\frac{4(2+k)}{3(5+k)} \pa U^{(\frac{3}{2})}
 \right](w)+ \cdots,
\nonu \\
\left(
\begin{array}{c}
\hat{G}_{22} \\
\hat{G}_{11} \\
\end{array} \right) (z) \, 
\left(
\begin{array}{c} 
U_{-}^{(2)} \\
V_{+}^{(2)} \\
\end{array} \right) (w) & = &
\frac{1}{(z-w)^2} \left[ \pm \frac{(3+2k)}{(5+k)} 
\left(
\begin{array}{c} 
\hat{G}_{12} \\
\hat{G}_{21} \\
\end{array} \right)-
\frac{4(2+k)}{(5+k)} T_{\mp}^{(\frac{3}{2})} \right](w)
\nonu \\
& + & \frac{1}{(z-w)} \left[  \pm W_{\mp}^{(\frac{5}{2})}
\pm \frac{(7+2k)}{3(5+k)} \pa 
\left(
\begin{array}{c} 
\hat{G}_{12} \\
\hat{G}_{21} \\
\end{array} \right)
-
\frac{4(3+k)}{3(5+k)} \pa T_{\mp}^{(\frac{3}{2})} \right.
\nonu
\\
&+& \left. \frac{2}{(5+k)} \left( -2i 
\hat{A}_3 \, 
\left(
\begin{array}{c} 
\hat{G}_{12} \\
\hat{G}_{21} \\
\end{array} \right)
 \pm 2i \hat{A}_3  \, T_{\mp}^{(\frac{3}{2})}
\mp i \hat{B}_{\pm} 
\left(
\begin{array}{c}
U^{(\frac{3}{2})} \\
V^{(\frac{3}{2})} \\
\end{array} \right)
\right) \right](w) \nonu \\
& + & \cdots.
\label{G22u-2}
\eea
As above,
the spin-$\frac{5}{2}$ fields in (\ref{G22u-2}) 
after subtracting these descendant 
fields (the derivative terms for the fields in the second-order pole with 
coefficient $\frac{1}{3}$) 
in the first-order pole are primary. 
The $U(1)$ charge conservation in these OPEs can be checked from Tables 
$2$ and $4$.

For the last current of spin-$\frac{5}{2}$ (\ref{U5half}), one also has 
the OPEs between the spin-$\frac{3}{2}$ currents and the 
spin-$\frac{5}{2}$ current
\bea
\hat{G}_{11}(z) \, U^{(\frac{5}{2})}(w) 
& = &
\frac{1}{(z-w)^2} \frac{4(5+3k)}{3(5+k)^2} 
\left[  \hat{A}_{+} \hat{B}_{-} \right](w)
\label{g11u5half}
\\
& + & \frac{1}{(z-w)} \left[ -\frac{2i}{(5+k)} 
\hat{A}_{+} U_{+}^{(2)} + \frac{2(-4+k)}{(5+k)^2}  \hat{A}_{+} \pa  \hat{B}_{-}  
\right. \nonu \\
& - & \left.   
  \frac{4i}{(5+k)} \hat{B}_{-} U_{-}^{(2)}
-  \frac{4}{3(5+k)} \hat{G}_{11}
\, \hat{G}_{11} + \frac{2}{(5+k)} \hat{G}_{11} 
\, U^{(\frac{3}{2})} \right](w) +\cdots,
\nonu \\
\hat{G}_{12}(z) \, U^{(\frac{5}{2})}(w) 
& = &  
\frac{1}{(z-w)^3} \frac{4(1+3k)}{(5+k)^2} \left[ i \hat{A}_{+} \right](w) 
\nonu \\
& + & \frac{1}{(z-w)^2} \left[ 
\frac{2(29+4k)}{3(5+k)} U_{-}^{(2)}
-\frac{12}{(5+k)^2} \hat{A}_{+}  \hat{A}_3 
\right. \nonu \\
& + &  \left. \frac{4(12+k)}{3(5+k)^2} \hat{A}_{+} 
\hat{B}_3  + \frac{6 i}{(5+k)^2} \pa  \hat{A}_{+}
 + \frac{2i}{(5+k)} T^{(1)} \hat{A}_{+}   \right](w) 
\nonu \\
& + & \frac{1}{(z-w)} \left[ 
\frac{4}{(5+k)} i \hat{A}_{+} \hat{T} -\frac{2}{(5+k)}
i \hat{A}_{+} W^{(2)}
-  \frac{10}{(5+k)^2}  \hat{A}_{+}  \pa \hat{A}_3
\right. \nonu \\
& - & \frac{2(-18+k)}{3(5+k)^2}  \hat{A}_{+}  \pa \hat{B}_3
-  \frac{4}{(5+k)} i \hat{A}_3  U_{-}^{(2)} -\frac{8}{(5+k)^2} \hat{A}_3 
\pa \hat{A}_{+}
\nonu \\
& + &  \frac{4(-3+k)}{3(5+k)^2} \hat{B}_3 
\pa \hat{A}_{+}
+\frac{2(11+k)}{3(5+k)} \pa  U_{-}^{(2)} -\frac{1}{(5+k)} i \pa T^{(1)}  
\hat{A}_{+}
\nonu \\
&- & \frac{2}{(5+k)} \hat{G}_{11} \hat{G}_{12} + \frac{2}{(5+k)} \hat{G}_{11}
T_{-}^{(\frac{3}{2})}
+\frac{4}{(5+k)^2} i \hat{A}_{+} \hat{A}_{+}
\hat{A}_{-}
\nonu \\
& -& \frac{4}{(5+k)^2} i \hat{A}_{+}  \hat{A}_3 \hat{A}_3
-\frac{8}{(5+k)^2} i \hat{A}_{+} \hat{A}_3 \hat{B}_3 
+  \frac{4}{(5+k)^2} i \hat{A}_{+} \hat{B}_{+}
\hat{B}_{-}
\nonu \\
& + & \frac{4}{(5+k)^2} i \hat{A}_{+} \hat{B}_3 \hat{B}_3
+   \frac{8}{(5+k)^2} i \hat{A}_3 \hat{A}_{+}  \hat{A}_3 
+\frac{2}{(5+k)} T^{(1)} \hat{A}_{+} \hat{A}_3
\nonu \\
&- & \left. 
\frac{2}{(5+k)} \hat{A}_3 T^{(1)} \hat{A}_{+}
 + \frac{4}{(5+k)} i \hat{B}_3  U_{-}^{(2)}
\right](w) + \cdots,
\label{g12u5half} \\
\hat{G}_{21}(z) \, U^{(\frac{5}{2})}(w) 
& = & 
-\frac{1}{(z-w)^3} \frac{16k}{3(5+k)^2} \left[ i \hat{B}_{-} \right](w) 
\nonu \\
& + & \frac{1}{(z-w)^2} \left[ 
-\frac{16(2+k)}{3(5+k)} U_{+}^{(2)}
-\frac{32(3+k)}{3(5+k)^2} \hat{A}_3 \hat{B}_{-} 
\right. \nonu \\
& + &  \left. \frac{8k}{(5+k)^2} \hat{B}_{-} 
\hat{B}_3  + \frac{4 k}{(5+k)^2} i \pa \hat{B}_{-}
 + \frac{4}{(5+k)} T^{(1)} i \hat{B}_{-}  \right](w) 
\nonu \\
& + & \frac{1}{(z-w)} 
\left[
-\frac{4(3+2k)}{3(5+k)^2} \hat{A}_3 \pa \hat{B}_{-} 
+\frac{12 k}{(3+7k)(5+k)} i \hat{B}_{-} \hat{T} 
\right.
\nonu \\
& - & \frac{2}{(5+k)} i \hat{B}_{-} W^{(2)}
-\frac{4(2+k)}{3(5+k)} \pa  U_{+}^{(2)}
+  \frac{2k}{3(5+k)^2} i \pa^2  \hat{B}_{-}
\nonu \\
& + & \frac{2(9+2k)}{3(5+k)^2} i \hat{A}_{-} 
\hat{A}_{+} \hat{B}_{-}
+  \frac{4k}{(5+k)^2} i 
\hat{B}_{-} \hat{B}_3 \hat{B}_3
-\frac{4k}{(5+k)^2} i \hat{B}_3 \hat{B}_{-} \hat{B}_3
\nonu \\
%\frac{1}{(5+k)} 
%\left[  \frac{12k}{(3+7k)} i ( \hat{B}_1  - i \hat{B}_2) \hat{T}  - 2i 
%( \hat{B}_1 - i\hat{B}_2) W^{(2)}
%  \right. \nonu \\
%& + &    \frac{4 i(3+2k)}{3(5+k)} \hat{A}_3 \, \pa 
%\hat{B}_2
%+ 2 i (\hat{B}_1 - i \hat{B}_2) \, T^{(2)}
%\nonu \\
%&+& \frac{4(2+k)}{3} \pa U_{+}^{(2)}
%+\frac{2k}{3(5+k)}  i \pa^2 ( \hat{B}_1 -   i \hat{B}_2)
%+2   T^{(1)} \, i \pa ( \hat{B}_1  - i \hat{B}_2)
%\nonu \\
%& +& 
% \frac{1}{(5+k)} \left( 4 i \hat{A}_1 \, \hat{A}_1 \, \hat{B}_1
%+ 4 \hat{A}_1 \, \hat{A}_1 \, \hat{B}_2
%- \frac{8(3+k)}{3}  \hat{A}_1 \, \hat{A}_2 \, \hat{B}_1
%\right. 
%\nonu \\
%& +&  \frac{8i(3+k)}{3}  \hat{A}_1 \, \hat{A}_2 \, \hat{B}_2 
%+ \frac{8(3+k)}{3} \hat{A}_2 \, \hat{A}_1 \, \hat{B}_1
%-\frac{8 i(3+k)}{3} \hat{A}_2 \, \hat{A}_1 \, \hat{B}_2
%+ 4 i \hat{A}_2 \, \hat{A}_2 \, \hat{B}_1
%\nonu \\
%&+ & 4  \hat{A}_2 \, \hat{A}_2 \, \hat{B}_2
%-\frac{4}{3} (3+2k) \hat{A}_3 \, \hat{B}_2 \, \hat{B}_3
%+ 4 i k \hat{B}_1 \, \hat{B}_3 \, \hat{B}_3 
%+ 4 k \hat{B}_2 \, \hat{B}_3 \, \hat{B}_3
%\nonu \\
%& + & \left. \left.
%\frac{4(3+2k)}{3}  \hat{B}_3 \, \hat{A}_3 \, \hat{B}_2
%- 
%  4 i k \hat{B}_3 \, \hat{B}_1 \, \hat{B}_3
%- 4 k \hat{B}_3 \, \hat{B}_2 \, \hat{B}_3
%\right) 
&- & 
\frac{2}{(5+k)} T^{(1)} \hat{B}_{-} \hat{B}_3
+\frac{2}{(5+k)} \hat{B}_3 T^{(1)}  \hat{B}_{-}
-\frac{2(3+2k)}{3(5+k)^2} i \hat{A}_{+} \hat{A}_{-} \hat{B}_{-}
\nonu \\
&+&  \left. \frac{2}{(5+k)} i \hat{B}_{-} T^{(2)}
\right](w) +  \cdots,
\label{g21u5half} \\
\hat{G}_{22}(z) \, U^{(\frac{5}{2})}(w) 
& = & 
\frac{1}{(z-w)^3} \left[ \frac{16}{(5+k)^2}  i (\hat{A}_3 -4
i \hat{B}_3)  - \frac{4(-5+2k)}{3(5+k)} T^{(1)} \right](w) 
\nonu \\
& + & \frac{1}{(z-w)^2} \left[ 
-\frac{4(18-9k+10k^2)}{3(3+7k)(5+k)} \hat{T}
+\frac{2(7+2k)}{(5+k)} T^{(2)}
\right. \nonu \\
& + &   \frac{4(3-4k)}{3(5+k)^2}  \hat{A}_1 \, 
\hat{A}_1  + \frac{4 (3-4k)}{3(5+k)^2}  \hat{A}_2  \, \hat{A}_2
 - \frac{8(3+2k)}{3(5+k)^2}  
\hat{A}_3 \, \hat{A}_3 
\nonu \\
& + &   \frac{8(3+2k)}{3(5+k)^2}  \hat{A}_3 \, 
\hat{B}_3  - \frac{8 (3+2k)}{3(5+k)^2}  \hat{B}_1  \, \hat{B}_1
 - \frac{8(3+2k)}{3(5+k)^2}  
\hat{B}_2 \, \hat{B}_2
\nonu \\
& + &  \left.  \frac{8(-3+k)}{3(5+k)^2}  \hat{B}_3 \, 
\hat{B}_3  + \frac{4 i}{(5+k)}  T^{(1)}  \, \hat{B}_3
+ \frac{2(-5+2k)}{3(5+k)} W^{(2)} 
\right](w) 
\nonu \\
&  + & \frac{1}{(z-w)} \left[ 
-\frac{1}{2} W^{(3)} +\frac{1}{2} P^{(3)} -\frac{8(-3+k)}{(19+23k)} T^{(1)}
\hat{T}
+\frac{1}{(5+k)} i  \hat{A}_{+}  V_{+}^{(2)}
\right. \nonu \\
&+ & \frac{1}{(5+k)} i   \hat{A}_{-}   U_{-}^{(2)}
+\frac{2(9-4k))}{3(5+k)^2}   \hat{A}_{-}  \pa 
 \hat{A}_{+} -\frac{3k}{(5+k)^2}   \hat{B}_{-} \pa
 \hat{B}_{+}
\nonu \\
& + & \frac{8(48+97k+29k^2)}{(3+7k)(19+23k)(5+k)} i \hat{A}_3 \hat{T}
-\frac{4}{(5+k)} i \hat{A}_3 T^{(2)}
+\frac{4(9+k)}{3(5+k)^2} \hat{A}_3 \pa \hat{B}_3
\nonu \\
&+ & \frac{1}{(5+k)} i  \hat{B}_{+}   U_{+}^{(2)}
+\frac{k}{3(5+k)^2}   \hat{B}_{+}  \pa
 \hat{B}_{-}
-  \frac{1}{(5+k)} i  \hat{B}_{-}   V_{-}^{(2)}
\nonu \\
 &- & \frac{8(-27+k)k}{(3+7k)(19+23k)(5+k)} i \hat{B}_3 \hat{T}
+\frac{4}{(5+k)} i \hat{B}_3 T^{(2)}
+\frac{4(-6+k)}{3(5+k)^2} \hat{B}_3 \pa \hat{A}_3
\nonu \\
& - & \frac{4}{(5+k)} i \hat{B}_3 W^{(2)}
-\frac{2k(-39+5k)}{3(3+7k)(5+k)} \pa \hat{T} + \pa T^{(2)}
+ \frac{(-7+k)}{3(5+k)} \pa  W^{(2)}
\nonu \\
& + & \frac{(9+4k)}{3(5+k)^2} i \pa^2 \hat{A}_3
+\frac{2k}{(5+k)^2} i \pa^2 \hat{B}_3 +\frac{2}{3(5+k)} \pa^2 T^{(1)}
-\frac{2}{(5+k)} i \pa T^{(1)} \hat{A}_3
\nonu \\
&+ & \frac{2}{(5+k)} i T^{(1)} \pa \hat{A}_3 +
 \frac{2}{(5+k)} i T^{(1)} \pa \hat{B}_3
-\frac{2}{(5+k)} \hat{G}_{12} \hat{G}_{21}
\nonu \\
&- & \frac{2}{(5+k)} \hat{G}_{22}  U^{(\frac{3}{2})}
+\frac{(15-8k)}{3(5+k)^2} i 
\hat{A}_{+} \hat{A}_{-} \hat{A}_3
+  \frac{4(3+k)}{3(5+k)^2} i \hat{A}_{-} 
\hat{A}_{+} \hat{A}_3
\nonu \\
&
+ & \frac{(-3+4k)}{3(5+k)^2} i \hat{A}_3 \hat{A}_{+}
\hat{A}_{-}
+  \frac{8}{(5+k)^2} i \hat{A}_3 \hat{A}_3 \hat{A}_3 
-\frac{16}{(5+k)^2}  i \hat{A}_3 \hat{A}_3 \hat{B}_3 
\nonu \\
& + & 
\frac{8}{(5+k)^2} i  \hat{A}_3 \hat{B}_3 \hat{B}_3   
+  \frac{8}{(5+k)^2} i  \hat{B}_{-} 
\hat{A}_3 \hat{B}_{+} 
-\frac{2k}{3(5+k)^2} i  \hat{B}_{-} 
\hat{B}_{+} \hat{B}_3 
\nonu \\
& + & \left. \frac{2k}{3(5+k)^2} i  \hat{B}_3 
\hat{B}_{+} 
\hat{B}_{-}
-\frac{2}{(5+k)} \hat{G}_{21} T_{-}^{(\frac{3}{2})}
\right](w) + \cdots, 
\label{g22u5half}
\eea
where the $W^{(3)}(w)$ is 
given by (\ref{w3}) and the other spin-$3$ field which is
primary can be written in terms of known currents as follows:
\bea
P^{(3)}(w) & = & -\frac{2i}{(5+k)} \left[ -\frac{i}{2} (5+k ) 
W^{(3)} + \hat{A}_{+} V_{+}^{(2)} -  \frac{i}{(5+k)} \hat{A}_{+} \pa 
\hat{A}_{-} \right.  
\nonu \\
&+& \hat{A}_{-} U_{-}^{(2)} - \frac{i}{(5+k)} \hat{A}_{-} \pa \hat{A}_{+}
- \frac{2i}{(5+k)} \hat{A}_3 \pa \hat{A}_3
+\frac{2i(-2+k)}{(5+k)} \hat{A}_3 \pa \hat{B}_3
\nonu \\
&- & \hat{B}_{+} U_{+}^{(2)} -\frac{i}{(5+k)} \hat{B}_{+} \pa 
\hat{B}_{-} -  \hat{B}_{-} V_{-}^{(2)}
\nonu \\
&- &  \frac{i}{(5+k)} \hat{B}_{-} \pa \hat{B}_{+} 
-\frac{2i(-4+k)}{(5+k)} \hat{B}_3 \pa \hat{A}_3
-\frac{2i}{(5+k)} \hat{B}_3 \pa \hat{B}_3
- i \pa \hat{T} + i \pa W^{(2)} 
\nonu \\
& -& \left. 
\pa T^{(1)} \hat{A}_3 + \pa T^{(1)} \hat{B}_3 - T^{(1)} \pa \hat{B}_3 
+\frac{i}{2} T^{(1)}  \hat{A}_{+}  \hat{A}_{-} 
-  \frac{i}{2}  \hat{A}_{-} T^{(1)}  \hat{A}_{+} \right](w).
\label{P3}
\eea
The $U(1)$ charge conservation in these OPEs can be checked from Tables 
$2, 3$ and $5$.

%%%%%%%%%%%%%%%%%%
%\subsection{The OPEs between four spin $\frac{3}{2}$ currents 
%and the higher spin 
%current of spins $(\frac{3}{2}, 2, 2, \frac{5}{2})$}
%%%%%%%%%%%%%%%%%%%%

For the spin-$2$ current (\ref{v+2}), one obtains
\bea
\hat{G}_{12}(z) \, V_{+}^{(2)}(w) & = &
\frac{1}{(z-w)^2} \left[ -\frac{(3+2k)}{(5+k)} \hat{G}_{22}+
\frac{4(2+k)}{(5+k)} V^{(\frac{3}{2})} \right](w)
\nonu \\
& + & \frac{1}{(z-w)} \left[ - V^{(\frac{5}{2})}
-\frac{(3+2k)}{3(5+k)} \pa \hat{G}_{22}+
\frac{4(4+k)}{3(5+k)} \pa V^{(\frac{3}{2})} \right.
\nonu
 \\
&+&  \left. \frac{2}{(5+k)} \left( i 
\hat{A}_{-} T_{-}^{(\frac{3}{2})}
+2 i \hat{A}_3 \, V^{(\frac{3}{2})} - i  \hat{B}_{+} (\hat{G}_{21} 
+
T_{+}^{(\frac{3}{2})}) - 2i \hat{B}_3 
\, V^{(\frac{3}{2})} \right) \right](w) \nonu \\
& + & \cdots.
\label{G12v+2} 
\eea
As above,
the spin-$\frac{5}{2}$ fields after subtracting the descendant 
fields (the derivative terms for the fields in the second-order pole with 
coefficient $\frac{1}{3}$) 
in the first-order pole are primary. 
The $U(1)$ charge conservation in these OPEs can be checked from Tables 
$2$ and $4$.

For the other spin-$2$ current (\ref{v-2}),
the following OPEs can be described as
\bea
\hat{G}_{21}(z) \, V_{-}^{(2)}(w) & = &
\frac{1}{(z-w)^2} \left[ -\frac{(6+k)}{(5+k)} \hat{G}_{22}+
\frac{2(7+k)}{(5+k)} V^{(\frac{3}{2})} \right](w)
\nonu \\
& + & \frac{1}{(z-w)}  \left[ V^{(\frac{5}{2})}
- \frac{(6+k)}{3(5+k)} \pa \hat{G}_{22}+
\frac{2(7+k)}{3(5+k)} \pa V^{(\frac{3}{2})}
 \right](w)+ \cdots.
\label{G21v-2} 
\eea
As above,
the spin-$\frac{5}{2}$ fields after subtracting these descendant 
fields (the derivative terms for the fields in the second-order pole with 
coefficient $\frac{1}{3}$) 
in the first-order pole are primary. 
The $U(1)$ charge conservation in these OPEs can be checked from Tables 
$2$ and $4$.

For the spin-$\frac{5}{2}$ current (\ref{v5half}), 
one has the following nontrivial OPEs between the spin-$\frac{3}{2}$ currents
and the spin-$\frac{5}{2}$ current
\bea
\hat{G}_{11}(z) \, V^{(\frac{5}{2})}(w) 
& = &
\frac{1}{(z-w)^3} \left[ \frac{8i(-1+3k)}{(5+k)^2} \hat{A}_3
- \frac{16ik}{3(5+k)^2} \hat{B}_3  + \frac{4(-7+2k)}{3(5+k)} T^{(1)}
\right](w)
\nonu \\
& + &
\frac{1}{(z-w)^2} \left[ -\frac{4(27+12k+10k^2)}{(9+21k)(5+k)}
\hat{T}+ \frac{2(7+2k)}{(5+k)} T^{(2)} 
+ \frac{2(-7+2k)}{3(5+k)} W^{(2)}
\right. \nonu \\
& - &   
 \frac{4(9+4k)}{3(5+k)^2} \hat{A}_1 \, \hat{A}_1- 
\frac{4(9+4k)}{3(5+k)^2} \hat{A}_2 \, \hat{A}_2 -
\frac{4(27+4k)}{3(5+k)^2}\hat{A}_3 \, \hat{A}_3 
+ \frac{40(3+k)}{3(5+k)^2}\hat{A}_3 \, \hat{B}_3 
\nonu \\
&-& \left. \frac{4(9+7k)}{3(5+k)^2} \hat{B}_1 \, \hat{B}_1
-\frac{4(9+7k)}{3(5+k)^2} \hat{B}_2 \, \hat{B}_2
-\frac{4(9+4k)}{3(5+k)^2} \hat{B}_3 \, \hat{B}_3 +
\frac{4i}{(5+k)} T^{(1)} \, \hat{A}_3 \right](w)
\nonu \\
& + & \frac{1}{(z-w)} \left[ 
\frac{3}{2} W^{(3)} +\frac{1}{2} P^{(3)}
+\frac{8(-3+k)}{(19+23k)} T^{(1)} \hat{T}
+\frac{3}{(5+k)} i  \hat{A}_{+} V_{+}^{(2)}
\right. \nonu \\  
&- & \frac{(3+4k)}{3(5+k)^2} \hat{A}_{+} \pa \hat{A}_{-}
 +\frac{1}{(5+k)} i \hat{A}_{-}
  U_{-}^{(2)}
-  \frac{(9+4k)}{3(5+k)^2}  \hat{A}_{-}
 \pa \hat{A}_{+} 
\nonu \\
& - & 
\frac{8(48+97k+29k^2)}{(3+7k)(19+23k)(5+k)} \hat{A}_3 \hat{T}
+  \frac{4}{(5+k)} i \hat{A}_3 T^{(2)}
+\frac{4}{(5+k)} i \hat{A}_3 W^{(2)} 
\nonu \\
& - & \frac{8(3+k)}{3(5+k)^2}
\hat{A}_3 \pa \hat{A}_3 
-  \frac{1}{(5+k)} i \hat{B}_{+}  U_{+}^{(2)}
-  \frac{(12+17k)}{3(5+k)^2}   \hat{B}_{+} \pa  \hat{B}_{-}
\nonu \\
& + &  \frac{k}{(5+k)^2}  \hat{B}_{-} \pa  \hat{B}_{+}
+\frac{8(-27+k)k}{(3+7k)(19+23k)(5+k)} i \hat{B}_3 \hat{T}
\nonu \\
& - & \frac{4}{(5+k)} i \hat{B}_3 T^{(2)}
+\frac{4(-3+4k)}{3(5+k)^2} \hat{B}_3 \pa \hat{A}_3 
-\frac{2(9-18k+5k^2)}{3(3+7k)(5+k)} \pa \hat{T} + \pa T^{(2)}
\nonu \\
& + & \frac{(-5+k)}{3(5+k)} \pa W^{(2)}
-\frac{2(3+8k)}{3(5+k)^2} i \pa^2 \hat{B}_3 +
\frac{2}{3(5+k)} \pa^2 T^{(1)}
 \nonu \\
& + & \frac{4}{(5+k)} i T^{(1)} 
\pa \hat{A}_3
-\frac{2}{(5+k)} \hat{G}_{11} V^{(\frac{3}{2})}
-\frac{2}{(5+k)} \hat{G}_{21} T_{-}^{(\frac{3}{2})}
\nonu \\
&- & \frac{4}{(5+k)^2} i  \hat{A}_{+}
 \hat{A}_{-} \hat{A}_3
-\frac{4}{(5+k)} i \hat{A}_3  \hat{A}_{+}
 \hat{A}_{-}
-  \frac{8}{(5+k)^2} i \hat{A}_3 \hat{A}_3 \hat{A}_3 
\nonu \\
& + & \frac{16}{(5+k)^2} i \hat{A}_3 \hat{A}_3 \hat{B}_3 
+\frac{2(9+k)}{3(5+k)^2} i \hat{A}_3  \hat{B}_{+}  \hat{B}_{-}
-  \frac{8}{(5+k)^2} i \hat{A}_3 \hat{B}_3 \hat{B}_3 
\nonu \\
& - &   
\frac{2(21+k)}{3(5+k)^2} i  \hat{B}_{-} \hat{A}_3 \hat{B}_{+}
+  \frac{2(3+2k)}{3(5+k)^2} i 
 \hat{B}_{-}  \hat{B}_{+} \hat{B}_3
-\frac{2(3+2k)}{3(5+k)^2} i 
\hat{B}_3 \hat{B}_{+}  \hat{B}_{-} 
\nonu \\
& - & \left. \frac{1}{(5+k)} i  \hat{B}_{-}  V_{-}^{(2)}
-\frac{2}{(5+k)} i \pa T^{(1)} \hat{A}_3
\right](w) +  \cdots,
\label{g11v5half} \\
\hat{G}_{12}(z) \, V^{(\frac{5}{2})}(w) 
& = & 
-\frac{1}{(z-w)^3} \frac{40k}{3(5+k)^2} \left[ i \hat{B}_{+} \right](w) 
\nonu \\
& + & \frac{1}{(z-w)^2} \left[ 
\frac{2(17+8k)}{3(5+k)} V_{-}^{(2)}
-\frac{4(9+2k)}{3(5+k)^2} \hat{A}_3  \hat{B}_{+} 
\right. \nonu \\
& + &  \left. \frac{4k}{(5+k)^2} \hat{B}_{+} 
\hat{B}_3  - \frac{2 k}{(5+k)^2} i \pa \hat{B}_{+}
 + \frac{2}{(5+k)^2} T^{(1)} i \hat{B}_{+}  \right](w) 
\nonu \\
& + & \frac{1}{(z-w)} 
\left[
\frac{4}{(5+k)} i  \hat{A}_3 V_{-}^{(2)}+ \frac{4(-9+k)}{3(5+k)^2}
\hat{A}_3 \pa \hat{B}_{+} 
-\frac{4}{(5+k)} i \hat{B}_{+} \hat{T} 
\right. \nonu \\
&+ & \frac{2}{(5+k)} i \hat{B}_{+} W^{(2)}
-\frac{4}{(5+k)} i \hat{B}_3   V_{-}^{(2)}
+\frac{4(4+k)}{3(5+k)} \pa  V_{-}^{(2)}
\nonu \\
&- & \frac{(-6+5k)}{6(5+k)^2} i \pa^2  \hat{B}_{+}
-\frac{1}{(5+k)} i \pa T^{(1)} \hat{B}_{+}
+ \frac{2}{(5+k)} i  T^{(1)} \pa \hat{B}_{+}
\nonu \\
& + & \frac{2}{(5+k)} \hat{G}_{22} T_{-}^{(\frac{3}{2})}
-\frac{(15+4k)}{3(5+k)^2} i \hat{A}_{+} 
\hat{A}_{-} \hat{B}_{+} 
+  \frac{(3+4k)}{3(5+k)^2} i 
\hat{A}_{-} 
\hat{A}_{+}  \hat{B}_{+}
\nonu \\
& - & \frac{4}{(5+k)^2} i \hat{A}_3 \hat{A}_3  \hat{B}_{+}
+  \frac{(-6+k)}{2(5+k)^2} i 
 \hat{B}_{+} 
\hat{B}_{+}  \hat{B}_{-}
-\frac{4}{(5+k)^2} i   \hat{B}_{+} \hat{B}_3 \hat{B}_3 
\nonu \\
&- & \left. \frac{(2+k)}{2(5+k)^2} i  \hat{B}_{-}
\hat{B}_{+} \hat{B}_{+} 
+\frac{8}{(5+k)^2} i \hat{B}_3 \hat{A}_3 \hat{B}_{+} 
%\frac{1}{(5+k)} 
%\left[  -4 i ( \hat{B}_1  + i \hat{B}_2) \hat{T}  + 2i 
%( \hat{B}_1 + i\hat{B}_2) W^{(2)}
%+ \frac{2i k}{(5+k)} \hat{B}_2 \, \pa \hat{B}_3
%  \right. \nonu \\
%& - &    \frac{2 (9+4k)}{3(5+k)}  ( \hat{B}_1 + i \hat{B}_2) \, \pa 
%\hat{A}_3 + \frac{2(-4+k)}{(5+k)} \hat{B}_1 \, \pa \hat{B}_3
%+ 4 i \hat{A}_3 \, V_{-}^{(2)} -4 i \hat{B}_3 \, V_{-}^{(2)}
%\nonu \\
%&+& \frac{4 i(-3 +k)}{3(5+k)} \hat{A}_3 \, \pa \hat{B}_2
%+\frac{4(4+k)}{3} \pa V_{-}^{(2)}
%- \frac{4 i k}{3(5+k)} \pa^2 \hat{B}_1 +\frac{2(-3+2k)}{3(5+k)}
%\pa^2 \hat{B}_2  
%\nonu \\
%& + &
%2 \hat{G}_{22} \, T_{-}^{(\frac{3}{2})}
%- \pa T^{(1)} i ( \hat{B}_1 + i \hat{B}_2)
%+ 2 T^{(1)} i \pa ( \hat{B}_1 + i \hat{B}_2)
%\nonu \\
%& +& 
% \frac{1}{(5+k)} \left( -4 i \hat{A}_1 \, \hat{A}_1 \, \hat{B}_1
%- 4 i \hat{A}_2 \, \hat{A}_2 \, \hat{B}_1
%-4 i  \hat{A}_3 \, \hat{A}_3 \, \hat{B}_1
%\right. 
%\nonu \\
%& +&  4  \hat{A}_3 \, \hat{A}_3 \, \hat{B}_2 
%+ 8 i  \hat{A}_3 \, \hat{B}_1 \, \hat{B}_3 +
%\frac{4 (-9+k)}{3} \hat{A}_3 \, \hat{B}_2 \, \hat{B}_3
%- 4 i \hat{B}_1 \, \hat{B}_1 \, \hat{B}_1
%\nonu \\
%&+ & 4  \hat{B}_1 \, \hat{B}_1 \, \hat{B}_2
%-4 i \hat{B}_1 \, \hat{B}_2 \, \hat{B}_2
%- 4 i  \hat{B}_1 \, \hat{B}_3 \, \hat{B}_3 
%+ 4  \hat{B}_2 \, \hat{A}_1 \, \hat{A}_1
%\nonu \\
%& + & \left. \left.
%4  \hat{B}_2 \, \hat{A}_2 \, \hat{A}_2
%+4 \hat{B}_2 \, \hat{B}_2 \, \hat{B}_2
%+ 4  \hat{B}_2 \, \hat{B}_3 \, \hat{B}_3
%- \frac{4(-3+k)}{3}  \hat{B}_3 \, \hat{A}_3 \, \hat{B}_2
%\right) 
\right](w) +  \cdots,
\label{g12v5half} \\
\hat{G}_{21}(z) \, V^{(\frac{5}{2})}(w) 
& = & 
\frac{1}{(z-w)^3} \frac{16k}{(5+k)^2} \left[ i \hat{A}_{-} \right](w) 
\nonu \\
& + & \frac{1}{(z-w)^2} \left[ 
-\frac{8(7+k)}{3(5+k)} V_{+}^{(2)}
-\frac{24}{(5+k)^2}  \hat{A}_{-} \, \hat{A}_3 
\right. \nonu \\
& + &  \left. \frac{16(9+k)}{3(5+k)^2} \hat{A}_{-} 
\hat{B}_3  - \frac{12 }{(5+k)^2} i \pa  \hat{A}_{-}
 + \frac{4}{(5+k)} T^{(1)} i 
\hat{A}_{-}  \right](w)
\nonu \\
& + & \frac{1}{(z-w)} 
\left[ 
-\frac{4(3+4k)}{(3+7k)(5+k)} i  \hat{A}_{-} \hat{T}
+\frac{2}{(5+k)} i \hat{A}_{-} T^{(2)}
\right. \nonu \\
&+ & \frac{12}{(5+k)^2}  \hat{A}_{-} \pa \hat{A}_3
+\frac{4(15+k)}{3(5+k)^2}  \hat{A}_{-} \pa \hat{B}_3
+  \frac{4}{(5+k)^2} \hat{A}_3 \pa  \hat{A}_{-}
\nonu \\
& - & \frac{2(7+k)}{3(5+k)} \pa   V_{+}^{(2)}
+\frac{2}{(5+k)} i T^{(1)} \pa  \hat{A}_{-}
-  \frac{4}{(5+k)^2} i \hat{A}_{+}
\hat{A}_{-} \hat{A}_{-}
\nonu \\
& + & 
\frac{4(12+k)}{3(5+k)^2} i  \hat{A}_{-} \hat{A}_3 \hat{B}_3
-\frac{8}{(5+k)^2} i \hat{A}_{-}
\hat{B}_{+} \hat{B}_{-}
- \frac{4}{(5+k)^2} i \hat{A}_{-}
\hat{B}_3 \hat{B}_3 \nonu \\
& + &  \frac{16}{(5+k)^2} i \hat{A}_3 
\hat{A}_{-} \hat{A}_3
%\frac{1}{(5+k)} 
%\left[  -\frac{4(3+4k)}{(3+7k)} i ( \hat{A}_1  - i \hat{A}_2) \hat{T}  
%+ 2 i 
%( \hat{A}_1 - i\hat{A}_2) T^{(2)}
%\right. \nonu \\
%&+ & 2   i 
%( \hat{A}_1 - i\hat{A}_2) W^{(2)}
%+ \frac{4(9+k)}{3(5+k)} \hat{A}_1 \, \pa \hat{B}_3
%+ \frac{8 i}{(5+k)} \hat{A}_2 \, \pa \hat{A}_3
%   \nonu \\
%& - &    \frac{12}{(5+k)}  \hat{A}_3 \, \pa 
%( \hat{A}_1 - i \hat{A}_2)  
%-\frac{2(7+k)}{3} \pa V_{+}^{(2)}
%+\frac{4 }{(5+k)} \pa^2 \hat{A}_2 
%+ 2 T^{(1)} i \pa ( \hat{A}_1 - i \hat{A}_2)
%\nonu \\
%& +& 
% \frac{1}{(5+k)} \left( -4 i \hat{A}_1 \, \hat{A}_1 \, \hat{A}_1
%- 12 i \hat{A}_1 \, \hat{A}_2 \, \hat{A}_2
%-4 i  \hat{A}_1 \, \hat{A}_3 \, \hat{A}_3
%\right. 
%\nonu \\
%& +&  \frac{4 i(12+k)}{3}  \hat{A}_1 \, \hat{A}_3 \, \hat{B}_3 
%- 8 i  \hat{A}_1 \, \hat{B}_1 \, \hat{B}_1 
%- 8 i \hat{A}_1 \, \hat{B}_2 \, \hat{B}_2
%- 4 i \hat{A}_1 \, \hat{B}_3 \, \hat{B}_3
%\nonu \\
%&- & 4  \hat{A}_2 \, \hat{A}_1 \, \hat{A}_1
%+ 8 i \hat{A}_2 \, \hat{A}_1 \, \hat{A}_2
%- 4   \hat{A}_2 \, \hat{A}_2 \, \hat{A}_2 
%- 4  \hat{A}_2 \, \hat{A}_3 \, \hat{A}_3
%\nonu \\
%& + & 
%\frac{4(12+k)}{3}  \hat{A}_2 \, \hat{A}_3 \, \hat{B}_3
%-8 \hat{A}_2 \, \hat{B}_1 \, \hat{B}_1
%- \frac{4i(9+k)}{3}  \hat{A}_2 \, \hat{B}_1 \, \hat{B}_2
%- 8  \hat{A}_2 \, \hat{B}_2 \, \hat{B}_2
%\nonu \\
%& - & \left. \left.
%4  \hat{A}_2 \, \hat{B}_3 \, \hat{B}_3
%-\frac{4i(6+k)}{3} \hat{A}_3 \, \hat{A}_1 \, \hat{B}_3
%- \frac{4(6+k)}{3}  \hat{A}_3 \, \hat{A}_2 \, \hat{B}_3
%+ \frac{4i(9+k)}{3}  \hat{B}_2 \, \hat{A}_2 \, \hat{B}_1
%\right) 
- 
\frac{4(6+k)}{3(5+k)^2} i \hat{A}_3 \hat{A}_{-}
\hat{B}_3
-\frac{20}{(5+k)^2} i \hat{A}_{-} \hat{A}_3 \hat{A}_3
\nonu \\
& + & \left. \frac{2}{(5+k)} i \hat{A}_{-} W^{(2)}
\right](w) +     \cdots,
\label{g21v5half} \\
\hat{G}_{22}(z) \, V^{(\frac{5}{2})}(w) 
& = & 
\frac{1}{(z-w)^2}  
\frac{56}{3(5+k)^2} \left[ -\hat{A}_{-}  
  \hat{B}_{+}    \right](w)
\nonu \\
 & + & \frac{1}{(z-w)}\left[ 
- \frac{4}{(5+k)}i 
\hat{A}_{-} \, V_{-}^{(2)}
-    \frac{2}{(5+k)}i 
\hat{B}_{+} \, V_{+}^{(2)} 
\right. 
\label{g22v5half} 
\\
&+ & \left. \frac{2}{(5+k)^2} \hat{B}_{+} \pa 
\hat{A}_{-} 
+ \frac{4}{3(5+k)} \hat{G}_{22} \, \hat{G}_{22}
+\frac{2}{(5+k)} \hat{G}_{22} V^{(\frac{3}{2})}
%- \frac{4}{(5+k)}i 
%(\hat{A}_1 - i \hat{A}_2) \, V_{-}^{(2)}    - \frac{2}{(5+k)^2} i 
%(\hat{A}_1 - i \hat{A}_2) \, \pa \hat{B}_2 
%\right. 
%\nonu \\
%& - &   \frac{2}{(5+k)}i 
%(\hat{B}_1 + i \hat{B}_2) \, V_{+}^{(2)}    + \frac{2}{(5+k)^2}  
%\hat{B}_1  \, \pa \hat{A}_1
%-\frac{1}{2(5+k)} \hat{G}_{21} \, \hat{G}_{21}
%\nonu \\
%& + &  \frac{5}{6(5+k)} \hat{G}_{22} \, \hat{G}_{22} +
%\frac{2}{(5+k)} \hat{G}_{22} \, V^{(\frac{3}{2})} 
%+\frac{2i}{(5+k)^2} \hat{A}_1 \, \hat{A}_3 \, \hat{B}_1
%\nonu \\
%& -& \left. \frac{2 i}{(5+k)^2} \hat{A}_3 \, \hat{A}_1 \, \hat{B}_1 
\right](w) +   \cdots,
\nonu
\eea
where $P_3(w)$ is given by Appendix (\ref{P3}).
The spin-$3$ field for the first three OPEs after subtracting the descendant 
fields (the derivative terms for the fields in the second-order pole with 
coefficient $\frac{1}{4}$) 
in the first-order pole are quasi-primary and those for the last OPE is 
primary. 
The $U(1)$ charge conservation in these OPEs can be checked from Tables 
$2, 3$ and $5$.

%%%%%%%%%%%%%%%%%%
\subsection{The OPEs between four spin $\frac{3}{2}$ currents 
and the higher spin 
current of spins $(2, \frac{5}{2}, \frac{5}{2}, 3)$}
%%%%%%%%%%%%%%%%%%%%

From the explicit expressions (\ref{nong11})-(\ref{nong21}),
and (\ref{w2simple}), 
the following four OPEs between the spin-$\frac{3}{2}$ currents and the 
spin-$2$ current can be obtained
\bea
\left(
\begin{array}{c}
\hat{G}_{11} \\
\hat{G}_{22} \\
\end{array} \right) (z) \, W^{(2)}(w)  & = & 
\frac{1}{(z-w)^2} \frac{1}{(5+k)} \left[ 
\left(
\begin{array}{c}
 \hat{G}_{11} \\
\hat{G}_{22} \\
\end{array} \right) +
(11+3k)  
\left(
\begin{array}{c}
-U^{(\frac{3}{2})} \\
V^{(\frac{3}{2})} \\
\end{array} \right)
\right](w) 
\nonu \\
& + &
\frac{1}{(z-w)} \frac{1}{(5+k)} \left[ 
-\pa  \left(
\begin{array}{c}
 \hat{G}_{11} \\
\hat{G}_{22} \\
\end{array} \right)  + (5+k) \pa 
\left(
\begin{array}{c}
-U^{(\frac{3}{2})} \\
V^{(\frac{3}{2})} \\
\end{array} \right)
\right. \nonu \\
& \pm  & \left. 2 i \hat{A}_{\pm}   
 \left(
\begin{array}{c}
 \hat{G}_{21} \\
\hat{G}_{12} \\
\end{array} \right)
+   2 i \hat{A}_{\pm} 
T_{\pm}^{(\frac{3}{2})}
-  2 i \hat{B}_{\mp} T_{\mp}^{(\frac{3}{2})}
\right](w) +\cdots,
\label{G11w2} \\
\left(
\begin{array}{c}
\hat{G}_{12} \\
\hat{G}_{21} \\
\end{array} \right) (z) \, W^{(2)}(w)  & = & 
\frac{1}{(z-w)^2} \frac{1}{(5+k)} \left[ 2(2+k) 
 \left(
\begin{array}{c}
 \hat{G}_{12} \\
\hat{G}_{21} \\
\end{array} \right)
 \mp
(-3+k)  T_{\mp}^{(\frac{3}{2})}\right](w) 
\nonu
\\
& + &
\frac{1}{(z-w)} \frac{1}{(5+k)} \left[ W_{\mp}^{(\frac{5}{2})} +
 \frac{1}{3} 2(2+k) \pa 
 \left(
\begin{array}{c}
 \hat{G}_{12} \\
\hat{G}_{21} \\
\end{array} \right)
\right. \nonu \\
 & \mp & \left. 
\frac{1}{3} (-3+k)  \pa T_{\mp}^{(\frac{3}{2})}
\right](w) 
+  \cdots.
\label{G12w2}
\eea
As above,
the spin-$\frac{5}{2}$ fields after subtracting the descendant 
fields (the derivative terms for the fields in the second-order pole with 
coefficient $\frac{1}{3}$) 
in the first-order pole are primary. 
The $U(1)$ charge conservation in these OPEs can be checked from Tables 
$2$ and $4$.

For the spin-$\frac{5}{2}$ current (\ref{w+5half}), 
one has the following OPEs between the spin-$\frac{3}{2}$ currents and the 
spin-$\frac{5}{2}$ current
\bea
\hat{G}_{11}(z) \, W_{+}^{(\frac{5}{2})}(w)  & = & 
\frac{1}{(z-w)^3} \frac{56k}{3(5+k)^2} \left[ i \hat{B}_{-} \right](w) 
\nonu \\
& + & \frac{1}{(z-w)^2} \left[ 
\frac{2(15+8k)}{3(5+k)} U_{+}^{(2)}
+\frac{4(-1+2k)}{3(5+k)^2}  \hat{A}_3 \,  \hat{B}_{-} 
\right. \nonu \\
& - &  \left. \frac{4k}{(5+k)^2} \hat{B}_{-}  
\hat{B}_3  - \frac{2k }{(5+k)^2} i \pa \hat{B}_{-} 
 - \frac{2}{(5+k)} T^{(1)} i \hat{B}_{-}  \right](w)
\nonu \\
& + & \frac{1}{(z-w)} 
\left[ \frac{4}{(5+k)} i \hat{A}_3 \, U_{+}^{(2)} 
+ \frac{12k}{(3+7k)(5+k)} i \hat{B}_{-}   \hat{T} 
+   \frac{2}{(5+k)} i \hat{B}_{-} T^{(2)}
\right. \nonu \\
& + & 
\frac{2(1+4k)}{3(5+k)^2} \hat{B}_{-}  \pa \hat{A}_3
+\frac{4(3+k)}{3(5+k)} \pa U_{+}^{(2)}
-  \frac{k}{3(5+k)^2} i \pa^2 \hat{B}_{-}
\nonu \\
& + & \frac{1}{(5+k)} i \pa T^{(1)}   \hat{B}_{-}
+\frac{ 2}{(5+k)} 
\hat{G}_{11}  \hat{G}_{21} +   
\frac{2}{(5+k)} \hat{G}_{11}  T_{+}^{(\frac{3}{2})}
\label{g11w+5half}
\\
& - & \frac{4}{(5+k)^2} i 
 \hat{A}_{+}  \hat{A}_{-}
 \hat{B}_{-} 
-  \frac{4(-5+k)}{3(5+k)^2} i \hat{A}_3  \hat{B}_{-}
\hat{B}_3 +\frac{k}{(5+k)^2} i   \hat{B}_{+}  
\hat{B}_{-}  \hat{B}_{-}
\nonu \\
&- & \frac{k}{(5+k)^2} i  \hat{B}_{-}  
\hat{B}_{+}  \hat{B}_{-}
-\frac{4k}{(5+k)^2} i  \hat{B}_{-}   
\hat{B}_3 \hat{B}_3
+  \frac{4(-5+k)}{3(5+k)^2} i \hat{B}_3 \hat{A}_3   
\hat{B}_{-}
\nonu \\
& + & \left. \frac{4k}{(5+k)^2} i \hat{B}_3   \hat{B}_{-}  \hat{B}_3
%\frac{1}{(5+k)} 
%\left[  \frac{12k}{(3+7k)} i ( \hat{B}_1  - i \hat{B}_2) \hat{T}  
%+ 2 i 
%( \hat{B}_1 - i\hat{B}_2) T^{(2)}
%\right. \nonu \\
%&- & 
% \frac{2 i(-5+4k)}{3(5+k)} \hat{B}_2 \, \pa \hat{A}_3
%- \frac{2 i k}{(5+k)} \hat{B}_2 \, \pa \hat{B}_3
%+ 4 i \hat{A}_3 \, U_{+}^{(2)} -\frac{4(-5+k)}{3(5+k)} \hat{A}_3 
%\, \pa \hat{B}_1  
%\nonu \\
%& + & \frac{2 i k}{3(5+k)} \pa^2 \hat{B}_1    
%+\frac{4(3+k)}{3} \pa U_{+}^{(2)}
%+\frac{k }{6(5+k)} \pa^2 \hat{B}_2 
%- 2 T^{(1)} i \pa ( \hat{B}_1 - i \hat{B}_2)
%\nonu \\
%& + & \pa T^{(1)} i \pa ( \hat{B}_1 - i \hat{B}_2)
%+ 2 \hat{G}_{11} \, \hat{G}_{21} + 2 \hat{G}_{11} \, T_{+}^{(\frac{3}{2})}
%\nonu \\
%& +& 
% \frac{1}{(5+k)} \left( -4 i \hat{A}_1 \, \hat{A}_1 \, \hat{B}_1
%- 4 \hat{A}_1 \, \hat{A}_1 \, \hat{B}_2
%-\frac{2(-5+4k)}{3}  \hat{A}_1 \, \hat{A}_2 \, \hat{B}_1
%\right. 
%\nonu \\
%& +&  \frac{2(5-4k)}{3}  \hat{A}_2 \, \hat{A}_1 \, \hat{B}_1 
%- 4 i  \hat{A}_2 \, \hat{A}_2 \, \hat{B}_1 
%- 4 \hat{A}_2 \, \hat{A}_2 \, \hat{B}_2
%- \frac{4 i(-5+k)}{3} \hat{A}_3 \, \hat{B}_1 \, \hat{B}_3
%\nonu \\
%&+ & k  \hat{B}_1 \, \hat{B}_1 \, \hat{B}_2
%- 4 i k \hat{B}_1 \, \hat{B}_3 \, \hat{B}_3
%- k   \hat{B}_2 \, \hat{B}_1 \, \hat{B}_1 
%- 4 k \hat{B}_2 \, \hat{B}_3 \, \hat{B}_3
%\nonu \\
%& + & \left. \left.
%\frac{4 i(-5+k)}{3}  \hat{B}_3 \, \hat{A}_3 \, \hat{B}_1
%+4 i k \hat{B}_3 \, \hat{B}_1 \, \hat{B}_3
%+ 4 k  \hat{B}_3 \, \hat{B}_2 \, \hat{B}_3
%\right) 
+  \frac{2}{(5+k)} T^{(1)}  \hat{B}_{-}  \hat{B}_3
-\frac{2}{(5+k)} \hat{B}_3 T^{(1)}  \hat{B}_{-} 
\right](w) +   \cdots,
\nonu \\
\hat{G}_{12}(z) \, W_{+}^{(\frac{5}{2})}(w)  & = & 
\frac{1}{(z-w)^3}  \left[ -\frac{8 i (1+3k)}{(5+k)^2} \hat{A}_3 -
\frac{80 i k}{3(5+k)^2} \hat{B}_3 - \frac{8(-3+k)}{3(5+k)} T^{(1)}
\right](w) 
\nonu \\
& + & \frac{1}{(z-w)^2} \left[ 
-\frac{4(15+65k+22k^2)}{3(3+7k)(5+k)} \hat{T}
+\frac{4(-3+k)}{3(5+k)} T^{(2)}
+\frac{4(4+k)}{(5+k)} W^{(2)}
\right.
\nonu \\
& - & \frac{16(-1+k)}{3(5+k)^2}  \hat{A}_1 \, \hat{A}_1 
-  \frac{16(-1+k)}{3(5+k)^2}  \hat{A}_2 \, \hat{A}_2 
+  \frac{4(13-4k)}{3(5+k)^2}  \hat{A}_3 \, \hat{A}_3
\nonu \\
& - &    \frac{4}{3(5+k)}  \hat{B}_1 \, \hat{B}_1 -
 \frac{4}{3(5+k)}  \hat{B}_2 \, \hat{B}_2 +
 \frac{4(-5+2k)}{3(5+k)^2}  \hat{B}_3 \, \hat{B}_3 -
 \frac{4 i}{(5+k)} T^{(1)} \, \hat{A}_3 \nonu \\
& + &  \left. 
 \frac{4 i }{(5+k)} T^{(1)} \, \hat{B}_3   
+ \frac{8(-4+k)}{3(5+k)^2}  \hat{A}_3 \, \hat{B}_3  
\right](w) \nonu \\
& + &  
\frac{1}{(z-w)}  \left[ \frac{1}{4} \pa \, \{ \hat{G}_{12} \, 
W_{+}^{(\frac{5}{2})}  \}_{-2}
 +\frac{24 i(1+3k)}{(19+23k)(5+k)} \left( \hat{T} \, \hat{A}_3-
\frac{1}{2} \pa^2 \hat{A}_3 \right) \right.
\nonu \\
& + &  \frac{80i k }{(19+23k)(5+k)}
 \left( \hat{T} \, \hat{B}_3-
\frac{1}{2} \pa^2 \hat{B}_3 \right)
\nonu \\
& + & \left. P^{(3)} 
+\frac{8(-3+k)}{(19+23k)}
 \left( \hat{T} \, T^{(1)}-
\frac{1}{2} \pa^2 T^{(1)} \right)
\right](w) +\cdots,
\label{g12w+5half} \\
\hat{G}_{21}(z) \, W_{+}^{(\frac{5}{2})}(w)  & = & 
 \frac{1}{(z-w)^2}  
\frac{4(19+3k)}{3(5+k)^2}  \left[ - \hat{A}_{-} \hat{B}_{-}  \right](w)
\nonu \\
& + & \frac{1}{(z-w)} 
\left[ 
-\frac{11}{3(5+k)}  \hat{G}_{21} \, \hat{G}_{21} 
+\frac{2}{(5+k)} i \hat{A}_{-} U_{+}^{(2)} 
\right. \nonu \\
& - & \left. \frac{2(-3+k)}{(5+k)^2} \hat{A}_{-} \pa 
 \hat{B}_{-}
-\frac{2}{(5+k)} i   \hat{B}_{-}  V_{+}^{(2)}
%\frac{1}{(5+k)} 
%\left[  2 i ( \hat{A}_1  - i \hat{A}_2) U_{+}^{(2)}  
%- 2 i 
%( \hat{B}_1 - i\hat{B}_2) V_{+}^{(2)}
%\right. \nonu \\
%&- & 
% \frac{2 (-3+k)}{(5+k)} \hat{A}_1 \, \pa ( \hat{B}_1 - i \hat{B}_2)
%- \frac{(-3+k)}{2} \hat{G}_{11} \, \hat{G}_{11} 
%- \frac{(13+3k)}{6} \hat{G}_{21} \, \hat{G}_{21} 
%\nonu \\
%& + & 
%\frac{1}{(5+k)} \left( 2 i(-3+k) \hat{A}_1 \, \hat{A}_3 \, \hat{B}_1
%+ 2 (-3+k) \hat{A}_1 \, \hat{A}_3 \, \hat{B}_2
%- 2 i (-3 +k)  \hat{A}_3 \, \hat{A}_1 \, \hat{B}_1
%\right.
%\nonu \\
%& - &   \left. \left. 2(-3+k)  \hat{A}_3 \, \hat{A}_1 \, \hat{B}_2 
%\right) 
\right](w) +   \cdots,
\label{g21w+5half} \\
\hat{G}_{22}(z) \, W_{+}^{(\frac{5}{2})}(w)  & = & 
\frac{1}{(z-w)^3} \frac{4(5+3k)}{(5+k)^2} \left[ i \hat{A}_{-} \right](w) 
\nonu \\
& + & \frac{1}{(z-w)^2} \left[ 
-\frac{2(27+4k)}{3(5+k)} V_{+}^{(2)}
-\frac{12}{(5+k)^2} \hat{A}_{-}  \hat{A}_3 
\right. \nonu \\
& + &  \left. \frac{4(2+k)}{3(5+k)^2} \hat{A}_{-} 
\hat{B}_3  - \frac{6 }{(5+k)^2} i \pa  \hat{A}_{-}
 + \frac{2}{(5+k)} T^{(1)} i \hat{A}_{-}  \right](w)
\nonu \\
& + & \frac{1}{(z-w)} 
\left[
 \frac{4(3+4k)}{(3+7k)(5+k)} i  \hat{A}_{-} \hat{T}  
- \frac{2}{(5+k)} i 
\hat{A}_{-} T^{(2)}
\right. \nonu \\
& - & \frac{2(-10+k)}{3(5+k)^2}  \hat{A}_{-} \pa \hat{B}_3 
-  \frac{4}{(5+k)^2} \hat{A}_3 
 \pa \hat{A}_{-} 
-\frac{4}{(5+k)} i \hat{B}_3  V_{+}^{(2)}
\nonu \\
&- & \frac{1}{(5+k)} i \pa T^{(1)} \hat{A}_{-} 
+ \frac{2}{(5+k)} i  T^{(1)} \pa \hat{A}_{-}
-\frac{2}{(5+k)} \hat{G}_{22} \, T_{+}^{(\frac{3}{2})}
\nonu \\
&+ & \frac{4}{(5+k)^2} i 
 \hat{A}_{+}
 \hat{A}_{-}  \hat{A}_{-}
- \frac{4}{(5+k)^2} i 
 \hat{A}_{-}
 \hat{A}_3   \hat{A}_3  
+  \frac{4(-7+k)}{3(5+k)^2} i  \hat{A}_{-} 
 \hat{A}_3   \hat{B}_3   
\nonu \\
& + &  \frac{4}{(5+k)^2} i  \hat{A}_{-}
 \hat{B}_3   \hat{B}_3   
%\frac{1}{(5+k)} 
%\left[  \frac{4(3+4k)}{(3+7k)} i ( \hat{A}_1  - i \hat{A}_2) \hat{T}  
%- 2 i 
%( \hat{A}_1 - i\hat{A}_2) T^{(2)}
%\right. \nonu \\
%&- & 
% \frac{26}{(5+k)} \hat{A}_1 \, \pa \hat{A}_3
%- \frac{2 (-10+ k)}{3(5+k)} \hat{A}_1 \, \pa \hat{B}_3
% -\frac{12}{(5+k)} \hat{A}_3 
%\, (\pa \hat{A}_1 - i \pa \hat{A}_2)  
%\nonu \\
%& - & 
%\frac{2(9+k)}{3} \pa V_{+}^{(2)} -4 i \hat{B}_3 \,  V_{+}^{(2)}
%+ 2 T^{(1)} i \pa ( \hat{A}_1 - i \hat{A}_2)
%\nonu \\
%& - & \pa T^{(1)} i  ( \hat{A}_1 - i \hat{A}_2)
% - 2 \hat{G}_{22} \, T_{+}^{(\frac{3}{2})}
%\nonu \\
%& +& 
% \frac{1}{(5+k)} \left( 4 i \hat{A}_1 \, \hat{A}_1 \, \hat{A}_1
%+ 16 \hat{A}_1 \, \hat{A}_1 \, \hat{A}_2
%- 2 i   \hat{A}_1 \, \hat{A}_2 \, \hat{A}_2
%\right. 
%\nonu \\
%& - & 6 i   \hat{A}_1 \, \hat{A}_3 \, \hat{A}_3 
%- \frac{4 i(-7+k)}{3}  \hat{A}_1 \, \hat{A}_3 \, \hat{B}_3 
%+ 4  i \hat{A}_1 \, \hat{B}_3 \, \hat{B}_3
%- 12 \hat{A}_2 \, \hat{A}_1 \, \hat{A}_1
%\nonu \\
%&+ & 6 i  \hat{A}_2 \, \hat{A}_1 \, \hat{A}_2
%+ 4 \hat{A}_2 \, \hat{A}_2 \, \hat{A}_2
%+ 4   \hat{A}_2 \, \hat{A}_3 \, \hat{A}_3 
%- \frac{4(-7+k)}{3} \hat{A}_2 \, \hat{A}_3 \, \hat{B}_3
%\nonu \\
%& + & 
%\frac{2 i(-10+k)}{3}  \hat{A}_2 \, \hat{B}_1 \, \hat{B}_2
%+4  \hat{A}_2 \, \hat{B}_3 \, \hat{B}_3
%+ 10 i  \hat{A}_3 \, \hat{A}_1 \, \hat{A}_3
%\nonu  \\
%& 
% - & \left. \left. 
%\frac{4 i(-1+k)}{3}  \hat{A}_3 \, \hat{A}_1 \, \hat{B}_3
%-\frac{4(-1+k)}{3}  \hat{A}_3 \, \hat{A}_2 \, \hat{B}_3
%- \frac{2i(-10+k)}{3}  \hat{B}_2 \, \hat{A}_2 \, \hat{B}_1
%\right) 
+  \frac{8}{(5+k)^2} i \hat{A}_3  \hat{A}_{-}
 \hat{A}_3  
-\frac{4(-1+k)}{3(5+k)^2} i
 \hat{A}_3  \hat{A}_{-}
 \hat{B}_3  
\nonu \\
 & - & \left. \frac{2(9+k)}{3(5+k)} \pa 
 V_{+}^{(2)}
- \frac{6}{(5+k)^2}  \hat{A}_{-} \pa \hat{A}_3
\right](w) +   \cdots,
\label{g22w+5half}
\eea
where $P_3(w)$ is given by Appendix (\ref{P3}).
The spin-$3$ field for the third OPE after subtracting the descendant 
fields (the derivative terms for the fields in the second-order pole with 
coefficient $\frac{1}{4}$) 
in the first-order pole is primary and those for the remaining OPEs are 
quasi-primary. 
The $U(1)$ charge conservation in these OPEs can be checked from Tables 
$2, 3$ and $ 5$.

For the other spin-$\frac{5}{2}$ current (\ref{w-5half}), 
the following OPEs between the spin-$\frac{3}{2}$ currents and the
spin-$\frac{5}{2}$ current can be obtained
\bea
\hat{G}_{11}(z) \, W_{-}^{(\frac{5}{2})}(w)  & = & 
-\frac{1}{(z-w)^3} \frac{4(5+3k)}{(5+k)^2} \left[ i \hat{A}_{+} \right](w) 
\nonu \\
& + & \frac{1}{(z-w)^2} \left[ 
\frac{2(27+4k)}{3(5+k)} U_{-}^{(2)}
-\frac{12}{(5+k)^2} \hat{A}_{+}  \hat{A}_3 
\right. \nonu \\
& + &  \left. \frac{4(2+k)}{3(5+k)^2} \hat{A}_{+} 
\hat{B}_3  + \frac{6 }{(5+k)^2} i \pa \hat{A}_{+} 
 + \frac{2}{(5+k)} T^{(1)} i 
\hat{A}_{+}  \right](w)
\nonu \\
& + & \frac{1}{(z-w)} 
\left[ 
 -\frac{4(3+4k)}{(3+7k)(5+k)} i \hat{A}_{+} \hat{T}  
+ \frac{2}{(5+k)} i 
\hat{A}_{+} T^{(2)}
\right. \nonu \\
& - & \frac{2(-10+k)}{3(5+k)^2}  \hat{A}_{+}   \pa \hat{B}_3
+  \frac{4}{(5+k)^2} \hat{A}_3 \pa \hat{A}_{+} 
-\frac{4}{(5+k)} i  \hat{B}_3  U_{-}^{(2)} 
\nonu \\
&-& \frac{1}{(5+k)} i \pa T^{(1)}  \hat{A}_{+}
+ \frac{2}{(5+k)} i  T^{(1)} \pa \hat{A}_{+}
+\frac{2}{(5+k)} \hat{G}_{11}  T_{-}^{(\frac{3}{2})}
\nonu \\
& + & \frac{12}{(5+k)^2} i  \hat{A}_{+} \hat{A}_3 \hat{A}_3
-  \frac{4(-7+k)}{3(5+k)^2} i  \hat{A}_{+}
\hat{A}_3 \hat{B}_3
-\frac{4}{(5+k)^2} i  \hat{A}_{+} \hat{B}_3 \hat{B}_3 
\nonu \\
%\frac{1}{(5+k)} 
%\left[  -\frac{4(3+4k)}{(3+7k)} i ( \hat{A}_1  + i \hat{A}_2) \hat{T}  
%+ 2 i 
%( \hat{A}_1 + i\hat{A}_2) T^{(2)}
%\right. \nonu \\
%&- & 
% \frac{2 (-10+ k)}{3(5+k)} \hat{A}_1 \, \pa \hat{B}_3
% +\frac{1}{(5+k)} \hat{A}_3 
%\, \pa \hat{A}_1  
%-\frac{12i}{(5+k)} \hat{A}_3 
%\, \pa \hat{A}_2  
%- 4 i \hat{B}_3 
%\, U_{-}^{(2)}  
%\nonu \\
%& - & 
%\frac{2(9+k)}{3} \pa U_{-}^{(2)} -\frac{5 i}{(5+k)} \pa^2 \hat{A}_1 
%+ 2 T^{(1)} i \pa ( \hat{A}_1 + i \hat{A}_2)
%\nonu \\
%& - & \pa T^{(1)} i \pa ( \hat{A}_1 + i \hat{A}_2)
% + 2 \hat{G}_{11} \, T_{-}^{(\frac{3}{2})}
%\nonu \\
%& +& 
% \frac{1}{(5+k)} \left( -4 i \hat{A}_1 \, \hat{A}_1 \, \hat{A}_1
%+ 3 \hat{A}_1 \, \hat{A}_1 \, \hat{A}_2
%+ 2 i   \hat{A}_1 \, \hat{A}_2 \, \hat{A}_2
%\right. 
%\nonu \\
%& - & 4 i   \hat{A}_1 \, \hat{A}_3 \, \hat{A}_3 
%- \frac{4 i(-7+k)}{3}  \hat{A}_1 \, \hat{A}_3 \, \hat{B}_3 
%- 4  i \hat{A}_1 \, \hat{B}_3 \, \hat{B}_3
%+ \hat{A}_2 \, \hat{A}_1 \, \hat{A}_1
%\nonu \\
%&- & 6 i  \hat{A}_2 \, \hat{A}_1 \, \hat{A}_2
%+ 4 \hat{A}_2 \, \hat{A}_2 \, \hat{A}_2
%-9   \hat{A}_2 \, \hat{A}_3 \, \hat{A}_3 
%+\frac{4(-7+k)}{3} \hat{A}_2 \, \hat{A}_3 \, \hat{B}_3
%\nonu \\
%& - & 
%\frac{2 i(-10+k)}{3}  \hat{A}_2 \, \hat{B}_1 \, \hat{B}_2
%+4  \hat{A}_2 \, \hat{B}_3 \, \hat{B}_3
%+ \frac{4 i(-1+k)}{3}  \hat{A}_3 \, \hat{A}_1 \, \hat{B}_3
%\nonu  \\
%& 
% + & \left. \left. 
%13  \hat{A}_3 \, \hat{A}_2 \, \hat{A}_3
%-\frac{4(-1+k)}{3}  \hat{A}_3 \, \hat{A}_2 \, \hat{B}_3
%+ \frac{2i(-10+k)}{3}  \hat{B}_2 \, \hat{A}_2 \, \hat{B}_1
%\right) 
&-&  \frac{16}{(5+k)^2} i
\hat{A}_3 \hat{A}_{+} \hat{A}_3
+\frac{4(-1+k)}{3(5+k)^2} i  \hat{A}_3 \hat{A}_{+}
\hat{B}_3
-  \frac{4}{(5+k)^2} i 
 \hat{A}_{+}  \hat{A}_{+}  
\hat{A}_{-}
\nonu \\
& + &  \left. \frac{10}{(5+k)^2} \hat{A}_{+}    \pa \hat{A}_3 
+ \frac{2(9+k)}{3(5+k)} \pa U_{-}^{(2)}
\right](w)
+    \cdots,
\label{g11w-5half} \\
\hat{G}_{12}(z) \, W_{-}^{(\frac{5}{2})}(w)  & = & 
\frac{1}{(z-w)^2}  
\frac{4(19+3k)}{3(5+k)^2}  \left[ - \hat{A}_{+} \hat{B}_{+}   \right](w)
\nonu \\
& + & \frac{1}{(z-w)} 
\left[ -\frac{11}{3(5+k)} \hat{G}_{12} \, \hat{G}_{12}
-\frac{2}{(5+k)} i \hat{B}_{+} U_{-}^{(2)}
\right. \nonu \\
& + & \left. \frac{2}{(5+k)} i  \hat{A}_{+}  V_{-}^{(2)}
-\frac{2(-3+k)}{(5+k)^2} \hat{A}_{+} 
 \pa \hat{B}_{+}
%\frac{1}{(5+k)} 
%\left[  2 i ( \hat{A}_1  + i \hat{A}_2) V_{-}^{(2)}  
%- 2 i 
%( \hat{B}_1 + i\hat{B}_2) U_{-}^{(2)}
%\right. \nonu \\
%&- & 
% \frac{2 (-3+k)}{(5+k)} \hat{A}_1 \, \pa ( \hat{B}_1 + i \hat{B}_2)
%- \frac{(13+3k)}{6} \hat{G}_{12} \, \hat{G}_{12} 
%- \frac{(-3+k)}{2} \hat{G}_{22} \, \hat{G}_{22} 
%\nonu \\
%& + & 
%\frac{1}{(5+k)} \left( -2 i(-3+k) \hat{A}_1 \, \hat{A}_3 \, \hat{B}_1
%+ 2 (-3+k) \hat{A}_1 \, \hat{A}_3 \, \hat{B}_2
%+ 2 i (-3 +k)  \hat{A}_3 \, \hat{A}_1 \, \hat{B}_1
%\right.
%\nonu \\
%& - &   \left. \left. 2(-3+k)  \hat{A}_3 \, \hat{A}_1 \, \hat{B}_2 
%\right)  
\right](w) 
+   \cdots,
\label{g12w-5half} \\
\hat{G}_{21}(z) \, W_{-}^{(\frac{5}{2})}(w)  & = & 
\frac{1}{(z-w)^3}  \left[ \frac{8 i (1+3k)}{(5+k)^2} \hat{A}_3 +
\frac{80 i k}{3(5+k)^2} \hat{B}_3 + \frac{8(-3+k)}{3(5+k)} T^{(1)}
\right](w) 
\nonu \\
& + & \frac{1}{(z-w)^2} \left[ 
-\frac{4(15+65k+22k^2)}{3(3+7k)(5+k)} \hat{T}
+\frac{4(-3+k)}{3(5+k)} T^{(2)}
+\frac{4(4+k)}{(5+k)} W^{(2)}
\right.
\nonu \\
& - & \frac{16(-1+k)}{3(5+k)^2}  \hat{A}_1 \, \hat{A}_1 
-  \frac{16(-1+k)}{3(5+k)^2}  \hat{A}_2 \, \hat{A}_2 
+  \frac{4(13-4k)}{3(5+k)^2}  \hat{A}_3 \, \hat{A}_3
\nonu \\
& - &    \frac{4}{3(5+k)}  \hat{B}_1 \, \hat{B}_1 -
 \frac{4}{3(5+k)}  \hat{B}_2 \, \hat{B}_2 +
 \frac{4(-5+2k)}{3(5+k)^2}  \hat{B}_3 \, \hat{B}_3 \nonu \\
& - & \left.
 \frac{4 i}{(5+k)} T^{(1)} \, \hat{A}_3  +  
 \frac{4 i }{(5+k)} T^{(1)} \, \hat{B}_3   
+ \frac{8(-4+k)}{3(5+k)^2}  \hat{A}_3 \, \hat{B}_3  
\right](w)
\nonu \\
& + &
\frac{1}{(z-w)}  \left[ \frac{1}{4} \pa \, \{ \hat{G}_{21} \, 
W_{-}^{(\frac{5}{2})} \}_{-2}
+\frac{24 i(1+3k)}{(19+23k)(5+k)} \left( \hat{T} \, \hat{A}_3-
\frac{1}{2} \pa^2 \hat{A}_3 \right) \right.
\nonu \\
& + &  \frac{80i k }{(19+23k)(5+k)}
 \left( \hat{T} \, \hat{B}_3-
\frac{1}{2} \pa^2 \hat{B}_3 \right)
\nonu \\
& + & \left. W^{(3)} 
+\frac{8(-3+k)}{(19+23k)}
 \left( \hat{T} \, T^{(1)}-
\frac{1}{2} \pa^2 T^{(1)} \right)
\right](w) +\cdots,
\label{g21w-5half} \\
\hat{G}_{22}(z) \, W_{-}^{(\frac{5}{2})}(w)  & = & 
\frac{1}{(z-w)^3} \frac{56k}{3(5+k)^2} \left[ -i \hat{B}_{+} \right](w) 
\nonu \\
& + & \frac{1}{(z-w)^2} \left[ 
-\frac{2(15+8k)}{3(5+k)} V_{-}^{(2)}
+\frac{4(-1+2k)}{3(5+k)^2}  \hat{A}_3 \, \hat{B}_{+} 
\right. \nonu \\
& - &  \left. \frac{4k}{(5+k)^2} \hat{B}_{+} 
\hat{B}_3  + \frac{2k }{(5+k)^2} i \pa \hat{B}_{+}  
 - \frac{2}{(5+k)} T^{(1)} i \hat{B}_{+}  \right](w)
\nonu \\
& + & \frac{1}{(z-w)} 
\left[
 \frac{4}{(5+k)} i \hat{A}_3 \, V_{-}^{(2)}
-\frac{4(-11+k)}{3(5+k)^2} \hat{A}_3 \pa  \hat{B}_{+}  
\right. \nonu \\
& - & \frac{2}{(5+k)} i \hat{B}_{+} T^{(2)}
+  \frac{2(13+4k)}{3(5+k)^2}   \hat{B}_{+}  \pa \hat{A}_3
-\frac{4(3+k)}{3(5+k)} \pa V_{-}^{(2)}
\nonu \\
&- & \frac{1}{(5+k)} i \pa T^{(1)}  \hat{B}_{+}
-\frac{2}{(5+k)} i T^{(1)}  \pa \hat{B}_{+}
-\frac{2}{(5+k)}  \hat{G}_{12}  \hat{G}_{22} 
\nonu \\
& - & \frac{2}{(5+k)}  \hat{G}_{22} \, T_{-}^{(\frac{3}{2})}
+\frac{4}{(5+k)^2} i   \hat{A}_{-}
 \hat{A}_{+}  \hat{B}_{+} 
+  \frac{k}{2(5+k)^2} i  \hat{B}_{+} 
 \hat{B}_{+}  \hat{B}_{-}
\nonu \\
& + & \left.  \frac{4k}{(5+k)^2} i  \hat{B}_{+} 
  \hat{B}_3 \hat{B}_3
-   \frac{k}{2(5+k)^2} i  \hat{B}_{-} 
 \hat{B}_{+}  \hat{B}_{+}
%\frac{1}{(5+k)} 
%\left[  -\frac{12k}{(3+7k)} i ( \hat{B}_1  + i \hat{B}_2) \hat{T}  
%- 2 i 
%( \hat{B}_1 + i\hat{B}_2) T^{(2)}
%\right. \nonu \\
%&+ & 
% \frac{2 i(7+4k)}{3(5+k)} \hat{B}_2 \, \pa \hat{A}_3
%+ 4 i \hat{A}_3 \, V_{-}^{(2)} -\frac{4(-11+k)}{3(5+k)} \hat{A}_3 
%\, \pa \hat{B}_1  
%\nonu \\
%& + & \frac{ i k}{3(5+k)} \pa^2 \hat{B}_1    
%-\frac{4(3+k)}{3} \pa V_{-}^{(2)}
%+\frac{k }{6(5+k)} \pa^2 \hat{B}_2 
%- 2 T^{(1)} i \pa ( \hat{B}_1 + i \hat{B}_2)
%\nonu \\
%& + & \pa T^{(1)} i  ( \hat{B}_1 + i \hat{B}_2)
%- 2 \hat{G}_{12} \, \hat{G}_{22} - 2 \hat{G}_{22} \, T_{-}^{(\frac{3}{2})}
%\nonu \\
%& +& 
% \frac{1}{(5+k)} \left( 4 i \hat{A}_1 \, \hat{A}_1 \, \hat{B}_1
%- 4 \hat{A}_1 \, \hat{A}_1 \, \hat{B}_2
%+\frac{2(7+4k)}{3}  \hat{A}_1 \, \hat{A}_2 \, \hat{B}_1
%\right. 
%\nonu \\
%& -&  \frac{2(7+4k)}{3}  \hat{A}_2 \, \hat{A}_1 \, \hat{B}_1 
%+ 4 i  \hat{A}_2 \, \hat{A}_2 \, \hat{B}_1 
%- 4 \hat{A}_2 \, \hat{A}_2 \, \hat{B}_2
%+ \frac{4 i(-11+k)}{3} \hat{A}_3 \, \hat{B}_1 \, \hat{B}_3
%\nonu \\
%&+ & k  \hat{B}_1 \, \hat{B}_1 \, \hat{B}_2
%+2 i k \hat{B}_1 \, \hat{B}_2 \, \hat{B}_2
%+4 i k   \hat{B}_1 \, \hat{B}_3 \, \hat{B}_3 
%-  k \hat{B}_2 \, \hat{B}_1 \, \hat{B}_1
%\nonu \\
%& - & 
%2 i k  \hat{B}_2 \, \hat{B}_1 \, \hat{B}_2
%-4  k \hat{B}_2 \, \hat{B}_3 \, \hat{B}_3
%- \frac{4 i (-11+k)}{3}  \hat{B}_3 \, \hat{A}_3 \, \hat{B}_1
%\nonu \\
%& -& \left. \left. 4 i  k \hat{B}_3 \, \hat{B}_1 \, \hat{B}_3
%+ 4 k  \hat{B}_3 \, \hat{B}_2 \, \hat{B}_3
%\right) 
-   \frac{4k}{(5+k)^2} i  \hat{B}_3
 \hat{B}_{+}   \hat{B}_3
\right.
\nonu \\
&- &  \left.  \frac{12k}{(3+7k)(5+k)} i \hat{B}_{+} \hat{T} 
-\frac{k}{6(5+k)^2} i \pa^2  \hat{B}_{+}
\right](w)
 +   \cdots.
\label{g22w-5half}
\eea
The spin-$3$ field for the second OPE after subtracting the descendant 
fields (the derivative terms for the fields in the second-order pole with 
coefficient $\frac{1}{4}$) 
in the first-order pole is primary and those for the remaining OPEs are 
quasi-primary. 
The $U(1)$ charge conservation in these OPEs can be checked from Tables 
$2, 3$ and $ 5$.

Finally, from the explicit for  the spin-$3$ current (\ref{w3}), 
the following complicated OPEs between the spin-$\frac{3}{2}$ currents 
and the spin-$3$ current can be obtained  
\bea
\hat{G}_{11}(z) \, W^{(3)}(w)  & = & 
\frac{1}{(z-w)^3} \left[-\frac{2(-3+k)(571+357k+110k^2)}{3(5+k)^2(19+23k)}
\hat{G}_{11} \right. \nonu \\
& - & \left.  \frac{2(-3+k)(1489+1225k+220k^2)}{3(5+k)^2(19+23k)}  
U^{(\frac{3}{2})}\right](w)
\nonu \\
& + & 
\frac{1}{(z-w)^2} \left[ 
-\frac{(19+5k)}{(5+k)} U^{(\frac{5}{2})} 
+ \frac{2(-10+k)}{(5+k)^2} i \hat{A}_{+} 
\hat{G}_{21}
\right.
\nonu \\
&- &   \frac{2(14+3k)}{(5+k)^2} i 
\hat{A}_{+} T_{+}^{(\frac{3}{2})}
-\frac{2(-5+211k+100k^2)}{(5+k)^2(19+23k)} i \hat{A}_3 \, \hat{G}_{11}
\nonu \\
&- &  \frac{6(7+3k)}{(5+k)^2} i \hat{A}_3 \, U^{(\frac{3}{2})}
-\frac{(-5+7k)}{(5+k)^2} i \hat{B}_{-}  
\hat{G}_{12} - \frac{18}{(5+k)^2} 
i \hat{B}_{-} T_{-}^{(\frac{3}{2})}
\nonu \\
& - &  \frac{2(76+411k+83k^2)}{(5+k)(19+23k)} i \hat{B}_3 \, \hat{G}_{11}
+\frac{2(-1+3k)}{(5+k)^2} i \hat{B}_3 \, U^{(\frac{3}{2})}
\nonu \\
& - & \frac{(-1047+16k+251k^2+12k^3)}{3(5+k)^2(19+23k)} \pa \hat{G}_{11}
\nonu \\
& - & \left.
\frac{4(-640-355k-73k^2+6k^3)}{3(5+k)^2(19+23k)} \pa U^{(\frac{3}{2})}  
%\right. \nonu \\
%&+ &  \frac{1}{(5+k)^2}\left(  4 (2+k) i( \hat{A}_1 +i \hat{A}_2)  
%\hat{G}_{21} + 8(2+k) i (\hat{A}_1 + i \hat{A}_2) T_{+}^{(\frac{3}{2})}
%\right. \nonu \\
%&+ & 
%2 (7+k) i( \hat{B}_1 -i \hat{B}_2)  
%\hat{G}_{12} - 4(7+k) i (\hat{B}_1 - i \hat{B}_2) T_{-}^{(\frac{3}{2})}
%\nonu \\
%&+ & \left. \left. 12 i \hat{A}_3 \, \hat{G}_{11}
%+ 4 i (3+k) \hat{A}_3 \, U^{(\frac{3}{2})}
%  -  4 i k \hat{B}_3 \, \hat{G}_{11}
%- 4 i (3+k) \hat{B}_3 \, U^{(\frac{3}{2})}
%\right) 
-  
\frac{(-199+k+12k^2)}{(5+k)(19+23k)} T^{(1)} \, \hat{G}_{11}
\right](w)
\nonu \\
& + &
\frac{1}{(z-w)}  \left[ 
\frac{4}{(5+k)^2}   \hat{A}_{-} \hat{A}_{+}
 U^{(\frac{3}{2})} 
+   
\frac{4}{(5+k)^2} \hat{B}_{-}
\hat{A}_{+}  V^{(\frac{3}{2})} 
-  \frac{2}{(5+k)}  i \hat{B}_{-} W_{-}^{(\frac{5}{2})}
\right.
\nonu \\
& + & \frac{8}{(5+k)^2} \hat{B}_{-} \hat{A}_3 \hat{G}_{12}
-\frac{8}{(5+k)^2} \hat{B}_{-} \hat{A}_3 T_{-}^{(\frac{3}{2})} 
- \frac{1}{(5+k)} 
\hat{B}_{-} \hat{B}_{+}  U^{(\frac{3}{2})}
\nonu \\
& + & \frac{2(7+2k)}{(5+k)^2} \hat{B}_{-} \hat{B}_3
T_{-}^{(\frac{3}{2})}   -  \frac{1}{3(5+k)} i \hat{B}_{-} \pa \hat{G}_{12} 
+ \frac{2 (9+4k)}{3(5+k)^2} i \hat{B}_{-} \pa T_{-}^{(\frac{3}{2})} 
\nonu \\
& - & 
\frac{4}{(5+k)^2} 
\hat{A}_{+} \hat{A}_{-} 
\hat{G}_{11} +\frac{1}{(5+k)}   \hat{B}_{+} \hat{B}_{-}
 U^{(\frac{3}{2})} 
-  
\frac{8}{(5+k)^2} \hat{A}_{+} \hat{A}_3 \hat{G}_{21}
\nonu \\
& + & \frac{2(4+k)}{(5+k)^2} i\hat{A}_{+}\pa \hat{G}_{21}
+   \frac{2 (6+k)}{(5+k)^2} i \hat{A}_{+}\pa 
T_{+}^{(\frac{3}{2})} 
-\frac{4}{(5+k)} i \hat{A}_3  U^{(\frac{5}{2})}
\nonu \\
& - & \frac{4}{(5+k)^2} \hat{A}_3 \hat{A}_3 \hat{G}_{11}
+ \frac{16}{(5+k)^2} \hat{A}_3 \hat{A}_3  U^{(\frac{3}{2})}
+ \frac{8}{(5+k)^2} \hat{A}_3 \hat{B}_3 \hat{G}_{11}
\nonu \\
& - &  \frac{16}{(5+k)^2} \hat{A}_3 \hat{B}_3  U^{(\frac{3}{2})}
 -\frac{2(33+181k+8k^2)}{3(5+k)^2(19+23k)} i \hat{A}_3 \pa \hat{G}_{11}
+ \frac{4}{(5+k)} i \hat{B}_3  U^{(\frac{5}{2})}
\nonu \\
& - & \frac{2(7+2k)}{(5+k)^2} \hat{B}_3 \hat{B}_{-} T_{-}^{(\frac{3}{2})}
 -   \frac{4}{(5+k)^2} \hat{B}_3 \hat{B}_3 \hat{G}_{11}
-\frac{2(-228+5k+37k^2)}{3(5+k)^2(19+23k)} i \hat{B}_3 \pa \hat{G}_{11}
\nonu \\
&+ &  \frac{2(1+5k)}{3(5+k)^2} i \hat{B}_3 \pa U^{(\frac{3}{2})}
-\frac{(-79-15k+4k^2)}{(5+k)(19+23k)} T^{(1)} \pa  \hat{G}_{11}
-  \frac{2(9+k)}{(5+k)^2} i \pa  \hat{A}_{+} \hat{G}_{21}
\nonu \\
& - & \frac{2(37+8k)}{3(5+k)^2} i \pa  \hat{A}_{+} T_{+}^{(\frac{3}{2})}
-  \frac{2(13+2k)}{(5+k)^2} i \pa \hat{A}_3   \hat{G}_{11}
-\frac{10}{(5+k)} i  \pa \hat{A}_3  U^{(\frac{3}{2})}
\nonu \\
&- & \frac{3(-3+k)}{(5+k)^2} i \pa \hat{B}_{-} \hat{G}_{12}
- \frac{2(6+k)}{(5+k)^2} i \pa \hat{B}_3  \hat{G}_{11}
-\frac{1}{(5+k)} \pa T^{(1)}  \hat{G}_{11}
\nonu \\
&-& \frac{(-9+2k)}{3(5+k)^2} \pa^2 \hat{G}_{11}  
-\pa  U^{(\frac{5}{2})}  + \frac{2(29+k)}{3(5+k)^2} i \hat{A}_3 \pa U^{(\frac{3}{2})}
+ \frac{2}{(5+k)} \hat{G}_{21}  U_{-}^{(2)}
\nonu \\
&- & \frac{2}{3(5+k)} i T_{+}^{(\frac{3}{2})} \pa  \hat{A}_{+}
-\frac{8(-21-64k+32k^2+7k^3)}{(3+7k)(19+23k)(5+k)} \hat{G}_{11} \hat{T}
\nonu \\
%+\frac{16}{(5+k)^2} 
%+ \frac{2(2+k)}{(5+k)^2} \hat{A}_3 (\hat{A}_1+i \hat{A}_2)
%\hat{G}_{21} 
%\nonu \\
%& - & \frac{4}{(5+k)^2}  \hat{A}_3 (\hat{A}_1+i \hat{A}_2)
%T_{+}^{(\frac{3}{2})}
%+ \frac{12}{(5+k)^2} i \hat{A}_3 \pa \hat{G}_{11}
%- \frac{4k}{(5+k)^2} i \hat{B}_3 \pa \hat{G}_{11}
%\nonu \\
%& - & \frac{4}{(5+k)} i \hat{B}_3 \pa U^{(\frac{3}{2})}
%+ \frac{2}{(5+k)} T^{(1)} \pa \hat{G}_{11}
%+\frac{2}{(5+k)^2} i \pa (\hat{B}_1-i \hat{B}_2) \hat{G}_{12}
%\nonu \\
%& - & 
%\frac{4}{(5+k)^2} i  \pa (\hat{B}_1-i \hat{B}_2)T_{-}^{(\frac{3}{2})}
%-\frac{2}{(5+k)} \pa T^{(1)}  \hat{G}_{11}
%-\frac{3}{(5+k)} \pa^2  U^{(\frac{3}{2})}
%\nonu \\
%&+ & \frac{2}{(5+k)} \hat{G}_{21}  U_{-}^{(2)}
%+\frac{2(3+k)}{(5+k)^2} i \hat{G}_{21} \pa  (\hat{A}_1+i \hat{A}_2)
%+\frac{2}{(5+k)} \hat{G}_{12} U_{+}^{(2)}
%\nonu \\
%& + & \left.
%\frac{4}{(5+k)} i U^{(\frac{3}{2})} \pa \hat{B}_3 
&+& \left. \frac{4}{(5+k)} \hat{G}_{11} T^{(2)} -\frac{16(-3+k)}{(19+23k)}
 U^{(\frac{3}{2})} \hat{T}
\right](w) +\cdots,
\label{g11w3} \\
\hat{G}_{12}(z) \, W^{(3)}(w)  & = & 
\frac{1}{(z-w)^3} \left[\frac{2(-2697-242k+1293k^2+110k^3)}{3(5+k)^2(19+23k)}
\hat{G}_{12} \right. \nonu \\
& - & \left.  \frac{2(-3+k)(1603+1363k+220k^2)}{3(5+k)^2(19+23k)}  
T_{-}^{(\frac{3}{2})}\right](w)
\nonu \\
& + & 
\frac{1}{(z-w)^2} \left[ 
\frac{(22+5k)}{(5+k)} W_{-}^{(\frac{5}{2})} 
+ \frac{2(7+k)}{(5+k)^2} i  \hat{A}_{+}  
\hat{G}_{22}
- \frac{8(-3+k)}{(19+23k)}  \pa T_{-}^{(\frac{3}{2})} 
\right. \nonu \\
&+ &  \frac{2(13+3k)}{(5+k)^2} i \hat{A}_{+} V^{(\frac{3}{2})}
- \frac{4(-31+52k+27k^2)}{(5+k)^2(19+23k)}  
i \hat{A}_3 \, \hat{G}_{12}
\nonu \\
&- &  \frac{2(11+3k)}{(5+k)^2} i \hat{A}_3 \, T_{-}^{(\frac{3}{2})}
+\frac{(17+k)}{(5+k)^2} i  \hat{B}_{+}  
\hat{G}_{11} - \frac{2(13+3k)}{(5+k)^2} 
i \hat{B}_{+} U^{(\frac{3}{2})}
\nonu \\
& -&  \frac{24(19+48k+5k^2)}{(5+k)^2(19+23k)} i  \hat{B}_3 \, \hat{G}_{12}
+  \frac{2(11+3k)}{(5+k)^2} i \hat{B}_3 \, T_{-}^{(\frac{3}{2})}
\nonu \\
&- & \left. \frac{12(-3+k)}{(19+23k)} T^{(1)} \hat{G}_{12}
+\frac{(349+2102k+405k^2+12k^3)}{3(5+k)^2(19+23k)} \pa \hat{G}_{12}
%\frac{2(-7+k)}{3(5+k)^2} \pa \hat{G}_{12}
%-\frac{4(-3+k)}{3(5+k)^2} \pa T_{-}^{(\frac{3}{2})} -
%\frac{2}{(5+k)} W_{-}^{(\frac{5}{2})}  
%\right. \nonu \\
%&+ &  \frac{1}{(5+k)^2}\left(  4 (2+k) i( \hat{A}_1 +i \hat{A}_2)  
%\hat{G}_{22} - 4(5+2k) i (\hat{A}_1 + i \hat{A}_2) V^{(\frac{3}{2})}
%\right. \nonu \\
%&+ & 
%2 (7+k) i( \hat{B}_1 +i \hat{B}_2)  
%\hat{G}_{11} + 4(8+k) i (\hat{B}_1 + i \hat{B}_2) U^{(\frac{3}{2})}
%\nonu \\
%&- & \left. \left. 8 i \hat{A}_3 \, \hat{G}_{12}
%+ 4 i (3+k) \hat{A}_3 \, T_{-}^{(\frac{3}{2})}
%  +  16 i  \hat{B}_3 \, \hat{G}_{12}
%- 4 i (3+k) \hat{B}_3 \, T_{-}^{(\frac{3}{2})}
%\right)
\right](w)
\nonu \\
& + &
\frac{1}{(z-w)}  \left[ \frac{4}{(5+k)^2} 
\hat{A}_{-} \hat{A}_{+}  T_{-}^{(\frac{3}{2})}
-  
\frac{8}{(5+k)^2} \hat{B}_{+}\hat{A}_{+}
\hat{G}_{21}
-\frac{2(15+7k)}{3(5+k)^2} 
\hat{A}_{-} \hat{A}_{+} \hat{G}_{12}
\right.
\nonu \\
& + & \frac{8}{(5+k)^2} \hat{B}_{+} \hat{A}_3 U^{(\frac{3}{2})}
+\frac{16}{(5+k)^2}
\hat{B}_{+} \hat{B}_{-} \hat{G}_{12} 
-  \frac{8(9+2k)}{3(5+k)^2} \hat{B}_{+} \hat{B}_{-}
   T_{-}^{(\frac{3}{2})}
\nonu \\
& - & \frac{(3+k)}{(5+k)^2} \hat{B}_{+} \hat{B}_3 \hat{G}_{11}
 +  \frac{4(15+4k)}{3(5+k)^2}
 \hat{B}_{-} \hat{B}_{+}  T_{-}^{(\frac{3}{2})}
-\frac{8}{(5+k)^2} \hat{B}_{+} \hat{B}_3 U^{(\frac{3}{2})}
\nonu \\
&+ & 
\frac{2(15+7k)}{3(5+k)^2} \hat{A}_{+} \hat{A}_{-}
\hat{G}_{12}  + \frac{1}{(5+k)} i \hat{B}_{+} \hat{B}_3 \pa
\hat{G}_{11} 
+  \frac{6}{(5+k)} \hat{A}_{+}
\hat{A}_3  V^{(\frac{3}{2})} \nonu \\
& + & \frac{2}{(5+k)} i \hat{B}_{+} \pa 
 U^{(\frac{3}{2})}
 -  \frac{8}{(5+k)^2} \hat{A}_{+}
\hat{B}_3  V^{(\frac{3}{2})}
+\frac{2(1+k)}{(5+k)^2} i \hat{A}_{+} \pa \hat{G}_{22}
\nonu \\
&- & \frac{2}{(5+k)} i  \hat{A}_{+} \pa   V^{(\frac{3}{2})}
-\frac{12}{(5+k)^2} \hat{B}_{-} \hat{B}_{+} \hat{G}_{12}
+  \frac{(3+k)}{(5+k)^2} \hat{B}_3  \hat{B}_{+} \hat{G}_{11} 
\nonu \\
& - & \frac{8(-19+2k+5k^2)}{(5+k)^2(19+23k)} i \hat{B}_3 \pa \hat{G}_{12} 
-\frac{2}{(5+k)} i \hat{B}_3 \pa  T_{-}^{(\frac{3}{2})}
+ \frac{4(27+7k)}{3(5+k)^2} i \hat{G}_{12} \pa \hat{A}_3 
\nonu \\
& - & \frac{6}{(5+k)} i T_{-}^{(\frac{3}{2})} \pa \hat{A}_3
-\frac{2k}{(5+k)^2} \hat{A}_{+} \hat{A}_3 \hat{G}_{22}
+  \frac{2}{(5+k)} \hat{G}_{11} V_{-}^{(2)} 
\nonu \\
& - & \frac{2}{(5+k)} \hat{G}_{22}  U_{-}^{(2)} 
+\frac{2k}{(5+k)^2} \hat{A}_3 \hat{A}_{+} \hat{G}_{22}
+  \frac{(6+k)}{(5+k)} \pa W_{-}^{(\frac{5}{2})} 
\nonu \\
& - & \frac{2(11+3k)}{(5+k)^2} \hat{A}_3 \hat{A}_{+}  V^{(\frac{3}{2})}
-\frac{16(-3+k)}{(19+23k)}  T_{-}^{(\frac{3}{2})} \hat{T}
-   
\frac{4(53+94k+9k^2)}{(5+k)^2(19+23k)}  \hat{A}_{+} \pa \hat{G}_{12}
\nonu \\
& + & \frac{2}{(5+k)} i  \hat{A}_{+} \pa  T_{-}^{(\frac{3}{2})}
- 
\frac{4(-3+k)}{(19+232k)} T^{(1)} \pa \hat{G}_{12} 
+ \frac{4(-27+23k+2k^2)}{(5+k)(19+23k)} \hat{G}_{12} \hat{T}
\nonu \\
& -& \left. \frac{2(27+7k)}{3(5+k)^2} i   T_{-}^{(\frac{3}{2})} \pa \hat{B}_3
-\frac{2(11+3k)}{(5+k)^2}  U^{(\frac{3}{2})}  \pa \hat{B}_{+}
\right](w) +\cdots,
\label{g12w3} \\
\hat{G}_{21}(z) \, W^{(3)}(w)  & = & 
\frac{1}{(z-w)^3} \left[-\frac{2(-2697-242k+1293k^2+110k^3)}{3(5+k)^2(19+23k)}
\hat{G}_{21} \right. \nonu \\
& - & \left.  \frac{2(-3+k)(1603+1363k+220k^2)}{3(5+k)^2(19+23k)}  
T_{+}^{(\frac{3}{2})}\right](w)
\nonu \\
& + & 
\frac{1}{(z-w)^2} \left[ 
-\frac{(24+5k)}{(5+k)} W_{+}^{(\frac{5}{2})}  
-\frac{2(-3+k)}{(5+k)^2} i \hat{A}_{-} 
\hat{G}_{11}
\right. \nonu \\
&- &  \frac{2(23+7k)}{(5+k)^2} 
i \hat{A}_{-} U^{(\frac{3}{2})}
-\frac{12(-23+2k+9k^2)}{(5+k)^2(19+23k)} i \hat{A}_3 \, \hat{G}_{21}
\nonu \\
&+ & \frac{2(17+5k)}{(5+k)^2} i \hat{A}_3 \, T_{+}^{(\frac{3}{2})}
-\frac{(-3+k)}{(5+k)^2} i\hat{B}_{-}  
\hat{G}_{22} 
+  \frac{2(29+5k)}{(5+k)^2} i \hat{B}_{-} V^{(\frac{3}{2})}
\nonu \\
& - & \frac{40(19+38k+3k^2)}{(5+k)^2(19+23k)} i \hat{B}_3 \, \hat{G}_{21}
- \frac{2(17+5k)}{(5+k)^2} 
i  \hat{B}_3 \, T_{+}^{(\frac{3}{2})}
\nonu \\
%\frac{2(-7+k)}{3(5+k)^2} \pa \hat{G}_{21}
%+\frac{4(-3+k)}{3(5+k)^2} \pa T_{+}^{(\frac{3}{2})} -
%\frac{2}{(5+k)} W_{+}^{(\frac{5}{2})}  
%\right. \nonu \\
%&+ &  \frac{1}{(5+k)^2}\left(  -4 (2+k) i( \hat{A}_1 -i \hat{A}_2)  
%\hat{G}_{11} - 4(5+2k) i (\hat{A}_1 - i \hat{A}_2) U^{(\frac{3}{2})}
%\right. \nonu \\
%&- & 
%2 (7+k) i( \hat{B}_1 -i \hat{B}_2)  
%\hat{G}_{22} + 4(8+k) i (\hat{B}_1 - i \hat{B}_2) V^{(\frac{3}{2})}
%\nonu \\
%&+ & \left. \left. 8 i \hat{A}_3 \, \hat{G}_{21}
%+ 4 i (3+k) \hat{A}_3 \, T_{+}^{(\frac{3}{2})}
%  -  16 i  \hat{B}_3 \, \hat{G}_{21}
%- 4 i (3+k) \hat{B}_3 \, T_{+}^{(\frac{3}{2})}
%\right) 
&-& 
\frac{(615+2386k+359k^2+12k^3)}{3(5+k)^2(19+23k)} \pa \hat{G}_{21}
-\frac{12(-3+k)}{(19+23k)} T^{(1)} \hat{G}_{21}
\nonu \\
& - & \left. 
\frac{4(-3+k)(131+37k+6k^2)}{3(5+k)^2(19+23k)} \pa T_{+}^{(\frac{3}{2})}
\right](w)
\nonu \\
& + &
\frac{1}{(z-w)}  \left[  
-\frac{4}{(5+k)^2} \hat{A}_{-} \hat{A}_{+}
\hat{G}_{21} -\frac{2}{(5+k)} i  \hat{A}_{-}
U^{(\frac{5}{2})}
 \right. 
\nonu \\
& + &  \frac{8}{(5+k)^2}  \hat{A}_{-}
 \hat{B}_{-} \hat{G}_{12}
-  \frac{8}{(5+k)^2} \hat{A}_{-} \hat{B}_3   
U^{(\frac{3}{2})}
+  \frac{2(-3+k)}{3(5+k)^2} i \hat{A}_{-} \pa \hat{G}_{11}
\nonu \\
&- & \frac{2(-7+k)}{3(5+k)^2} \hat{A}_{-}  \pa  U^{(\frac{3}{2})}
+\frac{4}{(5+k)^2}  \hat{B}_{+}
 \hat{B}_{-} \hat{G}_{21}
+  \frac{2}{(5+k)} i \hat{B}_{+}  V^{(\frac{5}{2})} 
\nonu \\
& + &  \frac{8}{(5+k)^2} \hat{B}_{+} \hat{A}_3  V^{(\frac{3}{2})}
-  \frac{8}{(5+k)^2}  \hat{B}_{-} \hat{B}_{+}
\hat{G}_{21} -\frac{6}{(5+k)}  \hat{B}_{-} \hat{B}_3  
V^{(\frac{3}{2})}
\nonu \\
& +& \frac{(-3+k)}{3(5+k)^2} i  \hat{B}_{-} \pa 
\hat{G}_{22} -\frac{2(1+k)}{3(5+k)^2} i \hat{B}_{-} \pa 
      V^{(\frac{3}{2})} 
+  \frac{8}{(5+k)^2} \hat{A}_{-} \hat{A}_3 
 U^{(\frac{3}{2})}
\nonu \\
&- & \frac{4(-23+2k+9k^2)}{(5+k)^2(19+23k)} i \hat{A}_3 \pa \hat{G}_{21}
+\frac{2}{(5+k)} i \hat{A}_3 \pa   T_{+}^{(\frac{3}{2})}
+\frac{2(11+3k)}{(5+k)^2} \hat{B}_3  \hat{B}_{-}  V^{(\frac{3}{2})}
\nonu \\
&- & \frac{8(19+48k+5k^2)}{(5+k)^2(19+23k)} i \hat{B}_3 \pa \hat{G}_{21}
-\frac{2}{(5+k)} i \hat{B}_3 \pa  T_{+}^{(\frac{3}{2})}
-\frac{4(-3+k)}{(19+23k)} T^{(1)} \pa \hat{G}_{21}
\nonu \\
& - & \frac{(4+k)}{(5+k)} \pa  W_{+}^{(\frac{5}{2})}
-\frac{2(1+k)}{(5+k)^2} i \pa  \hat{A}_{-} \hat{G}_{11} 
+ \frac{8}{(5+k)^2} i \pa \hat{A}_3 \hat{G}_{21}
\nonu \\
&- & \frac{2}{(5+k)} i \pa \hat{B}_3  T_{+}^{(\frac{3}{2})}
-\frac{(29+5k)}{6(5+k)^2} \pa^2 \hat{G}_{21} -\frac{4(9+2k)}{3(5+k)^2} \pa^2 
 T_{+}^{(\frac{3}{2})}
\nonu \\
&- & \frac{4(-27+23k+2k^2)}{(5+k)(19+23k)} \hat{G}_{21} \hat{T} 
-\frac{16(-3+k)}{(19+23k)}  T_{+}^{(\frac{3}{2})} \hat{T}
-\frac{2(11+3k)}{(5+k)^2} i  U^{(\frac{3}{2})} \pa  \hat{A}_{-} 
\nonu \\
&- & \left. \frac{1}{(5+k)} i \hat{G}_{22} \pa  \hat{B}_{-} 
\right](w) +\cdots,
\label{g21w3} \\
\hat{G}_{22}(z) \, W^{(3)}(w)  & = & 
\frac{1}{(z-w)^3} \left[\frac{2(-3+k)(571+357k+110k^2)}{3(5+k)^2(19+23k)}
\hat{G}_{22} \right. \nonu \\
& - & \left.  \frac{2(-3+k)(1489+1225k+220k^2)}{3(5+k)^2(19+23k)}  
V^{(\frac{3}{2})}\right](w)
\nonu \\
& + & 
\frac{1}{(z-w)^2} \left[ 
 \frac{(19+5k)}{(5+k)} V^{(\frac{5}{2})} 
- \frac{2(14+k)}{(5+k)^2} i \hat{A}_{-}  
\hat{G}_{12}   +  \frac{2(3+2k)}{(5+k)^2} 
i \hat{A}_{-} T_{-}^{(\frac{3}{2})}
\right. \nonu \\
& - & \frac{2(109+349k+100k^2)}{(5+k)^2(19+23k)} 
i\hat{A}_3 \, \hat{G}_{22}
+  \frac{2(-11+k)}{(5+k)^2} i \hat{A}_3 \, V^{(\frac{3}{2})} 
\nonu \\ 
& + & \frac{(29+k)}{(5+k)^2} i \hat{B}_{+}  
\hat{G}_{21} +   \frac{2(14+3k)}{(5+k)^2} 
i \hat{B}_{+} T_{+}^{(\frac{3}{2})}
-\frac{2(76+373k+37k^2)}{(5+k)^2(19+23k)} i \hat{B}_3 \, \hat{G}_{22}
\nonu \\
& + & \frac{2(33+5k)}{(5+k)^2} i  \hat{B}_3 \, V^{(\frac{3}{2})}
-\frac{(-161+47k+12k^2)}{(5+k)(19+23k)} T^{(1)} \hat{G}_{22}
\nonu \\
%-\frac{2(3+k)}{(5+k)^2} \pa \hat{G}_{22}
%+\frac{4(3+k)}{(5+k)^2} \pa V^{(\frac{3}{2})} 
%- \frac{2}{(5+k)} T^{(1)} \, \hat{G}_{22}
%\right. \nonu \\
%&+ &  \frac{1}{(5+k)^2}\left(  -4 (2+k) i( \hat{A}_1 -i \hat{A}_2)  
%\hat{G}_{12} + 8(2+k) i (\hat{A}_1 - i \hat{A}_2) T_{-}^{(\frac{3}{2})}
%\right. \nonu \\
%&- & 
%2 (7+k) i( \hat{B}_1 +i \hat{B}_2)  
%\hat{G}_{21} - 4(7+k) i (\hat{B}_1 + i \hat{B}_2) T_{+}^{(\frac{3}{2})}
%\nonu \\
%&- & \left. \left. 12 i \hat{A}_3 \, \hat{G}_{22}
%+ 4 i (3+k) \hat{A}_3 \, V^{(\frac{3}{2})}
%  +  4 i k \hat{B}_3 \, \hat{G}_{22}
%- 4 i (3+k) \hat{B}_3 \, V^{(\frac{3}{2})}
%\right) 
&+& 
\frac{(-1389-512k+113k^2+12k^3)}{3(5+k)^2(19+23k)} \pa \hat{G}_{22}
\nonu \\
& - & \left.
\frac{4(-89+445k+88k^2+6k^3)}{3(5+k)^2(19+23k)} \pa V^{(\frac{3}{2})} 
\right](w)
\nonu \\
& + &
\frac{1}{(z-w)}  \left[ 
\frac{8}{(5+k)^2} \hat{A}_{-} \hat{A}_{+}
\hat{G}_{22}
-\frac{2}{(5+k)} i \hat{A}_{-}  W_{-}^{(\frac{5}{2})}
+ \frac{2(10+k)}{(5+k)^2}  \hat{A}_{-}  \hat{A}_3
\hat{G}_{12} \right.
\nonu \\
& - & \frac{4}{(5+k)^2} \hat{A}_{-} 
\hat{B}_{+}  U^{(\frac{3}{2})}
+  \frac{8}{(5+k)^2}   \hat{A}_{-} \hat{B}_3
 T_{-}^{(\frac{3}{2})}
+\frac{2(2+k)}{3(5+k)^2} i \hat{A}_{-} 
 \pa \hat{G}_{12} \nonu \\
&- & \frac{2(15+2k)}{3(5+k)^2}
\hat{A}_{-} \pa  T_{-}^{(\frac{3}{2})}
 + \frac{(10+3k)}{(5+k)^2} \hat{B}_{+}
\hat{B}_{-} \hat{G}_{22}
 -  \frac{5}{(5+k)} 
\hat{B}_{+} \hat{B}_{-}  V^{(\frac{3}{2})}
\nonu \\
& - & \frac{8}{(5+k)^2}   \hat{B}_{+}  \hat{B}_3 
\hat{G}_{21}
+  \frac{(1+k)}{(5+k)^2} i \hat{B}_{+} \pa \hat{G}_{21}
-\frac{2(6+k)}{(5+k)^2} i \hat{B}_{+} \pa 
 T_{+}^{(\frac{3}{2})}
\nonu \\
& - & \frac{3(2+k)}{(5+k)^2}   
\hat{B}_{-} \hat{B}_{+} 
\hat{G}_{22} +\frac{(21+5k)}{(5+k)^2}   \hat{B}_{-}
\hat{B}_{+}  V^{(\frac{3}{2})}
-  \frac{4}{(5+k)} i \hat{A}_3  V^{(\frac{5}{2})}
\nonu \\
&-& \frac{2(10+k)}{(5+k)^2} \hat{A}_3  \hat{A}_{-}
\hat{G}_{12} 
- \frac{16(-3+k)}{(19+23k)}  V^{(\frac{3}{2})} \hat{T}
\nonu \\
& + & \frac{4}{(5+k)^2} \hat{A}_3 \hat{A}_3 \hat{G}_{22} 
-\frac{8}{(5+k)^2}  \hat{A}_3 \hat{B}_3 \hat{G}_{22}
+\frac{16}{(5+k)^2}   \hat{A}_3 \hat{B}_3  V^{(\frac{3}{2})}
\nonu \\
&- &  \frac{2(375+595k+8k^2)}{3(5+k)^2(19+23k)} 
i \hat{A}_3 \pa \hat{G}_{22} -
\frac{2(13+k)}{3(5+k)^2} i \hat{A}_3 \pa  V^{(\frac{3}{2})} 
\nonu \\
&+ & \frac{4}{(5+k)} i \hat{B}_3   V^{(\frac{5}{2})}
+\frac{4}{(5+k)^2} \hat{B}_3 \hat{B}_3 \hat{G}_{22}
-\frac{16}{(5+k)^2} \hat{B}_3 \hat{B}_3   V^{(\frac{3}{2})}
\nonu \\
&+ &  \frac{2(228+109k+101k^2) }{3(5+k)^2(19+23k)} 
i \hat{B}_3 \pa \hat{G}_{22}  - 
\frac{2(17+5k)}{3(5+k)^2} i \hat{B}_3 \pa  V^{(\frac{3}{2})} 
\nonu \\
&- &  \frac{(-41+31k+4k^2)}{(5+k)(19+23k)} T^{(1)} \pa \hat{G}_{22}
+ \pa  V^{(\frac{5}{2})} 
+\frac{2(7+2k)}{(5+k)^2} i
\pa \hat{A}_{-}    T_{-}^{(\frac{3}{2})}
\nonu \\
&- & \frac{2(7+2k)}{(5+k)^2}  i \pa \hat{A}_3  \hat{G}_{22}  - 
\frac{2}{(5+k)} i \pa \hat{A}_3   V^{(\frac{3}{2})} 
-\frac{(-29+7k)}{3(5+k)^2} i \pa  \hat{B}_{+}
\hat{G}_{21} 
\nonu \\
& + & \frac{2(37+8k)}{3(5+k)^2}  i \pa  \hat{B}_{+}
 T_{+}^{(\frac{3}{2})} +
\frac{1}{(5+k)}  \pa T^{(1)}  \hat{G}_{22}
-\frac{2}{(5+k)} \hat{G}_{21}   V_{-}^{(2)} 
\nonu \\
&+ & \frac{2(23+2k)}{3(5+k)^2} \hat{G}_{21} 
\pa   \hat{B}_{+}
+\frac{2}{3(5+k)} i  T_{+}^{(\frac{3}{2})} \pa  \hat{B}_{+}
\nonu \\
&+& \left. \frac{8}{(3+7k)(19+23k)(5+k)} \hat{G}_{22} \hat{T}
-\frac{4}{(5+k)} \hat{G}_{22}  T^{(2)} 
\right](w) +\cdots.
\label{g22w3}
\eea
The $U(1)$ charge conservation in these OPEs can be checked from Tables 
$2, 4, 6$ and $7$.

%%%%%%%%%%%%%%%%%%
\subsection{The OPEs between four spin $\frac{3}{2}$ currents 
and the higher spin 
currents in different basis
}
%%%%%%%%%%%%%%%%%%%%

Let us present the OPEs in different basis as we did in Appendix $(B.4)$.
The OPEs between the spin-$\frac{3}{2}$ currents 
$\hat{G}_a(z)$ and the  spin-$1$ current $\hat{T}^{(1)}(w)=T^{(1)}(w)$
given by (\ref{G11t1}) and (\ref{G12t1})
can be rewritten in terms of hatted higher spins in (\ref{16new})
as follows:
\bea
\left(
\begin{array}{c}
\hat{G}_{11} \\
\hat{G}_{22}
\end{array} \right)(z) \, \hat{T}^{(1)}(w) & = &
\frac{1}{(z-w)} 2 \left(
\begin{array}{c}
  \hat{U}^{(\frac{3}{2})}
\\
  \hat{V}^{(\frac{3}{2})}
\end{array}
\right)(w) +
\cdots, 
%\label{newG11t1} 
\nonu \\
\left(
\begin{array}{c}
\hat{G}_{12} \\
\hat{G}_{21} 
\end{array} \right) (z) \, \hat{T}^{(1)}(w) & = &
\frac{1}{(z-w)} 2 \left(
\begin{array}{c} 
\hat{T}_{-}^{(\frac{3}{2})} \\
\hat{T}_{+}^{(\frac{3}{2})}
\end{array}
\right)(w) +
\cdots. 
\label{newnewG12t1} 
\eea
There are no $\hat{G}_a(w)$ dependences in the right hand side of
(\ref{newnewG12t1}).
 The spin-$1$ current $\hat{T}^{(1)}(z)$
acting on the four spin-$\frac{3}{2}$ currents of large ${\cal N}=4$
nonlinear algebra leads to 
the four currents with same spins but they are located at higher 
spin multiplet. 

The OPEs between 
 the spin-$\frac{3}{2}$ currents 
$\hat{G}_a(z)$ and the  four spin-$\frac{3}{2}$ currents
in the right hand side of 
(\ref{newnewG12t1}) or (\ref{16new})
can be described by
\bea
\left(
\begin{array}{c}
\hat{G}_{11} \\
\hat{G}_{22}
\end{array} \right)(z) \, 
\hat{T}_{\pm}^{(\frac{3}{2})} (w) & = & 
-\frac{1}{(z-w)}  \left(
\begin{array}{c} 
 \hat{U}_{+}^{(2)} +\frac{i}{3} \hat{T}^{(1)} \hat{B}_{-} \\
V_{-}^{(2)} -\frac{i}{3} \hat{T}^{(1)} \hat{B}_{+} 
\end{array}
\right)(w) +\cdots,
 \nonu \\
\left(
\begin{array}{c}
\hat{G}_{12} \\
\hat{G}_{21} 
\end{array} \right)(z) \, \hat{T}_{\pm}^{(\frac{3}{2})}(w) & = & 
\frac{1}{(z-w)^2} \, \hat{T}^{(1)}(w) +
\frac{1}{(z-w)} 
 \left(
\begin{array}{c} 
\{  \hat{G}_{12} \, \hat{T}_{+}^{(\frac{3}{2})}\}_{-1} \\
\{  \hat{G}_{21} \, \hat{T}_{-}^{(\frac{3}{2})}\}_{-1}
\end{array}
\right)(w)   + \cdots,
%\label{newG12t+3half} 
\nonu \\
\left(
\begin{array}{c}
\hat{G}_{22} \\
\hat{G}_{11} \\
\end{array} \right)(z) \, \hat{T}_{\pm}^{(\frac{3}{2})}(w) & = & 
- \frac{1}{(z-w)} 
\left(
\begin{array}{c} 
\hat{V}_{+}^{(2)} +\frac{i}{k} \hat{T}^{(1)} \hat{A}_{-} \\
\hat{U}_{-}^{(2)} -\frac{i}{k} \hat{T}^{(1)} \hat{A}_{+} \\
\end{array} \right)(w) +\cdots,
%\label{newG22t+3half}
\nonu \\
\left(
\begin{array}{c}
\hat{G}_{12} \\
\hat{G}_{21} \\
\end{array} \right)(z) \, 
\left(
\begin{array}{c}
\hat{U}^{(\frac{3}{2})} \\
\hat{V}^{(\frac{3}{2})} \\
\end{array} \right) (w) & = & 
\frac{1}{(z-w)}  
\left(
\begin{array}{c} 
\hat{U}_{-}^{(2)} -\frac{i}{k} \hat{T}^{(1)} \hat{A}_{+} \\
\hat{V}_{+}^{(2)} +\frac{i}{k} \hat{T}^{(1)} \hat{A}_{-} \\
\end{array} \right)(w) +\cdots,
%\label{newG12u3half} 
\nonu \\
\left(
\begin{array}{c}
\hat{G}_{21} \\
\hat{G}_{12} \\
\end{array} \right) (z) \, 
\left(
\begin{array}{c} 
\hat{U}^{(\frac{3}{2})} \\
\hat{V}^{(\frac{3}{2})} \\
\end{array} \right)(w) & = &
\frac{1}{(z-w)}   \left( 
\begin{array}{c}
 \hat{U}_{+}^{(2)} +\frac{i}{3} \hat{T}^{(1)} \hat{B}_{-} \\
  \hat{V}_{-}^{(2)} -\frac{i}{3} \hat{T}^{(1)} \hat{B}_{+} \\
\end{array} \right) (w) +\cdots,
%\label{G21u3half} 
\nonu
\\
\left(
\begin{array}{c}
\hat{G}_{22} \\
\hat{G}_{11} \\
\end{array} \right)(z) \, 
\left(
\begin{array}{c}
\hat{U}^{(\frac{3}{2})} \\
\hat{V}^{(\frac{3}{2})} \\
\end{array} \right) (w) & = &
\frac{1}{(z-w)^2} \, \hat{T}^{(1)}(w) +
\frac{1}{(z-w)}  
 \left(
\begin{array}{c} 
\{  \hat{G}_{22} \, \hat{U}^{(\frac{3}{2})}\}_{-1} \\
\{  \hat{G}_{11} \, \hat{V}^{(\frac{3}{2})}\}_{-1}
\end{array}
\right)(w) 
 \nonu \\
& + &   \cdots.
\label{newnewG22u3half} 
\eea
All the third-order poles appeared in previous basis 
are disappeared in this
new basis (\ref{newnewG22u3half}).
From Appendix (\ref{newnewG22u3half}), one sees 
the six higher spin-$2$ currents appear in the right hand side.
Note that the fields $\hat{T}^{(1)} \hat{A}_{\pm}(w)$
and $\hat{T}^{(1)} \hat{B}_{\pm}(w)$
are primary fields under the stress energy tensor $\hat{T}(z)$ 
because $\hat{T}^{(1)}(z)$
commutes with both $\hat{A}_{\pm}(w)$ and $\hat{B}_{\pm}(w)$ and these five 
spin-$1$ currents are primary.
The spin-$1$ current $\hat{T}^{(1)}(w)$ appears in the second-order poles
of second and sixth OPEs in Appendix (\ref{newnewG22u3half}).

The first-order poles in Appendix (\ref{newnewG22u3half}) are given by
\bea
\left(
\begin{array}{c}
\{  \hat{G}_{12} \, \hat{T}_{+}^{(\frac{3}{2})}\}_{-1} \\
\{  \hat{G}_{21} \, \hat{T}_{-}^{(\frac{3}{2})}\}_{-1} \\
\end{array} \right)(w) & = &  
\mp  
\frac{1}{2}
(\hat{T}^{(2)} 
+ \hat{W}^{(2)})(w) \mp   \frac{1}{2}
(\hat{T}^{(2)} - \hat{W}^{(2)})(w)
\nonu \\
& \pm & \frac{i}{k} \hat{T}^{(1)} \hat{A}_3(w) 
\pm \frac{i}{3} \hat{T}^{(1)} \hat{B}_3(w)
+\frac{1}{2} \pa \hat{T}^{(1)}(w)
\pm   \frac{(k+3)}{ (7 k+3)} \tilde{ \hat{T}}(w), 
\nonu \\
\left(
\begin{array}{c}
\{  \hat{G}_{22} \, \hat{U}^{(\frac{3}{2})}\}_{-1} \\
\{  \hat{G}_{11} \, \hat{V}^{(\frac{3}{2})}\}_{-1} \\
\end{array} \right)(w) & = &  
\mp\frac{1}{2}
(\hat{T}^{(2)} 
+ \hat{W}^{(2)})(w) \pm   \frac{1}{2}
(\hat{T}^{(2)} - \hat{W}^{(2)})(w)
\nonu \\
& \mp & \frac{i}{k} \hat{T}^{(1)} \hat{A}_3(w) 
\pm \frac{i}{3} \hat{T}^{(1)} \hat{B}_3(w)
+\frac{1}{2} \pa \hat{T}^{(1)}(w)
\pm \tilde{\hat{T}}(w). 
\label{pole2exp}
\eea
In (\ref{pole2exp}), each five term 
is primary field 
 except the descendant field 
$ \frac{1}{2} \pa \hat{T}^{(1)}(w)$ which should be present due to the 
second-order pole term, $\hat{T}^{(1)}(w)$,
and moreover $\tilde{\hat{T}}(w)$ was defined in (\ref{tildehatt})
previously.

Now one calculates the OPEs between
 the spin-$\frac{3}{2}$ currents 
$\hat{G}_a(z)$ and the  above six spin-$2$ currents
in the right hand side of  Appendix (\ref{newnewG22u3half}) or (\ref{16new}) 
\bea
\hat{G}_{11}(z) \, 
\left(
\begin{array}{c} 
\hat{U}_{-}^{(2)} \\
\hat{T}^{(2)} -\hat{W}^{(2)} \\
\hat{V}_{+}^{(2)} \\
\end{array} \right) (w) & = &
 \frac{1}{(z-w)^2} \left[ 
\frac{2(-1+k)(5+2k)}{k(5+k)} \right] 
\left(
\begin{array}{c}
0 \\
\hat{U}^{(\frac{3}{2})}\\
-\hat{T}_{+}^{(\frac{3}{2})} \\
\end{array} \right)(w)
\nonu \\
& + & \frac{1}{(z-w)} 
\left(
\begin{array}{c}
\frac{2 i (2 k+5)}{k (k+5)} \hat{A}_{+} \hat{U}^{(\frac{3}{2})} \\
\{ \hat{G}_{11} \, ( 
\hat{T}^{(2)} -\hat{W}^{(2)}) \}_{-1}
 \\
\{ \hat{G}_{11} \, 
\hat{V}_{+}^{(2)} \}_{-1}
 \\
\end{array} \right)(w)
 +  \cdots,
\nonu \\
\hat{G}_{12}(z) \, 
\left(
\begin{array}{c} 
\hat{U}_{-}^{(2)} \\
\hat{T}^{(2)} -\hat{W}^{(2)} \\
\hat{V}_{+}^{(2)} \\
\end{array} \right) (w) & = &
 \frac{1}{(z-w)^2} \left[ 
\frac{2(-1+k)(5+2k)}{k(5+k)} \right] 
\left(
\begin{array}{c}
0 \\
\hat{T}_{-}^{(\frac{3}{2})}\\
\hat{V}^{(\frac{3}{2})} \\
\end{array} \right)(w)
\nonu \\
& + & \frac{1}{(z-w)} 
\left(
\begin{array}{c}
\frac{2 i (2 k+5)}{k (k+5)} \hat{A}_{+} \hat{T}_{-}^{(\frac{3}{2})} \\
\{ \hat{G}_{12} \, ( 
\hat{T}^{(2)} -\hat{W}^{(2)}) \}_{-1}
\\
\{ \hat{G}_{12} \, 
\hat{V}_{+}^{(2)} \}_{-1}
 \\
\end{array} \right)(w)
 +  \cdots,
\nonu \\
\hat{G}_{21}(z) \, 
\left(
\begin{array}{c} 
\hat{U}_{-}^{(2)} \\
\hat{T}^{(2)} -\hat{W}^{(2)} \\
\hat{V}_{+}^{(2)} \\
\end{array} \right) (w) & = &
 \frac{1}{(z-w)^2} \left[ 
\frac{2(-1+k)(5+2k)}{k(5+k)} \right] 
\left(
\begin{array}{c}
\hat{U}^{(\frac{3}{2})}\\
-\hat{T}_{+}^{(\frac{3}{2})} \\
 0 \\
\end{array} \right)(w)
\nonu \\
& + & \frac{1}{(z-w)} 
\left(
\begin{array}{c} 
\{ \hat{G}_{21} \, 
\hat{U}_{-}^{(2)} \}_{-1}
\\
\{ \hat{G}_{21} \, ( 
\hat{T}^{(2)} -\hat{W}^{(2)}) \}_{-1}
  \\
   -\frac{2 i (5+2k)}{k(5+k)} \hat{A}_{-} \hat{T}_{+}^{(\frac{3}{2})}  \\
\end{array} \right)(w)
 +  \cdots,
\nonu \\
\hat{G}_{22}(z) \, 
\left(
\begin{array}{c} 
\hat{U}_{-}^{(2)} \\
\hat{T}^{(2)} -\hat{W}^{(2)} \\
\hat{V}_{+}^{(2)} \\
\end{array} \right) (w) & = &
 \frac{1}{(z-w)^2} \left[ 
\frac{2(-1+k)(5+2k)}{k(5+k)} \right] 
\left(
\begin{array}{c}
-\hat{T}_{-}^{(\frac{3}{2})}\\
-\hat{V}^{(\frac{3}{2})} \\
 0 \\
\end{array} \right)(w)
\nonu \\
& + & \frac{1}{(z-w)} 
\left(
\begin{array}{c}
\{ \hat{G}_{22} \, 
\hat{U}_{-}^{(2)} \}_{-1}
\\
\{ \hat{G}_{22} \, ( 
\hat{T}^{(2)} -\hat{W}^{(2)}) \}_{-1}
\\ 
 -\frac{2 i (5+2k)}{k(5+k)} \hat{A}_{-} \hat{V}^{(\frac{3}{2})} 
\\
\end{array} \right)(w)
 +  \cdots,
\nonu \\
\hat{G}_{11}(z) \, 
\left(
\begin{array}{c} 
\hat{V}_{-}^{(2)} \\
\hat{T}^{(2)} +\hat{W}^{(2)} \\
\hat{U}_{+}^{(2)} \\
\end{array} \right) (w) & = &
 \frac{1}{(z-w)^2} \left[ 
\frac{4(8+k)}{3(5+k)} \right] 
\left(
\begin{array}{c}
-\hat{T}_{-}^{(\frac{3}{2})}\\
-\hat{U}^{(\frac{3}{2})} \\
 0 \\
\end{array} \right)(w)
\nonu \\
& + & \frac{1}{(z-w)} 
\left(
\begin{array}{c}
\{ \hat{G}_{11} \, 
\hat{V}_{-}^{(2)} \}_{-1} 
 \\
\{ \hat{G}_{11} \, ( 
\hat{T}^{(2)} + \hat{W}^{(2)}) \}_{-1}
 \\
-\frac{2i (8+k)}{3(5+k)}  \hat{B}_{-} \hat{U}^{(\frac{3}{2})} \\
\end{array} \right)(w)
 +  \cdots,
\nonu \\
\hat{G}_{12}(z) \, 
\left(
\begin{array}{c} 
\hat{V}_{-}^{(2)} \\
\hat{T}^{(2)} +\hat{W}^{(2)} \\
\hat{U}_{+}^{(2)} \\
\end{array} \right) (w) & = &
 \frac{1}{(z-w)^2} \left[ 
\frac{4(8+k)}{3(5+k)} \right] 
\left(
\begin{array}{c}
0 \\
\hat{T}_{-}^{(\frac{3}{2})}\\
\hat{U}^{(\frac{3}{2})} \\
\end{array} \right)(w)
\nonu \\
& + & \frac{1}{(z-w)} 
\left(
\begin{array}{c}
\frac{2 i (8+k)}{3(5+k)} \hat{B}_{+} T_{-}^{(\frac{3}{2})} \\
\{ \hat{G}_{12} \, ( 
\hat{T}^{(2)} + \hat{W}^{(2)}) \}_{-1}
 \\
\{ \hat{G}_{12} \, \hat{U}_{+}^{(2)} \}_{-1}
 \\
\end{array} \right)(w)
 +  \cdots,
\nonu \\
\hat{G}_{21}(z) \, 
\left(
\begin{array}{c} 
\hat{V}_{-}^{(2)} \\
\hat{T}^{(2)} +\hat{W}^{(2)} \\
\hat{U}_{+}^{(2)} \\
\end{array} \right) (w) & = &
 \frac{1}{(z-w)^2} \left[ 
\frac{4(8+k)}{3(5+k)} \right] 
\left(
\begin{array}{c}
\hat{V}^{(\frac{3}{2})}\\
-\hat{T}_{+}^{(\frac{3}{2})} \\
0 \\
\end{array} \right)(w)
\nonu \\
& + & \frac{1}{(z-w)} 
\left(
\begin{array}{c}
\{ \hat{G}_{21} \, 
\hat{V}_{-}^{(2)} \}_{-1} \\
\{ \hat{G}_{21} \, ( 
\hat{T}^{(2)} + \hat{W}^{(2)}) \}_{-1}
 \\
-\frac{2 i (8+k)}{3(5+k)} \hat{B}_{-} 
\hat{T}_{+}^{(\frac{3}{2})} \\
\end{array} \right)(w)
 +  \cdots,
\nonu \\
\hat{G}_{22}(z) \, 
\left(
\begin{array}{c} 
\hat{V}_{-}^{(2)} \\
\hat{T}^{(2)} +\hat{W}^{(2)} \\
\hat{U}_{+}^{(2)} \\
\end{array} \right) (w) & = &
 \frac{1}{(z-w)^2} \left[ 
\frac{4(8+k)}{3(5+k)} \right] 
\left(
\begin{array}{c}
 0 \\
\hat{V}^{(\frac{3}{2})}\\
-\hat{T}_{+}^{(\frac{3}{2})} \\
\end{array} \right)(w)
\nonu \\
& + & \frac{1}{(z-w)} \left(
\begin{array}{c}
 \frac{2 i (8+k)}{3(5+k)} \hat{B}_{+} \hat{V}^{(\frac{3}{2})}  \\
\{ \hat{G}_{22} \, ( 
\hat{T}^{(2)} + \hat{W}^{(2)}) \}_{-1}
 \\
\{ \hat{G}_{22} \,  
\hat{U}_{+}^{(2)}  \}_{-1}
 \\
\end{array} \right)(w) +  \cdots.
\label{newnewG22u-2}
\eea
The $k$-dependent 
structure constants appearing in the second-order pole (in the 
right hand side) of 
Appendix (\ref{newnewG22u-2}) for the $({\bf 3}, 
{\bf 1})$ are common up to the signs and those for the $({\bf 1}, {\bf 3})$
have same value up to the signs.
The nonlinear quadratic terms ($\hat{A}_{+} \hat{U}^{(\frac{3}{2})}(w), \cdots$) 
appearing in the first-order poles in
Appendix (\ref{newnewG22u-2}) are primary fields because the OPEs between the 
corresponding spin-$1$ currents and the corresponding spin-$\frac{3}{2}$
currents do not have any singular terms.
For example, $\hat{A}_{+}(z) \, \hat{U}^{(\frac{3}{2})}(w) = + \cdots$.

The nonlinear spin-$\frac{5}{2}$ fields appearing in (\ref{newnewG22u-2}) 
are given by as follows:
\bea
\left(
\begin{array}{c}
\{ \hat{G}_{21} \, 
\hat{U}_{-}^{(2)} \}_{-1} \\
\{ \hat{G}_{12} \, 
\hat{V}_{+}^{(2)} \}_{-1} \\
\end{array} \right)(w) & = & 
\pm
\left(
\begin{array}{c}
\hat{U}^{(\frac{5}{2})} \\
\hat{V}^{(\frac{5}{2})} \\
\end{array} \right)  (w) 
+
\frac{2(-1+k)(5+2k)}{3k(5+k)} \pa 
\left(
\begin{array}{c}
\hat{U}^{(\frac{3}{2})} \\
\hat{V}^{(\frac{3}{2})} \\
\end{array} \right)(w) 
\nonu \\
& + &
\frac{(2 k+5)}{k(k+5) 
(38 k^2+41 k-25)} \left[ \mp 2 i  (8 k^2-51 k+25) 
\hat{A}_{\pm} \hat{T}_{\pm}^{(\frac{3}{2})}
\right. 
\nonu \\
& \mp & 4 i k (23 k-5)
\hat{A}_3 
\left(
\begin{array}{c}
\hat{U}^{(\frac{3}{2})} \\
\hat{V}^{(\frac{3}{2})} \\
\end{array} \right)
 \mp  12 i k (k-1) (k+1) 
\hat{B}_{\mp} \hat{T}_{\mp}^{(\frac{3}{2})}
\nonu \\
& \mp & 
12 i k (k-1) (k+1) 
\hat{B}_3 
\left(
\begin{array}{c}
\hat{U}^{(\frac{3}{2})} \\
\hat{V}^{(\frac{3}{2})} \\
\end{array} \right)
\nonu \\
& - &   5 (k-1) (k+1) (k+5)
\hat{T}^{(1)} \left(
\begin{array}{c} 
\hat{G}_{11} \\
\hat{G}_{22} \\
\end{array} \right)
\nonu \\
& - & \left. \frac{4}{3} (k-1)  (7 k^2+4 k+25)
 \pa 
\left(
\begin{array}{c}
\hat{U}^{(\frac{3}{2})} \\
\hat{V}^{(\frac{3}{2})} \\
\end{array} \right)
 \right](w),
\nonu \\
\left(
\begin{array}{c}
\{ \hat{G}_{22} \, 
\hat{U}_{-}^{(2)} \}_{-1} \\
\{ \hat{G}_{11} \, 
\hat{V}_{+}^{(2)} \}_{-1} \\
\end{array} \right)
(w) & = & \pm \hat{W}_{\mp}^{(\frac{5}{2})}(w) 
-
\frac{2(-1+k)(5+2k)}{3k(5+k)} 
\pa \hat{T}_{\mp}^{(\frac{3}{2})}(w)
\nonu \\
& + &
\frac{(2k +5)}{k (k+5) (38 k^2+41 k-25)}
\left[ \right. \nonu \\
& \mp & 
2 i  (8 k^2-51 k+25)
\hat{A}_{\pm} 
\left(
\begin{array}{c}
\hat{V}^{(\frac{3}{2})} \\
 \hat{U}^{(\frac{3}{2})} \\
\end{array} \right)
\nonu \\
& \pm & 
4 i  k (23 k-5)
\hat{A}_3 \hat{T}_{\mp}^{(\frac{3}{2})}
\pm 12 i k (k-1) (k+1) 
\hat{B}_{\pm} 
\left(
\begin{array}{c}
\hat{U}^{(\frac{3}{2})} \\
\hat{V}^{(\frac{3}{2})} \\
\end{array} \right)
\nonu \\
& \mp & 
12 i k (k-1) (k+1)
\hat{B}_3 \hat{T}_{\mp}^{(\frac{3}{2})}
\nonu \\
& + & 5 (k-1) (k+1)(k+5) 
\hat{T}^{(1)} 
\left(
\begin{array}{c}
\hat{G}_{12} \\
\hat{G}_{21} \\
\end{array} \right)
\nonu \\
& + & \left. 
\frac{4}{3} (k-1) (7 k^2+4 k+25)
 \pa \hat{T}_{\mp}^{(\frac{3}{2})} \right](w),
\nonu \\
\left(
\begin{array}{c}
\{ \hat{G}_{11} \, ( 
\hat{T}^{(2)} -\hat{W}^{(2)}) \}_{-1} \\
\{ \hat{G}_{22} \, ( 
\hat{T}^{(2)} -\hat{W}^{(2)}) \}_{-1} \\
\end{array} \right)
(w) 
& = &
\left(
\begin{array}{c}
\hat{U}^{(\frac{5}{2})} \\
 \hat{V}^{(\frac{5}{2})}  \\
\end{array} \right)(w) 
\pm \frac{2(-1+k)(5+2k)}{3k(5+k)} \pa 
\left(
\begin{array}{c}
\hat{U}^{(\frac{3}{2})} \\
\hat{V}^{(\frac{3}{2})} \\
\end{array} \right)(w)
\nonu \\
& + &
\frac{(2 k +5)}{k (k+5) (38 k^2+41 k-25)}
\left[ -
4 i  k (23 k-5)
\hat{A}_{\pm} \hat{T}_{\pm}^{(\frac{3}{2})}
\right. 
\nonu \\
& + & 
4 i  
(15 k^2+46 k-25)
\hat{A}_3 \left( 
\begin{array}{c}
\hat{U}^{(\frac{3}{2})} \\
\hat{V}^{(\frac{3}{2})} \\
\end{array} \right)
\nonu \\
& - &    
12 i k (k-1) (k+1)
\hat{B}_{\mp} \hat{T}_{\mp}^{(\frac{3}{2})} 
\nonu \\
& - &  12 i k (k-1) (k+1)
\hat{B}_3 
 \left( 
\begin{array}{c}
\hat{U}^{(\frac{3}{2})} \\
 \hat{V}^{(\frac{3}{2})} \\
\end{array} \right)
\nonu \\
& \mp &    5 (k-1) (k+1) (k+5)
\hat{T}^{(1)} 
 \left( 
\begin{array}{c}
\hat{G}_{11} \\
\hat{G}_{22} \\
\end{array} \right)
\nonu \\
&\mp  & \left.  \frac{4}{3} (k-1) 
(7 k^2+4 k+25)
\pa 
\left( 
\begin{array}{c}
\hat{U}^{(\frac{3}{2})} \\
 \hat{V}^{(\frac{3}{2})} \\
\end{array} \right)
\right](w)
\nonu \\
& + &
\frac{6 k}{(7k +3)} \left[
\mp\frac{ i }{(k+2)}
\hat{A}_{\pm} 
\left( 
\begin{array}{c}
\hat{G}_{21} \\
\hat{G}_{12} \\
\end{array} \right)
\mp  
 \frac{ i }{(k+2)}
\hat{A}_3 
\left( 
\begin{array}{c}
\hat{G}_{11} \\
\hat{G}_{22} \\
\end{array} \right) \right.
\nonu \\
& \pm &  \frac{ i }{5}
\hat{B}_{\mp} 
\left( 
\begin{array}{c}
\hat{G}_{12} \\
\hat{G}_{21} \\
\end{array} \right)
\pm  \frac{ i }{5}
\hat{B}_3 
\left( 
\begin{array}{c}
\hat{G}_{11} \\
\hat{G}_{22} \\
\end{array} \right)
\nonu \\
& + & \left.  \pa 
\left( 
\begin{array}{c}
\hat{G}_{11} \\
\hat{G}_{22} \\
\end{array} \right)
+  \frac{3 (9 k+13)}{5 (k+2)}
\hat{T}^{(1)} 
\left( 
\begin{array}{c}
\hat{U}^{(\frac{3}{2})} \\
\hat{V}^{(\frac{3}{2})}  \\
\end{array} \right)
\right](w),
\nonu \\
\left(
\begin{array}{c}
\{ \hat{G}_{12} \, ( 
\hat{T}^{(2)} -\hat{W}^{(2)}) \}_{-1} \\
\{ \hat{G}_{21} \, ( 
\hat{T}^{(2)} -\hat{W}^{(2)}) \}_{-1} \\
\end{array} \right)
(w) 
& = &
-\hat{W}_{\mp}^{(\frac{5}{2})}(w) 
 \pm 
\frac{2(-1+k)(5+2k)}{3k(5+k)} \pa \hat{T}_{\mp}^{(\frac{3}{2})}(w)
\nonu \\
& + &
\frac{(2 k +5)}{k (k+5) (38 k^2+41 k-25)}
\left[ 
4 i  k (23 k-5)
\hat{A}_{\pm} 
\left( 
\begin{array}{c}
\hat{V}^{(\frac{3}{2})} \\
\hat{U}^{(\frac{3}{2})}  \\
\end{array} \right)
\right. 
\nonu \\
& + & 
4 i  
(15 k^2+46 k-25)
\hat{A}_3 \hat{T}_{\mp}^{(\frac{3}{2})}
\nonu \\
& - &     
12 i k (k-1) (k+1)
\hat{B}_{\pm} 
\left( 
\begin{array}{c}
\hat{U}^{(\frac{3}{2})} \\
 \hat{V}^{(\frac{3}{2})}  \\
\end{array} \right)
\nonu \\
& + &  12 i k (k-1) (k+1)
\hat{B}_3 \hat{T}_{\mp}^{(\frac{3}{2})} 
\nonu \\
& \mp &    5 (k-1) (k+1) (k+5)
\hat{T}^{(1)} 
\left(
\begin{array}{c}
\hat{G}_{12} \\
\hat{G}_{21} \\
\end{array} \right)
\nonu \\
&\mp  & \left.  \frac{4}{3} (k-1)  
(7 k^2+4 k+25)
\pa \hat{T}_{\mp}^{(\frac{3}{2})} \right](w)
\nonu \\
& + &
\frac{6 k}{(7k +3)} \left[ \pm
\frac{ i }{(k+2)}
\hat{A}_{\pm} 
\left(
\begin{array}{c}
\hat{G}_{22} \\
\hat{G}_{11} \\
\end{array} \right)
\mp  
 \frac{ i }{(k+2)}
\hat{A}_3 
\left(
\begin{array}{c}
\hat{G}_{12} \\
\hat{G}_{21} \\
\end{array} \right) \right.
\nonu \\
& \pm & \frac{ i }{5}
\hat{B}_{\pm} 
\left(
\begin{array}{c}
\hat{G}_{11} \\
\hat{G}_{22} \\
\end{array} \right)
 \mp  \frac{ i }{5}
\hat{B}_3 
\left(
\begin{array}{c}
\hat{G}_{12} \\
\hat{G}_{21} \\
\end{array} \right)+ \pa 
\left(
\begin{array}{c}
\hat{G}_{12} \\
\hat{G}_{21}
\end{array} \right)
\nonu \\
& + & \left.  \frac{3 (9 k+13)}{5 (k+2)}
\hat{T}^{(1)} \hat{T}_{\mp}^{(\frac{3}{2})}
\right](w),
\nonu \\
\left(
\begin{array}{c}
\{ \hat{G}_{11} \, 
\hat{V}_{-}^{(2)} \}_{-1} \\
\{ \hat{G}_{22} \, 
\hat{U}_{+}^{(2)} \}_{-1} \\
\end{array} \right)
(w) & = & \mp \hat{W}_{\mp}^{(\frac{5}{2})}(w) 
-\frac{4(8+k)}{9(5+k)} \pa \hat{T}_{\mp}^{(\frac{3}{2})}(w)
\nonu \\
& + &
\frac{(k +8)}{ (k+5) (38 k^2+41 k-25)}
\left[
\pm 32 i k  
\hat{A}_{\pm} 
\left(
\begin{array}{c}
\hat{V}^{(\frac{3}{2})} \\
 \hat{U}^{(\frac{3}{2})} \\
\end{array} \right)
\right.
\nonu \\
& \mp & 
32 i  k 
\hat{A}_3 \hat{T}_{\mp}^{(\frac{3}{2})}
\pm \frac{2}{3} i (2k+1)(5+k) 
\hat{B}_{\pm} 
\left(
\begin{array}{c}
\hat{U}^{(\frac{3}{2})} \\
\hat{V}^{(\frac{3}{2})} \\
\end{array} \right)
\nonu \\
& \pm & 
4 i (-5+5k+6k^2)
\hat{B}_3 \hat{T}_{\mp}^{(\frac{3}{2})}
+ \frac{8}{3} (k+2)(k+5) 
\hat{T}^{(1)} 
\left(
\begin{array}{c}
\hat{G}_{12} \\
\hat{G}_{21} \\
\end{array} \right)
\nonu \\
& - & \left. 
\frac{8}{9} (2 k^2-31 k-25)
 \pa \hat{T}_{\mp}^{(\frac{3}{2})} \right](w),
\nonu \\
\left(
\begin{array}{c}
\{ \hat{G}_{21} \, 
\hat{V}_{-}^{(2)} \}_{-1} \\
\{ \hat{G}_{12} \, 
\hat{U}_{+}^{(2)} \}_{-1} \\
\end{array} \right)
(w) & = &
\pm 
\left(
\begin{array}{c}
\hat{V}^{(\frac{5}{2})} \\
\hat{U}^{(\frac{5}{2})} \\
\end{array}
\right)(w) 
+\frac{4(8+k)}{9(5+k)} \pa 
\left(
\begin{array}{c}
\hat{V}^{(\frac{3}{2})} \\
\hat{U}^{(\frac{3}{2})} \\
\end{array}
\right)(w)
\nonu \\
& + &
\frac{(k +8)}{ (k+5) (38 k^2+41 k-25)}
\left[ \mp
32 i  k 
\hat{A}_{\mp} \hat{T}_{\mp}^{(\frac{3}{2})}
\right. 
\nonu \\
& \mp & 
32 i  k
\hat{A}_3 
\left(
\begin{array}{c}
\hat{V}^{(\frac{3}{2})} \\
\hat{U}^{(\frac{3}{2})} \\
\end{array}
\right)
\pm    
\frac{2}{3} i  (k+5) (2k+1)
\hat{B}_{\pm} \hat{T}_{\pm}^{(\frac{3}{2})} 
\nonu \\
& \mp &  4 i (-5+5k+6k^2)
\hat{B}_3 
\left(
\begin{array}{c}
\hat{V}^{(\frac{3}{2})} \\
 \hat{U}^{(\frac{3}{2})} \\
\end{array}
\right)
\nonu \\
& - &   \frac{8}{3} (k+2) (k+5)
\hat{T}^{(1)} 
\left(
\begin{array}{c}
\hat{G}_{22} \\
\hat{G}_{11} \\
\end{array} \right)
\nonu \\
& + & \left.    \frac{8}{9} (-25-31k+2k^2) 
\pa 
\left(
\begin{array}{c}
\hat{V}^{(\frac{3}{2})} \\
\hat{U}^{(\frac{3}{2})} \\
\end{array}
\right)
\right](w),
\nonu \\
\left(
\begin{array}{c}
\{ \hat{G}_{11} \, ( 
\hat{T}^{(2)} + \hat{W}^{(2)}) \}_{-1} \\
\{ \hat{G}_{22} \, ( 
\hat{T}^{(2)} + \hat{W}^{(2)}) \}_{-1} \\
\end{array} \right)
(w) 
& = &
\left(
\begin{array}{c}
\hat{U}^{(\frac{5}{2})} \\
\hat{V}^{(\frac{5}{2})} \\
\end{array}
\right)
(w) 
\mp \frac{4(8+k)}{9(5+k)} \pa 
\left(
\begin{array}{c}
\hat{U}^{(\frac{3}{2})} \\
\hat{V}^{(\frac{3}{2})} \\
\end{array}
\right)(w)
\nonu \\
& + &
\frac{(k +8)}{ (k+5) (38 k^2+41 k-25)}
\left[ -
32 i  k 
\hat{A}_{\pm} \hat{T}_{\pm}^{(\frac{3}{2})}
\right. 
\nonu \\
& - & 
32 i k  
\hat{A}_3 
\left(
\begin{array}{c}
\hat{U}^{(\frac{3}{2})} \\
\hat{V}^{(\frac{3}{2})} \\
\end{array} \right)
-    
4 i (-5+5k+6k^2)
\hat{B}_{\mp} \hat{T}_{\mp}^{(\frac{3}{2})} 
\nonu \\
& + &  \frac{8}{3} i (-5+13k+10k^2)
\hat{B}_3 
\left(
\begin{array}{c}
\hat{U}^{(\frac{3}{2})} \\
 \hat{V}^{(\frac{3}{2})} \\
\end{array} \right)
\nonu \\
&\pm &   \frac{8}{3} (2+k) (k+5)
\hat{T}^{(1)} 
\left(
\begin{array}{c}
\hat{G}_{11} \\
\hat{G}_{22} \\
\end{array} \right)
\nonu \\
&\mp  & \left.  \frac{8}{9}   
(2 k^2-31 k-25)
\pa 
\left(
\begin{array}{c}
\hat{U}^{(\frac{3}{2})} \\
\hat{V}^{(\frac{3}{2})} \\
\end{array} \right) 
\right](w)
\nonu \\
& + &
\frac{(3+4k)}{(7k +3)} \left[
\pm \frac{ 2 i }{(k+2)}
\hat{A}_{\pm} 
\left(
\begin{array}{c}
\hat{G}_{21} \\
\hat{G}_{12} \\
\end{array} \right)
\pm  
 \frac{ 2 i }{(k+2)}
\hat{A}_3 
\left(
\begin{array}{c}
\hat{G}_{11} \\
\hat{G}_{22} \\
\end{array} \right)
\right. \nonu \\
& \mp &   \frac{ 2 i }{5}
\hat{B}_{\mp} 
\left(
\begin{array}{c}
\hat{G}_{12} \\
\hat{G}_{21} \\
\end{array} \right)
 \mp  \frac{ 2i }{5}
\hat{B}_3 
\left(
\begin{array}{c}
\hat{G}_{11} \\
\hat{G}_{22} \\
\end{array} \right) 
- 2 \pa 
\left(
\begin{array}{c}
\hat{G}_{11} \\
\hat{G}_{22} \\
\end{array} \right)
 \nonu \\
& - & \left.  \frac{6  (9 k+13)}{5 (k+2)}
\hat{T}^{(1)} 
\left(
\begin{array}{c}
\hat{U}^{(\frac{3}{2})} \\
\hat{V}^{(\frac{3}{2})} \\
\end{array} \right)
\right](w),
\nonu \\
\left(
\begin{array}{c}
\{ \hat{G}_{12} \, ( 
\hat{T}^{(2)} + \hat{W}^{(2)}) \}_{-1} \\
\{ \hat{G}_{21} \, ( 
\hat{T}^{(2)} + \hat{W}^{(2)}) \}_{-1} \\
\end{array} \right)
(w) 
& = &
\hat{W}_{\mp}^{(\frac{5}{2})}(w) 
\pm \frac{4(8+k)}{9(5+k)} \pa \hat{T}_{\mp}^{(\frac{3}{2})}(w)
\nonu \\
& + &
\frac{(k +8)}{ (k+5) (38 k^2+41 k-25)}
\left[- 
32 i  k 
\hat{A}_{\pm} 
\left(
\begin{array}{c}
\hat{V}^{(\frac{3}{2})} \\
\hat{U}^{(\frac{3}{2})} \\
\end{array} \right) 
\right. 
\nonu \\
& + & 
32 i k  
\hat{A}_3 \hat{T}_{\mp}^{(\frac{3}{2})}
+    
4 i (-5 +5k+6k^2)
\hat{B}_{\pm} 
\left(
\begin{array}{c} 
\hat{U}^{(\frac{3}{2})} \\
 \hat{V}^{(\frac{3}{2})} \\
\end{array} \right) 
\nonu \\
& + &  \frac{8}{3} i (-5 +13k+10k^2)
\hat{B}_3 \hat{T}_{\mp}^{(\frac{3}{2})} 
\mp   \frac{8}{3} (k+2) (k+5)
\hat{T}^{(1)} 
\left(
\begin{array}{c}
\hat{G}_{12} \\
\hat{G}_{21} \\
\end{array} \right)
\nonu \\
&\pm  &  \left. \frac{8}{9}  
(2 k^2-31 k-25)
\pa \hat{T}_{\mp}^{(\frac{3}{2})} \right](w)
\nonu \\
& + &
\frac{(3+4 k)}{(7k +3)} \left[
\mp \frac{ 2i }{(k+2)}
\hat{A}_{\pm} 
\left(
\begin{array}{c}
\hat{G}_{22} \\
\hat{G}_{11} \\
\end{array} \right)
\pm  
 \frac{ 2i }{(k+2)}
\hat{A}_3 
\left(
\begin{array}{c}
\hat{G}_{12} \\
\hat{G}_{21} \\
\end{array} \right) \right.
\nonu \\
& \mp &  \frac{ 2i }{5}
\hat{B}_{\pm} 
\left(
\begin{array}{c}
\hat{G}_{11} \\
\hat{G}_{22} \\
\end{array} \right)
\pm  \frac{ 2i }{5}
\hat{B}_3 
\left(
\begin{array}{c}
\hat{G}_{12} \\
\hat{G}_{21} \\
\end{array} \right) 
\nonu \\
&
\pm & \left.  2 \pa 
\left(
\begin{array}{c}
\hat{G}_{12} \\
\hat{G}_{21} \\
\end{array} \right)
 -  \frac{6  (9 k+13)}{5 (k+2) }
\hat{T}^{(1)} \hat{T}_{\mp}^{(\frac{3}{2})}(w)
\right](w).
\label{opecomplicated}
\eea
In each OPE of Appendix (\ref{opecomplicated}), 
the first line contains spin-$\frac{5}{2}$ current and  
the derivative (descendant) term 
from the spin-$\frac{3}{2}$ current in the second-order 
pole. The next lines in the OPE consist of a primary field
which is a quadratic expression between the higher spin currents and
the currents from large ${\cal N}=4$ nonlinear algebra. 
The other
primary field which is also a quadratic in $\hat{A}_i(w)$, 
$\hat{B}_i(w)$ and 
$\hat{G}_a(w)$ (as well as  a derivative term and 
a product between the higher spin currents)
appears in other OPEs.  

For example, in Table $4$, the fourth row describes the spin-$\frac{5}{2}$
current $\hat{W}_{+}^{(\frac{5}{2})}(w)$ and other $12$ (hatted) composite fields
corresponding to the first-order pole $\{ \hat{G}_{21} \, ( 
\hat{T}^{(2)} -\hat{W}^{(2)}) \}_{-1}(w)$ in Appendix (\ref{opecomplicated}).
Among those composite fields, six of them participate in one 
spin-$\frac{5}{2}$ primary field where the OPE between 
the spin-$1$ and the spin-$\frac{3}{2}$ currents 
contains the first-order pole term $\hat{T}_{+}^{(\frac{3}{2})}(w)$
(therefore the ordering between these currents involves the derivative 
term $\pa \hat{T}_{+}^{(\frac{3}{2})}(w)$) 
and the remaining ones do in other 
spin-$\frac{5}{2}$ primary field containing the product of two higher 
spin currents $\hat{T}^{(1)} 
\hat{T}_{+}^{(\frac{3}{2})}(w)$.
One can easily check that the OPE   $\hat{T}^{(1)}(z) \, 
\hat{T}_{+}^{(\frac{3}{2})}(w)$ provides the first-order pole term 
$-\frac{1}{2} \hat{G}_{21}(w)$ and therefore the ordering between two 
currents can give the derivative term $\pa \hat{G}_{21}(w)$. 
One can see the similar features in the first-order poles of 
$\{ \hat{G}_{11} \, \hat{V}_{+}^{(2)} \}_{-1}(w)$,  
$\{ \hat{G}_{22} \, \hat{U}_{+}^{(2)} \}_{-1}(w)$ or 
$\{ \hat{G}_{21} \, ( 
\hat{T}^{(2)} +\hat{W}^{(2)}) \}_{-1}(w)$.
Furthermore, the Table $4$ describes the remaining 
spin-$\frac{5}{2}$ currents, $\hat{W}_{-}^{(\frac{5}{2})}(w), 
\hat{U}^{(\frac{5}{2})}(w)$ and $\hat{V}^{(\frac{5}{2})}(w)$.
Similar analysis corresponding to the remaining $12$ OPEs of 
Appendix (\ref{opecomplicated}) can be done without any difficulty.

The OPEs between
 the spin-$\frac{3}{2}$ currents 
$\hat{G}_a(z)$ and the  above four spin-$\frac{5}{2}$ currents
in Appendix (\ref{newnewG22u-2}) or (\ref{16new}) 
can be obtained by
\bea
%\hat{G}_{11}(z) \, \hat{U}^{(\frac{5}{2})}(w) & = & 
%\frac{1}{(z-w)^2} \, \left[ \right](w) +
%\frac{1}{(z-w)} \, \left[ \right](w) + \cdots====0,
%\nonu \\
\left(
\begin{array}{c}
\hat{G}_{12} \\
\hat{G}_{21} \\
\end{array} \right) (z)  \, \left( 
\begin{array}{c} 
\hat{U}^{(\frac{5}{2})} \\
\hat{V}^{(\frac{5}{2})} \\
\end{array} \right) (w) & = & 
\pm \frac{1}{(z-w)^2} \, \left[ \frac{64 k (k+1) (k+8)}{(k+5) 
(38 k^2+41 k-25)} \right] \left(
\begin{array}{c} 
\hat{U}_{-}^{(2)} \\
\hat{V}_{+}^{(2)} \end{array} 
\right) (w) \nonu \\
& + & 
\frac{1}{(z-w)} \, \left[ \mbox{spin-$3$ composite 
fields} \right](w) + \cdots,
\nonu \\
\left(
\begin{array}{c}
\hat{G}_{21} \\
\hat{G}_{12} \\
\end{array} \right) (z) \, 
\left(
\begin{array}{c} 
\hat{U}^{(\frac{5}{2})} \\
\hat{V}^{(\frac{5}{2})} \\
\end{array} \right)(w) & = & 
\mp \frac{1}{(z-w)^2} \, \left[ \frac{96 (k-1) (k+1) (2 k+5)}{
(k+5) (38 k^2+41 k-25)} \right]  \left(
\begin{array}{c}  
\hat{U}_{+}^{(2)} \\
\hat{V}_{-}^{(2)} \\
\end{array} \right)(w) \nonu \\
& + & 
\frac{1}{(z-w)} \, \left[ \mbox{spin-$3$ composite fields} \right](w) + \cdots,
\nonu \\
\left(
\begin{array}{c}
\hat{G}_{22} \\
\hat{G}_{11} \\
\end{array} \right) (z) \, 
\left(
\begin{array}{c}
\hat{U}^{(\frac{5}{2})} \\
\hat{V}^{(\frac{5}{2})} \\
\end{array} \right)(w) & = & 
\frac{1}{(z-w)^2} \, \frac{(k+1)}{(k+5)  (38 k^2+41 k-25)}
\nonu \\
& \times & \left[
-\frac{96  (-5+3k)(3+4k+2k^2)}
{ (7 k+3)}
\tilde{\hat{T}} \right.   \nonu \\
& + &  
48 (k-1)  (2 k+5)
(\hat{T}^{(2)} 
+ \hat{W}^{(2)}) \nonu \\
& + &  \left. 32 k  (k+8)
(\hat{T}^{(2)} - \hat{W}^{(2)}) \right](w)
\nonu \\
&+ &  
\frac{1}{(z-w)} \, \left[\mp \hat{W}^{(3)} +  
\mbox{spin-$3$ composite  fields}
 \right](w) + \cdots,
\nonu \\
%\hat{G}_{11}(z) \, \hat{V}^{(\frac{5}{2})}(w) & = & 
%\frac{1}{(z-w)^2} \, \left[ \right](w) +
%\frac{1}{(z-w)} \, \left[ \hat{W}^{(3)} + \cdots  \right](w) + \cdots,
%\nonu \\
%\hat{G}_{12}(z) \, \hat{V}^{(\frac{5}{2})}(w) & = & 
%\frac{1}{(z-w)^2} \, \left[ \right](w) +
%\frac{1}{(z-w)} \, \left[ \frac{96 (k-1) (k+1) (2 k+5)}{(k+5) 
%(38 k^2+41 k-25)} \right]  \hat{V}_{-}^{(2)}(w) + \cdots,
%\nonu \\
%\hat{G}_{21}(z) \, \hat{V}^{(\frac{5}{2})}(w) & = & 
%\frac{1}{(z-w)^2} \, \left[ \right](w) +
%\frac{1}{(z-w)} \, \left[ -\frac{64 k (k+1) (k+8)}{(k+5) 
%(38 k^2+41 k-25)}\right]\hat{V}_{+}^{(2)} (w) + \cdots,
%\nonu \\
%\hat{G}_{22}(z) \, \hat{V}^{(\frac{5}{2})}(w) & = & 
%\frac{1}{(z-w)^2} \, \left[ \right](w) +
%\frac{1}{(z-w)} \, \left[ \right](w) + \cdots====0,
%\nonu \\
\left(
\begin{array}{c}
\hat{G}_{11} \\
\hat{G}_{22} \\
\end{array} \right)(z) \, 
\hat{W}_{\pm}^{(\frac{5}{2})}
(w) & = & 
\pm \frac{1}{(z-w)^2} \, \left[\frac{96 (k-1) (k+1) (2 k+5)}{(k+5) 
(38 k^2+41 k-25)} \right] \left(
\begin{array}{c} 
\hat{U}_{+}^{(2)} \\
\hat{V}_{-}^{(2)} \\
\end{array} \right)(w) \nonu \\
& + & 
\frac{1}{(z-w)} \, \left[\mbox{spin-$3$ composite 
fields} \right](w) + \cdots,
\nonu \\
\left(
\begin{array}{c}
\hat{G}_{12} \\
\hat{G}_{21} \\
\end{array} \right) (z) \, \hat{W}_{\pm}^{(\frac{5}{2})}(w) & = & 
\frac{1}{(z-w)^2} \, \frac{(k+1)}{(k+5)  (38 k^2+41 k-25)}
\nonu \\
& \times & \left[
-\frac{96  (10 k^3+34 k^2-11 k-15)}
{ (7 k+3)}
\tilde{\hat{T}} \right.   \nonu \\
& + &  
48 (k-1)  (2 k+5)
(\hat{T}^{(2)} 
+ \hat{W}^{(2)}) \nonu \\
& - &  \left. 32 k  (k+8)
(\hat{T}^{(2)} - \hat{W}^{(2)}) \right](w)
\nonu \\
& + & 
\frac{1}{(z-w)} \, \left[\mp \hat{W}^{(3)} + \mbox{spin-$3$ composite 
fields} 
\right](w) + \cdots,
\nonu \\
%\hat{G}_{21}(z) \, \hat{W}_{+}^{(\frac{5}{2})}(w) & = & 
%\frac{1}{(z-w)^2} \, \left[ \right](w) +
%\frac{1}{(z-w)} \, \left[ \right](w) + \cdots====0,
%\nonu \\
\left(
\begin{array}{c}
\hat{G}_{22} \\
\hat{G}_{11} \\
\end{array} \right) (z) \, \hat{W}_{\pm}^{(\frac{5}{2})}(w) & = & 
\mp \frac{1}{(z-w)^2} \, \left[\frac{64 k (k+1) (k+8)}{(k+5) 
(38 k^2+41 k-25)} \right] \left( 
\begin{array}{c} \hat{V}_{+}^{(2)} \\
\hat{U}_{-}^{(2)} \\
\end{array} \right) (w) \nonu \\
&+ &
\frac{1}{(z-w)} \, \left[ \mbox{spin-$3$ composite 
fields} \right](w) + \cdots,
%\hat{G}_{11}(z) \, \hat{W}_{-}^{(\frac{5}{2})}(w) & = & 
%\frac{1}{(z-w)^2} \, \left[\frac{64 k (k+1) (k+8)}{(k+5) 
%(38 k^2+41 k-25)} \right]\hat{U}_{-}^{(2)} (w) +
%\frac{1}{(z-w)} \, \left[ \right](w) + \cdots,
%\nonu \\
%\hat{G}_{12}(z) \, \hat{W}_{-}^{(\frac{5}{2})}(w) & = & 
%\frac{1}{(z-w)^2} \, \left[ \right](w) +
%\frac{1}{(z-w)} \, \left[ \right](w) + \cdots====0,
%\nonu \\
%\hat{G}_{21}(z) \, \hat{W}_{-}^{(\frac{5}{2})}(w) & = & 
%\frac{1}{(z-w)^2} \, \left[ \right](w) +
%\frac{1}{(z-w)} \, \left[ \hat{W}^{(3)} + \cdots \right](w) + \cdots.
%\nonu \\
%\hat{G}_{22}(z) \, \hat{W}_{-}^{(\frac{5}{2})}(w) & = & 
%\frac{1}{(z-w)^2} \, \left[-\frac{96 (k-1) (k+1) (2 k+5)}{
%(k+5) (38 k^2+41 k-25)} \right]\hat{V}_{-}^{(2)} (w) \nonu \\
%& + &
%\frac{1}{(z-w)} \, \left[ \right](w) + \cdots.
\label{opeopeope}
\eea
where the primary field $\tilde{\hat{T}}(w)$ is given by 
Appendix 
(\ref{tildehatt}) and we do not present the complete expressions for the 
spin-$3$ composite fields consisting of 
$11$ currents and $16$ higher spin currents 
appearing in the first-order poles \cite{Ahn2014}.
All the third-order poles appeared in previous basis 
are disappeared in this
new basis (\ref{opeopeope}).

Finally,  the OPEs between
the spin-$\frac{3}{2}$ currents 
$\hat{G}_a(z)$ and the  above four spin-$3$ currents
in Appendix (\ref{opeopeope}) or (\ref{16new}) 
\bea
\left( 
\begin{array}{c} 
\hat{G}_{11} \\
\hat{G}_{22} \\
\end{array} \right) (z) \, \hat{W}^{(3)}(w) & = & 
%\frac{1}{(z-w)^3} \, \left[ \right](w) +
\mp \frac{1}{(z-w)^2} \, \left[ \frac{(327 k+859) 
(38 k^2+41 k-25)}{4 (786 k^3+3727 k^2+2920 k-1925)}
\right] 
\left(
\begin{array}{c}
\hat{U}^{(\frac{5}{2})} \\
\hat{V}^{(\frac{5}{2})} \\
\end{array} \right) (w) \nonu \\
& + & 
\frac{1}{(z-w)} \, \left[ \mbox{spin-$\frac{7}{2}$ composite
fields} \right](w) + \cdots,
\nonu \\
\left( 
\begin{array}{c} 
\hat{G}_{12} \\
\hat{G}_{21} \\
\end{array} \right) (z) \, \hat{W}^{(3)}(w) & = & 
%\frac{1}{(z-w)^3} \, \left[ \right](w) +
\pm \frac{1}{(z-w)^2} \, \left[ \frac{(327 k+859) 
(38 k^2+41 k-25)}{4 (786 k^3+3727 k^2+2920 k-1925)}
\right] 
\hat{W}_{\mp}^{(\frac{5}{2})}(w) \nonu \\
& + & 
\frac{1}{(z-w)} \, \left[\mbox{spin-$\frac{7}{2}$ composite 
fields} \right](w) + \cdots.
%\nonu \\
%\hat{G}_{12}(z) \, \hat{W}^{(3)}(w) & = & 
%\frac{1}{(z-w)^3} \, \left[ \right](w) +
%\frac{1}{(z-w)^2} \, \left[ \right](w) +
%\frac{1}{(z-w)} \, \left[ \right](w) + \cdots,
%\nonu \\
%\hat{G}_{21}(z) \, \hat{W}^{(3)}(w) & = & 
%\frac{1}{(z-w)^3} \, \left[ \right](w) +
%\frac{1}{(z-w)^2} \, \left[ \right](w) +
%\frac{1}{(z-w)} \, \left[ \right](w) + \cdots,
%\nonu \\
%\hat{G}_{22}(z) \, \hat{W}^{(3)}(w) & = & 
%\frac{1}{(z-w)^3} \, \left[ \right](w) +
%\frac{1}{(z-w)^2} \, \left[ \right](w) +
%\frac{1}{(z-w)} \, \left[ \right](w) + \cdots.
\label{ope3half3}
\eea
Note that the structure constants appearing in the second-order poles 
have same values in Appendix (\ref{ope3half3}) and   
we do not present the complete expressions for the 
spin-$\frac{7}{2}$ composite fields \cite{Ahn2014}.
All the third-order poles appeared in previous basis 
are disappeared in this
new basis (\ref{ope3half3}).

%%%%%%%%%%%%%%%%%%%%%%%%%%%%%%%%%%%%%%%%%%%%%%%%%%%%%%%%%%%%%%%%%%

\end{document}